%% file: wrapper_s.tex
\documentclass[11pt]{cernrepdbg}
\usepackage{graphicx,epsf,epsfig,here}

\newcommand{\gsim}{\buildrel > \over {_\sim}}
\newcommand{\lsim}{\buildrel < \over {_\sim}}
\newcommand{\lqcd}{\Lambda_{\rm QCD}}
\newcommand{\order}[1]{${\cal O}(#1)$}
\newcommand{\morder}[1]{{\cal O}(#1)}
\newcommand{\bQ}{\overline{Q}}

\def\NJ{N_{\J}}
\def\J{J/\psi}
\def\Nc{N_c}
\def\Ncbar{N_{\overline c}}
\def\ccbar{c \overline c}
\def\Nccbar{N_{\ccbar}}
\newcommand{\pt}{\ensuremath{p_T}}
\newcommand{\TeV}{\mathrm{TeV}}

\newcommand{\be}{\begin{eqnarray}}
\newcommand{\ee}{\end{eqnarray}}

\def\BE{\begin{equation}}
\def\EE{\end{equation}}

\def\lsim{\mathrel{\raise.3ex\hbox{$<$\kern-.75em\lower1ex\hbox{$\sim$}}}}
\def\gsim{\mathrel{\raise.3ex\hbox{$>$\kern-.75em\lower1ex\hbox{$\sim$}}}}
\def\to{\rightarrow}

\def\fun#1#2{\lower3.6pt\vbox{\baselineskip0pt\lineskip.9pt}}
\makeatletter

\def\vereq#1#2{\lower3pt\vbox{\baselineskip1.5pt \lineskip1.5pt
\ialign{$\m@th#1\hfill##\hfil$\crcr#2\crcr\sim\crcr}}}

\makeatother
\newcommand{\lgl}{$\langle$}                      
\newcommand{\rgl}{$\rangle$}                                     

\newcommand{\cbarc    }{$c \bar c$} 
\newcommand{\bbarb    }{$b \bar b$} 

\newcommand{\mup     }{$\mu^+$}                        
\newcommand{\mum     }{$\mu^-$}

\newcommand{\kaon  }{$K$}

\newcommand{\jpsi}{$J/\psi$}                      
\newcommand{\psip}{$\psi^{\prime}$}
\newcommand{\ups}{$\Upsilon$}                 
\newcommand{\upsp}{$\Upsilon^{\prime}$}
\newcommand{\upspp}{$\Upsilon^{\prime\prime}$}

\newcommand{\zo   }{$Z^0$}                            

\newcommand{\ccb   }{$c\overline c$}              
\newcommand{\bbb   }{$b\overline b$}              
              
\newcommand{\pp}{$pp$}           
\newcommand{\aacol}{$AA$}          
\newcommand{\PbPb}{Pb+Pb}           
\newcommand{\SnSn}{Sn+Sn}           
\newcommand{\ArAr}{Ar+Ar}           
\newcommand{\KrKr}{Kr+Kr}           
\newcommand{\etta}{$\eta$}
\newcommand{\abseta}{$\vert\eta\vert$}

\newcommand{\ptmu}{$p_{\rm T}^{\mu}$}                    
\newcommand{\Gonc}{GeV/$c$}                      
\newcommand{\mass}{GeV/$c^2$}                      
\begin{document}
\title{HARD PROBES IN HEAVY ION COLLISIONS AT THE LHC:\\
HEAVY FLAVOUR PHYSICS}
\author{
{\bf Convenors}: M.~Bedjidian$^{1}$, P.~Crochet$^{7}$, S.~Frixione$^{9}$, 
D.~Kharzeev$^{11}$, R.~Vogt$^{22,23}$\\
{\bf Editors}: R.~Vogt$^{22,23}$, S.~Frixione$^{9}$\\
{\bf Contributing authors}: 
M.~Bedjidian$^{1}$, D.~Blaschke$^{2,3}$, G.T.~Bodwin$^{4}$,
N.~Carrer$^{5}$, B.~Cole$^{6}$,
P.~Crochet$^{7}$, A.~Dainese$^{8}$, A.~Deandrea$^{1}$,
S.~Frixione$^{9}$, P.~Hoyer$^{10}$,
D.~Kharzeev$^{11}$, O.L.~Kodolova$^{12}$, R.~Kvatadze$^{13}$, Jungil Lee$^{4}$,
I.P.~Lokhtin$^{12}$,
M.~Mangano$^{14}$, N.~Marchal$^{15}$,
M.~Nardi$^{16}$, G.~Nardulli$^{17}$, H.~Niemi$^{18}$, S.~Peign\'e$^{15}$,
P.~Petreczky$^{11}$, A.D.~Polosa$^{14}$,
H.~Satz$^{19}$, H.~Takai$^{11}$, S.~Tapprogge$^{20}$,
R.L.~Thews$^{21}$, E.~Vercellin$^{16}$, R.~Vogt$^{22,23}$
}
\institute{
$^{1}$Institut de Physique Nucl\'eaire de Lyon, Lyon, France \\
$^{2}$Fachbereich Physik, Universit\"at Rostock, Rostock, Germany \\
$^{3}$Bogoliubov Laboratory of Theoretical Physics,
Joint Institute of Nuclear Research, Dubna, Russia\\
$^{4}$High Energy Physics Division, Argonne National Laboratory, Argonne, IL,
USA \\
$^{5}$EP Division, CERN, Gen\`eve, Switzerland\\
$^{6}$Department of Physics, Columbia University, New York, NY, USA\\
$^{7}$Laboratoire de Physique Corpusculaire, CNRS/IN2P3, Clermont-Ferrand, 
France \\
$^{8}$INFN Padova, Universit\`a di Padova, Padova, Italy \\
$^{9}$INFN, Sezione di Genova, Genova, Italy \\
$^{10}$Department of Physical Sciences, University of Helsinki, Helsinki, 
Finland \\
$^{11}$Physics Department, Brookhaven National Laboratory, Upton, NY, USA\\
$^{12}$Institute of Nuclear Physics, Moscow State University, Moscow, Russia\\
$^{13}$High Energy Physics Institute, Tbilisi, Georgia\\
$^{14}$Theory Division, CERN, Gen\`eve, Switzerland\\
$^{15}$LAPTH, Annecy-le-Vieux, France \\
$^{16}$INFN Torino, Universit\`a di Torino, Torino, Italy \\
$^{17}$INFN Bari, Universit\`a di Bari, Bari, Italy \\ 
$^{18}$Department of Physics, University of Jyv\"asky\"a, 
Jyv\"asky\"a, Finland \\
$^{19}$Fakult\"at f\"ur Physik, Universit\"at Bielefeld, Bielefeld, Germany\\
$^{20}$Helsinki Institute of Physics, University of Helsinki, Helsinki, 
Finland \\
$^{21}$Department of Physics, University of Arizona, Tucson, AZ, USA \\
$^{22}$Nuclear Science Division, Lawrence Berkeley National 
Laboratory, Berkeley, CA, USA\\
$^{23}$Physics Department, University of California at Davis, Davis, CA, USA
}
\maketitle
\begin{abstract}
We present the results from the heavy quarks and quarkonia working group.
This report gives benchmark heavy quark and quarkonium cross
sections for $pp$ and $pA$ collisions at the LHC against which the $AA$ rates
can be compared in the study of the quark-gluon plasma.  We also provide an
assessment of the theoretical uncertainties in these benchmarks.  We then
discuss some of the cold matter effects on quarkonia production, including
nuclear absorption, scattering by produced hadrons, and energy loss in the
medium.  Hot matter effects that could reduce the observed quarkonium rates
such as color screening and thermal activation are then discussed.  Possible
quarkonium enhancement through coalescence of uncorrelated heavy quarks and
antiquarks is also described.  Finally, we discuss the capabilities of the LHC
detectors to measure heavy quarks and quarkonia as well as the Monte Carlo
generators used in the data analysis.
\end{abstract}

\newpage
\tableofcontents
\setcounter{footnote}{0}
%

\include{intro}
\include{open}
\include{quarkonium}

\include{energyloss}

\include{kinetic}

\include{hotqcd}
\include{montecarlo}
\include{alice_s}
\include{cms}

\include{atlas_s}
\include{ack}

\include{biblio}
\end{document}

%% file: intro.tex
\section[INTRODUCTION]{INTRODUCTION~\protect
\footnote{Author: R.~Vogt.}}

Heavy quarks are a sensitive probe of the
collision dynamics at both short and long timescales.  On 
one hand, perturbative heavy-quark production takes place at
$\tau \propto 1/m_Q$. On the other hand, heavy quarks decay weakly so that
their lifetime is greater than that of the medium created in heavy ion
collisions.  In addition, quarkonium states, bound $Q \overline Q$ pairs,
have binding energies of the order of a few hundred MeV, comparable to the
plasma screening mass.  In a quark-gluon plasma, interactions with hard gluons
can overcome this threshold, leading to a large
probability for quarkonium breakup.

However, for heavy quarks and quarkonium to be effective plasma probes, 
the baseline production
cross sections should be well established since one would
ideally like to normalize the $AA$ rates to $pp$ collisions.  
The following two
sections of this chapter specifically address the baseline rates 
for open heavy 
flavors and quarkonia in turn.  
We emphasize that the baseline rates in $pp$, $pA$ and $AA$ collisions are
large enough for high statistics studies of all quarkonium and heavy quark
states.  Thus a complete physics program can be carried out.

The heavy flavor cross
sections will be calculated to next-to-leading order (NLO) in perturbative QCD.
No data are available on the $c \overline c$
total cross sections at collider energies to better fix production parameters
such as the mass, $m_c$, and the factorization and renormalization scales,
$\mu_F$ and $\mu_R$ respectively, at high energies.  However, it should be
possible to reliably interpolate from $pp$ measurements at 14 TeV if no 
$pp$ run at 5.5 TeV is available, as discussed in section~\ref{sec:open}.
   
There is still some uncertainty in the quarkonium production mechanism.
The quarkonium baseline rates are calculated using the
color evaporation model (CEM) at NLO for the total cross sections.  
The $p_T$ distributions from the CEM are compared to those calculated in
nonrelativistic QCD (NRQCD).  The comover enhancement scenario for quarkonium 
production is also introduced although no predictions for the rates are given.

Once the $pp$ baselines are discussed, the effects of modifications of the
parton distribution functions in nuclei, referred to here as shadowing, 
are discussed in sections \ref{sec:open} and \ref{sec:quarkon}.  
At small momentum
fractions, the nuclear gluon distribution is expected to be substantially
reduced relative to that in the nucleon.  We show how both the total rates
and the rapidity and $p_T$ distributions could be affected by
these modifications.  Although the nuclear gluon distribution is not well
known, it can be measured in $pA$ and $Ap$ relative to $pp$
interactions at the LHC.
The $AA$ baseline rates without any other nuclear effects can then be more
reliably predicted.  Final state effects on the quarkonium rates are discussed
in sections~\ref{sec:dima}-\ref{sec:lattice}.

It is well known that other effects in cold matter can change the expected
quarkonium rates.  Nuclear absorption and secondary scattering with produced
particles (comover scattering)
are both effects that can cause quarkonium states to break up. In 
section~\ref{sec:dima}, the absorption cross section is postulated to increase
with energy.  Its value is extrapolated to LHC
energies and the effects on the expected $J/\psi$ and $\Upsilon$ rates in Pb+Pb
and Ar+Ar collisions are discussed.
The cross sections for quarkonium interactions with comovers have been studied
in a number of models.  Many estimates of the comover cross section exist.
A discussion of one of the most recent is also
presented in section~\ref{sec:dima}.  The last topic discussed in this section
is energy loss in the medium.
It is shown that for angles less than $m_Q/E$, soft gluon radiation is
suppressed, reducing the energy loss of $c$ and $b$ quarks
relative to massless partons.  The effects
on the $D/\pi$ ratio are considered for both cold matter and `hot' plasma.

Quarkonium suppression in a quark-gluon plasma was first
predicted by Matsui and
Satz \cite{Matsui:1986dk}. No bound states should exist at temperatures
$T > T_D$ when the screening radius, $1/\mu_D(T)$, is smaller than the size of
the bound state even though the
bound state may still exist above the critical temperature for deconfinement,
{\it i.e.}\ $T_c < T_D$.  Early estimates of the dissociation of
quarkonium by color screening suggested that the charmonium states above the
$J/\psi$ would break up near $T_c$, as do hadrons made up of light quarks,
while the $J/\psi$ would need a somewhat
higher temperature to break up.  Similar
results were predicted for the $\Upsilon$ family except that the tightly bound
$\Upsilon(1S)$ appeared to break up only at temperatures considerably above
$T_c$.  Later, it was realized that changes in the gluon momentum distributions
near $T_c$ as well as thermal activation of the quarkonium states 
can contribute significantly to quarkonium suppression, leading to
breakup for $T < T_D$.
More recent lattice calculations suggest that screening
may not be an efficient mechanism near $T_c$ so that gluon dissociation may
dominate quarkonium suppression.  In this case, there is no specific transition
temperature.  These results are summarized
in section~\ref{sec:lattice}. 

The most natural prediction is then that the per nucleon
quarkonium rates in $AA$
collisions would be substantially reduced relative to those in $pp$ and $pA$
collisions.  However, if the large number of $Q \overline Q$ pairs produced
at LHC energies
is taken into account, the initially produced uncorrelated heavy quark pairs
could provide an important source of final-state quarkonium, particularly
charmonium.  This coalescence mechanism could lead to quarkonium enhancement
at the LHC instead of suppression.  Some predictions for the LHC are given
in section~\ref{sec:thews}.  The RHIC $J/\psi$
data will place important constraints on this
type of mechanism even though the number of uncorrelated $c \overline c$ pairs
is much smaller than at the LHC.  Since the $b \overline b$ rate is high enough
for uncorrelated $b \overline b$ production in heavy ion collisions at the LHC,
a smaller effect on $\Upsilon$ production may also be expected.

Both the baseline distributions and the in-medium effects introduced here
need to be simulated for data analysis.  This is typically done with a standard
set of Monte Carlo tools.  In section~\ref{sec:monte}, heavy quark production
with the $pp$ generators 
\textsc{Pythia} and \textsc{Herwig} are compared to LO and NLO calculations.
Calculations in $AA$ generators are also briefly discussed.

Finally, we turn to the experimental capabilities for measurements of these
states in heavy ion collisions at the LHC.  The ALICE and CMS collaborations
have made detailed studies of quarkonium detection, shown in 
sections~\ref{sec:alice} and \ref{sec:cms} respectively.  ALICE may be able to 
reconstruct heavy flavor mesons.  
This effort is also discussed in section~\ref{sec:alice}. CMS
is less likely to reconstruct heavy flavor mesons through their decay channels
but can focus on their contributions to the dilepton continuum, as discussed
in section~\ref{sec:cms}.  Since the ATLAS heavy ion effort is more recent, 
only a short
summary of their potential capabilities is included in section~\ref{sec:atlas}.

It is clear that systematic efforts in this area will provide a rich array
of data.  These data will
provide important information about the gluon distribution
in the nucleus in cold matter and the properties of the medium in the
hot matter produced in $AA$ collisions.

%% file: open.tex
\section[BENCHMARK HEAVY QUARK CROSS SECTIONS]
{BENCHMARK HEAVY QUARK CROSS SECTIONS~\protect
\footnote{Authors: S.~Frixione, M.~L.~Mangano.}}
\label{sec:open}

In this Section we collect some benchmark results for the rates 
and kinematics features of charm and bottom quark production. We shall
consider $pp$, $p$Pb, Pb$p$ and \PbPb\ collisions. We shall
always assume a beam energy of 2.75~TeV per nucleon for Pb
and 7~TeV for protons.  We shall also study $pp$ collisions
with 2.75~TeV proton beams. 

The main goal of this Section is to establish to which extent $pp$ and $pA$
measurements can be used to infer
the expected Pb+Pb rates in the absence of high-density effects specific
to the nuclear-nuclear environment. We shall show
that, while the prediction of absolute rates at the LHC energies is
affected by large theoretical uncertainties, the extrapolation of the
rates from one energy to another, or from $pp$ to $p$Pb, is under much
better theoretical control.  Our results stress the importance of $pA$
and lower energy $pp$ control runs.

Our calculations
were performed in the framework of next-to-leading order, NLO,
QCD~\cite{Nason:1987xz,Nason:1989zy,Beenakker:1990ma,Mangano:jk}.  We
expect that
NLO QCD properly describes the main features of the production
mechanism  in $pp$ and $pA$ collisions
(see e.g. Refs.~\cite{Frixione:1994nb,Frixione:1997ma}) and,
in particular, can well describe the extrapolation of these features from
one energy or beam type to others.  We note that data are available
for bottom and, most recently, charm production in $p\bar{p}$
collisions at energies up to almost 2~TeV. The absolute rates
predicted by NLO QCD are affected by a theoretical uncertainty of
approximately a factor of 2.  Large corrections also arise from the
inclusion of resummation contributions and from the
nonperturbative fragmentation of the bare heavy quarks into
hadrons. Recent studies~\cite{Cacciari:2002pa,Cacciari:2003zu,Frixione:2003ei}
indicate nevertheless that, once all known and calculable effects
(higher-order logarithmically-enhanced
emissions, fragmentation effects, and initial-state radiation)
which complement the plain NLO calculation are
included, the picture
which emerges form the comparison with the Tevatron data is very
satisfactory. Indeed, comparison of the theory with charm production
data~\cite{Acosta:2003ax,Korn:2003pt} is very good even though the 
small charm quark mass might have suggested that reliable rate estimates were
not possible.  While including effects beyond
NLO is crucial for reliable absolute predictions,
accounting for them here would not affect the thrust of our analysis.
We show that, once we can normalize the production properties
against data from $pp$ and $p$Pb
collisions, the extrapolation to \PbPb\ is well constrained.
It should thus be possible to safely isolate quark-gluon plasma effects.

The theoretical systematics we shall explore include:
\begin{itemize}
\item the heavy quark masses;
\item the choice of factorization and renormalization scales ($\mu_F$
  and $\mu_R$);
\item the choice of parton distribution functions (PDFs).
\end{itemize}
We explore the uncertainties due to the PDF choice at the level
of the proton parton densities, using the MRS99~\cite{Martin:1999ww} set
as a default, and the more recent sets MRST2001~\cite{Martin:2001es} 
and CTEQ6M~\cite{Pumplin:2002vw} for comparisons. We use the
EKS98~\cite{Eskola:1998iy,Eskola:1998df} parameterization of the
modifications of the parton distribution functions in nuclei. Although 
originally fitted with GRV LO proton PDFs, as argued in 
Ref.~\cite{Eskola:1998df}, these ratios are to a good extent 
independent of the PDF choice.

In the case of charm production we consider the following
mass and scale input parameter sets ${\cal P}$:
\begin{center}
\begin{tabular}{|l|l|c|c|c|c|} \hline
${\cal P}$ & PDF & $m_c$ (GeV) & $\mu_F/\mu_0$ & $\mu_R/\mu_0$  \\ \hline
I   &  MRS99     & 1.5  &   2  &   1  \\ \hline
II  &  MRS99     & 1.2  &   2  &   1  \\ \hline
III &  MRS99     & 1.8  &   2  &   1  \\ \hline
IV  &  MRS99     & 1.5  &   2  &   2  \\ \hline
V   &  MRST2001  & 1.5  &   2  &   1  \\ \hline
VI  &  CTEQ6M    & 1.5  &   2  &   1  \\ \hline
\end{tabular}
\end{center}
where
\be
\mu_0 = \sqrt{(p_{T,c}^2+p_{T,\overline{c}}^2)/2+m_c^2} \; .
\ee
In the case of bottom quarks, we have instead:
\begin{center}
\begin{tabular}{|l|l|c|c|c|c|} \hline
${\cal P}$ & PDF & $m_b$ (GeV) & $\mu_F/\mu_0$ & $\mu_R/\mu_0$  \\ \hline
I   & MRS99    & 4.75  &   1    &   1   \\ \hline
II  & MRS99    & 4.5   &   1    &   1   \\ \hline
III & MRS99    & 5     &   1    &   1   \\ \hline
IV  & MRS99    & 4.75  &   2    &   0.5 \\ \hline
V   & MRS99    & 4.75  &   0.5  &   2   \\ \hline
VI  & MRST2001 & 4.75  &   1    &   1   \\ \hline
VII & CTEQ6M   & 4.75  &   1    &   1   \\ \hline
\end{tabular}
\end{center}
with $\mu_0$ defined as above but now $b$ replaces $c$.

\subsection{$x$ coverage in charm production} 
\label{xcov}

We first study the range of $x$ values that contribute to charm
production. This question is relevant for understanding
whether the current parameterizations of nuclear densities allow safe
extrapolations.  We also want to establish whether the constraints set by charm
measurements in $p$Pb collisions cover $x$ ranges which are 
relevant for \PbPb. As a reference, we shall use parameter set ${\cal P} =
{\rm I}$ here and limit ourselves to LO predictions.

We shall consider the full pseudorapidity range as well as limited
pseudorapidity regions defined by the acceptance coverage typical of
ALICE. In particular, we consider the central region,
$\vert \eta \vert < 0.9$,  and the forward region, $2.5 < \eta < 4$.

\begin{figure}
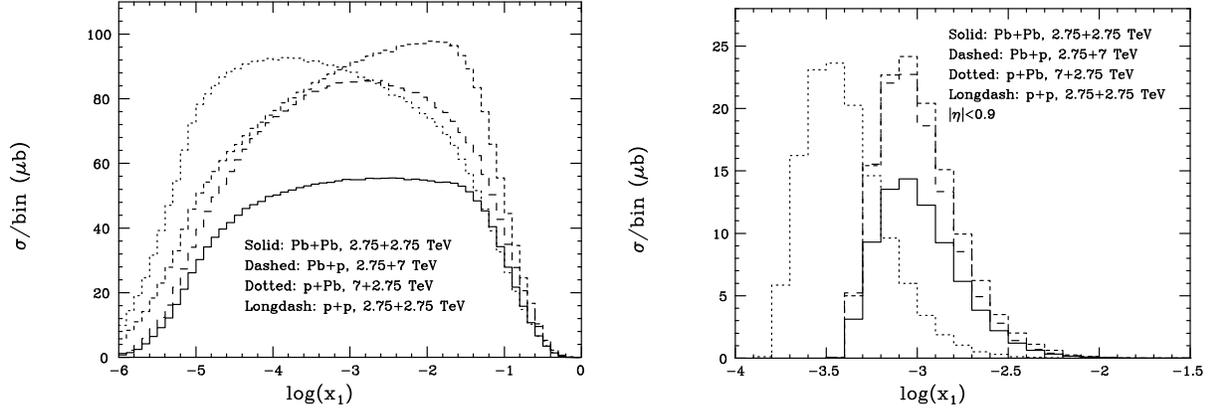
 
\begin{center} 
\centerline{
\includegraphics[width=0.48\textwidth,clip]{x1rng_all.eps} \hfill
\includegraphics[width=0.48\textwidth,clip]{x1rng_cen.eps}}
\vskip -0.4cm 
\caption{Parton-$x_1$ distributions for charm production in the full
pseudorapidity range (left) and in the central region (right). 
The cross section per nucleon is given.
\label{fig:xrng}}
\end{center} 
\end{figure}

\begin{figure} 
\begin{center} 
\centerline{
\includegraphics[width=0.8\textwidth,clip]{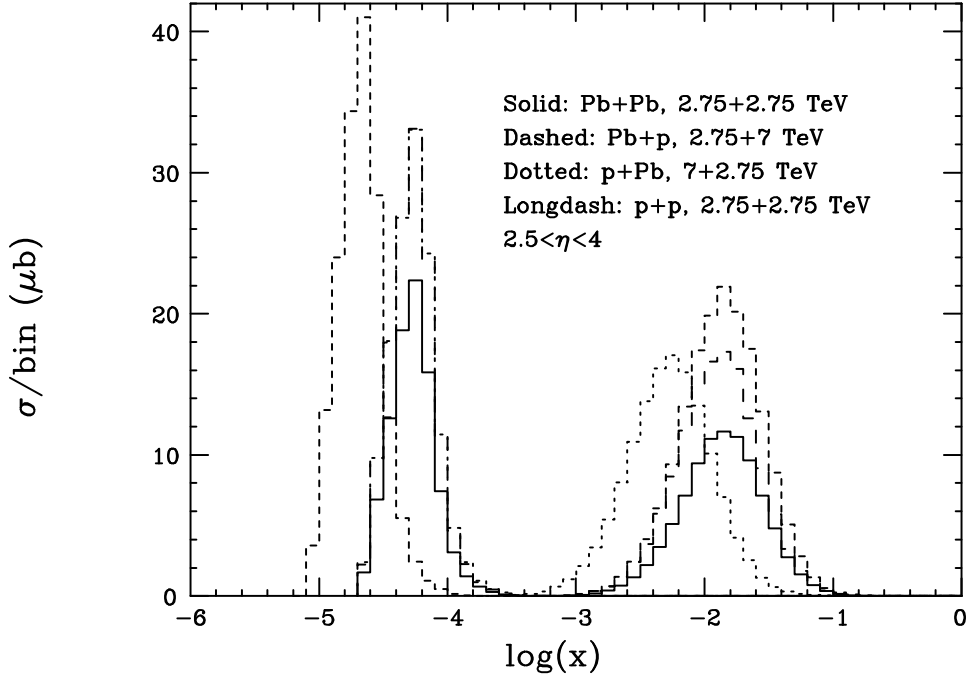}}
\vskip -0.4cm 
\caption{Parton-$x$ distributions for charm production in the forward
pseudorapidity range $2.5<\eta<4$. The cross section per nucleon is given.
\label{fig:xrng_fwd}}
\end{center} 
\end{figure}

The left (right) plot in Fig.~\ref{fig:xrng} shows the differential
rate distribution for $c\overline{c}$ production over the full (central)
pseudorapidity range as a function of $x_1$, the momentum fraction of
partons in the beam traveling in the positive $\eta$ direction. The $x_2$
distribution is identical for $pp$ and \PbPb\
collisions, while the $x_1$ and $x_2$ distributions are interchanged
for $p$Pb and Pb$p$ collisions.
The same distributions, obtained for charm pairs in the
forward pseudorapidity regions are given in Fig.~\ref{fig:xrng_fwd},
where the two peak structures refer to the $x_2$ (left peak) and
$x_1$ (right peak) variables.
In the case of central production, we note that the bulk of the
cross section comes from $x$ values peaked at $10^{-3}$. The gluon
density of the proton in this region is well known, thanks to the HERA
data. No data are, however, available for the nuclear densities in this
$x$ region, at least not in a range of $Q^2$ relevant for charm
production.  We see that the $x$ distributions for Pb$p$ and
\PbPb\ collisions are very similar in shape. This suggests that a determination
of nuclear corrections to the nucleon PDFs extracted from a Pb$p$ run
would be sufficient to properly predict the \PbPb\ behaviour.

In the case of forward production, the $x$ ranges probed inside the
two beams are clearly asymmetric. The selected $x$ ranges have
almost no overlap with the domains probed in central $c\overline{c}$
production. In particular, the range relevant for the beam traveling
in the negative $\eta$ direction peaks below $10^{-4}$, a region
where data for $Q^2 \sim 10$~GeV$^2$ are not now available. Assuming
a reliable extrapolation of the HERA data to this domain for the
proton beam, the determination of the nuclear corrections for Pb will
therefore require a $p$Pb run. A Pb$p$ run would probably have lower
priority since the peak of the $x_1$ distribution is at
$x_1 > 10^{-2}$ where current data are more reliable.

\subsection{Total cross sections}

We study here the predictions for total cross sections,
starting with charm production. Table~\ref{tab:chrxs} gives results for
$pp$ collisions at both $\sqrt{S}=14$~TeV and 5.5~TeV.  The rates vary
over a large range and the $K$ factors, quantifying the size of NLO
corrections, are very large. As anticipated previously, these factors make
absolute rate estimates quite uncertain. Nevertheless, the
predicted extrapolation from 14 to 5.5 TeV is independent of the
chosen parameter combination at the level of few percent, as shown in
the last two columns of Table~\ref{tab:chrxs}.
Similar results and conclusions are obtained for bottom
quarks, detailed in Table~\ref{tab:botxs}. 
\begin{table}
\begin{center}
\caption{Production cross sections for charm pairs in $pp$ collisions
at 14 TeV and 5.5 TeV. The penultimate column gives the ratios of
cross sections  at the two energies for the various parameter
sets. The last column gives the ratios, normalized to the ratio
obtained with ${\cal P} = {\rm I}$.}
\label{tab:chrxs}
\vspace*{0.1cm}
\begin{tabular}{|l|c|c|} \hline
${\cal P}$ & $\sigma_{14}$ (mb) & 
$K = \sigma_{\rm NLO}/\sigma_{\rm LO}$
 \\ \hline
I &   10.4 &  1.7
 \\ \hline
II &  16.7 & 1.7
 \\ \hline
III &  6.8 &  1.7
 \\ \hline
IV &  7.3 &  2.1
 \\ \hline
V & 8.57  & 1.8
 \\ \hline
VI &  10.6 &  1.8
\\\hline
\end{tabular}
\nolinebreak
\begin{tabular}{|c|c|c|} \hline
$\sigma_{5.5}$ (mb) & $R({\cal P})=\sigma_{5.5}/\sigma_{14}$ &
 $R({\cal P})/R({\cal P}={\mathrm I})$
 \\ \hline
 5.4  & 0.52 &  1
 \\ \hline
 9.2 &  0.55 &  1.06
 \\ \hline
 3.4 &  0.50 &  0.96
 \\ \hline
 3.7 &  0.51 &  0.98
 \\ \hline
 4.2  & 0.49 & 0.94
 \\ \hline
  5.3  & 0.50 & 0.96
 \\ \hline
\end{tabular}                                                                 
\end{center}
\end{table}

\begin{table}
\begin{center}
\caption{Production cross sections for bottom pairs in $pp$ collisions
at 14 TeV and 5.5 TeV. The penultimate column gives the ratios of
cross sections  at the two energies for the various parameter
sets. The last column gives the ratios, normalized to the ratio
obtained with ${\cal P} = {\rm I}$.}
\label{tab:botxs}
\vspace*{0.1cm}
\begin{tabular}{|l|c|c|} \hline
${\cal P}$ & $\sigma_{14}$ (mb) 
& $K = \sigma_{\rm NLO}/\sigma_{\rm LO}$
 \\ \hline
I &   0.43 &  2.3
 \\ \hline
II &   0.51 & 2.4
 \\ \hline
III & 0.37 & 2.3
 \\ \hline
IV & 0.66 & 1.4
 \\ \hline
V &  0.20 & 3.2
 \\ \hline
VI & 0.40 & 2.4
 \\ \hline
VII & 0.45  & 2.4
 \\ \hline
\end{tabular}                                                                 
\nolinebreak
\begin{tabular}{|c|c|c|} \hline
$\sigma_{5.5}$ (mb) & $R({\cal P})=\sigma_{5.5}/\sigma_{14}$ &
 $R({\cal P})/R({\cal P}={\mathrm I})$
 \\ \hline
 0.17 &  0.40 & 1
 \\ \hline
 0.20 &   0.39 &   0.98
 \\ \hline
 0.15 &  0.41 & 1.03
 \\ \hline
 0.26 &  0.39 &  0.98
 \\ \hline
 0.088 &  0.44 & 1.1
 \\ \hline
 0.17 & 0.43 & 1.08
 \\ \hline
 0.18 & 0.40  & 1
 \\ \hline
\end{tabular}                                                                 
\end{center}
\end{table}

We next consider the extrapolation of
cross sections within the central, Table~\ref{tab:chrxs_cen}, and
forward, Table~\ref{tab:chrxs_fwd}, acceptance regions.
We also consider the
extrapolation of the $p$Pb, Pb$p$ and $pp$ rates to \PbPb.  As in the
case of the total rates, the extrapolations appear to have a limited
dependence on the parameter set, contrary to the large variations
of the absolute rates. We do not explicitly quote the PDF dependence 
since it is very small.
 
While in the case of central production there is clearly no difference
between Pb$p$ and $p$Pb, the forward acceptance has an intrinsic
asymmetry. In is interesting to note that the extrapolation of the
forward rate appears to be more stable when using a Pb$p$ normalization 
rather than a $p$Pb one. 

We find qualitatively similar results for bottom production.

\begin{table}
\begin{center}
\caption{Single inclusive charm production cross sections per nucleon in the
central region. The $pp$ cross section is calculated at $\sqrt{S}=5.5$~TeV.}
\label{tab:chrxs_cen}
\vspace*{0.1cm}
\begin{tabular}{|l|c|c|c|c|c|} \hline
  ${\cal P}$ & $\sigma_{pp}$ (mb) & $\sigma_{{\rm Pb}p}$ (mb) &
  $\sigma_{\rm PbPb}$ (mb)   
  & $\sigma_{\rm PbPb}/\sigma_{pp}$
  & $\sigma_{\rm PbPb}/\sigma_{{\rm Pb}p}$
 \\ \hline
I &  0.88 & 1.00  & 0.59  & 0.67  & 0.59  
 \\ \hline
II & 1.5  & 1.6   & 0.98  & 0.65  & 0.61  
 \\ \hline
III & 0.55 & 0.64  & 0.38  & 0.69  & 0.59 
 \\ \hline
IV & 0.58 & 0.66   &  0.40 & 0.69  & 0.61 
 \\ \hline
\end{tabular}
\end{center}
\end{table}

\begin{table}
\begin{center}
\caption{Single inclusive charm production cross sections per nucleon in the
forward region. The $pp$ cross section is calculated at $\sqrt{S}=5.5$~TeV.}
\label{tab:chrxs_fwd}
\vspace*{0.1cm}
\begin{tabular}{|l|c|c|c|c|c|c|c|} \hline
  ${\cal P}$ & $\sigma_{pp}$ (mb) & $\sigma_{{\rm Pb}p}$ (mb) &
  $\sigma_{p{\rm Pb}}$ (mb)  &
  $\sigma_{\rm PbPb}$ (mb)   
  & $\sigma_{\rm PbPb}/\sigma_{pp}$
  & $\sigma_{\rm PbPb}/\sigma_{{\rm Pb}p}$
  & $\sigma_{\rm PbPb}/\sigma_{p{\rm Pb}}$
 \\ \hline
I &  0.75 & 0.92 & 0.79  & 0.50  & 0.67  & 0.54 & 0.63
 \\ \hline
II & 1.3  & 1.5  & 1.3  & 0.83  &  0.64 & 0.55 & 0.64
 \\ \hline
III & 0.47 & 0.60  & 0.51  & 0.33  & 0.70  & 0.55 & 0.65
 \\ \hline
IV & 0.51 & 0.65  & 0.56  & 0.36  & 0.71  & 0.55 & 0.64
 \\ \hline
\end{tabular}
\end{center}
\end{table}

\subsection{Inclusive \pt\ spectra}

\begin{figure}
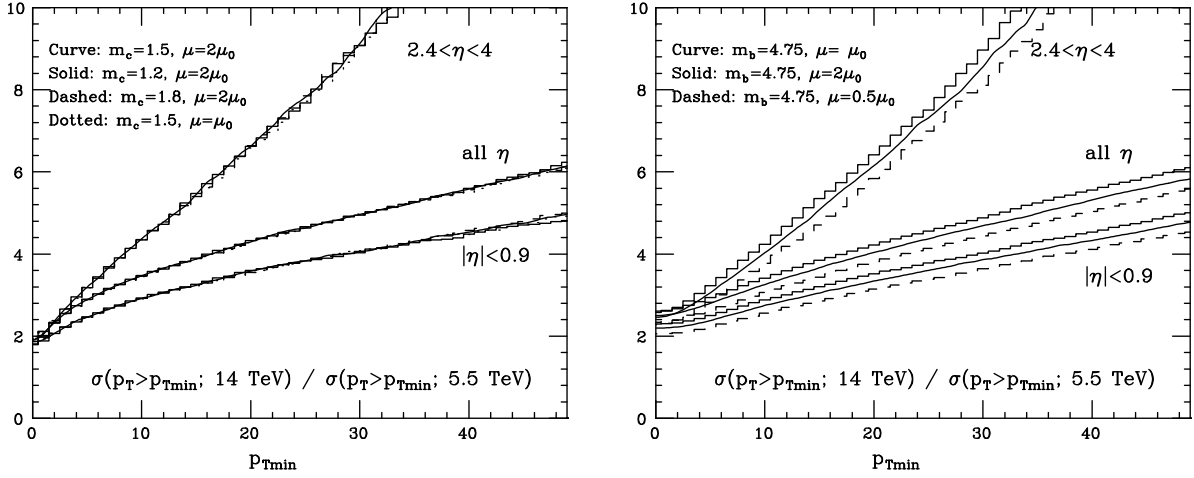
 
\begin{center} 
\centerline{
\includegraphics[width=0.47\textwidth,clip]{cratio.eps}
\hfill
\includegraphics[width=0.47\textwidth,clip]{bratio.eps}
 }
\vskip -0.4cm 
\caption{Charm (left) and bottom (right) ratios at
14~TeV/5.5~TeV. 
\label{fig:cratio}}
\end{center} 
\end{figure}

Here we present more detailed studies of the 
reliability of extrapolating the charm transverse momentum spectra.
Figure~\ref{fig:cratio} shows the ratios of the charm and bottom quark
\pt\ spectra in $pp$
collisions at 14 and 5.5~TeV. Rather than the differential spectra, we present
the rates integrated above a given \pt\ threshold, $p_{T {\rm min}}$.  
The curves correspond to different parameter sets. In spite of the
strong sensitivity of the ratios to the pseudorapidity range and to $p_{T {\rm
min}}$, we see once again that the dependence on the input
parameter set is very small, ensuring a rather safe extrapolation. 

\begin{figure}
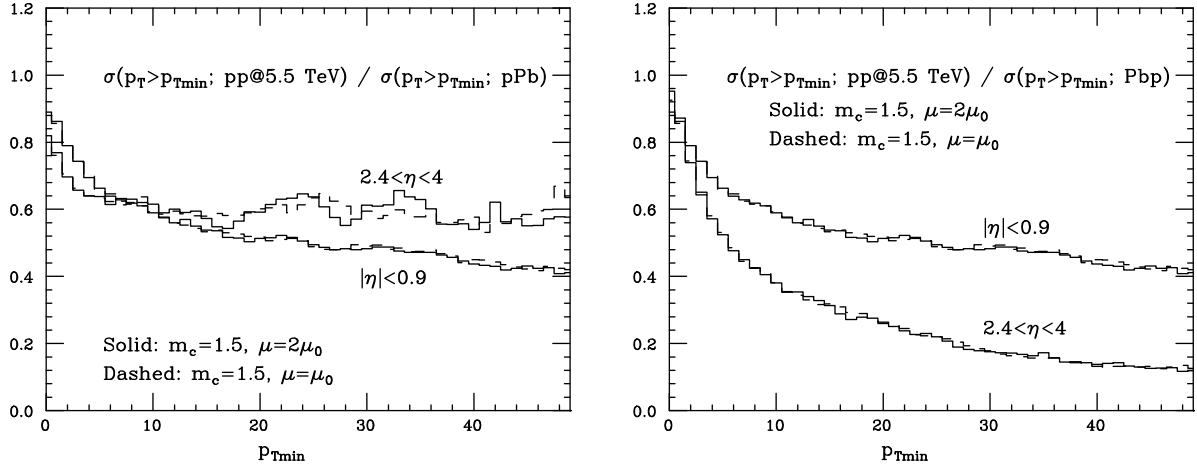
 
\begin{center} 
\centerline{
\includegraphics[width=0.47\textwidth,clip]{cratiopPb.eps}
\hfill
\includegraphics[width=0.47\textwidth,clip]{cratioPbp.eps}
}
\vskip -0.4cm 
\caption{Charm \pt\ ratios per nucleon for $pp$(5.5~TeV)/($p$Pb) (left) and 
$pp$(5.5~TeV)/(Pb$p$) (right). \label{fig:cratiopPb}}
\end{center} 
\end{figure}

In Fig.~\ref{fig:cratiopPb} we present similar ratios for charm
production, comparing the spectra in 5.5~TeV $pp$ collisions with
$p$Pb and Pb$p$ interactions. We consider here only predictions obtained 
with $m_c = 1.5$ GeV but different values of the renormalization
scale, ${\cal P} = {\rm I}$ and IV.

\begin{figure}
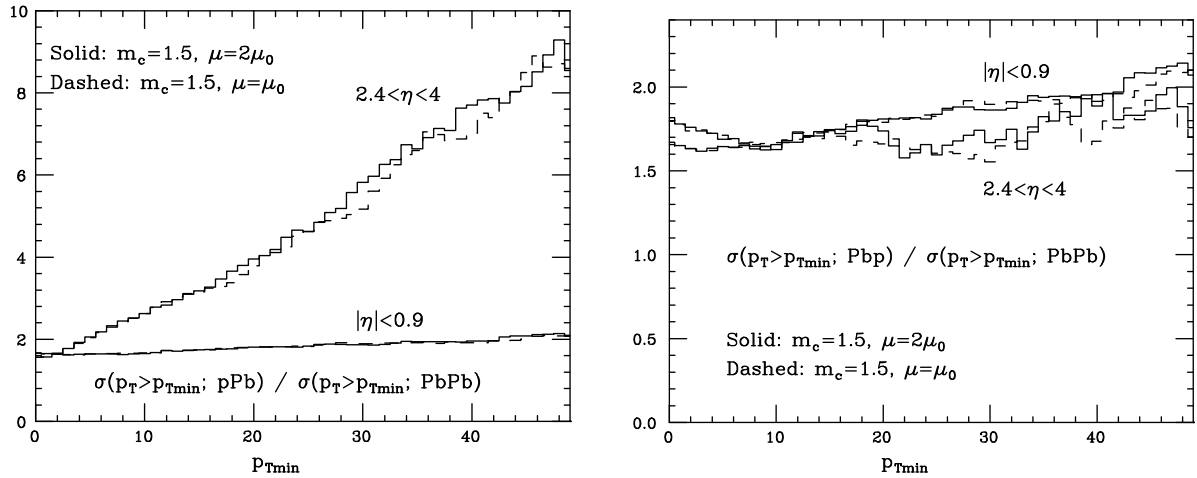
 
\begin{center} 
\centerline{
\includegraphics[width=0.47\textwidth,clip]{cratioPbPb.eps}
\hfill
\includegraphics[width=0.47\textwidth,clip]{cratioPbPbPb.eps}
 }
\vskip -0.4cm 
\caption{Charm \pt\ ratios per nucleon for ($p$Pb)/(\PbPb) (left) and 
(Pb$p$)/(\PbPb) (right). \label{fig:cratioPbPb}}
\end{center} 
\end{figure}

In Fig.~\ref{fig:cratioPbPb} we consider the ratio of
the spectra in $p$Pb and Pb$p$ collisions to
\PbPb. We note that, while the $p$Pb/PbPb ratio shows 
a strong \pt\ and $\eta$ dependence, the Pb$p$/PbPb ratios 
are quite similar up to an overall
normalization factor. This is because production in the forward
proton region is enhanced by the larger energy of the proton relative
to the nucleon energy in the Pb nucleus. When the charm is instead
detected in the forward region, in the direction of the Pb nucleus,
the spectrum is much less sensitive to whether the ``target'' is a
proton or another Pb beam.  In this case, therefore, a Pb$p$ run 
provides a better normalization benchmark compared to a $p$Pb run. 
This is opposite the conclusion reached in subsection~\ref{xcov} for the
determination of the small $x$ gluon density in Pb. 

\subsection{$\Delta\phi$ azimuthal correlations}

Azimuthal correlations between heavy quark pairs could provide useful
information on the mechanisms of heavy quark
production in central nuclear-nuclear collisions.  A large
fraction of these correlations could be washed out by the fact that
the average number of charm pairs produced in \PbPb\ collisions is
large and the detection of two charm quarks in the same event is no
guarantee that they both originate from the same nucleon-nucleon
interaction.  However, the probability of tagging correlated pairs increases at
large \pt\ since these rates are smaller.  

\begin{figure} 
\begin{center} 
\centerline{
\includegraphics[width=0.8\textwidth,clip]{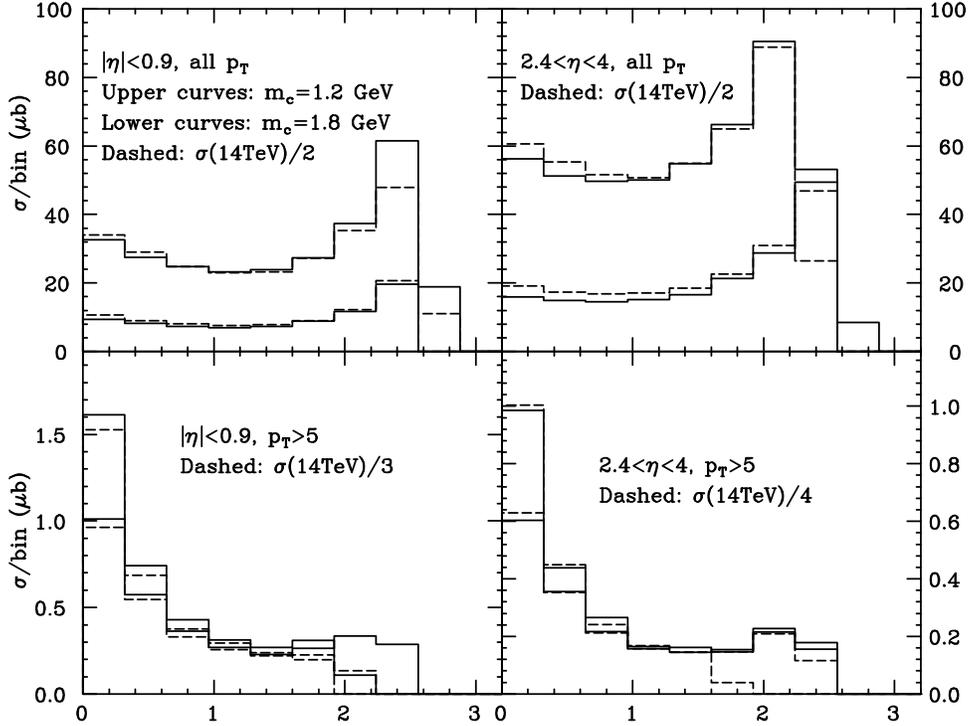}}
\vskip -0.4cm 
\caption{Charm  azimuthal correlations in $pp$ collisions. The solid
lines correspond to $\sqrt{S}=5.5$~TeV, the dashed lines to 
$\sqrt{S}=14$~TeV, rescaled by the amount indicated.
 \label{fig:cdphi}}
\end{center} 
\end{figure}

The plots in Fig.~\ref{fig:cdphi} represent the distribution of the
azimuthal difference, $\Delta\phi$, between the $Q$ and $\overline{Q}$, 
defined as:
\be
\Delta \phi=\vert \phi_Q - \phi_{\overline Q}\vert \; ,
\ee
in $pp$ collisions at 14 and 5.5 TeV.
The upper plots refer to distributions for pairs in the central and
forward regions without any \pt\ cut. The lower plots require a
minimum \pt\ of 5~GeV for both the $c$ and $\overline{c}$. Note both the very
strong suppression of back-to-back production over all $p_T$,
indicating very strong Sudakov suppression effects, and the
enhancement of production in the same hemisphere at high $p_T$, a result of the
dominant contribution from final-state gluon splitting into
$c\overline{c}$.  Also note the similar shape 
predicted for 14 and 5.5 TeV collisions after rescaling the
overall normalization as well 
as the mild dependence on the input parameters. 

\begin{figure} 
\begin{center} 
\centerline{\includegraphics[width=0.8\textwidth,clip]{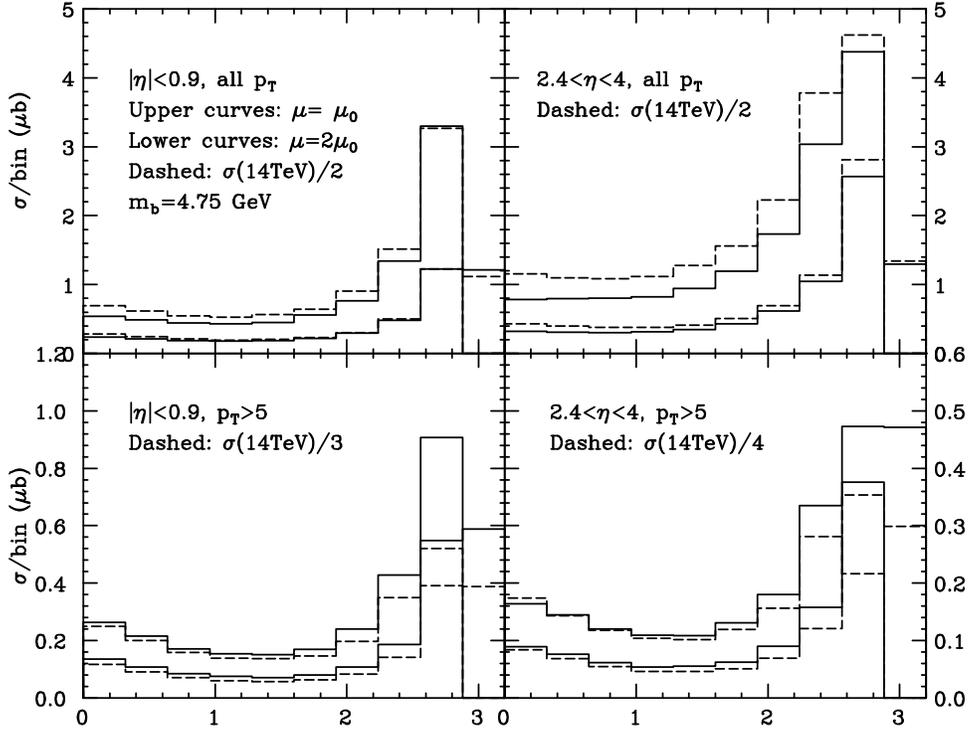} }
\vskip -0.4cm 
\caption{Bottom  azimuthal correlations in $pp$ collisions. The solid
lines correspond to $\sqrt{S}=5.5$~TeV, the dashed lines to 
$\sqrt{S}=14$~TeV, rescaled by the amount indicated.
 \label{fig:bdphi}}
\end{center} 
\end{figure}

The results for bottom pair production are given in
Fig.~\ref{fig:bdphi}. In this case, the dominant contributions are
in the back-to-back region, both without and with the $p_T$ cut.
The distributions are similar when Pb beams are considered.

\subsection{Conclusions}

We summarize here the findings of this study. We find that, in spite
of the large uncertainties involved in the absolute predictions for
charm and bottom quarks at the energies of relevance for the LHC, the
correlation between the rates expected at different energies or with
different beams is very strong. Using data extracted in $pp$
collisions at 14 TeV it is possible to predict
the rates expected in $pp$ collisions at 5.5 TeV to within few percent. 
A comparison of
these predicted rates with $p$Pb and Pb$p$ measurements will therefore
allow a solid extraction of the nuclear modifications of the parton PDFs
in Pb. In spite of the different energy configurations for the $p$Pb
and the \PbPb\ runs, it is possible to extrapolate to 
\PbPb\ collisions in the absence of strong
medium-dependent modifications. Differences with respect to these
predictions could then used to infer properties of the production and
propagation of heavy quarks in the dense medium resulting from 
high energy \PbPb\ collisions. We stress once more the value of $p$Pb and 
Pb$p$ control runs, and possibly also lower energy $pp$ runs. 
We showed examples indicating the need of both $p$Pb and
Pb$p$ runs.  The $p$Pb runs are useful to determine the
small-$x$ gluon density of Pb while Pb$p$ runs provide a better
normalization benchmark for inclusive charm \pt\ spectra.
A concrete plan of measurements to benchmark the predictions
against real data and to determine the residual systematic
uncertainties in the extrapolations remains to be outlined.

%% file: quarkonium.tex
\section[QUARKONIUM BASELINE PREDICTIONS]
{QUARKONIUM BASELINE PREDICTIONS~\protect
\footnote{Section coordinator: R. Vogt.}}\label{sec:quarkon}

\subsection{Introduction}
\label{quarkon.intro}

In this section, we discuss quarkonium production in $pp$, $pA$, and $AB$ 
collisions.  Here we only consider the effects of nuclear shadowing on
quarkonium production.  Other possible consequences of nuclear collisions
such as absorption and energy loss in nuclear matter
and finite temperature effects are discussed later in this chapter. 

Early studies of high-$p_T$ quarkonium production revealed that direct
$J/\psi$ production in the color singlet model was inadequate to describe
charmonium hadroproduction.  However, it may be able to describe charmonium
production in cleaner environments such as photoproduction.  Given this
difficulty with hadroproduction, several other approaches have been developed.
The Color Evaporation Model (CEM), 
discussed in section~\ref{quarkon.cem}, treats
heavy flavor and quarkonium production on an equal footing, the color being
`evaporated' through an unspecified process which does not change the momentum.
Nonrelativistic QCD (NRQCD), sections~\ref{quarkon.nrqcd1} 
and~\ref{quarkon.nrqcd2}, is an effective field theory
in which the short distance partonic interactions produce $Q\bQ$ 
pairs in color singlet or color octet states and nonperturbative matrix 
elements describe the evolution of the $Q\bQ$ pair into a quarkonium
state. The color singlet model essentially corresponds to dropping all 
but the first term in the NRQCD expansion 
(see sect.~\ref{NRQCD-factorization}).
Finally, the Comover Enhancement Scenario (CES), section~\ref{quarkon.ces}, 
bases its predictions on the color singlet model enhanced by absorption 
of gluons from the medium.

\subsection[Quarkonium production in the Color Evaporation
Model]{Quarkonium production in the Color Evaporation
Model~\protect\footnote{Author: R. Vogt.}}
\label{quarkon.cem}

To better understand quarkonium suppression, it is necessary to have a good
estimate of the expected yields.  However, there are still a number of 
unknowns about quarkonium production in the primary
nucleon-nucleon interactions.  In this section, we discuss quarkonium 
production in the color evaporation model (CEM) and give predictions for
production in $pp$ and $AA$ interactions at the LHC. 

The CEM was first discussed a long time ago~\cite{Barger:1979js,Barger:1980mg}
and has enjoyed considerable phenomenological success. 
In the CEM, the quarkonium production cross section is some fraction $F_C$ of 
all $Q\bQ$ pairs below the $H \overline H$ threshold where $H$ is
the lowest mass heavy flavor hadron.  Thus the CEM cross section is
simply the $Q\bQ$ production cross section with a cut on the pair mass
but without any constraints on the 
color or spin of the final state.  The produced $Q\bQ$ 
pair then neutralizes its color by
interaction with the collision-induced color field---``color evaporation''.
The $Q$ and the $\bQ$ either combine with light
quarks to produce heavy-flavored hadrons or bind with each other 
to form quarkonium.  The additional energy needed to produce
heavy-flavored hadrons when the partonic center of mass energy, 
$\sqrt{\hat s}$, is less than $2m_H$, the heavy hadron
threshold, is obtained nonperturbatively from the
color field in the interaction region.
Thus the yield of all quarkonium states
may be only a small fraction of the total $Q\overline 
Q$ cross section below $2m_H$.
At leading order, the production cross section of quarkonium state $C$ in
an $AB$ collision is
\begin{eqnarray}
\sigma_C^{\rm CEM} = F_C \sum_{i,j} \int_{4m_Q^2}^{4m_H^2} d\hat{s}
\int dx_1 dx_2~f_{i/A}(x_1,\mu^2)~f_{j/B}(x_2,\mu^2)~ 
\hat\sigma_{ij}(\hat{s})~\delta(\hat{s} - x_1x_2s)\, 
\, , \label{sigtil}
\end{eqnarray} 
where $A$ and $B$ can be any hadron or nucleus,
$ij = q \overline q$ or $gg$, $\hat\sigma_{ij}(\hat s)$ is the
$ij\rightarrow Q\bQ$ subprocess cross section and $f_{i/A}(x,\mu^2)$
is the parton density in the hadron or nucleus.  The total $Q \overline
Q$ cross section takes $\hat{s} \rightarrow s$.
In a collision where either or both $A$ and $B$ is a nucleus, 
we use the EKS98 parameterization of the nuclear modifications 
of the parton densities \cite{Eskola:1998iy,Eskola:1998df} 
to model shadowing effects.  
Since quarkonium production is gluon dominated, 
isospin is negligible.  However, because shadowing effects on the gluon
distribution may be large, they could strongly influence the results.

The fraction $F_C$ must be universal so that, once it is fixed by data, the
quarkonium production ratios should be constant as a function of $\sqrt{s}$,
$y$ and $p_T$.  The actual value of $F_C$ depends on the heavy quark mass, 
$m_Q$, the scale, $\mu^2$, the parton densities and 
the order of the calculation.
It was shown in Ref.~\cite{Gavai:1994in} that the quarkonium production ratios
were indeed constant, as expected by the model.

Of course the leading order calculation in Eq.~(\ref{sigtil}) is insufficient
to describe high $p_T$ quarkonium production since the $Q\bQ$ pair
$p_T$ is zero at LO.  Therefore, the CEM was 
taken to NLO \cite{Gavai:1994in,Schuler:1996ku} using the exclusive 
$Q\bQ$ hadroproduction code of Ref.~\cite{Mangano:jk}.  
At NLO in the CEM, the process $gg \rightarrow g Q\bQ$
is included, providing a good
description of the quarkonium $p_T$ distributions at the Tevatron
\cite{Schuler:1996ku}.  In the exclusive NLO calculation of
Ref.~\cite{Mangano:jk}, both the $Q$ and $\bQ$ variables
are available at each point in the phase space over which the
integration is carried out. Thus, one is not forced to choose
the renormalization and factorization scales proportional to
the heavy-quark mass. A popular choice is for example
$\mu\propto m_T \equiv\sqrt{m_Q^2 + p_T^2}$, with 
$p_T^2 = 0.5(p_{T_Q}^2 + p_{T_{\bQ}}^2)$.  
For simplicity, we refer to $\mu \propto m_Q$ in the
following but the proportionality to $m_T$ is implied.

\begin{figure}[htb]
\setlength{\epsfxsize=0.95\textwidth}
\setlength{\epsfysize=0.4\textheight}
\centerline{\epsffile{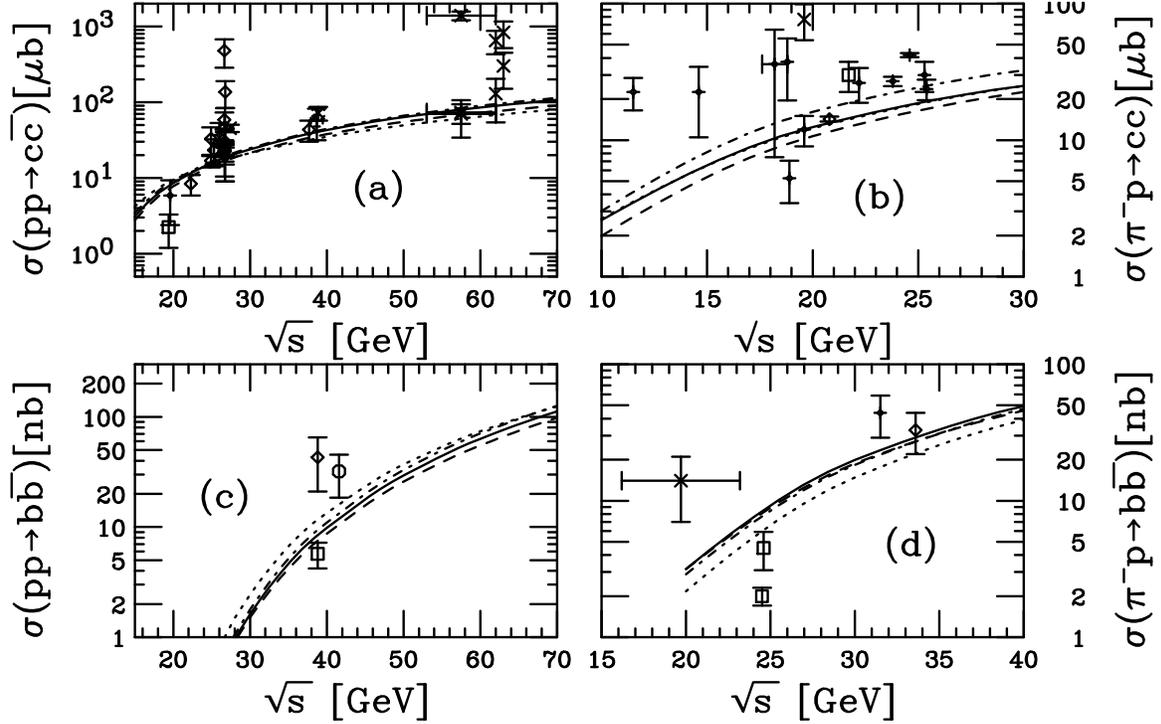}}
\caption{The $c \overline c$, (a) and (b), and $b \overline b$, (c) and (d), 
total cross section data in $pp$ and $\pi^- p$ interactions compared to NLO
calculations.  In (a) and (b), we show the results obtained with parameter
choices $\psi1$ (solid), $\psi2$ (dashed), $\psi3$ (dot-dashed) and $\psi4$ 
(dotted).  In (c) and (d), we show those obtained with 
$\Upsilon1$ (solid), $\Upsilon2$ (dashed), $\Upsilon3$ (dot-dashed) and 
$\Upsilon4$ (dotted).}
\label{qqbfig}
\end{figure}

We use the same parton densities and parameters that agree with 
the $Q\bQ$ total cross section data, given in Table~\ref{qqbparams},
to determine $F_C$ for $J/\psi$ and $\Upsilon$ production.  
The fit parameters \cite{Vogt:2002vx,Vogt:2002ve} for the parton densities 
\cite{Martin:1998sq,Lai:1999wy,Gluck:1998xa}, quark
masses and scales are given in
Table~\ref{qqbparams} while the $Q\bQ$ cross sections calculated with
these parameters are 
compared to $pp \rightarrow Q\bQ$ and $\pi^- p \rightarrow Q \overline
Q$ data in Fig.~\ref{qqbfig}.

\begin{table}[htb]
\begin{center}
\caption{Parameters used to obtain the `best' agreement to the $Q\bQ$
cross sections.}
\begin{tabular}{|cccc||cccc|} \hline
\multicolumn{4}{|c||}{$c \overline c$} & \multicolumn{4}{c|}{$b 
\overline b$} \\ \hline
Label & PDF & $m_c$ (GeV) & $\mu/m_c$ & Label & PDF & $m_b$ (GeV) & 
$\mu/m_b$ \\ \hline
$\psi1$ & MRST HO & 1.2 & 2   & $\Upsilon1$ & MRST HO & 4.75 & 1   \\
$\psi2$ & MRST HO & 1.4 & 1   & $\Upsilon2$ & MRST HO & 4.5  & 2   \\
$\psi3$ & CTEQ 5M & 1.2 & 2   & $\Upsilon3$ & MRST HO & 5.0  & 0.5 \\
$\psi4$ & GRV 98 HO & 1.3 & 1 & $\Upsilon4$ & GRV 98 HO & 4.75 & 1 \\ \hline
\end{tabular}
\label{qqbparams}
\end{center}
\end{table}

\begin{figure}[htb]
\setlength{\epsfxsize=0.95\textwidth}
\setlength{\epsfysize=0.4\textheight}
\centerline{\epsffile{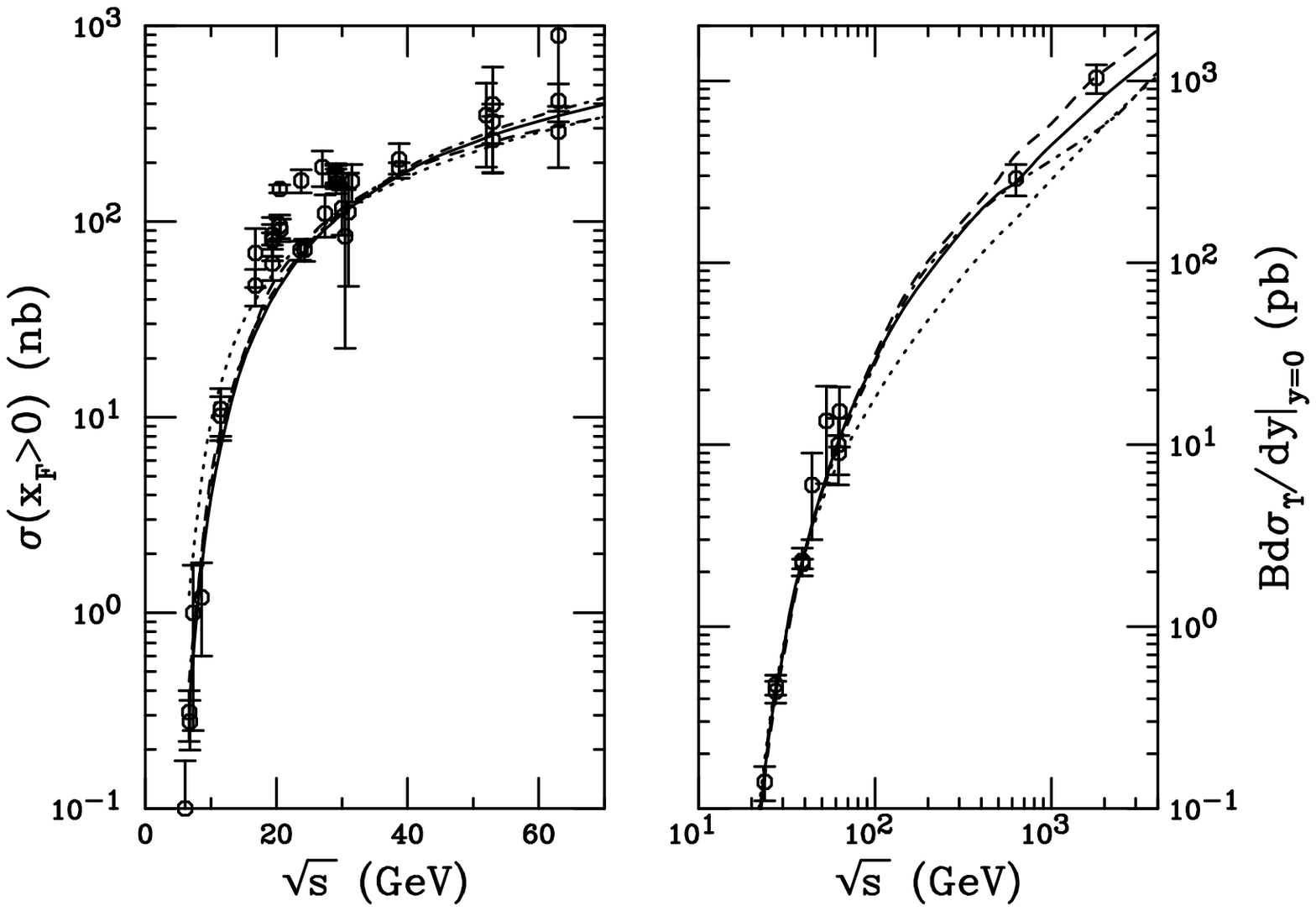}}
\caption{Forward $J/\psi$ (left) and combined $\Upsilon + \Upsilon' +
\Upsilon''$ inclusive (right) cross sections calculated to NLO in the CEM.  On
the left-hand side, we show the results obtained with parameter choices 
$\psi1$ (solid), $\psi2$ (dashed), $\psi3$ (dot-dashed) and $\psi4$ (dotted).  
On the right-hand side, we show those obtained using $\Upsilon1$ (solid), 
$\Upsilon2$ (dashed), $\Upsilon3$ (dot-dashed) and 
$\Upsilon4$ (dotted).}
\label{psiupsfixt}
\end{figure}

We now describe the extraction of $F_C$ for the individual quarkonium states.
The $J/\psi$ has been measured in $pp$ and $pA$ interactions up to $\sqrt{s} =
63$ GeV.  The data are of two types: the forward cross section, $\sigma(x_F >
0)$, and the cross section at zero rapidity, $d\sigma/dy|_{y=0}$.  All the
cross sections are inclusive with feeddown from $\chi_c$ and $\psi'$
decays.  To obtain $F_{J/\psi}$ for inclusive $J/\psi$ production, the 
normalization of Eq.~(\ref{sigtil}) is fit for the $c \overline c$ parameters
in Table~\ref{qqbparams}.  The comparison of $\sigma_{J/\psi}^{\rm CEM}$ to the
$x_F > 0$ data for all four fits is shown on the left-hand side of 
Fig.~\ref{psiupsfixt}.  The ratios of the direct production cross sections to
the inclusive $J/\psi$ cross section can be determined from data on inclusive
cross section ratios and branching fractions.  These direct
ratios, $R_C$, given 
in Table~\ref{ratios}, are multiplied by the inclusive fitted $F_{J/\psi}$ to
obtain the direct production fractions, $F^{\rm dir}_C = F_{J/\psi} R_C$.

\begin{table}[htb]
\begin{center}
\caption{Direct quarkonium production ratios, $R_C = \sigma^{\rm 
dir}_C/\sigma_{C'}^{\rm inc}$ where $C' = J/\psi$ and $\Upsilon$.
From Ref.~\protect \cite{Digal:2001ue}.}
\begin{tabular}{|cccccccccc|} \hline
& $J/\psi$ & $\psi'$ & $\chi_{c1}$ & $\chi_{c2}$ & $\Upsilon$ & $\Upsilon'$
& $\Upsilon''$ & $\chi_b(1P)$ & $\chi_b(2P)$ \\ \hline
$R_C$ & 0.62 & 0.14 & 0.60 & 0.99 & 0.52 & 0.33 & 0.20 & 1.08 & 0.84 \\ \hline
\end{tabular}
\label{ratios}
\end{center}
\end{table}

The same procedure, albeit somewhat more complicated due to the larger number 
of bottomonium states below the $B \overline B$ threshold, is followed for the 
bottomonium.  For most data below $\sqrt{s} = 100$ GeV, 
the three bottomonium $S$ states were either not separated or their sum was 
reported.  No  $x_F$-integrated cross sections were available so that we fit 
the CEM $\Upsilon$ cross
section to the effective lepton pair cross section at $y=0$ for the
three $\Upsilon(nS)$ states.  The extracted fit fraction is labeled $F_{\sum
\Upsilon}$.  The comparison of $\sigma_\Upsilon^{\rm CEM}$ with 
$F_{\sum \Upsilon}$ to 
the data for all parameter sets in Table~\ref{qqbparams} is shown on the
right-hand side of Fig.~\ref{psiupsfixt}.  Using the individual branching
ratios of the $\Upsilon$, $\Upsilon'$ and $\Upsilon''$ to lepton pairs and the 
total cross sections reported by CDF \cite{Affolder:1999wm}, 
it is possible to extract 
the inclusive $\Upsilon$ fit fraction, $F_\Upsilon$.  The direct production
ratios obtained in Ref.~\cite{Gunion:1996qc} have been updated in 
Ref.~\cite{Digal:2001ue} using recent CDF $\chi_b$ data.  
The resulting direct to inclusive $\Upsilon$ ratios, $R_C$, are also
given in Table~\ref{ratios}.  The subthreshold $b \overline b$ cross section
is then multiplied by $F_C^{\rm dir} = F_\Upsilon R_C$ to obtain the direct
bottomonium cross sections.

The energy dependence shown in Fig.~\ref{psiupsfixt} for both states is well
reproduced by the NLO CEM.  All the fits are equivalent for $\sqrt{s} \leq 100$
GeV but differ by up to a factor of two at 5.5 TeV.  Since the $p_T$-integrated
$\Upsilon$ cross sections have been measured at the Tevatron, the $\sqrt{s}$
range of the extrapolation to the LHC is rather small for the $\Upsilon$.
The high energy $\Upsilon$ data seem to agree best with the energy dependence 
obtained with the parameter sets $\Upsilon1$ and $\Upsilon2$.
A similar check cannot be made for the $J/\psi$ because the high lepton $p_T$
cut excludes $J/\psi$ acceptance for $p_T =0$ at the Tevatron.  However, the 
good agreement with the lower energy data results in less than a factor of 
two difference between the four cases at $\sqrt{s} = 5.5$ TeV.
\begin{table}[htbp]
\caption[]{The inclusive $J/\psi$ production fractions obtained from data for
the cases given in Table~\protect \ref{qqbparams}.
The direct charmonium cross sections for $pp$ collisions at 5.5 TeV are also 
given.}
\begin{center}
\begin{tabular}{|cc||cccc|} \hline
& & \multicolumn{4}{c|}{$\sigma^{\rm
dir}$ ($\mu$b)} \\ \hline
Case & $F_{J/\psi}$ & $J/\psi$ & $\chi_{c1}$ & $\chi_{c2}$ & $\psi'$ \\ \hline
$\psi1$ & 0.0144 & 19.0 & 18.3 & 30.2 & 4.3 \\
$\psi2$ & 0.0248 & 12.4 & 12.0 & 19.8 & 2.8 \\
$\psi3$ & 0.0155 & 22.2 & 21.6 & 35.6 & 5.0 \\
$\psi4$ & 0.0229 & 19.8 & 19.3 & 31.8 & 4.5 \\ \hline
\end{tabular}
\label{psitab}
\end{center}
\end{table}

The $pp$ cross sections obtained for the individual states at 5.5 TeV
are shown in Tables~\ref{psitab} and \ref{upstab} along with the values of the
inclusive $F_C$.
We give both $F_{\sum \Upsilon}$ and $F_\Upsilon$ for bottomonium.
Only the direct cross sections are given in the tables.  To obtain 
$\sigma^{\rm dir}$, we use the 
production ratios in Table~\ref{ratios}.

\begin{table}[htbp]
\caption[]{The inclusive $\Upsilon$ production fractions obtained from data
for the cases given in Table~\protect \ref{qqbparams}.  The direct 
bottomonium cross sections 
for $pp$ collisions at 5.5 TeV are also given.}
\begin{center}
\begin{tabular}{|ccc||ccccc|} \hline
& & & \multicolumn{5}{c|}{$\sigma^{\rm
dir}$ (nb)} \\ \hline
Case & $F_{\sum \Upsilon}$ & $F_\Upsilon$ & $\Upsilon$ & $\Upsilon'$  
& $\Upsilon''$ & $\chi_b(1P)$ & $\chi_b(2P)$ \\ \hline
$\Upsilon1$ & 0.000963 & 0.0276 & 188 & 119 & 72 & 390 & 304  \\
$\Upsilon2$ & 0.000701 & 0.0201 & 256 & 163 & 99 & 532 & 414  \\
$\Upsilon3$ & 0.001766 & 0.0508 & 128 &  82 & 49 & 267 & 208  \\
$\Upsilon4$ & 0.000787 & 0.0225 & 145 &  92 & 56 & 302 & 235  \\ 
\hline
\end{tabular}
\label{upstab}
\end{center}
\end{table}

The range of the fit parameters allows us to explore the dependence of
$F_{J/\psi}$ and $F_\Upsilon$ on $m_Q$ and $\mu$.  The range of fit parameters
is limited because we only choose parameters that are in relatively
good agreement with the $Q\bQ$ total cross sections.  The $F_{J/\psi}$ 
obtained with the parameter sets $\psi1$ and $\psi3$, employing the same 
mass and 
scale, are rather similar but $\sigma^{\rm dir}$ differs by 20\% at 5.5 TeV.
For the same PDFs but with a lower scale, case $\psi2$ relative to $\psi1$, 
$F_{J/\psi}$ is about a factor of two larger.  However, the cross sections at 
5.5 TeV differ only by 50\%.  Changing the mass and PDF at the same scale,
cases $\psi3$ and $\psi4$, does not change $F_{J/\psi}$ substantially.

Since the $b$ quark is more massive, the scale dependence can be more sensibly
explored.  The $b \overline b$ cross sections for cases $\Upsilon1$, 
$\Upsilon2$ and $\Upsilon3$ are essentially equivalent.  However, the resulting
$F_\Upsilon$ differs by a factor of 2.5 with the highest $m_b$ giving the
largest $F_\Upsilon$ and the lowest $\sigma^{\rm dir}$.  There is a factor of
two between the corresponding $\Upsilon$ cross sections.  For different
PDFs but the same mass and scale, $\Upsilon1$ and $\Upsilon4$, the fitted
$F_\Upsilon$'s differ by only $\sim 10$\%. 

\begin{table}[htbp]
\caption[]{The direct cross section per nucleon pair and the dilepton cross
section per nucleon multiplied by $A^2$ for the minimum bias lepton pair
cross section.  The results are given for $\psi1$.  
We compare $pp$ to $pA$ and $AA$ interactions.}
\begin{center}
\begin{tabular}{|cc||cccc||cc|} \hline
 & & \multicolumn{4}{c||}{$\sigma^{\rm dir}$/nucleon pair ($\mu$b)} &
\multicolumn{2}{c|}{$B \sigma^{\rm inc} A^2$ ($\mu$b)} \\ \hline
System & $\sqrt{s}$ (TeV) & $J/\psi$ & $\chi_{c1}$ & 
$\chi_{c2}$ & $\psi'$ & $J/\psi$ & $\psi'$
\\ \hline
$pp$   & 14   & 32.9 & 31.8 & 52.5 & 7.43 & 3.18    & 0.057 \\
$pp$   & 8.8  & 25.0 & 24.2 & 39.9 & 5.65 & 2.42    & 0.044 \\
$p$Pb  & 8.8  & 19.5 & 18.9 & 31.1 & 4.40 & 392.3   & 7.05  \\
$pp$   & 7    & 21.8 & 21.1 & 34.9 & 4.93 & 2.11    & 0.038 \\
O+O    & 7    & 17.6 & 17.0 & 28.1 & 3.98 & 436.2   & 7.84  \\
$pp$   & 6.3  & 20.5 & 19.9 & 32.8 & 4.63 & 1.99    & 0.036 \\
Ar+Ar  & 6.3  & 15.0 & 14.5 & 23.9 & 3.38 & 2321    & 41.7  \\
$pp$   & 6.14 & 20.2 & 19.6 & 32.3 & 4.56 & 1.96    & 0.035 \\
Kr+Kr  & 6.14 & 13.7 & 13.2 & 21.8 & 3.08 & 9327    & 167.6 \\
$pp$   & 5.84 & 19.6 & 19.0 & 31.3 & 4.42 & 1.90    & 0.034 \\
Sn+Sn  & 5.84 & 12.8 & 12.4 & 20.4 & 2.89 & 17545   & 315.2 \\
$pp$   & 5.5  & 18.9 & 18.3 & 30.2 & 4.26 & 1.83    & 0.033 \\
Pb+Pb  & 5.5  & 11.7 & 11.3 & 18.7 & 2.64 & 48930   & 879   \\ \hline
\end{tabular}
\end{center}
\label{psisigs}
\end{table}

We now show the effects of nuclear shadowing on the total cross sections for
one particular set of parameters. We choose parameter set $\psi1$ for 
charmonium and set
$\Upsilon1$ for bottomonium.  The results are given in Tables~\ref{psisigs}
and \ref{upssigs}.  In the middle of each table, the direct cross sections
per nucleon pair are given for all states.  For every applicable energy, the
$pp$ and $AA$, or for 8.8 TeV, the $pp$ and 
$pA$ cross sections are compared to 
directly show the shadowing effects.  We can see that the effects are largest
for charmonium and for the heaviest nuclei even though these are at the lowest
energies and thus the highest $x$.  Shadowing effects on the gluon 
distributions do change significantly at this low $x$ 
\cite{Eskola:1998iy,Eskola:1998df}.
In all cases, the effect is less than a factor of two.

On the right-hand side of the tables, the inclusive cross sections are 
multiplied by the lepton pair branching ratios.  They are also multiplied by
$A^2$ to reproduce the minimum bias cross sections.  The reduction due to
shadowing is then the only nuclear dependence included.  We have not added
in nuclear absorption effects in cold matter, discussed in section
\ref{section:abs}. 

\begin{table}[htbp]
\caption[]{The direct cross section per nucleon pair and the dilepton cross
section per nucleon multiplied by $A^2$ for the minimum bias lepton pair
cross section.  The results are given for $\Upsilon1$.
We compare $pp$ to $pA$ and $AA$ interactions.}
\begin{center}
\begin{tabular}{|cc||ccccc||ccc|} \hline
 & & \multicolumn{5}{c||}{$\sigma^{\rm dir}$/nucleon pair ($\mu$b)} &
\multicolumn{3}{c|}{$B \sigma^{\rm inc} A^2$ ($\mu$b)} \\ \hline
System & $\sqrt{s}$ (TeV) & $\Upsilon$ & $\Upsilon'$ & 
$\Upsilon''$ & $\chi_b(1P)$ & $\chi_b(2P)$ & 
$\Upsilon$ & $\Upsilon'$ & $\Upsilon''$ 
\\ \hline
$pp$   & 14   & 0.43 & 0.27 & 0.16  & 0.89 & 0.69 & 0.020 & 0.0050 & 0.0030 \\
$pp$   & 8.8  & 0.29 & 0.18 & 0.11  & 0.60 & 0.47 & 0.014 & 0.0040 & 0.0020 \\
$p$Pb  & 8.8  & 0.25 & 0.16 & 0.097 & 0.52 & 0.41 & 2.51  & 0.65   & 0.37   \\
$pp$   & 7    & 0.23 & 0.15 & 0.090 & 0.48 & 0.38 & 0.011 & 0.0029 & 0.0016 \\
O+O    & 7    & 0.21 & 0.13 & 0.081 & 0.44 & 0.34 & 2.57  & 0.66   & 0.38   \\
$pp$   & 6.3  & 0.21 & 0.14 & 0.082 & 0.44 & 0.34 & 0.010 & 0.0026 & 0.0015 \\
Ar+Ar  & 6.3  & 0.18 & 0.12 & 0.070 & 0.38 & 0.29 & 13.8  & 3.59   & 2.02   \\
$pp$   & 6.14 & 0.21 & 0.13 & 0.080 & 0.43 & 0.33 & 0.0099& 0.0026 & 0.0014 \\
Kr+Kr  & 6.14 & 0.17 & 0.11 & 0.066 & 0.35 & 0.28 & 57.4  & 14.8   & 8.38   \\
$pp$   & 5.84 & 0.20 & 0.12 & 0.076 & 0.41 & 0.32 & 0.0094& 0.0024 & 0.0014 \\
Sn+Sn  & 5.84 & 0.16 & 0.10 & 0.062 & 0.33 & 0.26 & 108.1 & 28.0   & 15.8   \\
$pp$   & 5.5  & 0.19 & 0.12 & 0.070 & 0.39 & 0.30 & 0.0090& 0.0020 & 0.0013 \\
Pb+Pb  & 5.5  & 0.15 & 0.094& 0.057 & 0.31 & 0.24 & 304   & 78.8   & 44.4   \\
\hline
\end{tabular}
\end{center}
\label{upssigs}
\end{table}

\begin{figure}[htb] 
\setlength{\epsfxsize=0.75\textwidth}
\setlength{\epsfysize=0.4\textheight}
\centerline{\epsffile{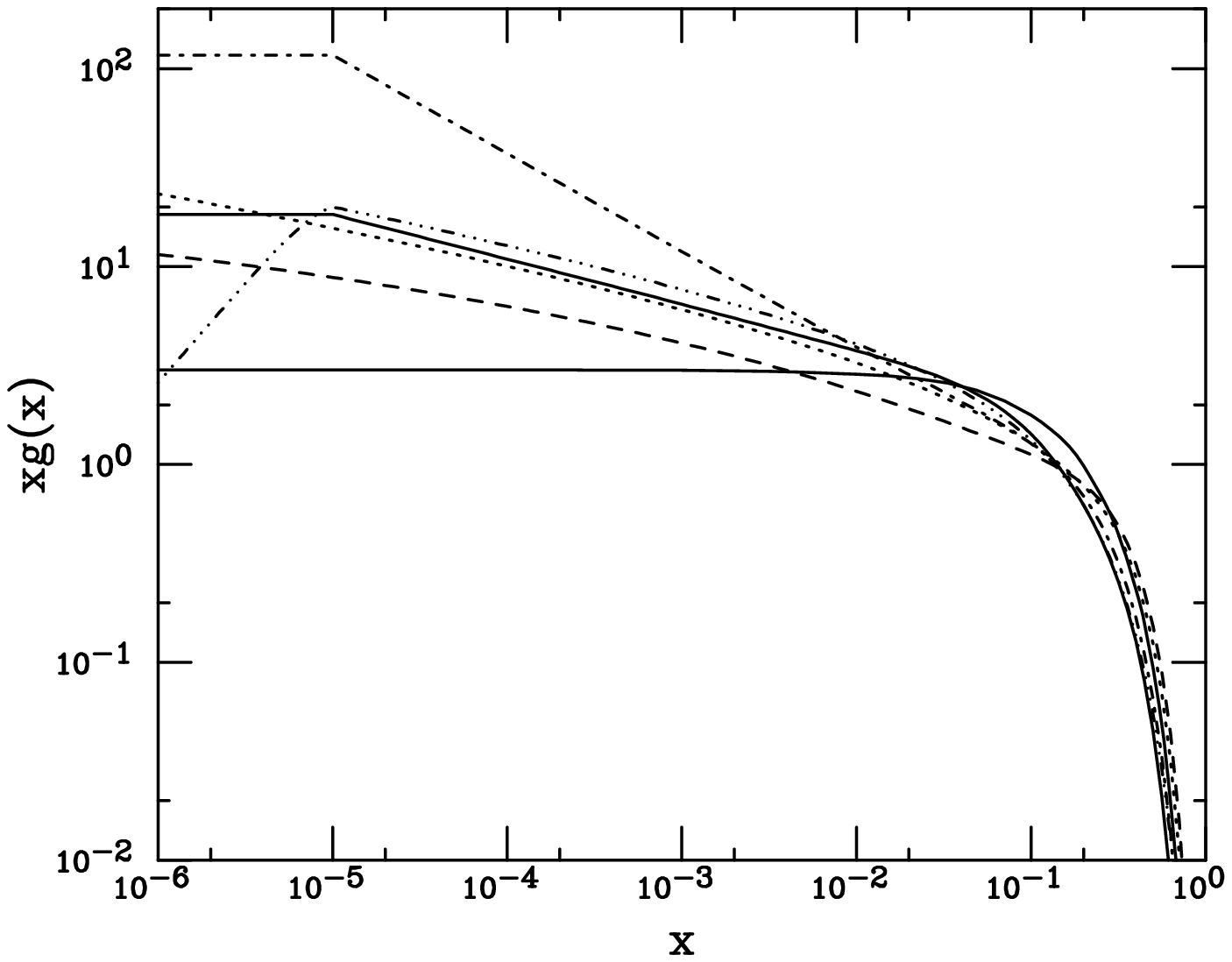}}
\caption[]{Gluon distribution functions in the proton at the scale of the
charmonium calculations. The lower solid curve is the scale independent
$(1-x)^5$, the other
solid curve employs the MRST HO distributions with $\mu = 2.4$ GeV,
the dashed, GRV 98 HO with $\mu = 1.3$ GeV, the dot-dashed, MRSD-' with
$\mu= 2.4$ GeV, the dotted, GRV HO with $\mu = 1.3$ GeV and the
dot-dot-dot-dashed, CTEQ 5M with $\mu = 2.4$ GeV.}
\label{pdfglue}
\end{figure}

The $J/\psi$ cross sections reported here are a factor of two or more lower
than those calculated in Ref.~\cite{Gavai:1994in}.  This should not be a 
surprise because the PDFs used in those calculations were available before
the first low-$x$ HERA data and generally overestimated the increase at low 
$x$.  At $x \sim 10^{-3}$, the MRS D-$^\prime$ 
gluon density~\cite{Martin:1992zi} is 
nearly a factor of five greater than the MRST gluon density 
\cite{Martin:1998sq}
based on more recent HERA data that better constrain the low-$x$ gluon 
density.  The differences between the GRV HO \cite{Gluck:1991ng} and GRV 98 HO
\cite{Gluck:1998xa} are smaller.  The gluon densities used in this study are
compared to those used in Ref.~\cite{Gavai:1994in} in Fig.~\ref{pdfglue}.

\begin{figure}[htb]
\setlength{\epsfxsize=0.95\textwidth}
\setlength{\epsfysize=0.5\textheight}
\centerline{\epsffile{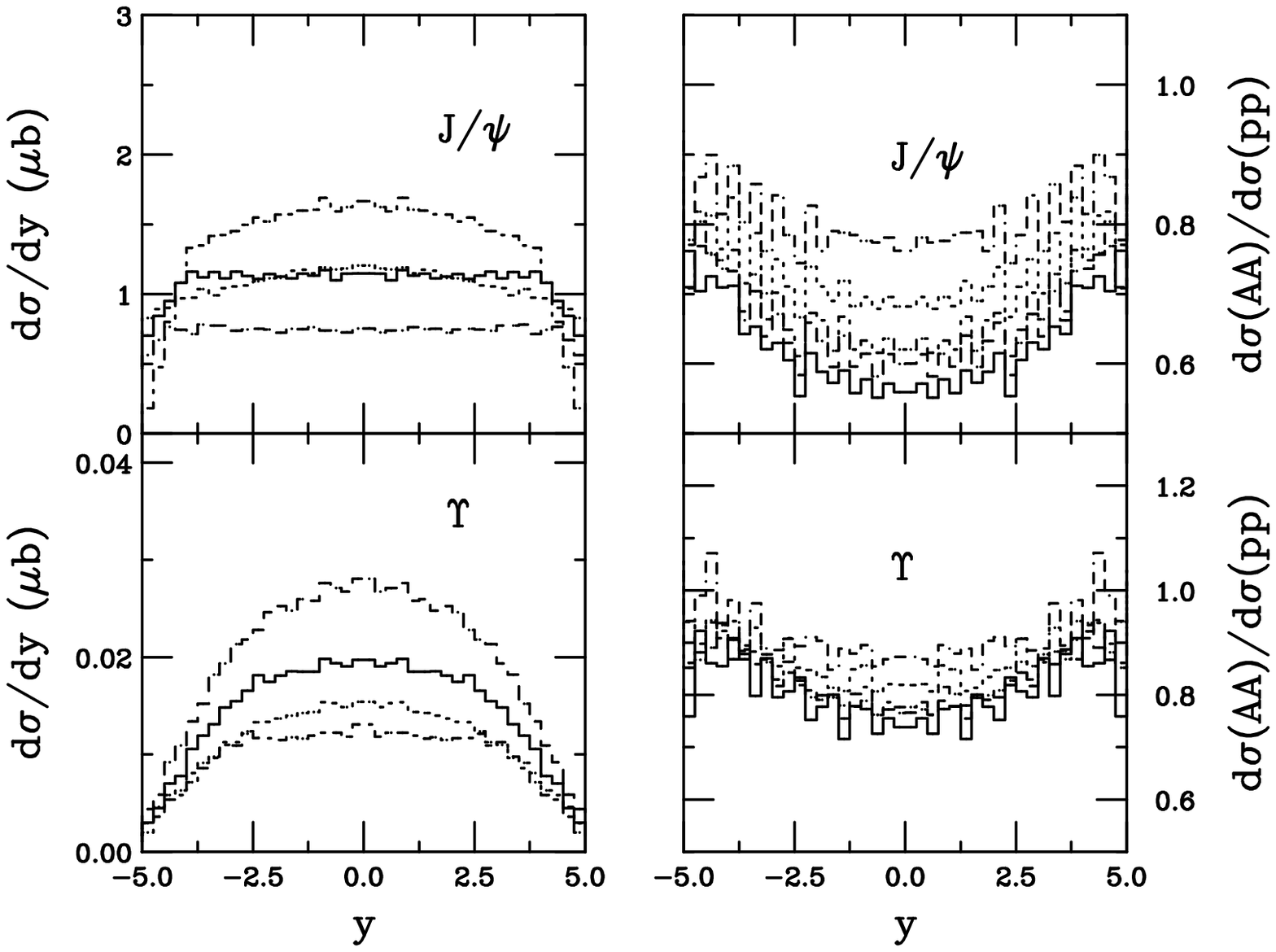}}
\caption{The direct $J/\psi$ and $\Upsilon$ 
rapidity distributions in Pb+Pb collisions
(left) and the $AA$/$pp$ ratios (right).  All the plots are normalized per
nucleon. On the left-hand
side, the $J/\psi$ calculations are 
$\psi1$ (solid), $\psi2$ (dashed),
$\psi3$ (dot-dashed) and $\psi4$ (dotted) while 
the $\Upsilon$ calculations are 
$\Upsilon1$ (solid), $\Upsilon2$ (dashed), $\Upsilon3$ (dot-dashed) and 
$\Upsilon4$ (dotted).  
The ratios on the right-hand side use $\psi1$ for the $J/\psi$ and $\Upsilon1$
for the $\Upsilon$.  The ratios are given
for the maximum $AA$ energy: $A =$ Pb (solid), Sn (dashed), Ar
(dot-dashed) and O (dot-dash-dash-dashed).
}
\label{psiupsydep}
\end{figure}

The direct $J/\psi$ and $\Upsilon$ rapidity distributions in Pb+Pb interactions
at 5.5 TeV/nucleon for all the parameter choices 
are compared in the left-hand side of 
Fig.~\ref{psiupsydep}.  The rapidity distributions reflect the differences
in the total cross sections quite well.  The `corners' in the $J/\psi$ rapidity
distributions at $|y|\sim 4$ occur at $x \sim 10^{-5}$, the lowest
$x$ for which the MRST and CTEQ5 densities are valid.  The behavior of the
parton gluon densities for $x \leq 10^{-5}$ varies significantly, as shown in
Fig.~\ref{pdfglue}.  The 
minimum $x$ of the GRV 98 densities is $10^{-9}$ so that no problems are
encountered for this set.  The MRST densities below $x = 10^{-5}$ are fixed
to the density at this minimum value and are thus constant for lower values of
$x$.  On the other hand, the CTEQ5M distributions turn over and decrease for
$x < 10^{-5}$, causing the steep drop in the rapidity distributions at high
$|y|$. Only the GRV98 HO distributions are smooth over all $y$.

On the right-hand side of Fig.~\ref{psiupsydep}, the $AA/pp$ ratios are 
compared for all the $AA$ combinations given in Tables~\ref{psisigs} and
\ref{upssigs}.  Since the rapidity distributions are not smooth due to the
Monte Carlo integration of Ref.~\cite{Mangano:jk}, the ratios enhance the 
fluctuations.  The `corners' do not appear in the ratios because the change
in slope occurs at the same point for $pp$ and $AA$ at the same energy.  The 
biggest effect of shadowing is at midrapidity when both $x$ values are small.
As the rapidity increases, the ratios also increase since, {\it e.g.}\ at large
$y$, $x_1$ is large and in the antishadowing region, reducing the shadowing
effect in the product.  The effect on the $J/\psi$ is the largest, from a
$\sim 45$\% effect on Pb+Pb at $y=0$ to a 20\% effect at $y=0$ for O+O.  The
overall effect on the $\Upsilon$ is lower since the $x$ values probed, as well
as the scale, are larger.  The evolution of the shadowing parameterization
decreases the effect.  Thus the $\Upsilon$ 
result is $\sim 25$\% for Pb+Pb and only $\sim 10$\% for O+O.

\begin{figure}[htb]
\setlength{\epsfxsize=0.95\textwidth}
\setlength{\epsfysize=0.5\textheight}
\centerline{\epsffile{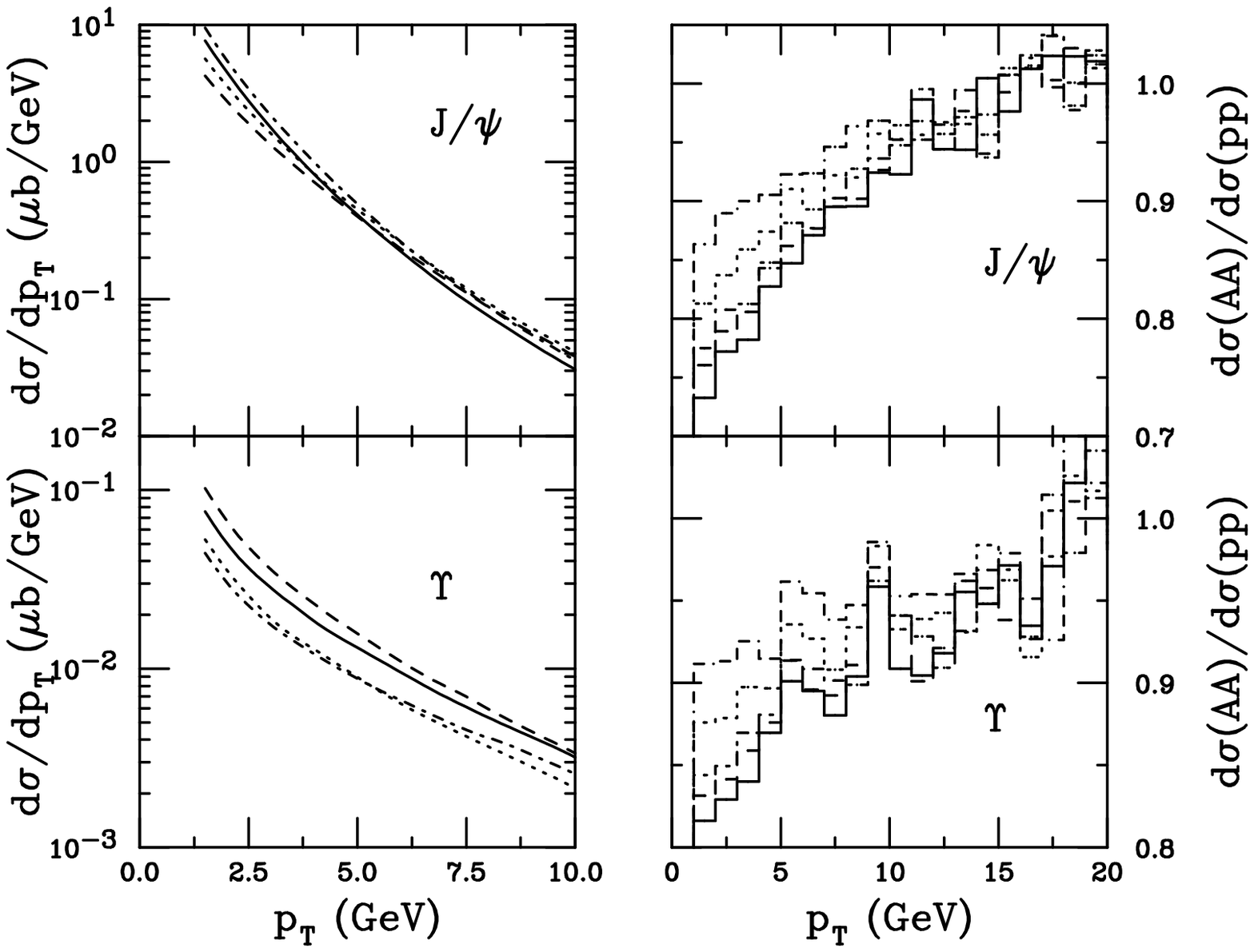}}
\caption{The direct $J/\psi$ and $\Upsilon$ $p_T$ distributions in Pb+Pb 
collisions (left) and the $AA$/$pp$ ratios (right).  All the plots are
normalized per nucleon.  On the left-hand
side, the $J/\psi$ calculations are 
$\psi1$ (solid), $\psi2$ (dashed),
$\psi3$ (dot-dashed) and $\psi4$ (dotted) while the 
$\Upsilon$ calculations are 
$\Upsilon1$ (solid), $\Upsilon2$ (dashed), $\Upsilon3$ (dot-dashed) and 
$\Upsilon4$ (dotted).  
The ratios on the right-hand side use $\psi1$ for the $J/\psi$ and $\Upsilon1$
for the $\Upsilon$.  The ratios are given
for the maximum $AA$ energy: $A =$ Pb (solid), Sn (dashed), Ar
(dot-dashed) and O (dot-dash-dash-dashed).
}
\label{psiupsptdep}
\end{figure}

The rapidity-integrated $p_T$ distributions of direct quarkonium production are
compared for all fit parameters in Pb+Pb collisions at 5.5 TeV in
Fig.~\ref{psiupsptdep}.  The cross
sections are not shown all the way down to $p_T = 0$ because we have not
included any intrinsic $k_T$ broadening.  Broadening effects on quarkonium
are discussed in the $pA$ chapter of this report.  The distributions are
all fairly similar but changing the mass and scale has an effect on the slope,
as is particularly obvious for the $\Upsilon$.  The highest $m_b$, 
$\Upsilon3$, has the hardest slope while that of the lowest $m_b$, $\Upsilon2$,
decreases the fastest with $p_T$.  

The $AA/pp$ ratios are also shown in Fig.~\ref{psiupsptdep}.  The fluctuations
are again large but the general trend is clear.  The ratios at low $p_T$
are similar to those at midrapidity and increase to unity around $p_T \sim 15
- 20$ GeV for both the $J/\psi$ and $\Upsilon$.

\subsection[Quarkonium production in Non-Relativistic QCD]
{Quarkonium production in Non-Relativistic QCD~\protect
\footnote{Authors: G.T. Bodwin, Jungil Lee and R. Vogt.}}\label{quarkon.nrqcd1}

\subsubsection{The NRQCD Factorization Method}
\label{NRQCD-factorization}%

In both heavy-quarkonium decays and hard-scattering production, large
energy-momentum scales appear. The heavy-quark mass $m_Q$ is much larger than
$\Lambda_{\rm QCD}$, and, in the case of production, the transverse
momentum $p_T$ can be much larger than $\Lambda_{\rm QCD}$ as well. 
Thus, the associated values of $\alpha_s$
are much less than one: $\alpha_s(m_c)\approx 0.25$ and
$\alpha_s(m_b)\approx 0.18$. Therefore, one might hope that it would be
possible to calculate the rates for heavy quarkonium production and decay 
accurately in perturbation theory. However, there are clearly
low-momentum, nonperturbative effects associated with the dynamics of
the quarkonium bound state that invalidate the direct application of 
perturbation theory. 

In order to make use of perturbative methods, one must first separate
the short-distance/high-momentum, perturbative effects from the
long-distance/low-momentum, nonperturbative effects---a process which is
known as ``factorization.'' One convenient way to carry out this
separation is through the use of the effective field theory
Nonrelativistic QCD (NRQCD)
\cite{Caswell:1985ui,Thacker:1990bm,Bodwin:1994jh}. NRQCD reproduces
full QCD accurately at momentum scales of order $m_Qv$ and smaller,
where $v$ is heavy-quark velocity in the bound state in the
center-of-mass (CM) frame, with $v^2\approx 0.3$ for charmonium and
$v^2\approx 0.1$ for bottomonium. Virtual processes involving momentum
scales of order $m_Q$ and larger can affect the lower-momentum
processes.  Their effects are taken into account through the
short-distance coefficients of the operators that appear in the NRQCD
action.

Because $Q\overline Q$ production occurs at momentum scales of order $m_Q$ or
larger, it manifests itself in NRQCD through contact interactions. As a
result, the quarkonium production cross section can be written as a sum
of the products of NRQCD matrix elements and short-distance coefficients:
\begin{equation}
\sigma(H)=\sum_n {F_n(\Lambda)\over m_Q^{d_n-4}}\langle 0|
{\cal O}_n^H(\Lambda)|0\rangle \, \, .
\label{prod-fact}
\end{equation}
Here, $H$ is the quarkonium state, $\Lambda$ is the ultraviolet cutoff of
the effective theory, the $F_n$ are short-distance coefficients, and the 
${\cal O}_n^H$ are four-fermion operators, whose mass dimensions are 
$d_n$. A formula similar to Eq.~(\ref{prod-fact}) exists for the inclusive
quarkonium annihilation rate \cite{Bodwin:1994jh}.

The short-distance coefficients $F_n(\Lambda)$ are essentially the
process-dependent partonic cross sections to make a $Q\overline Q$ pair. 
The $Q\overline Q$ pair can be
produced in a color-singlet state or in a color-octet state. The
short-distance coefficients are determined by matching the square of
the production amplitude in NRQCD to full QCD. Because the 
$Q\overline Q$ production scale is of order $m_Q$ or greater, 
this matching can be carried out in perturbation theory.

The four-fermion operators in Eq.~(\ref{prod-fact}) create a $Q\overline Q $
pair, project it onto an intermediate state that consists of a heavy
quarkonium plus anything, and then annihilate the $Q\overline Q$ pair. The
vacuum matrix element of such an operator is the probability for a
$Q\overline Q$ pair to form a quarkonium plus anything. These matrix elements
are somewhat analogous to parton fragmentation functions. They contain
all of the nonperturbative physics that is associated with evolution of the
$Q\overline Q$ pair into a quarkonium state.

Both color-singlet and color-octet four-fermion operators appear in
Eq.~(\ref{prod-fact}). They correspond, respectively, to the evolution of
a $Q\overline Q$ pair in a relative color-singlet state or a relative
color-octet state into a color-singlet quarkonium. If we drop all of the
color-octet contributions in Eq.~(\ref{prod-fact}), then we have the
color-singlet model \cite{Schuler:1994hy}. In contrast, NRQCD is not a
model, but a rigorous consequence of QCD in the limit $\Lambda_{\rm
QCD}/m_Q\rightarrow 0$.

The NRQCD decay matrix elements can be calculated in lattice simulations
\cite{Bodwin:1993wf,Bodwin:2001mk} or determined from phenomenology.
However, at present, the production matrix elements must be obtained
phenomenologically, as it is not yet known how to formulate the
calculation of production matrix elements in lattice simulations. In
general, the production matrix elements are different from the decay
matrix elements. However, in the color-singlet case, the production and
decay matrix elements can be related through the vacuum-saturation
approximation, up to corrections of relative order $v^4$
\cite{Bodwin:1994jh}.

An important property of the matrix elements, which greatly increases
the predictive power of NRQCD, is the fact that they are universal, {\it
i.e.}, process independent. NRQCD $v$-power-counting rules 
organize the sum over operators in Eq.~(\ref{prod-fact}) as an
expansion in powers of $v$. Through a given order in $v$, only a limited
number of operator matrix elements contribute. Furthermore, at leading
order in $v$, there are simplifying relations between operator matrix
elements, such as the heavy-quark spin symmetry \cite{Bodwin:1994jh} and the
vacuum-saturation approximation \cite{Bodwin:1994jh}, that reduce the number of
independent phenomenological parameters.  In contrast, the CEM ignores
the hierarchy of matrix elements in the $v$ expansion.

The proof of the factorization formula (\ref{prod-fact}) relies both on
NRQCD and on the all-orders perturbative machinery for proving
hard-scattering factorization.  A detailed proof does not yet exist, but
work is in progress \cite{qiu-sterman}. Corrections to the
hard-scattering part of the factorization are thought to be of order
$(m_Q v/p_T)^2$, not $(m_Q/p_T)^2$, in the unpolarized case and of order
$m_Q v/p_T$, not $m_Q/p_T$, in the polarized case. It is not known if there
is a factorization formula at low $p_T$ or for the $p_T$-integrated
cross section. The presence of soft gluons in the quarkonium 
binding process makes the application of the standard factorization 
techniques problematic at low $p_T$.

In the decay case, the color-octet matrix elements can be
interpreted as the probability to find the quarkonium in a Fock state
consisting of a $Q\overline Q$ pair plus some gluons. It is a common
misconception that color-octet production proceeds, like color-octet
decay, through a higher Fock state. However, in color-octet production,
the gluons that neutralize the color are in the final state, not the
initial state. There {\it is} a higher-Fock-state process, but it
requires the production of gluons that are nearly collinear to the
$Q\overline Q$ pair, and it is, therefore, suppressed by additional
powers of $v$.

In practical theoretical calculations of the quarkonium production 
and decay rates, a number of significant uncertainties arise. In many
instances, the series in $\alpha_s$ and in $v$ of Eq.~(\ref{prod-fact}) 
converge slowly, and the uncertainties from their
truncation are large---sometimes of order 100\%. In addition, the matrix
elements are often poorly determined, either from phenomenology or
lattice measurements, and the important linear combinations of matrix
elements vary from process to process, making tests of universality
difficult. There are also large
uncertainties in the heavy-quark masses (approximately 10\% for $m_c$
and 5\% for $m_b$, for the mass ranges used in the calculations) that 
can be very significant for
quarkonium rates proportional to a large power of the mass.

\subsubsection{Experimental Tests of NRQCD Factorization}

Here, we give a brief review of some of the successes of NRQCD, as well
as some of the open questions.  We concentrate on hadroproduction
results for both unpolarized and polarized production.  We also discuss
briefly some recent two-photon, $e^+ e^-$, and photoproduction results.

\begin{figure}
\begin{center}
\epsfig{figure=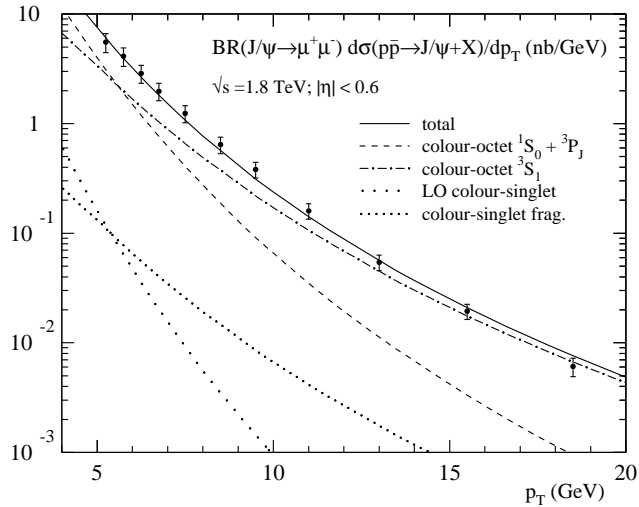,width=10cm}
\caption{$J/\psi$ cross section as a function of $p_T$. The data points 
are from the CDF measurement \cite{Abe:1997jz}. The solid curve is 
the NRQCD-factorization fit to the data given in Ref.~\cite{Kramer:2001hh}. 
The other curves give various contributions to the NRQCD-factorization 
fit. From Ref.~\cite{Kramer:2001hh}.}
\label{fig-tevatron}
\end{center}
\end{figure}

Using the NRQCD-factorization approach, one can obtain a good fit to the
high-$p_T$ CDF data \cite{Abe:1997jz}, while the color-singlet model
under predicts the data by more than an order of magnitude. (See
Fig.~\ref{fig-tevatron}.)  The $p_T$ dependence of the unpolarized
Tevatron charmonium data has been studied under a number of model
assumptions, including LO collinear factorization, parton-shower
radiation, $k_T$ smearing, and $k_T$ factorization. (See
Ref.~\cite{Kramer:2001hh} for a review.)

Several uncertainties in the theoretical predictions affect the
extraction of the NRQCD char\-mo\-ni\-um-production matrix elements from
the data. There are large uncertainties in the theoretical predictions
that arise from the choices of the factorization scale, the
renormalization scale, and the parton distributions. The  extracted
values of the octet matrix elements are very sensitive to the
small-$p_T$ behavior of the cross section and this, in turn, leads to a
sensitivity to the behavior of the small-$x$ gluon distribution.
Furthermore, the effects of multiple soft-gluon emission are important,
and their omission in the fixed-order perturbative calculations leads to
overestimates of the matrix elements. Effects of higher-order
corrections in $\alpha_s$ are a further uncertainty in the theoretical
predictions. Similar theoretical uncertainties arise in the extraction
of the NRQCD production matrix elements for the $\Upsilon$
\cite{Braaten:2000cm} states, but, owing to large statistical
uncertainties, they are less significant for the fits than in the
charmonium case.

\begin{figure}
\begin{tabular}{cc}
\includegraphics[width=8cm]{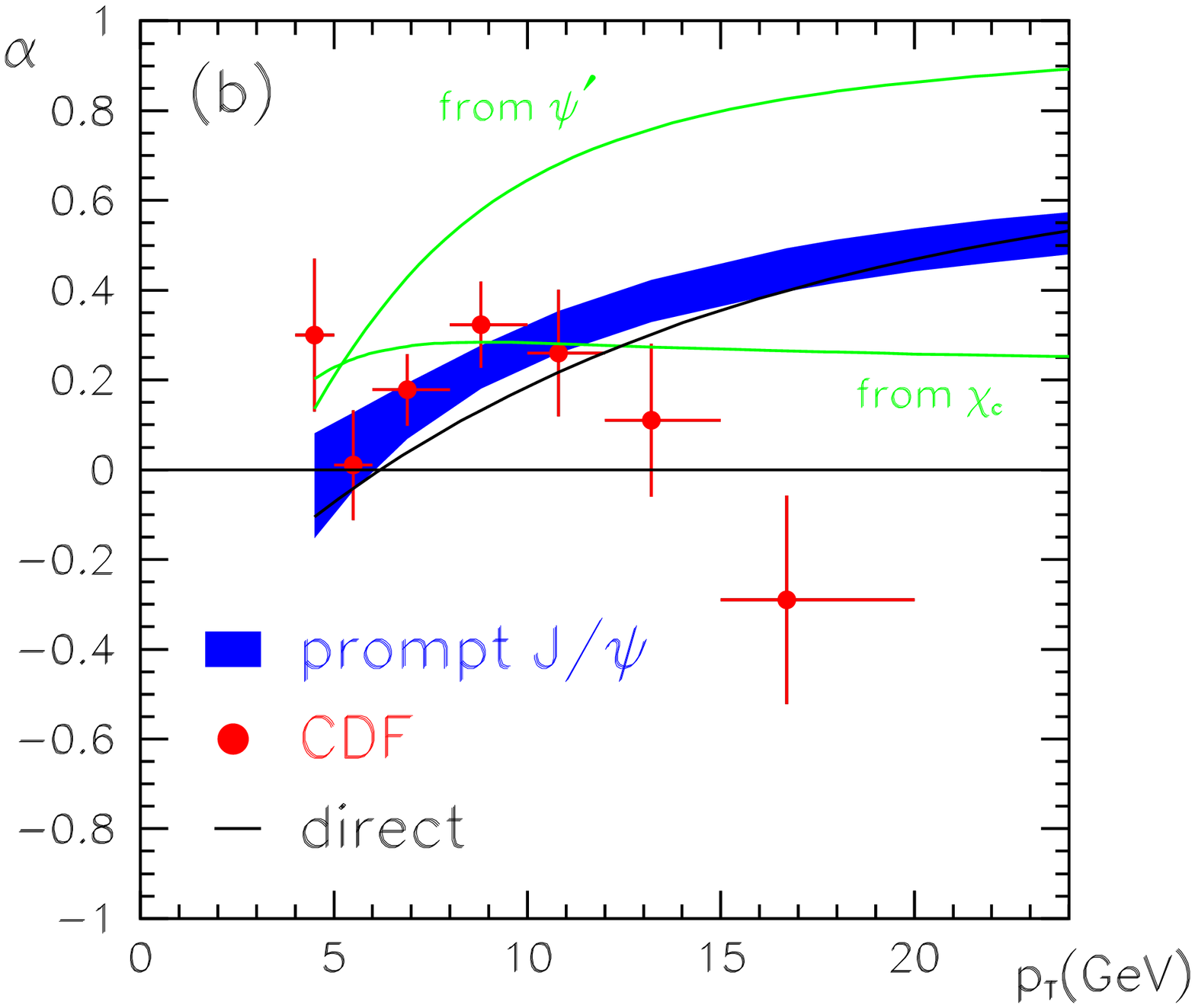}&
\includegraphics[width=8cm]{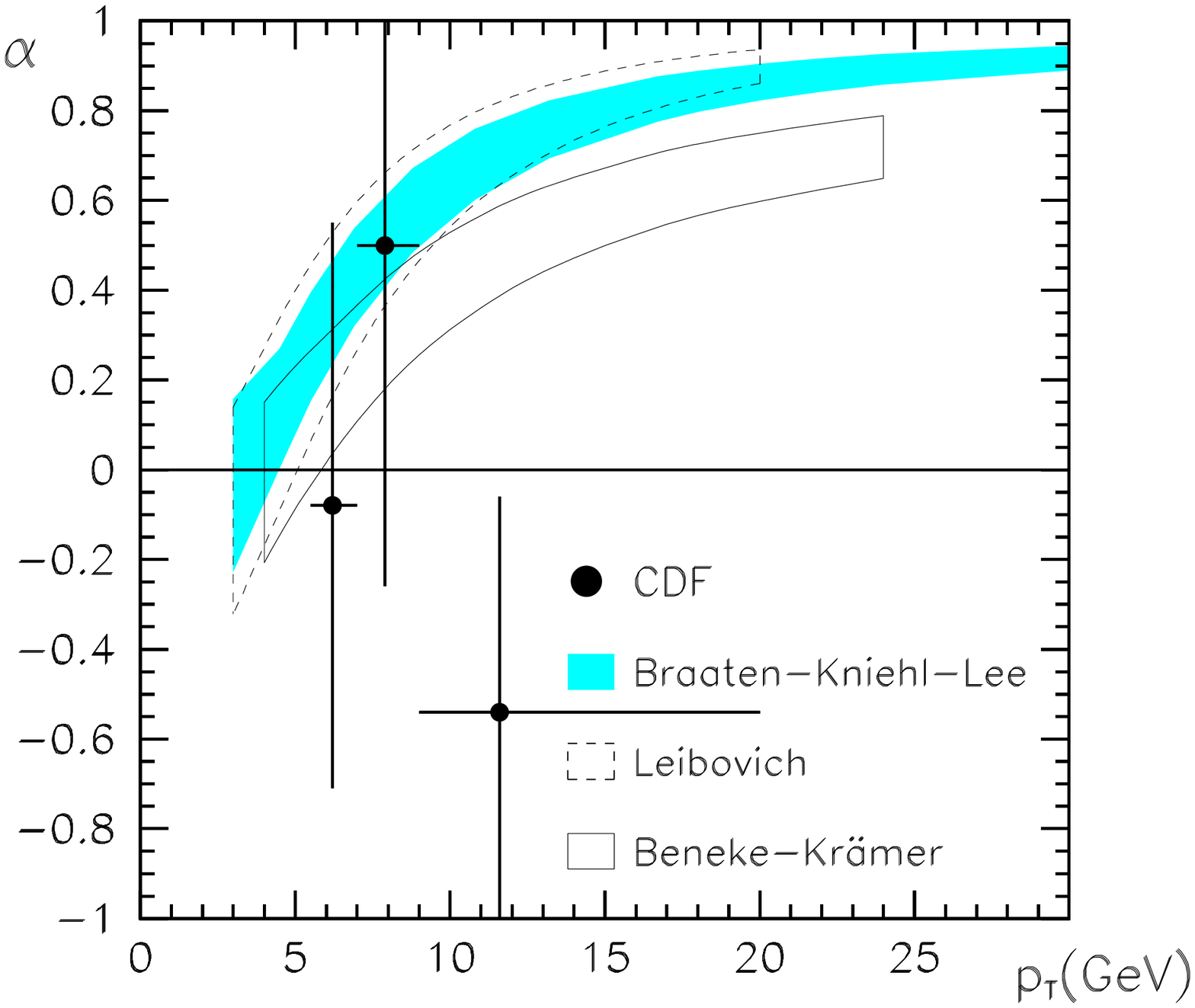}
\end{tabular}
\caption{Left-hand side: $J/\psi$ polarization at the Tevatron. 
The band is the total NRQCD-factorization prediction. The other curves 
give the contributions from feeddown from higher charmonium states. 
Right-hand side: $\psi'$ polarization at the Tevatron. 
The bands give various NRQCD-factorization predictions. The data 
points are from the CDF measurement \cite{Affolder:2000nn}. From 
Ref.~\cite{Braaten:1999qk}.}
\label{fig-pol}
\end{figure}

At large $p_T$ ($p_T\gsim 4m_c$ for the $J/\psi$) the dominant
quarkonium-production mechanism is gluon fragmentation into a
$Q\overline Q$ pair in a ${}^3S_1$ color-octet state.
The fragmenting gluon is nearly on mass shell
and is, therefore, transversely polarized. Furthermore, the velocity-scaling
rules predict that the color-octet $Q\overline Q$ state retains its
transverse polarization as it evolves into $S$-wave quarkonium
\cite{Cho:1994ih}, up to corrections of relative order $v^2$. Radiative
corrections, color-singlet production, and feeddown from higher states
can dilute the quarkonium polarization
\cite{Beneke:1995yb,Leibovich:1996pa,Beneke:1996yw,Braaten:1999qk,Kniehl:2000nn}. Despite this 
dilution, a substantial polarization is expected at large $p_T$.  Its
detection would be a ``smoking gun'' for the presence of color-octet
production. In contrast, the color-evaporation model predicts no
quarkonium polarization. The CDF measurement of the $J/\psi$ and $\psi'$
polarization as a function of $p_T$ \cite{Affolder:2000nn} is shown in
Fig.~\ref{fig-pol}, along with the NRQCD factorization prediction
\cite{Leibovich:1996pa,Beneke:1996yw,Braaten:1999qk}. 
The analysis of $\psi'$ polarization is simpler
than for the $J/\psi$, since feeddown does not play a r\^ole. However, the
statistics are not as good for the $\psi'$.
The degree of polarization is $\alpha=(1-3\xi)/(1+\xi)$, where $\xi$ is 
the fraction of events with longitudinal polarization.
$\alpha=1$ corresponds to 100\% transverse polarization, and
$\alpha=-1$ corresponds to 100\% longitudinal polarization. The observed
polarization is in relatively good agreement with the prediction, except
in the highest $p_T$ bin, although the prediction of increasing
polarization with increasing $p_T$ is not in evidence.

Because the polarization depends on a ratio of matrix elements, some of
the theoretical uncertainties are reduced compared with those in the
production cross section, and, so, the polarization is probably not
strongly affected by multiple soft-gluon emission or $K$ factors.
Contributions of higher order in $\alpha_s$ could conceivably change the
rates for the various spin states by a factor of two. Therefore, it is
important to carry out the NLO calculation, which involves significant
computational difficulties. It is known that order-$v^2$ corrections to
parton fragmentation into quarkonium can be quite large
\cite{Bodwin:2003wh}. If spin-flip corrections to the NRQCD matrix
elements, which are nominally suppressed by powers of $v$, are also
large, perhaps because the velocity-scaling rules need to be modified,
then spin-flip contributions could significantly dilute the $J/\psi$
polarization. Nevertheless, in the context of NRQCD, it is difficult to
see how there could not be substantial charmonium polarization for
$p_T>4m_c$.

Compared to the $J/\psi$-polarization prediction, the
$\Upsilon$-polarization prediction has smaller $v$-ex\-pan\-sion
uncertainties. However, because of the higher $\Upsilon$ mass, it is
necessary to go to higher $p_T$ to insure that fragmentation dominates
and that there is substantial polarization. Unfortunately, the current
Tevatron data run out of statistics in this high-$p_T$ region.
CDF finds that $\alpha=-0.12\pm 0.22$ for $8 <p_T<20\hbox{ GeV}$
\cite{Acosta:2001gv}, which is consistent with both
the NRQCD-factorization prediction \cite{Braaten:2000gw} and the
zero-polarization prediction of the CEM. There are also discrepancies
between the polarizations observed in fixed-target experiments and the
NRQCD predictions.

Calculations of inclusive $J/\psi$ and $\Upsilon$ production in $\gamma
\gamma$ collisions \cite{Klasen:2001cu,Cho:1995vv} have been compared
with LEP data
\cite{Todorova-Nova:2001pt,Chapkine:2002,Alexander:1995vh}. Both the
$J/\psi$ and $\Upsilon$ measurements favor the NRQCD predictions over
those of the color-singlet model.

Belle \cite{Abe:2001za} and BaBar \cite{Aubert:2001pd} have also
measured the $J/\psi$ total cross sections in $e^+e^- \rightarrow J/\psi
X$. The results of the two experiments are incompatible with each other,
but they both seem to favor NRQCD over the color-singlet model. A
surprising new result from Belle \cite{Abe:2002rb} is that most of the
produced $J/\psi$'s are accompanied by an additional $c\overline c$
pair: $\sigma(e^+e^-\rightarrow J/\psi\,c\overline c)
/\sigma(e^+e^-\rightarrow J/\psi\, X)=0.59^{+0.15}_{-0.13}\pm 0.12$.
Perturbative QCD plus the color-singlet model predict that this ratio
should be about $0.1$ \cite{Cho:1996cg}. There seems to be a major
discrepancy between theory and experiment. However, the
order-$\alpha_s^2$ calculation lacks color-octet contributions, including
those that produce $J/\psi\, c\overline c$. Although these contributions
are suppressed by $v^4 \approx 0.1$, it is possible that the
short-distance coefficients are large.  In other results, the angular
distributions favor NRQCD, but the polarization measurements show no
evidence of the transverse polarization that would be expected in
color-octet production. However, the center-of-mass momentum is rather
small, and, hence, one would not expect the polarization to be large.

Quarkonium production has also been measured in inelastic photoproduction
\cite{Merkel:qu,Bertolin:2000xz} and deep-inelastic scattering (DIS)
\cite{Meyer:1998fw,Mohrdieck:2000mk} at HERA. The NRQCD calculation
deviates from the data near large photon-momentum fractions, owing to
the large LO color-octet contribution. The NLO color-singlet result
agrees with the data over all momentum fractions, as well as with the
data as a function of $p_T$. See Ref.~\cite{Kramer:2001hh} for a more
complete review.  In the case of deep-inelastic scattering, the $Q^2$
and $p_T$ dependences are in agreement with NRQCD, but the results are
more ambiguous for the dependence on the longitudinal momentum fraction.

\subsubsection{Quarkonium Production in Nuclear Matter}

The existing factorization ``theorems'' for quarkonium production in
hadronic collisions are for cold hadronic matter. These theorems predict
that nuclear matter is ``transparent'' for $J/\psi$ production at large
$p_T$. That is, at large $p_T$, all of the nuclear effects are contained
in the nuclear parton distributions. The corrections to this
transparency are of order $(m_Q v/p_T)^2$ for unpolarized cross
sections and of order $m_Qv/p_T$ for polarized cross sections.

The effects of transverse-momentum kicks from multiple elastic
collisions between active partons and spectators in the nucleons are
among those effects that are suppressed by $(m_Q v/p_T)^2$. Nevertheless,
these multiple-scattering effects can be important because the
production cross section falls steeply with $p_T$ and because the number
of scatterings grows linearly with the path length through
nuclear matter. Such elastic interactions can be expressed in terms of
eikonal interactions \cite{Bodwin:1988fs} or higher-twist matrix
elements \cite{Qiu:2001hj}.

Inelastic scattering of quarkonium by nuclear matter is also an
effect of higher order in $(m_Q v/p_T)^2$. However, it can become
dominant when the amount of nuclear matter that is traversed by the
quarkonium is sufficiently large. Factorization breaks down when
\begin{equation}
L\gsim {{\rm min}(z_Q,z_{\overline Q})P_H^2\over 
M_A k_T^2({\rm tot})} \, \, ,
\end{equation}
where $L$ is the length of the quarkonium path in the nucleus, $M_A$ is
the mass of the nucleus, $z$ is the parton longitudinal momentum
fraction, $P_H$ is the momentum of the quarkonium in the
parton CM frame, and $k_T({\rm tot})$ is the accumulated transverse
momentum ``kick'' from passage through the nuclear matter. This
condition for the breakdown of factorization is similar to
``target-length condition'' in Drell-Yan production
\cite{Bodwin:1981fv,Bodwin:1984hc}. Such a breakdown 
is observed in the Cronin effect at low $p_T$ and in
Drell-Yan production at low $Q^2$, where the cross section is
proportional to $A^\alpha$, and $\alpha < 1$.

It is possible that multiple-scattering effects may be
larger for color-octet production than for color-singlet production.
In the case of color-octet production, the pre-quarkonium $Q\overline Q$
system carries a nonzero color charge and, therefore, has a larger 
amplitude to exchange soft gluons with spectator partons.

At present, there is no complete, rigorous theory to account for all of
the effects of multiple scattering and we must resort to
``QCD-inspired'' models. A reasonable requirement for models is that
they be constructed so that they are compatible with the factorization
result in the large-$p_T$ limit. Many models treat interactions of the
pre-quarkonium with the nucleus as on-shell (Glauber) scattering. 
This assumption should be examined carefully, as on-shell
scattering is known, from the factorization proofs, not to be a valid
approximation in leading order in $(m_Q v/p_T)^2$.

\subsubsection{NRQCD Predictions for the LHC
}
\label{quarkon.nrqcd2}

In this section, we shall use the formalism of NRQCD to give predictions
for quarkonium production in the LHC energy range. We rewrite
the cross section in Eq.~(\ref{prod-fact}) for the inclusive production of a 
charmonium state $H$ as follows:
\begin{equation} 
\sigma(H) \;=\; \sum_n
\sigma^{(Q\overline Q)_n} \;
        \langle {\cal O}^{H}_n \rangle \, \, ,
\label{sig-fact}
\end{equation}
where $\sigma^{(Q \overline Q)_n} = F_n(\Lambda)/m_Q^{d_n-4}$,
$\langle {\cal O}_n^H \rangle = \langle 0| {\cal O}_n^H|0\rangle$, 
and $n$ runs over
all the color and angular momentum states of the $Q \overline Q$ pair.
The cross sections $\sigma^{(Q \overline Q)_n}$
can be calculated in perturbative QCD.
All dependence on the final state $H$ is contained in the 
nonperturbative NRQCD matrix elements 
$\langle {\cal O}^{H}_n \rangle$.

The most important matrix elements for
$J/\psi=\psi(1S)$ and $\psi'=\psi(2S)$ production
can be reduced to the color-singlet parameter 
$\langle {\cal O}^{\psi(nS)}_1(^3S_1) \rangle$ 
and the three color-octet parameters
$\langle {\cal O}^{\psi(nS)}_8(^3S_1) \rangle$,
$\langle {\cal O}^{\psi(nS)}_8(^1S_0) \rangle$, and
$\langle {\cal O}^{\psi(nS)}_8(^3P_0) \rangle$.
Two of the three color-octet matrix elements only appear in the linear 
combination
\begin{eqnarray}
M_k^{\psi(nS)} = (k/m_c^2)\langle {\cal O}^{\psi(nS)}_8(^3P_0) \rangle 
+ \langle
{\cal O}^{\psi(nS)}_8(^1S_0)\rangle \, \, . \label{mk}
\end{eqnarray}  
The value of $k$ is sensitive to the $p_T$
dependence of the fit.  At the Tevatron, $k \approx 3$.  Fits to fixed-target
total cross sections give larger values, $k \approx \hbox{6--7}$ 
\cite{Beneke:1996tk}.
The most important matrix elements for $\chi_{cJ}$ production
can be reduced to a color-singlet parameter
$\langle {\cal O}^{\chi_{c0}}_1(^3P_0) \rangle$
and a single color-octet parameter
$\langle {\cal O}^{\chi_{c0}}_8(^3S_1) \rangle$.
These matrix elements are sufficient to
calculate the prompt $J/\psi$ cross section to leading order in 
$\alpha_s$ and to order $v^4$ relative to 
the color-singlet contribution.  
\begin{table}
\begin{center}
\caption{Matrix elements for charmonium production. Note that here
$\langle {\cal O}_1^H \rangle = \langle {\cal
O}_1^{\psi(nS)}(^3S_1) \rangle$ for $J/\psi$ and $\psi'$, but $\langle
{\cal O}_1^H \rangle=\langle {\cal O}_1^{\chi_{c0}}(^3P_0) \rangle$ for
$\chi_{c0}$.  Uncertainties are statistical only. From
Ref.~\cite{Braaten:1999qk}.}
\label{tab:me-1}
\renewcommand{\arraystretch}{1.5}
$$
\begin{array}{|c|cccc|}
\hline\hline
 H & \langle {\cal{O}}_1^{H} \rangle  & \langle
 {\cal{O}}_8^{H}(^3S_1) \rangle  & k &
 M_k^{H} \\ \hline
J/\psi & 1.3 \pm 0.1~{\rm GeV^3} & (4.4 \pm 0.7)\times 10^{-3}~{\rm GeV}^3 &  
3.4 & (8.7 \pm 0.9)\times 10^{-2}~{\rm GeV}^3 \\[-1mm] 
\psi' & 0.65 \pm 0.06~{\rm GeV^3} & (4.2 \pm 1.0)\times 10^{-3}~{\rm GeV}^3 & 
3.5 & (1.3 \pm 0.5)\times 10^{-2}~{\rm GeV}^3 
 \\[-1mm]
\chi_{c0} & (8.9 \pm 1.3)\times 10^{-2}~{\rm GeV^5} & 
(2.3 \pm 0.3)\times 10^{-3}~{\rm GeV}^3 & & \\[1mm] \hline \hline
\end{array}
$$
\renewcommand{\arraystretch}{1.0}
\end{center}
\end{table}

In $pp$ collisions, different partonic processes for 
$Q \overline Q$ production dominate in different $p_T$ ranges.
If $p_T$ is of order $m_Q$, fusion processes dominate, and, so, 
the $Q \overline Q$ pair is produced in the hard-scattering process.
These contributions can be written in the form
\begin{equation}
\sigma_{\rm Fu}(H) = \sum_{i,j} \int dx_1 \, dx_2 \, 
f_{i/A}(x_1,\mu^2) f_{j/B}(x_2,\mu^2)
\hat \sigma_{ij}^{(Q \overline Q)_n} \;
\langle {\cal O}^{H}_n \rangle \, \,  ,
\label{sig-fusion}
\end{equation}
where $A$ and $B$ are the incoming hadrons or nuclei.  
In Eq.~(\ref{sig-fusion}),
we include the parton processes $i j \rightarrow Q \overline Q \, X$,
where $ij=gg,q \overline q, qg$ and $\overline q g$, and $q=u,d,s$.
The relevant partonic cross sections $\hat \sigma_{ij}^{(Q \overline Q)_n}$
are given in Refs.~\cite{Cho:1995vh,Cho:1995ce}.

For $p_T\gg m_Q$, the dominant partonic process is gluon 
fragmentation through the color-octet ${}^3S_1$ channel.
This contribution can be expressed as
\begin{equation}
\sigma_{\rm Fr}(H) = \sum_{i,j}\int dx_1 \, dx_2 \, dz
f_{i/A} (x_1,\mu^2) f_{j/B} (x_2,\mu^2)
\hat \sigma_{ij}^{g}
D_g^{(Q \overline Q)_8({}^3S_1)} (z,\mu_{\rm Fr}^2) \; 
\langle {\cal O}^{H}_n \rangle \,
\, ,
\label{sig-frag}
\end{equation}
where $D_g^{(Q \overline Q)_n({}^3S_1)}(z,\mu_{\rm Fr}^2)$ is the
fragmentation function for a gluon fragmenting into a $Q\overline Q$ pair,
$P/z$ is the momentum of the fragmenting gluon, $P$ is the momentum of
the $Q\overline Q$ pair, and $\mu_{\rm Fr}$ is the fragmentation scale. The
fragmentation process scales as $d \hat \sigma/dp_T^2 \sim
1/p_T^4$~\cite{Braaten:1993rw,Braaten:1994vv}. The fragmentation process
is actually included in the fusion processes of Eq.~(\ref{sig-fusion}).
In the limit $p_T\gg m_Q$, the fusion processes that proceed through
$g^*\to (Q\overline Q)_8({}^3S_1)$ are well-approximated by the expression
(\ref{sig-frag}).  At large $p_T$, one can evolve the fragmentation
function in the scale $\mu_{\rm Fr}$, thereby resumming large logarithms
of $p_T^2/m_Q^2$. Such a procedure leads to a smaller short-distance
factor \cite{Beneke:1996yw} and a more accurate prediction at large
$p_T$ than would be obtained by using the fusion cross section
(\ref{sig-fusion}). However, in our calculations, we employ the fusion
cross section (\ref{sig-fusion}), which leads to systematic
over-estimation of the cross section at large $p_T$.

In order to predict the cross section for prompt $J/\psi$ production
(including $\chi_c$ and $\psi'$ feeddown) at the LHC, we need the values
of the NRQCD matrix elements.  There have been several previous
extractions of the color-octet matrix elements
\cite{Beneke:1996yw,Braaten:1999qk,%
Cho:1995vh,Cho:1995ce,Cacciari:1995yt,Kniehl:1998qy,Kniehl:1999vf} from
the CDF $J/\psi$, $\chi_c$ and $\psi'$ $p_T$ distributions
\cite{Abe:1997jz,Abe:1997yz}. We use the matrix elements given in
Ref.~\cite{Braaten:1999qk}, which are shown in
Table~\ref{tab:me-1}. Our calculations are based on the MRST LO
parton distributions \cite{Martin:1998np}. 
\begin{figure}
\begin{center}
\includegraphics[width=8cm]{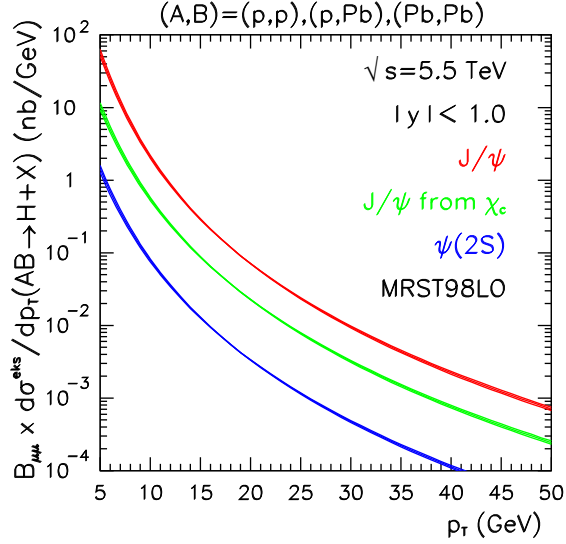}
\end{center}
\caption{\label{lee.fig1} Differential cross sections per nucleon
multiplied by leptonic branching fractions
for prompt $J/\psi$ (upper curves),
$J/\psi$ from $\chi_c$ (middle curves), and
prompt $\psi(2S)$ (lower curves) in $pp$, $p$Pb, and Pb+Pb collisions
at $\sqrt{s}=5.5$~TeV. 
The EKS98 parameterization~\cite{Eskola:1998iy,Eskola:1998df} 
is employed for $p$Pb and Pb+Pb collisions.
}
\end{figure}
In calculating the cross section per nucleon
for prompt $J/\psi$ production in $pA$ or $AA$ collisions, we take
$f_{i/A}=f_{i/p} R^A_i$.  We employ the EKS98 
parameterization~\cite{Eskola:1998iy,Eskola:1998df}
for the nuclear shadowing ratio $R^A_i$.
We evolve $\alpha_s$ at one-loop accuracy, and 
we set $\mu = (4 m_c^2 + p_T^2)^{1/2}$ and $m_c=1.5$ GeV.

There are several sources of uncertainty in our predictions for the cross
sections. There are large uncertainties in the NRQCD matrix elements
themselves. The errors shown in Table \ref{tab:me-1} are 
statistical only. There are additional large uncertainties in the matrix 
elements that arise from truncations of the series in $\alpha_s$ and 
$v$ in the theoretical expressions that
are used to extract the matrix elements. The matrix elements $\langle
{\cal O}_8(^1S_0) \rangle$ and $\langle {\cal O}_8(^3P_0) \rangle$ are
fixed by the data only in the linear combination $M_k^H$. In the present
calculation, we take $\langle {\cal O}_8(^1S_0) \rangle=xM_k^H$ and
$\langle {\cal O}_8(^3P_0) \rangle/m_c^2=(1-x) M_k^H/k$, use the
values of $k$ given in Table \ref{tab:me-1}, and choose $x=1/2$.
Variation of $x$ between $0$ and $1$ affects the cross sections at 
low $p_T$ by amounts on the order of 5\%. There are
additional uncertainties in the predicted cross sections that arise from
the choices of the parton distributions, the charm-quark mass $m_c$, and
the scale $\mu$. Because they affect the matrix-element fits,
these uncertainties are highly correlated with those of the matrix 
elements. We have not tried to estimate their effects on the
predicted cross sections.
        
\begin{figure}
\begin{tabular}{cc}
\includegraphics[width=8cm]{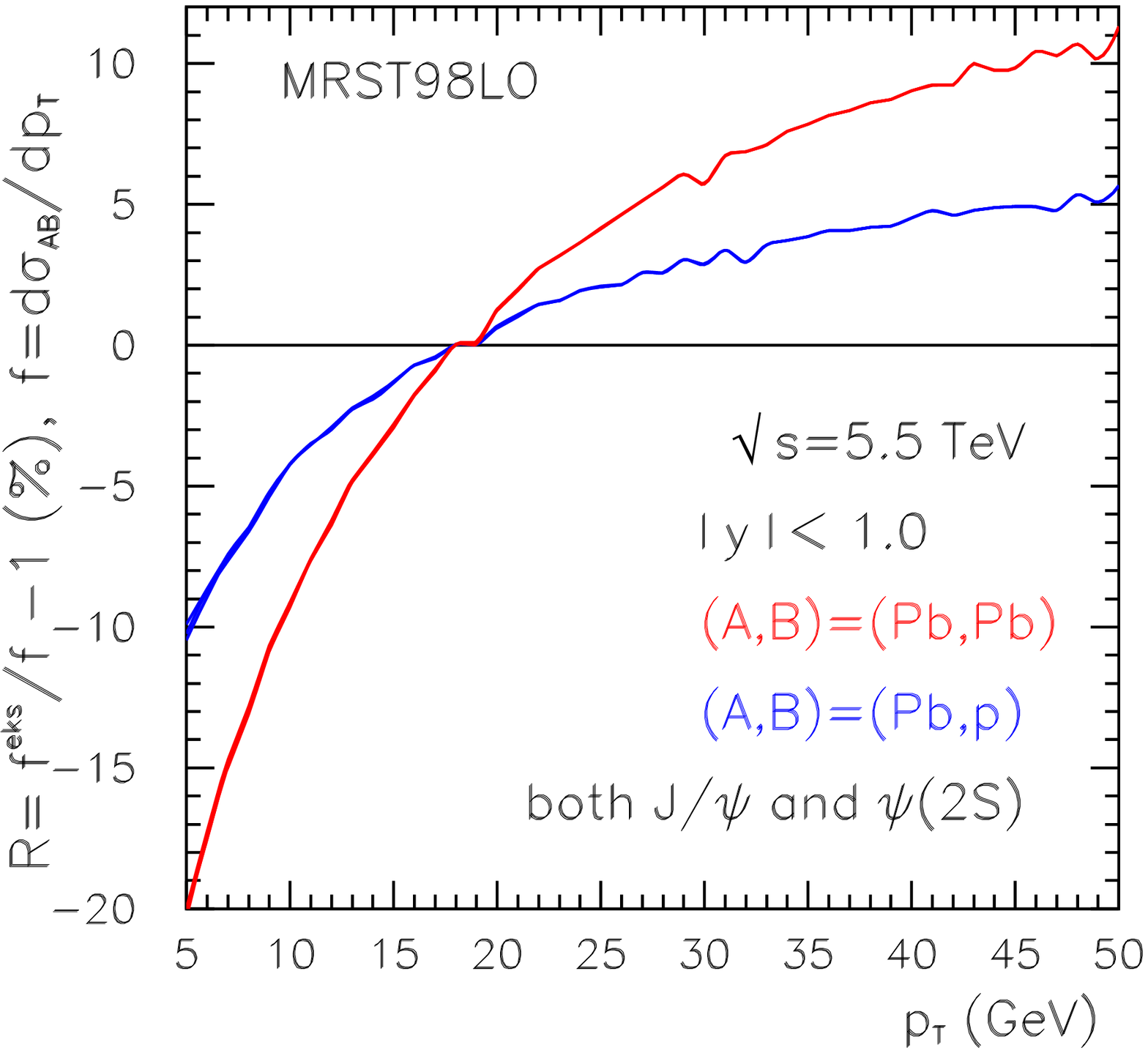}&
\includegraphics[width=8cm]{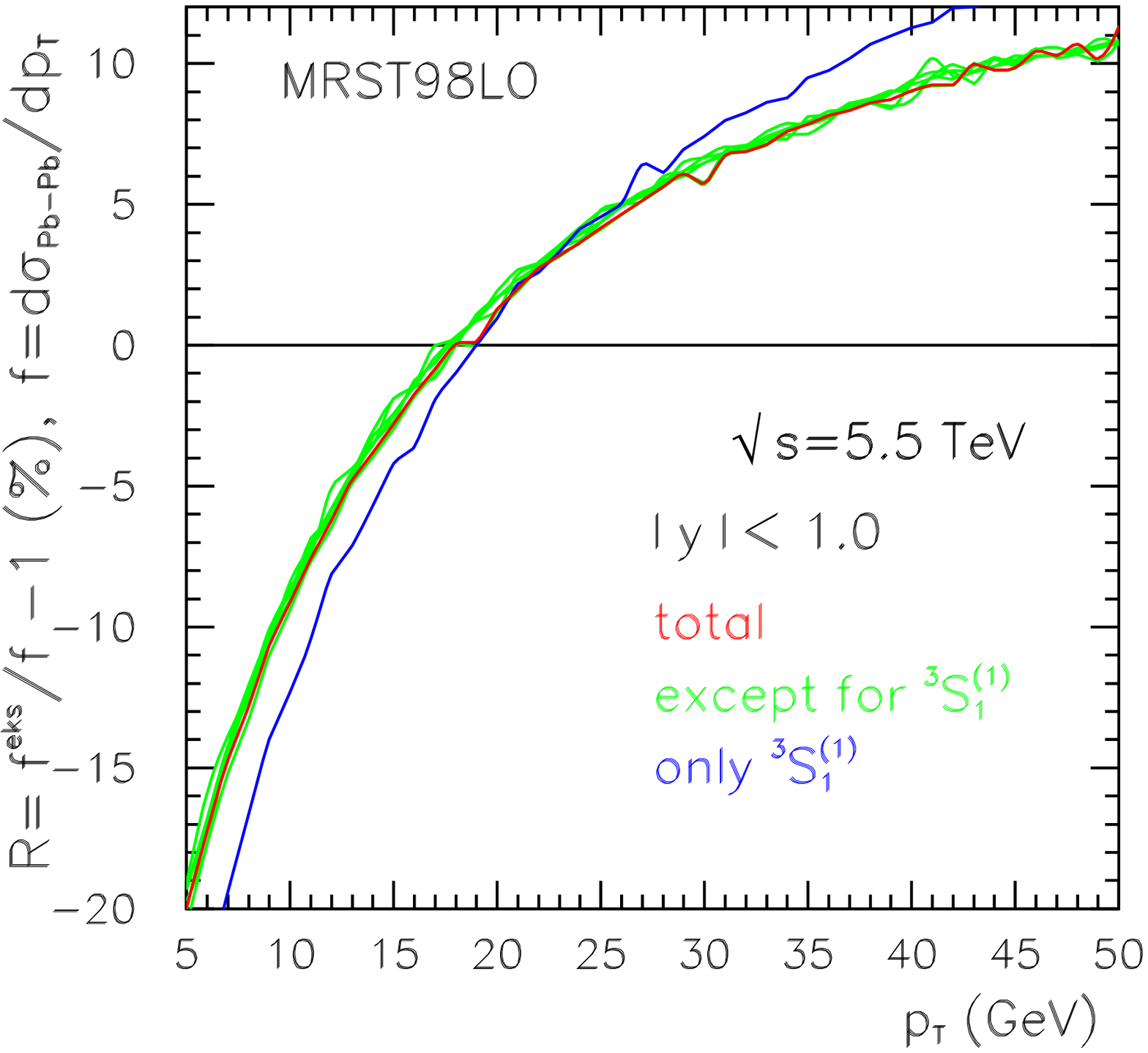}
\end{tabular}
\caption{\label{lee.fig2} 
The $p_T$ dependence of $R_{AB}$ [Eq.~(\ref{R})].
(a) We compare the results for $p$Pb and Pb+Pb collisions.  (The $p_T$ 
dependence is stronger in the Pb+Pb result.)
(b) We show the dependence of $R_{AB}$ on the various production channels
in Pb+Pb collisions at $\sqrt{s}=5.5$~TeV.
}
\end{figure}

In Fig.~\ref{lee.fig1}, we show the $p_T$ distributions per nucleon
multiplied by the dilepton branching fractions for prompt $J/\psi$
(upper curves), $J/\psi$ from $\chi_c$ decays (middle curves), and
prompt $\psi'$ (lower curves) at $\sqrt{s}=5.5$~TeV. For $p$Pb and Pb+Pb
collisions, we use the EKS98
parameterization~\cite{Eskola:1998iy,Eskola:1998df} to account for the
effect of nuclear shadowing. The $pp$, $p$Pb, and Pb+Pb results
essentially lie on top of each other in Fig.~\ref{lee.fig1}, owing to the
many decades covered in the plot.

In order to display small differences between the distributions, we
define the function $R_{AB}$:
\begin{eqnarray}
R_{AB}(p_T)=\frac{d\sigma_{AB}/dp_T - d\sigma_{pp}/dp_T}
                 {d\sigma_{pp}/dp_T} \, \, .
\label{R}
\end{eqnarray}
In Fig.~\ref{lee.fig2}, we present $R_{AB}$ as a function of $p_T$. As
is shown in Fig.~\ref{lee.fig2}(a), nuclear shadowing increases the
cross section at large $p_T$ and decreases it at small $p_T$. The
deviation of the Pb+Pb cross section from the $pp$ cross section is
twice as large as that seen in the case of $p$Pb collisions. In order to
investigate the dependence of the shadowing effect on the short-distance
cross sections that arise in hadroproduction of $S$-wave charmonium
states in Pb+Pb collisions, we plot $R_{AB}$ for all channels separately.
[See Fig.~\ref{lee.fig2}(b).] Even though the $p_T$ dependence the
contribution to the cross section of the color-octet $^3S_1$ channel is
quite different from those of the color-octet $^1S_0$ and $^3P_J$
channels, all three channels show the same nuclear effect. The only
channel that shows a slightly different behavior is the color-singlet
channel, which gives a negligible contribution to the cross section.
While the differential cross sections in Fig.~\ref{lee.fig1} are
strongly dependent on the nonperturbative NRQCD matrix elements,
$R_{AB}$ is almost independent of the matrix elements, making it a good
observable for studying nuclear shadowing at the LHC.

The $\Upsilon$ rates are somewhat more difficult to calculate because of
the many feeddown contributions.  The matrix elements are also not
particularly well known. Since it is unlikely that all the different
contributions can be disentangled, we follow the approach of
Ref.~\cite{Braaten:2000cm} and compute the inclusive $\Upsilon(nS)$
production cross section
\begin{eqnarray}
d\sigma(\Upsilon(nS))_{\rm inc} & = & 
d \sigma^{(b \overline b)_1(^3S_1)} 
        \langle {\cal O}_1^{\Upsilon(nS)}(^3S_1) \rangle_{\rm inc}
+ \sum_J d \sigma^{(b \overline b)_1(^3P_J)} 
        \langle {\cal O}_1^{\Upsilon(nS)}(^3P_J) \rangle_{\rm inc}
\nonumber
\\ 
&& + \mbox{} d \sigma^{(b \overline b)_8(^3S_1)}
        \langle {\cal O}_8^{\Upsilon(nS)}(^3S_1) \rangle_{\rm inc}
+ d \sigma^{(b \overline b)_8(^1S_0)}
        \langle {\cal O}_8^{\Upsilon(nS)}(^1S_0) \rangle_{\rm inc}
\nonumber
\\
&& 
+ \left( \sum_J (2J+1) d \sigma^{(b \overline b)_8(^3P_J)} \right)
        \langle {\cal O}_8^{\Upsilon(nS)}(^3P_0) \rangle_{\rm inc} \, \, ,
\label{sig-total}
\end{eqnarray}
where the last term makes use of heavy-quark spin symmetry to relate all
of the octet ${}^3P_J$ matrix elements to the octet ${}^3P_0$ matrix element.
The ``inclusive'' matrix elements are defined by
\begin{equation}
\langle {\cal O}_i^{\Upsilon(nS)}(n) \rangle_{\rm inc} =
\sum_H B_{H \to \Upsilon(nS)} \langle {\cal O}_i^H(n) \rangle \, \, ,
\label{O-total}
\end{equation}
where $i=1$ or 8 for singlet or octet, respectively.
\begin{table}
\begin{center}
\caption{Inclusive color-singlet matrix elements for bottomonium
production.  The errors on the $^3S_1$ matrix elements come from
estimates of the $\Upsilon(nS)$ decay rate to lepton pairs.  The errors
on the $^3P_J$ states come from an average over potential-model
estimates.  The inclusive matrix elements are a linear combination of
branching ratios, as in Eq.~(\protect\ref{O-total}).  The $S$-state
matrix elements are in units of GeV$^3$ while the $P_J$-state matrix
elements are in units of GeV$^5$.  From Ref.~\cite{Braaten:2000cm}.}
\label{upssing}
\renewcommand{\arraystretch}{1.5}
$$
\begin{array}{|c|cccc|}
\hline\hline
 H & \langle {\cal{O}}_1^{H}(^3S_1) \rangle_{\rm inc} & \langle
 {\cal{O}}_1^H(^3P_0) \rangle_{\rm inc}  & \frac{1}{3} \langle 
{\cal{O}}_1^H(^3P_1) \rangle_{\rm inc} & \frac{1}{5} \langle 
{\cal{O}}_1^H(^3P_2) \rangle_{\rm inc}  \\ \hline
\Upsilon(3S) & 4.3 \pm 0.9 & 0 & 0 & 0 \\[-1mm] 
\Upsilon(2S) & 5.0 \pm 0.7 & 0.12 \pm 0.06 & 0.55 \pm 0.15 & 0.42 \pm 0.10
 \\[-1mm]
\Upsilon(1S) & 12.8 \pm 1.6 & < 0.2 & 1.23 \pm 0.25 & 0.84 \pm 0.15
 \\[-1mm] \hline \hline
\end{array}
$$
\renewcommand{\arraystretch}{1.0}
\end{center}
\end{table}
The sum over $H$ includes the $\Upsilon(nS)$ as well as all higher states
that can decay to $\Upsilon(nS)$.  The branching ratio for $H \to H'$
decays is $B_{H \to H'}$ with $B_{H \to H} \equiv 1$.  Only $\chi_b(1P)$
and $\chi_b(2P)$ decays are included; the possibility of feeddown from
the as-yet unobserved $\chi_b(3P)$ states is neglected.  In the linear
combination $M_k^{\Upsilon(nS)}$, the color-octet matrix element from
the $^3P_0$ state is neglected, and, so, $M_k^{\Upsilon(nS)} = \langle
{\cal O}_8^{\Upsilon(nS)}(^1S_0) \rangle_{\rm inc}$.  We use $m_b =
4.77$ GeV and the MRST LO parton distributions. The values of the
inclusive color-singlet matrix elements are given in
Table~\ref{upssing}, and the values of the inclusive color-octet matrix
elements, from Ref.~\cite{Braaten:2000cm}, are given in
Table~\ref{upsoct}.

\begin{table}
\begin{center}
\caption{Inclusive color-octet matrix elements for bottomonium
production. The matrix elements were fit using the MRSTLO parton 
distributions. The first set of error bars is from $\chi^2$ fits to the
$\Upsilon$ $p_T$ distributions in the region $p_T>8$ GeV.  The second
set is associated with the variation of the scales and corresponds to
multiplying $\mu = \sqrt{m_b^2 + p_T^2}$ by 2 (upper error) and 0.5
(lower error).  The matrix elements are in units of $10^{-2}$ GeV$^3$.
From Ref.~\cite{Braaten:2000cm}.}
\label{upsoct}
\renewcommand{\arraystretch}{1.5}
$$
\begin{array}{|c|ccc|}
\hline\hline
 H & \langle {\cal{O}}_8^{H}(^3S_1) \rangle_{\rm inc} & \langle
 {\cal{O}}_8^H(^1S_0) \rangle_{\rm inc}  & \frac{5}{m_b^2} \langle
{\cal{O}}_1^H(^3P_0) \rangle_{\rm inc}  \\ \hline
\Upsilon(3S) & 3.7 \pm 1.7 ^{+1.7}_{-1.3} & 7.5 \pm 4.9 ^{+3.4}_{-2.5} & 0
 \\[-1mm] 
\Upsilon(2S) & 19.6 \pm 6.3 ^{+8.9}_{-6.5} & -8.7 \pm 11.1 ^{-2.4}_{+1.8} & 0
 \\[-1mm]
\Upsilon(1S) & 11.7 \pm 3.0 ^{+5.7}_{-4.2} & 18.1 \pm 7.2 ^{+11.4}_{-8.1} & 0
 \\[-1mm] \hline \hline
\end{array}
$$
\renewcommand{\arraystretch}{1.0}
\end{center}
\end{table}

\begin{figure}
\begin{center}
\includegraphics[width=8cm]{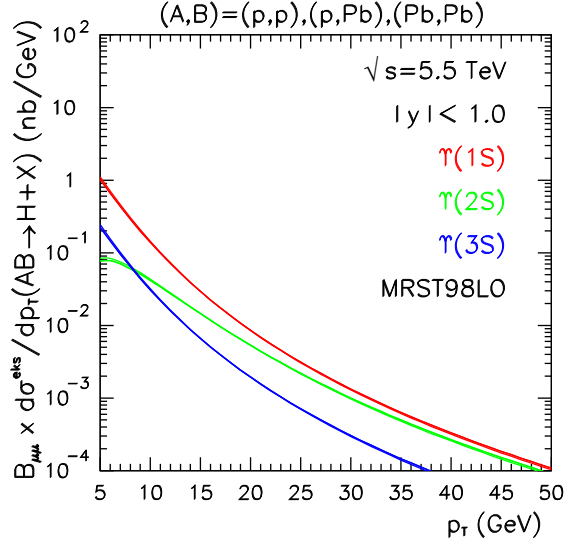}
\end{center}
\caption{\label{lee.fig3} Differential cross sections per nucleon
multiplied by leptonic branching fractions
for inclusive $\Upsilon(1S)$ (upper curves),
$\Upsilon(2S)$ (middle curves), and
prompt $\Upsilon(3S)$ (lower curves) in $pp$, $p$Pb, and Pb+Pb collisions
at $\sqrt{s}=5.5$~TeV. 
The EKS98 parameterization~\cite{Eskola:1998iy,Eskola:1998df} 
is employed for $p$Pb and Pb+Pb collisions.
}
\end{figure}

In Fig.~\ref{lee.fig3}, we show the $p_T$ distributions per nucleon
multiplied by the dilepton branching fractions for the 3 $\Upsilon$ $S$
states at $\sqrt{s}=5.5$~TeV.  The feeddown contributions are included
as in Eq.~(\ref{sig-total}).  For $p$Pb and Pb+Pb collisions, we use the
EKS98 parameterization~\cite{Eskola:1998iy,Eskola:1998df} in order to
account for the effects of nuclear shadowing. The $pp$, $p$Pb, and Pb+Pb
results lie essentially on top of each other in Fig.~\ref{lee.fig3}.

The unusual relative behavior of the $\Upsilon(2S)$ and $\Upsilon(3S)$
states at both low and high $p_T$ is due to the fact that the
bottomonium matrix elements are not very well determined.  For $p_T <
10$ GeV, the $\Upsilon(2S)$ cross section drops below the $\Upsilon(3S)$
cross section because the $\Upsilon(2S)$ has a large negative
color-octet matrix element. (See Table~\ref{upsoct}.) The short-distance
coefficients multiplying $M_k^{\Upsilon(nS)}$ are significant at low
$p_T$. Thus, there is a large cancellation between the octet $^3S_1$
matrix element and $M_k^{\Upsilon(nS)}$, which reduces the
$\Upsilon(2S)$ cross section in this region, causing it to drop below
the $\Upsilon(3S)$ cross section at low $p_T$. At the high-$p_T$ end of
the spectrum, the large value of the $^3S_1$ $\Upsilon(2S)$ color-octet
matrix element (Table~\ref{upsoct}) causes the $\Upsilon(2S)$ cross
section to approach that of the $\Upsilon(1S)$.  In this region, the
color-octet $^3S_1$ contribution dominates the other channels. Its large
matrix element gives the $\Upsilon(2S)$ an unreasonably large cross
section relative to that of the $\Upsilon(1S)$.  The $\Upsilon(2S)$ rate
at $p_T \approx 15$ GeV is more reasonable because the large and positive
$^3S_1$ contribution, and the large and negative $M_k^{\Upsilon(2S)}$
contribution nearly cancel each other.

\begin{figure}
\begin{tabular}{cc}
\includegraphics[width=8cm]{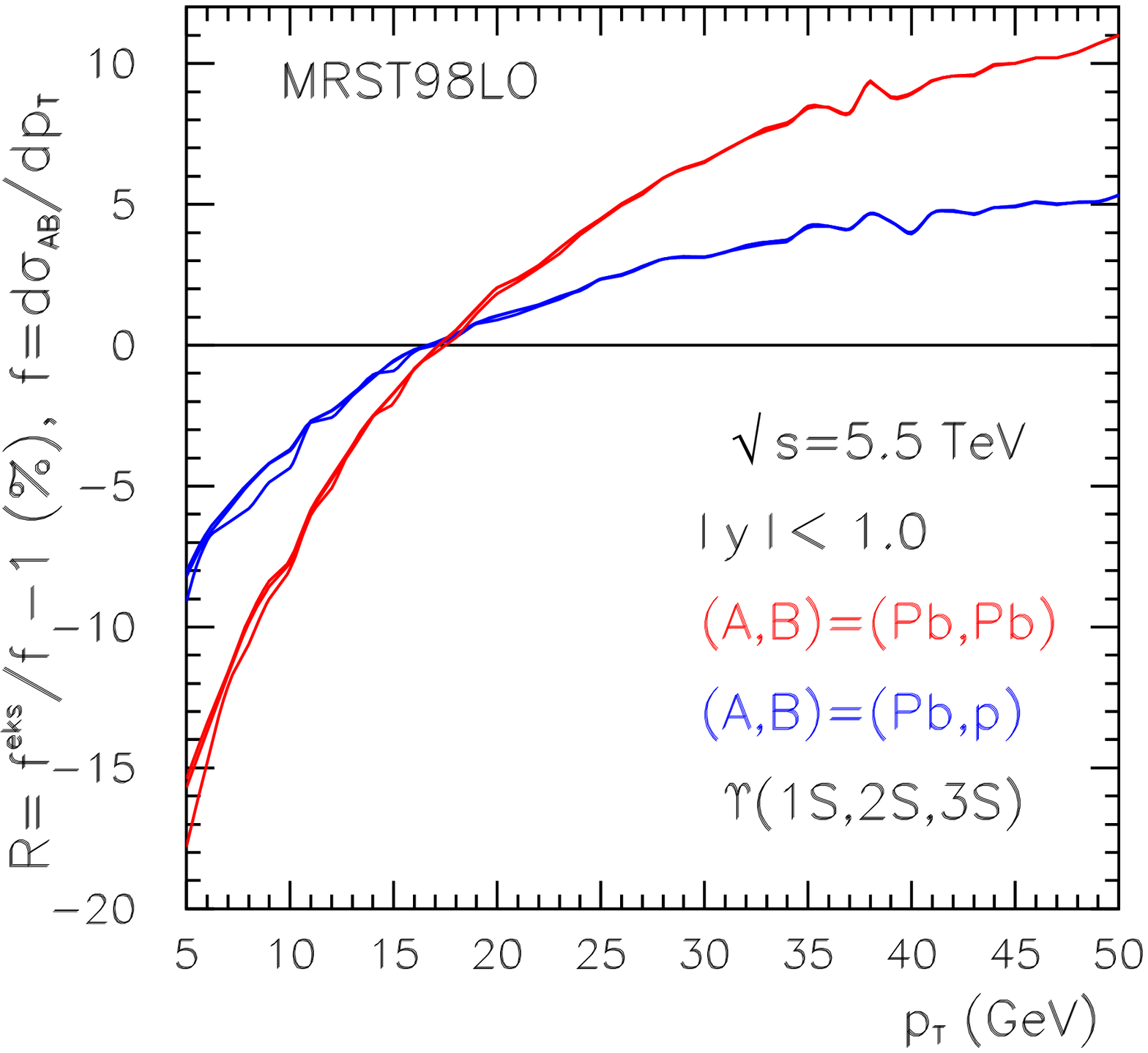}&
\includegraphics[width=8cm]{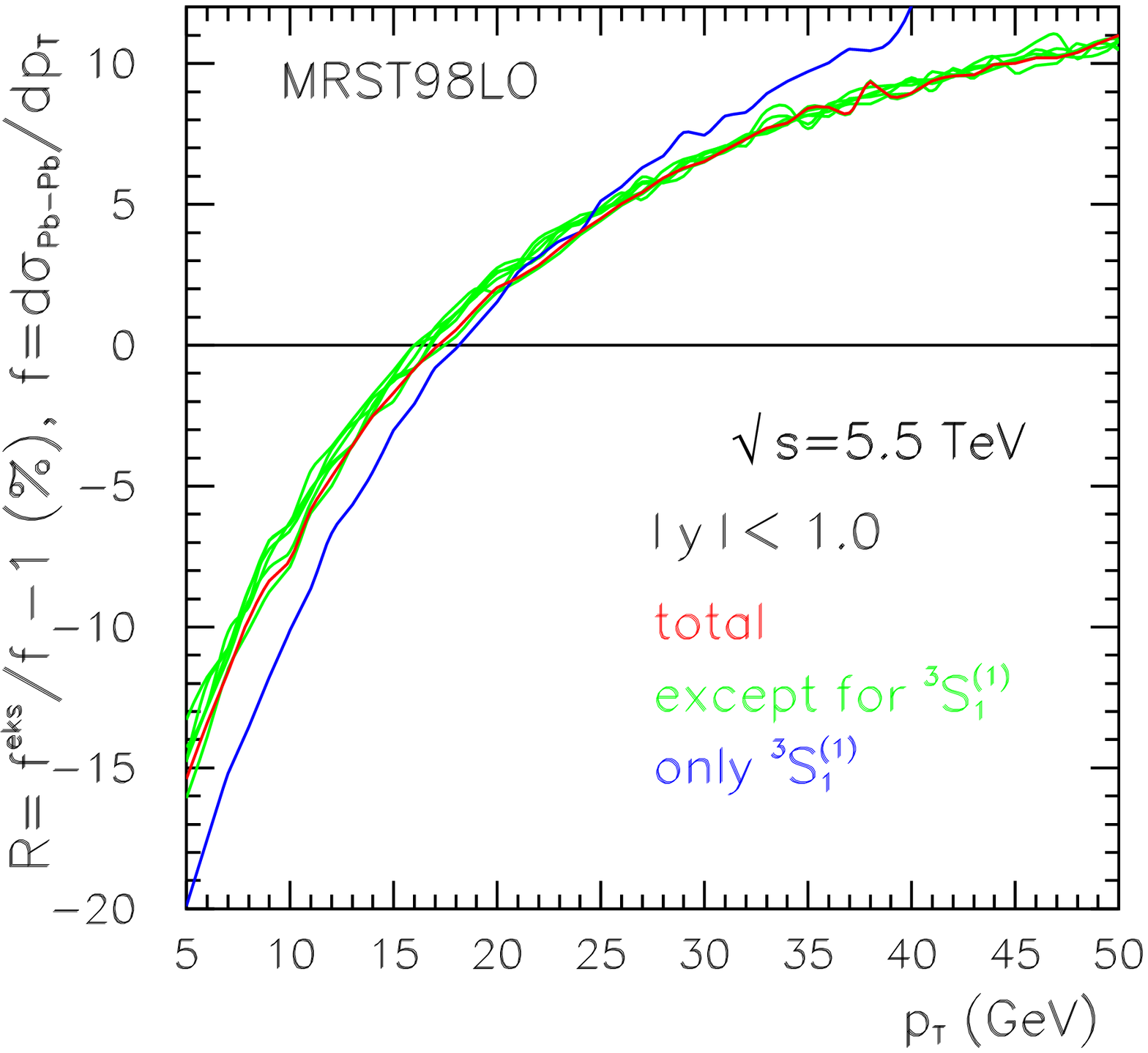}
\end{tabular}
\caption{\label{lee.fig4} 
The $p_T$ dependence of $R_{AB}$ [Eq.~(\ref{R})] for $\Upsilon$ 
production.
(a) We compare the results for $p$Pb and Pb+Pb collisions.  (The $p_T$ 
dependence is stronger in the Pb+Pb result.)
(b) We show the dependence of $R_{AB}$ on the various production channels
in Pb+Pb collisions at $\sqrt{s}=5.5$~TeV.
}
\end{figure}

Better determinations of the $\Upsilon$ matrix elements are required in
order to make more accurate predictions of the NRQCD
$\Upsilon$-production rates at the LHC. As is shown in the
$\Upsilon$-polarization analysis in Ref.~\cite{Braaten:2000gw}, some
theoretical predictions have quite large uncertainties even at Tevatron
energies, owing to our poor knowledge of the matrix elements. However,
as is shown in Fig.~\ref{lee.fig4}, the ratio $R_{AB}$ is still a good
measure of the effect of shadowing on $\Upsilon$ production.  The ratio
is independent of the $\Upsilon$ state and is quite similar to the
$J/\psi$ ratio in Fig.~\ref{lee.fig2}.  The shadowing effect in Pb+Pb
interactions may be somewhat less for the $\Upsilon$ at $p_T \approx 5$
GeV than for the $J/\psi$, but the difference is small.  Note also, from
Fig.~\ref{lee.fig4}(b), that $R_{AB}$ is essentially independent of the
matrix elements and is, therefore, largely unaffected by their
uncertainties.

\subsection[Comparison of CEM and NRQCD Results]
{Comparison of CEM and NRQCD Results~\protect \footnote{Author: R. Vogt.}}
\label{quarkon.comp}

Here we briefly compare the $p_T$ distributions of inclusive $J/\psi$ and
$\Upsilon$ production in Pb+Pb collisions at 5.5 TeV calculated in the CEM and
NRQCD approaches.  Neither calculation includes any intrinsic transverse
momentum effects which could alter the slopes of the $p_T$ distributions.

\begin{figure}
\begin{tabular}{cc}
\includegraphics[width=8cm]{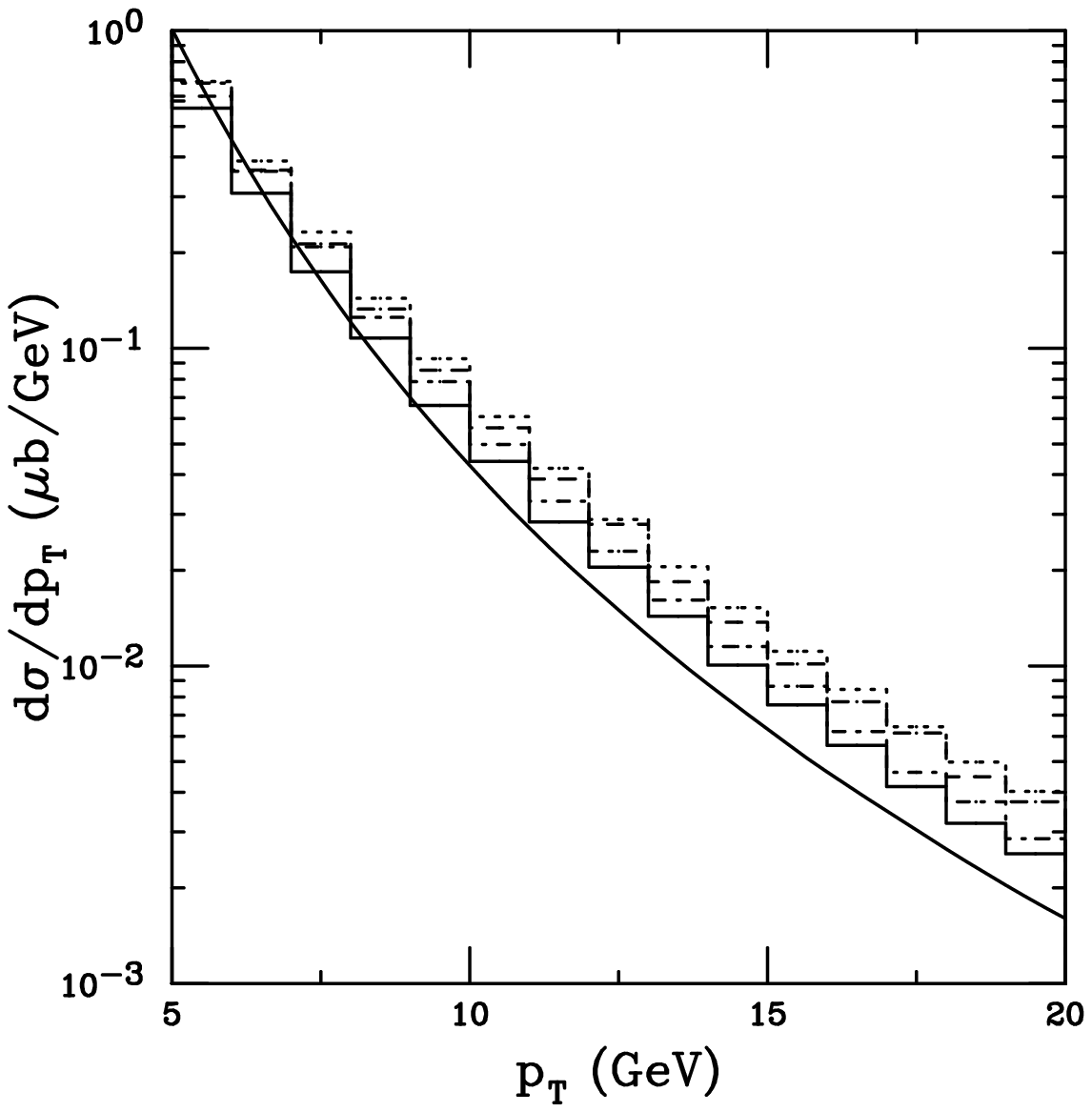}&
\includegraphics[width=8cm]{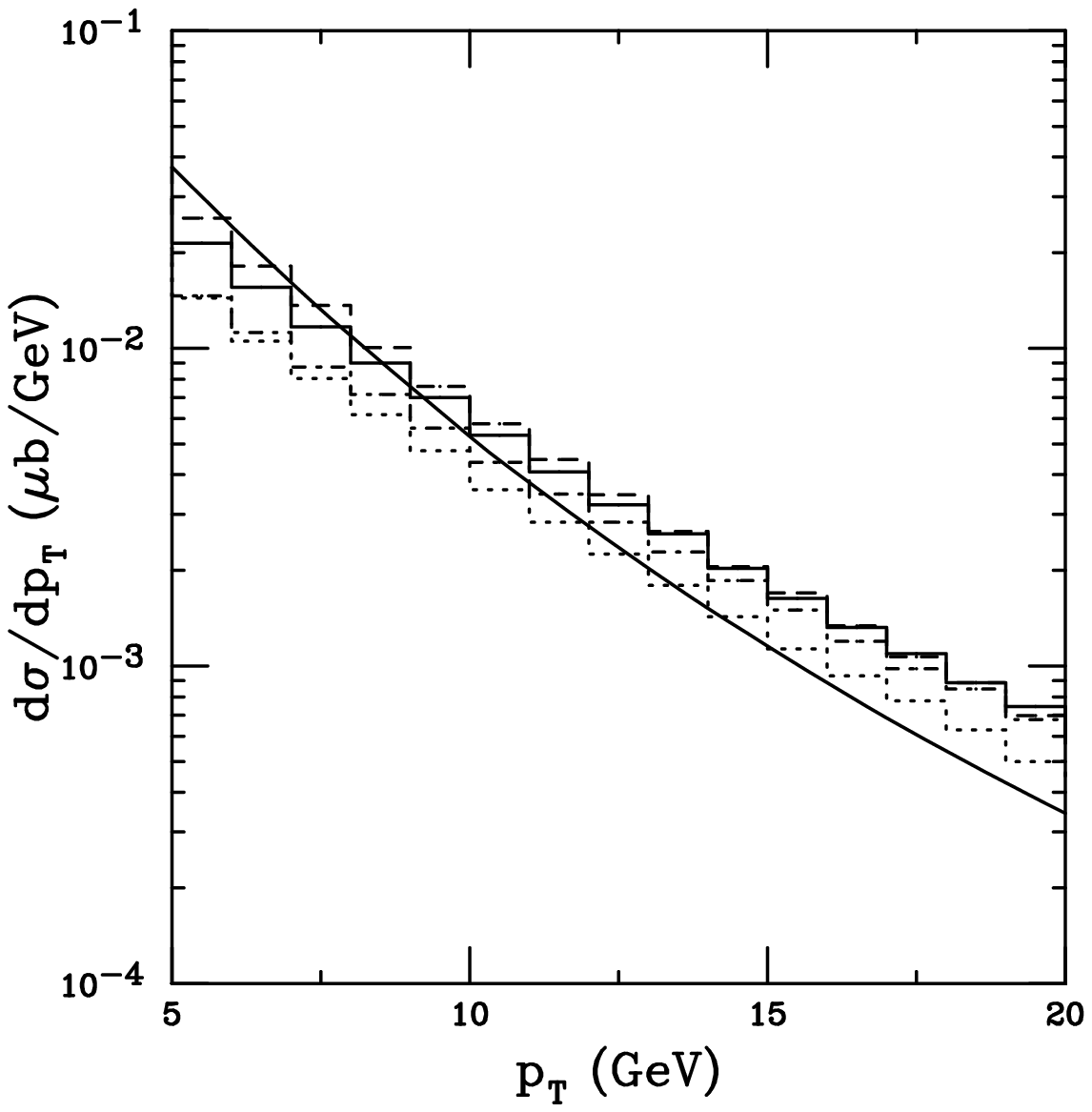}
\end{tabular}
\caption{\label{upsicomp} 
The $p_T$ dependence of the NRQCD and CEM results for inclusive $J/\psi$ (left)
and $\Upsilon$ (right) production in 5.5 TeV Pb+Pb collisions
is compared.  The curves are the NRQCD
predictions while the histograms are the CEM predictions.  On the left-hand
side, the $J/\psi$ CEM predictions are 
$\psi1$ (solid), $\psi2$ (dashed),
$\psi3$ (dot-dashed) and $\psi4$ (dotted).  On the right-hand side the CEM 
$\Upsilon$ calculations are 
$\Upsilon1$ (solid), $\Upsilon2$ (dashed), $\Upsilon3$ (dot-dashed) and 
$\Upsilon4$ (dotted).  
}
\end{figure}

The $J/\psi$ distributions are compared on the left-hand side of
Fig.~\ref{upsicomp}.  The NRQCD
result from Fig.~\ref{lee.fig1} is given in the solid curve.  The branching
ratio to lepton pairs has been removed and the cross section converted to
$\mu$b. The CEM results from Fig.~\ref{psiupsptdep} are shown in the
histograms. In this case, the direct $J/\psi$ cross section has been converted
to the inclusive cross section.  Considering the difference in mass, scale and
parton densities in the two approaches, $m_c = 1.5$ GeV and $\mu = 
\sqrt{4m_c^2 + p_T^2}$ with MRST98LO for NRQCD and the parameters in
Table~\ref{qqbparams} for the CEM, the agreement is rather good over the $p_T$
range shown.  However, the $p_T$ slopes appear to be somewhat different.  
Note also that the NRQCD calculations are made in the
rapidity interval $|y|<1$ while the CEM results are integrated over all
rapidity, affecting the relative normalization.  
We can expect a comparable level of agreement for the
other charmonium states which have a $p_T$ dependence similar to that of 
the inclusive $J/\psi$ in Fig.~\ref{lee.fig1}. 

The inclusive $\Upsilon (1S)$ distributions are compared on the right-hand side
of Fig.~\ref{upsicomp}.  Here also we have converted the NRQCD result to $\mu$b
and divided out the branching ratio to lepton pairs.
We again see a relatively good agreement of the
calculations in the two approaches, despite some differences in the masses,
scales and parton densities used, as in the case of the $J/\psi$.  The rapidity
$|y|<1$ cut on the NRQCD result is a smaller relative factor for $\Upsilon$
production due to the narrower $\Upsilon$ rapidity distribution, see
Fig.~\ref{psiupsydep}.  We note that the agreement of the two approaches
for the $\Upsilon (2S)$ and $\Upsilon (3S)$ states would not be as good, 
primarily due to the poorly determined matrix elements for these states, as
previously discussed.  

\subsection[The Comover Enhancement Scenario (CES)]
{The Comover Enhancement Scenario (CES)~\protect
\footnote{Authors: P. Hoyer, N. Marchal and S. Peign\'e.}}
\label{quarkon.ces}

\subsubsection{The Quarkonium Thermometer}

The production of heavy $Q\bQ$ quarkonia may
offer valuable insights into QCD dynamics, complementary to those
given by open heavy flavor production. In both cases, the creation of
the heavy quark pair requires an initial parton collision of hardness
\order{m_Q}. Most of the time the heavy quarks hadronize
independently of each other and are incorporated into separate
hadrons. The QCD factorization theorem exploits the conservation of
probability in the hadronization process to express the total heavy
quark production cross section in terms of target and projectile
parton distributions and a perturbative subprocess cross section such
as $\sigma(gg\rightarrow Q\bQ)$.

The quarkonium cross section is a small fraction of the open flavor
one and is thus not constrained by the standard QCD factorization
theorems. 
Nevertheless, it is plausible that the initial $Q\bQ$ production 
is governed by the usual parton distributions and
hard subprocess cross sections with the invariant mass of the
$Q\bQ$ pair constrained to be close to threshold. Before the
quarkonium emerges in the final state there can, however, be further
interactions which, due to the relatively low binding energy, can
either ``make or break'' the bound state. Quarkonium studies can thus
give new information about the environment of hard production, from
the creation of the heavy quark pair until its ``freeze-out''. The
quantum numbers of the quarkonium state furthermore impose
restrictions on its interactions. Thus states with negative charge
conjugation, $\psi(nS)$, or total spin $J=1$, $\chi_{c1}$,
require the $Q\bQ$ pair to interact at least once after its
creation via $gg\to Q\bQ$.

Despite an impressive amount of data on the production of several
quarkonium states with a variety of beams and targets we still have a
poor understanding of the underlying QCD dynamics. Thus quarkonia
cannot yet live up to their potential as `thermometers' of $AB$
collisions, where $A,B = \gamma^{(*)}$, hadron or nucleus. Rather, it
appears that we need simultaneous studies and comparisons of several
processes to gain insight into the production dynamics.

We will now summarize the successes and failures of the
Color Singlet Model \cite{Berger:1980ni,Baier:1981uk},
which we consider as a guideline for understanding the
nature of the quarkonium production dynamics.

\subsubsection{Successes and Failures of the Color Singlet Model} 
\label{csmsect}

In the Color Singlet Model (CSM), the $Q\bQ$ pair is directly prepared
with the proper quantum numbers in the initial hard subprocess and
further interactions are assumed to be absent. The quarkonium production
amplitude is then given by the overlap of the non-relativistic wave
function with that of the $Q\bQ$ pair.

This model at NLO correctly predicts the
normalization and momentum dependence of the $J/\psi$ {\em photoproduction}
rate \cite{Kramer:2001hh,Adloff:2002ex}. While the absolute
normalization of the CSM prediction is uncertain by a factor of $2 -
3$ there appears to be no need for any additional production
mechanism for longitudinal momentum fractions $0.3 \leq x_F \leq 0.9$
and $1 \leq p_T^2 \leq 60$ GeV$^2$. The comparison with
leptoproduction data \cite{Adloff:2002ey} is less conclusive since
only LO CSM calculations exist.

The CSM underestimates the directly produced $J/\psi$ and
$\psi'$ {\em hadroproduction} rates by more than an order of magnitude.
This is true both at low $p_{\perp} \lsim m_c$ (fixed target)
\cite{Kaplan:1996tb} and at high
$p_{\perp} \gg m_c$ (collider)  \cite{Kramer:2001hh}. Similar
discrepancies for the $\Upsilon$
states~\cite{Affolder:1999wm,Abe:1995an,Alexopoulos:1995dt}
indicate that the anomalous
enhancement does not decrease quickly with increasing quark mass.

The
inelastic cross section ratio $\sigma(\psi')/\sigma_{\rm dir}(J/\psi)$ is
similar in photoproduction \cite{Bertolin:1999yj} and hadroproduction
\cite{Alexopoulos:1997yd,Lourenco:1996wn} and consistent with the value
$\simeq 0.24$ expected in the CSM~\cite{Vanttinen:1994sd}.
The ratio does not depend on $x_F$
in the projectile fragmentation region and is independent of the
nuclear target size in $hA$ collisions. The CSM thus underestimates
the $J/\psi$ and $\psi'$
hadroproduction cross sections, as well as that of the
$\chi_{c1}$ \cite{Vanttinen:1994sd},
by similar large factors. The quantum numbers of these charmonium
states require final-state gluon emission in the CSM, $gg \to J/\psi g$.
This emission is not required for the $\chi_{c2}$ where the 
CSM cross section $\sigma(gg \to \chi_{c2})$ is only a factor $\sim 2$
below the hadroproduction data \cite{Vanttinen:1994sd}.

In the CSM, $\chi_c$ photoproduction is suppressed by a power of $\alpha_s$
compared to the $J/\psi$ and $\psi'$ production rates.
One indeed observes a smaller value
of the $\sigma (\chi_{c2})/\sigma ( J/ \psi)$ ratio in
photoproduction~\cite{Roudeau:1988wb} than in
hadroproduction~\cite{Antoniazzi:1992iv,Antoniazzi:1993yf}.

\subsubsection{Description of the CES and its generic timescales}

The analysis of agreements and discrepancies between the CSM and quarkonium
data led to the comover enhancement scenario of quarkonium
production~\cite{Hoyer:1998ha,Hoyer:1998dr}.
Hadroproduced $Q\bQ$ pairs are created within a comoving color field
and form $J/\psi$, $\psi'$ and $\chi_{c1}$ through gluon {\em
absorption} rather than emission, enhancing the cross
section relative to the CSM since the pair gains rather than loses
energy and momentum. The $\chi_{c2}$ cross section is not as strongly 
influenced since no gluon needs to be absorbed or emitted. Most importantly,
such a mechanism
is consistent with the success of the CSM in photoproduction
since no color fields are expected in the photon fragmentation region,
$x_F \gsim 0.3$.

\begin{figure}[thb]
\begin{center}
\includegraphics[width=6cm]{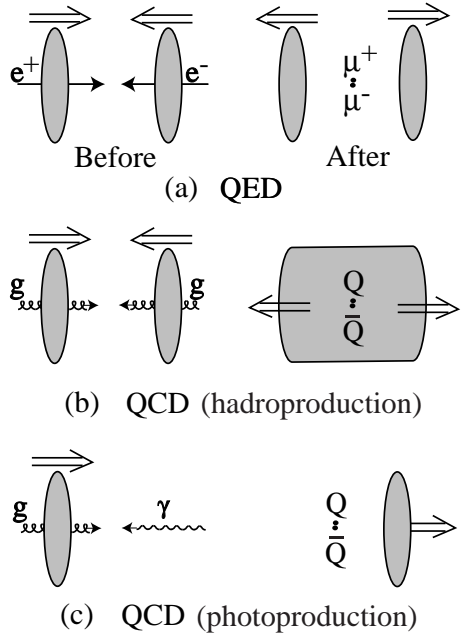}
\caption{Schematic scenarios of gauge field interactions are compared:
(a) $e^+ e^- \rightarrow \mu^+ \mu^-$ in QED; (b) hadroproduction of a 
$Q\bQ$ pair, {\it e.g.}\ $gg \rightarrow Q\bQ$; and (c)
photoproduction of a $Q\bQ$ pair, $\gamma g \rightarrow Q \overline
Q$.  The creation of a comoving color field is specific to hadroproduction,
(b).}
\label{comofield}
\end{center}
\end{figure}

The origin of a comoving color field in hadroproduction is illustrated 
in Fig.~\ref{comofield}. Light charged particles carry
gauge fields which are radiated in high energy annihilations
into a heavy particle pair. In $e^+e^- \to \mu^+\mu^-$ annihilations, the
photon fields pass through each other and materialize as forward
bremsstrahlung, Fig.~\ref{comofield}(a). 
In $gg \to Q\bQ$, on the other hand, the
self-interaction of the color field can also result in the creation of a
gluon field at intermediate rapidities, Fig.~\ref{comofield}(b). Hadroproduced
$Q\bQ$ pairs thus find themselves surrounded by a color field. We
postulate that interactions between the $Q\bQ$ pair and this
comoving field are important in quarkonium hadroproduction. In direct
photoproduction, the incoming photon does not carry any color field
and the $Q\bQ$ pair is left in a
field-free environment after the collision, Fig.~\ref{comofield}(c). 
The proposed rescattering thus does not affect (non-resolved) photoproduction.

The importance of rescattering effects in hadroproduction as compared
to photoproduction is also suggested by data on open charm
production. Hadroproduced $D \overline D$ pairs are almost uncorrelated in
azimuthal angle \cite{Aitala:1998kh}, at odds with
standard QCD descriptions. Photoproduced pairs on the other hand,
emerge nearly back-to-back \cite{Frabetti:1996vi}, following the
charm quarks of the underlying $\gamma g\to c\overline c$ process.

Since the hardness of the gluons radiated in the
creation process increases with quark mass, the rescattering effect
persists for bottomonium. Due to the short timescale of the radiation
the heavy quark pair remains in a compact configuration during
rescattering and overlaps with the quarkonium wave function at the
origin. The successful CSM result for
$\sigma(\psi')/\sigma_{\rm dir}(J/\psi)$~\cite{Vanttinen:1994sd} 
is thus preserved.

The $Q\bQ$ pair may also interact with the more distant projectile
spectators after it has expanded and formed quarkonium. Such
spectator interactions are more frequent for nuclear projectiles and
can cause the breakup (absorption) of the bound state. This
conventional mechanism of quarkonium suppression in nuclei is thus
fully compatible with, but distinct from, 
interactions with the comoving color field.

We have investigated the consequences of the CES using pQCD to
calculate the interaction between the $Q\bQ$ and the comoving
field. While we find consistency with data, quantitative predictions
depend on the structure of the comoving field. Hence tests of the CES 
must rely on its generic features, especially the proper timescales over which
the $Q\bQ$ proceeds from production through hadronization.

The CES distinguishes three proper timescales in quarkonium production:
\begin{itemize}
\item $\tau_Q \sim 1/m_Q$, the $Q\bQ$ pair production time;
\item $\tau_{AP}$,
the DGLAP scale over which the comoving field is created and
interacts with the $Q\bQ$ pair;
\item $\tau_\Lambda \sim 1/\lqcd$, while rescattering with
comoving spectators may occur.
\end{itemize}
In the following we will consider quarkonium production
at $p_{\perp} \lsim m_Q$.
In quarkonium production at $p_T \gg m_Q$, a large $p_T$
parton is first created on a timescale $1/p_T$, typically through
$gg \to gg^*$. The virtual gluon then fragments, $g^* \to Q\bQ$, in
proper time $\tau_Q$.
Thus high $p_T$ quarkonium production is also describable with the CES 
\cite{Peigne:2000wd}.

\paragraph{Timescale $\tau_Q \sim 1/m_Q$: creation of the 
$Q\bQ$ pair --- }

The $Q\bQ$ pair is created in a standard parton subprocess,
typically $gg \to Q\bQ$, at a time scale $\tau_Q \sim 1/m_Q$.
This first stage is common to other
theoretical approaches such as the 
CEM~\cite{Fritzsch:1977ay,Halzen:1977rs,Gluck:1977zm} and 
NRQCD~\cite{Bodwin:1994jh,Beneke:1996tk}. The 
momentum distribution of the
$Q\bQ$ is determined by the product of projectile $(A)$ and target
$(B)$ parton distributions, such as
$g_A(x_1) g_B(x_2)$ where $A$, $B$ may be a hadron, nucleus or resolved real or
virtual photon.  (In the case of direct photoproduction or DIS, when the
projectile is a photon or lepton, the process depends on the single
distribution $g_B(x_2)$.)  Such production
is consistent with the quarkonium data.

According to pQCD, the $Q\bQ$ is dominantly produced 
{\it close to threshold} in a color octet, 
$S=L=0$ configuration. Such a state can
obtain the quarkonium quantum numbers through a further interaction which
flips a heavy quark spin and turns the pair into a color singlet. The
amplitude for processes of this type are suppressed by the factor
$k/m_Q$ where $k$ is the momentum scale of the interaction. The
various theoretical approaches differ in the scale assumed for $k$.
\begin{description}
\item{CSM:} Here $k=\morder{m_Q}$. Thus $J/\psi$
production proceeds via the emission of a hard gluon in the primary
process, $gg \to Q\bQ+g$. The $\chi_{c2}$ is produced
without gluon emission, 
$gg \to \chi_{c2}$, through a subdominant $S=L=1$ color singlet
production amplitude.
\item{NRQCD:} The $Q\bQ$ quantum numbers are
changed via gluon emission at the bound state momentum scale
$k=\morder{\alpha_s m_Q}$. This corresponds to an expansion in powers of
the bound state velocity $v=k/m_Q$, introducing nonperturbative
matrix elements that are fit to data.
\item{CEM:} Here $k=\morder{\lqcd}$. Soft
interactions are postulated to change the $Q\bQ$ quantum numbers
with probabilities that are specific for each quarkonium state but
independent of kinematics, projectile and target.
\item{CES:} The quantum numbers of the $Q\bQ$ are changed
in perturbative interactions with a comoving field at scale
$k=\morder{1/\tau_{AP}}$, as described below.
\end{description}

\paragraph{Timescale $\tau_{AP}$: interactions with the comoving field --- }

The scale $\tau_{AP}$ refers to the time in which collinear
bremsstrahlung, the source of QCD scaling violations, is emitted in
the heavy quark creation process \cite{Hoyer:1998ha}.
Thus $1/\tau_{AP}$ characterizes
the effective hardness of logarithmic integrals of the type
$\int_{\mu_F}^{m_Q} dk/k$ where $\mu_F \ll m_Q$ is the factorization
scale. We stress that $1/\tau_{AP}$ is an intermediate but still
perturbative scale, $\tau_Q \ll \tau_{AP} \ll \tau_{\Lambda}$, which
grows with $m_Q$.

The fact that the $Q\bQ$ pair acquires the quarkonium quantum
numbers over the per\-tur\-ba\-ti\-ve time\-scale $\tau_{AP}$ is a feature
of the CES and distinguishes it from other approaches. At this
time, the pair is still compact and couples to quarkonia
via the bound state wavefunction at the origin
or its derivative(s). Thus no
new parameters are introduced in this transition. However, the
interactions of the $Q\bQ$ pair depend on the properties
of the comoving color field such as its intensity and polarization.
Quantitative predictions in the CES are only possible
when the dependence on the comoving field is weak.

Ratios of radially excited quarkonia, such as
$\sigma(\psi')/\sigma_{\rm dir}(J/\psi)$, are insensitive to the comoving
field and are thus expected to be process-independent when
absorption on spectators at later times can be ignored, see below.
The fact that this ratio is observed to be roughly universal
\cite{Alexopoulos:1997yd,Lourenco:1996wn} is one of the
main motivations for the CES. Even the measured variations of the
ratio in different reactions agree with expectations, see 
Ref.~\cite{Hoyer:fq} for a discussion of its
systematics in elastic and inelastic photoproduction, leptoproduction
and hadroproduction at low and high $p_T$.

The ratio $\sigma(\chi_{c1})/\sigma(\chi_{c2})$ is
measured to be $0.6 \pm 0.3$ in pion-induced~\cite{Koreshev:1996wd}
and $0.31 \pm 0.14$ in proton-induced~\cite{Alexopoulos:1999wp}
reactions. The CSM underestimates
this ratio 
as well
as that of~$ \sigma ( J/ \psi )/\sigma (\chi_2 )$ \cite{Vanttinen:1994sd}.
The rescattering contribution
increases $\sigma ( J/\psi )$ and $\sigma (\chi_1 )$, enhancing
the above ratios. 

\paragraph{Nuclear target dependence --- }

The quarkonium cross section can be influenced by rescattering effects
in both the target and projectile fragmentation regions. For
definiteness, we assume the charmonium is produced in the
projectile fragmentation region, $x_F > 0$.

The nuclear target dependence is usually parameterized as $\sigma(hA \to
J/\psi+X) \propto A^\alpha$. Data show that  $\alpha < 1$ and
obeys Feynman scaling: $\alpha$ depends on (and decreases with) $x_F$ 
rather than on the momentum fraction~$x_2$ of the target
parton~\cite{Hoyer:1990us,Leitch:1999ea}. The comparison with lepton pair
production in the Drell-Yan process shows that the $J/\psi$ nuclear suppression
cannot be attributed to shadowing of parton distributions in the
nucleus~\cite{Lourenco:1996wn}. The $A$ dependence is thus difficult 
to explain in the CSM, NRQCD and
CEM approaches.

In the Feynman scaling regime, we may assume that the $Q\bQ$ pair
energy is high enough to remain compact while traversing the target.
The relative transverse momentum of the $Q$ and $\bQ$ could
increase as a result of rescattering in the target, thus suppressing
the binding probability. However, this
explanation is unlikely in view of the absence of nuclear
suppression in photoproduction~\cite{Amaudruz:1991sr}.

In the CES, the nuclear target suppression is ascribable
to absorption of the comoving color field in the target nucleus. This
field is emitted by a projectile parton with transverse size
$\tau_{AP}$, larger than the size, $\sim 1/m_Q$, of the
$Q\bQ$ pair. Due to Lorentz time dilation, the field is emitted long
before reaching the target and reinteracts with the $Q\bQ$ long after 
passing the
target. Absorption of the comoving field in the target implies
suppression of $J/\psi$ production in the CES. At high energies,
we have $x_1 \simeq x_F$, which explains the Feynman scaling of this
effect.  Moreover, as $x_F$ increases,
less energy is available to be radiated to the gluon field which
therefore becomes softer, further increasing its absorption in the
target and thus the nuclear suppression.

This explanation is consistent with the fact that the nuclear target
suppression of the $J/\psi$ and the $\psi'$ is found to be the same
for $x_F \gsim 0.2$~\cite{Leitch:1999ea}. It also predicts that the 
suppression will be
similar for $\chi_{c1}$ production. On the other hand, the nuclear
target suppression should be {\em reduced} for $\chi_{c2}$ since a
substantial fraction is directly produced 
without gluon absorption from the comoving field. A measurement of
$\sigma(hA \to \chi_{c1},\chi_{c2}+X)$ in the projectile fragmentation region
would thus constitute an important test of the CES.

In a Glauber picture of the nuclear suppression, a relatively large value
for the absorption cross section is required, 
$\sigma_{\rm abs} \sim 5 \ {\rm mb}$.
We interpret this value as the joint cross section of the $Q\bQ$ pair
and the comoving field, thus of order $\tau_{AP}^2 \gg 1/m_Q^2$.
Since $1/\tau_{AP}$ scales with $m_Q$,
we expect less nuclear absorption for the
$\Upsilon$ states than for the $J/ \psi$, as observed
experimentally~\cite{Alde:1991sw,Leitch:di}.

\paragraph{Timescale $\tau_\Lambda \sim 1/\lqcd$: interactions with
comoving spectators --- }

By the time, $\tau_\Lambda \sim 1$ fm, that the $Q\bQ$ pair encounters
comoving projectile spectators, the pair has already expanded and is
distributed according to the quarkonium wave function. The spectator
rescattering effects are thus independent of the quarkonium formation
process. Larger and more loosely bound charmonia are
more easily broken up by secondary scattering. Hence the $\chi_c$
and $\psi'$ cross sections should be depleted compared to that of the
$J/\psi$. Likewise bottomonium is generally less affected by spectator
interactions than charmonium.

Spectator interactions at large enough $x_F$
are likely to be unimportant for hadron
projectiles judging from the approximate universality of the
$\sigma(\psi')/\sigma_{\rm dir}(J/\psi)$ ratio in photo- and
hadroproduction. The lower ratio seen in nucleus-nucleus
collisions~\cite{Abreu:1999nn},
on the other hand, is most naturally explained by absorption on
spectators. It would be important to confirm this by also measuring the
ratio in $hA$ scattering in the nuclear fragmentation region.

\subsubsection{Summary}

Data on quarkonium production have proved challenging for QCD models.
The richness of the observed phenomena indicates that quarkonium cross
sections are indeed sensitive to the environment of the hard QCD
scattering. We may, however, only be able to decipher its message
through systematic experimental and theoretical studies of several
species of quarkonia produced with a variety of beams and targets in
a range of kinematic conditions.

The rates of quarkonium production at the LHC are large enough for high
statistics studies of the production mechanism.  Differences in $pp$ rates and
$p_T$ distributions may help distinguish between the CEM, NRQCD and CES
approaches. 


\noindent
{\em Acknowledgments} 
S.~P. would like to thank Cristina Volpe for very stimulating discussions.

%% file: energyloss.tex
\section[ABSORPTION AND ENERGY LOSS]{ABSORPTION AND ENERGY LOSS~\protect
\footnote{Section coordinator: D.~Kharzeev.}}
\label{sec:dima}

In this section, we first discuss $J/\psi$ absorption in matter by nucleons 
(sect.~\ref{section:abs}) and comoving secondaries (sect.~\ref{section:como}).
We then present a brief review of heavy quark energy loss 
(sect.~\ref{section:eloss}).

The effective 
nucleon absorption cross sections in section~\ref{section:abs}
are obtained by extrapolating the values extracted from the fixed-target 
measurements to higher energies. 
The resulting charmonium suppression can be 
used as a baseline in the search for deconfinement.  

In section~\ref{section:como}, an evaluation of the couplings of the
$J/\psi$ to 
$D^{(*)}D^{(*)}$ and $D^{(*)}D^{(*)}\pi$ are presented in the Constituent
Quark Model. These couplings are a crucial ingredient in the
calculation of cross sections for the processes $\pi J/\psi\to
D^{(*)}\overline{D}^{(*)}$, an  important background for the $J/\psi$
suppression signal in quark-gluon plasma. While final results relevant
to the cross section for $J/\psi$ absorption by comovers are not yet available,
the finished calculation should provide some of the most reliable results
for these processes.

Energy loss has been much discussed in the context of jet quenching.  While
some calculations have addressed quarkonium, this aspect of energy loss is
still too unsettled to be summarized.  Therefore, only energy loss by heavy
quarks in medium is evaluated in section~\ref{section:eloss}.

\subsection[Charmonium Suppression in Nuclear Collisions at the 
LHC: the baseline]{Charmonium Suppression in Nuclear Collisions at the 
LHC: the baseline~\protect
\footnote{Authors: D. Kharzeev, M. Nardi and H. Satz.}}
\label{section:abs}


Suppression of the $J/\psi$ in nucleus-nucleus collisions was proposed as 
a signal of color deconfinement a long time ago~\cite{Matsui:1986dk}. 
Shortly after, such a suppression was 
observed in O+U and, subsequently, in S+U collisions 
\cite{Baglin:iv,Baglin:bj,Baglin:wi}. It was promptly noted that at least part 
of the observed effect was already present in $pA$ collisions  
\cite{Capella:1988ha,Gerschel:uh}. 
A quantitative analysis of charmonium suppression in nuclear matter has been 
performed in Ref.~\cite{Kharzeev:1996yx}.  Here we extend that analysis to 
RHIC and LHC energies, $\sqrt{s} = 200$ GeV and 5.5 TeV respectively.  
In this extension, we assume that the increased
energy does not change the nature of the nuclear absorption
process. In principle, interference between successive scatterings
could lead to modifications at higher energies. The forthcoming 
$p(d)A$ data from RHIC could clarify this point.

Apart from the independently-determined nuclear matter distributions, the only 
parameter entering the calculation is the effective absorption cross section 
of the quarkonium precursor in nuclear matter, determined from $pA$ data.  
The latest data indicate $\sigma_{\rm abs} \simeq 5$ mb
\cite{Abreu:2002hc,Abreu:rm}. We assume that  
$\sigma_{\rm abs}$ increases with charmonium-nucleon center of mass 
energy as $s_{J/\psi N}^\Delta$, reflecting the growth of the small $x$
gluon density in nucleons.  A massive $Q \overline Q$ state produced at zero 
rapidity will interact with nucleons at center of mass energy squared
$s_{J/\psi N} \simeq 2 m_{J/\psi} ( \sqrt{s}/2)$ where $\sqrt{s}$ is the 
center of mass energy per nucleon pair of the heavy ion collision. Therefore, 
the effective absorption cross section will increase with $\sqrt{s}$ as 
\begin{equation}
\sigma_{\rm abs}(\sqrt{s}) = \sigma_{\rm abs}(\sqrt{s_0})\ \left({s \over 
s_0}\right)^{\Delta/2} \, \, .
\label{endep1}
\end{equation}
\begin{table}[htb]
\begin{center}
\caption{Effective charmonium--nucleon absorption cross sections at 
SPS, RHIC, and LHC energies.  The middle row is the result for the central
value of $\sigma_{\rm abs}(\sqrt{s_0})$ while the upper and lower rows show the
errors on $\sigma_{\rm abs}$.}
\begin{tabular}{|c|c|c|}
\hline
\multicolumn{3}{|c|}{$\sigma_{\rm abs}$ (mb)} \\ \hline
SPS & RHIC & LHC \\ \hline
4.5  &  6.1  &  9.3 \\ 
5.0  &  6.8  &  10.3 \\
5.5  &  7.5  &  11.3 \\ \hline
\end{tabular}
\label{kharzeev:tab1}
\end{center}
\end{table}
We choose $\sqrt{s_0} \simeq 17.3$ GeV (the CERN SPS energy), 
$\sigma_{\rm abs}(\sqrt{s_0}) = 5 \pm 0.5$ mb for charmonium, 
and $\Delta = 0.125$.  Note that since $\sigma_{\rm abs}$ is the absorption
cross section of the precursor charmonium state, it is assumed to be the 
same for all charmonium resonances.  With these parameters, the values of
the absorption cross sections at RHIC and LHC are given in 
Table~\ref{kharzeev:tab1}.  
\begin{figure}[htb]
\begin{center}
\epsfig{file=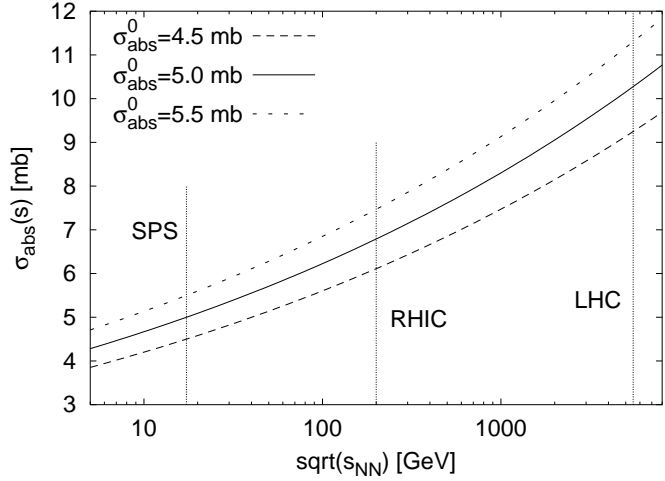, width=10cm}
\caption{The energy dependence of $\sigma_{\rm abs}$.}
\label{kharzeev:fig1}
\end{center}
\end{figure}
The energy dependence of the absorption cross 
section is shown in Fig.~\ref{kharzeev:fig1}. 

The quarkonium survival probability in an $AB$ collision is given by, see 
Eq.~(6) of Ref.~\cite{Kharzeev:1996yx},
\begin{eqnarray*}
S_{J/\psi}(b)& = & \int d^2 s \int d z_A d z_B \rho_A(\vec{s},z_A) 
\rho_B(\vec{b}-\vec{s},z_B) \\*
&&\times \exp \left\{-(A-1)\int_{z_A}^\infty d z_A^\prime
        \rho_A(\vec{s},z_A^\prime) \sigma_{\rm abs} \right\} \\*
&&\times \exp \left\{-(B-1)\int_{z_B}^\infty d z_B^\prime
\rho_B(\vec{b}-\vec{s},z_B^\prime) \sigma_{\rm abs} \right\} \, \, .
\end{eqnarray*}
\begin{table}[htb]
\begin{center}
\caption{$J/\psi$ survival probability in Pb+Pb collisions as a function of 
impact parameter for SPS, RHIC, and LHC energies.  The central results are
calculated with the value of $\sigma_{\rm abs}$ given in the middle row of
Table~\protect\ref{kharzeev:tab1}.  The errors represent the uncertainty in
$\sigma_{\rm abs}$.}
\label{kharzeev:tab2}
\begin{tabular}{|c|ccc|}
\hline
& \multicolumn{3}{|c|}{$S_{J/\psi}$} \\
$b$ (fm)    &      SPS         &        RHIC        &       LHC       \\
\hline
$\langle b \rangle$ & 0.52 $\pm$ 0.03  & 0.42 $\pm$ 0.03  & 0.29 $\pm$ 0.03 \\
\hline
 0 & 0.47 $\pm$ 0.03 & 0.37 $\pm$ 0.04 & 0.25 $\pm$ 0.03 \\
 1 & 0.47 $\pm$ 0.03 & 0.37 $\pm$ 0.04 & 0.25 $\pm$ 0.03 \\
 2 & 0.47 $\pm$ 0.03 & 0.37 $\pm$ 0.04 & 0.25 $\pm$ 0.03 \\
 3 & 0.48 $\pm$ 0.03 & 0.38 $\pm$ 0.04 & 0.25 $\pm$ 0.03 \\
 4 & 0.48 $\pm$ 0.03 & 0.38 $\pm$ 0.04 & 0.25 $\pm$ 0.03 \\
 5 & 0.49 $\pm$ 0.03 & 0.39 $\pm$ 0.04 & 0.26 $\pm$ 0.03 \\
 6 & 0.50 $\pm$ 0.03 & 0.40 $\pm$ 0.04 & 0.27 $\pm$ 0.03 \\
 7 & 0.52 $\pm$ 0.03 & 0.42 $\pm$ 0.04 & 0.29 $\pm$ 0.03 \\
 8 & 0.54 $\pm$ 0.03 & 0.44 $\pm$ 0.03 & 0.31 $\pm$ 0.03 \\
 9 & 0.57 $\pm$ 0.03 & 0.47 $\pm$ 0.03 & 0.34 $\pm$ 0.03 \\
10 & 0.60 $\pm$ 0.03 & 0.51 $\pm$ 0.03 & 0.38 $\pm$ 0.03 \\
11 & 0.65 $\pm$ 0.03 & 0.57 $\pm$ 0.03 & 0.44 $\pm$ 0.03 \\
12 & 0.71 $\pm$ 0.02 & 0.63 $\pm$ 0.03 & 0.51 $\pm$ 0.03 \\
\hline
\end{tabular}
\end{center}
\end{table}
The resulting $J/\psi$ survival probability as a function of impact parameter 
in a Pb+Pb collision is given in Table~\ref{kharzeev:tab2} for SPS, RHIC and 
LHC energies.  
\begin{figure}[htb]
\begin{center}
\epsfig{file=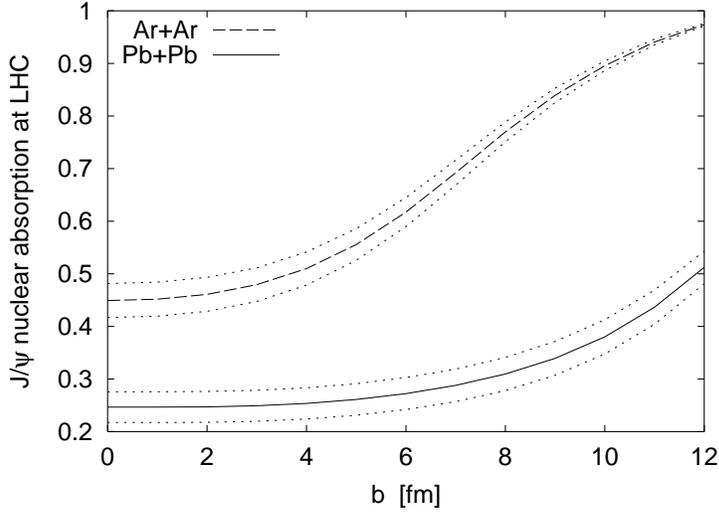, width=10cm}
\caption{The impact parameter dependence of $S_{J/\psi}$ in 
Pb+Pb and Ar+Ar collisions at the LHC.  We assume $\sqrt{s} = 5.5$ TeV for
both systems.  In both cases, the survival probabilities in the
central curves are calculated with the median $\sigma_{\rm abs}$ extrapolated
to LHC energies while the upper and lower curves give the uncertainty in 
$S_{J/\psi}$ due to the absorption cross section.}
\label{kharzeev:fig2}
\end{center}
\end{figure}
The LHC result is also shown in Fig.~\ref{kharzeev:fig2}.
\begin{table}[htb]
\begin{center}
\caption{The $J/\psi$ survival probability in Ar+Ar collisions at the LHC as a 
function of impact parameter.  The central results are
calculated with the value of $\sigma_{\rm abs}$ given in the middle row of
Table~\protect\ref{kharzeev:tab1}.  The errors represent the uncertainty in
$\sigma_{\rm abs}$.}
\begin{tabular}{|c|c|}
\hline
$b$ (fm)    &  $S_{J/\psi}$      \\
\hline
$\langle b \rangle$   & 0.52 $\pm$ 0.03   \\
\hline
 0 & 0.45 $\pm$ 0.03 \\
 1 & 0.45 $\pm$ 0.03 \\ 
 2 & 0.46 $\pm$ 0.03 \\ 
 3 & 0.48 $\pm$ 0.03 \\ 
 4 & 0.51 $\pm$ 0.03 \\ 
 5 & 0.56 $\pm$ 0.03 \\ 
 6 & 0.62 $\pm$ 0.03 \\ 
 7 & 0.69 $\pm$ 0.02 \\ 
 8 & 0.77 $\pm$ 0.02 \\ 
\hline
\end{tabular}
\label{kharzeev:tab3}
\end{center}
\end{table}

To illustrate the $A$ dependence, we have also calculated 
the survival probability in Ar+Ar collisions at the LHC.
The results are shown in Table~\ref{kharzeev:tab3} and
Fig.~\ref{kharzeev:fig2}. 

\begin{figure}[htb]
\epsfxsize=10truecm \centerline{\epsffile{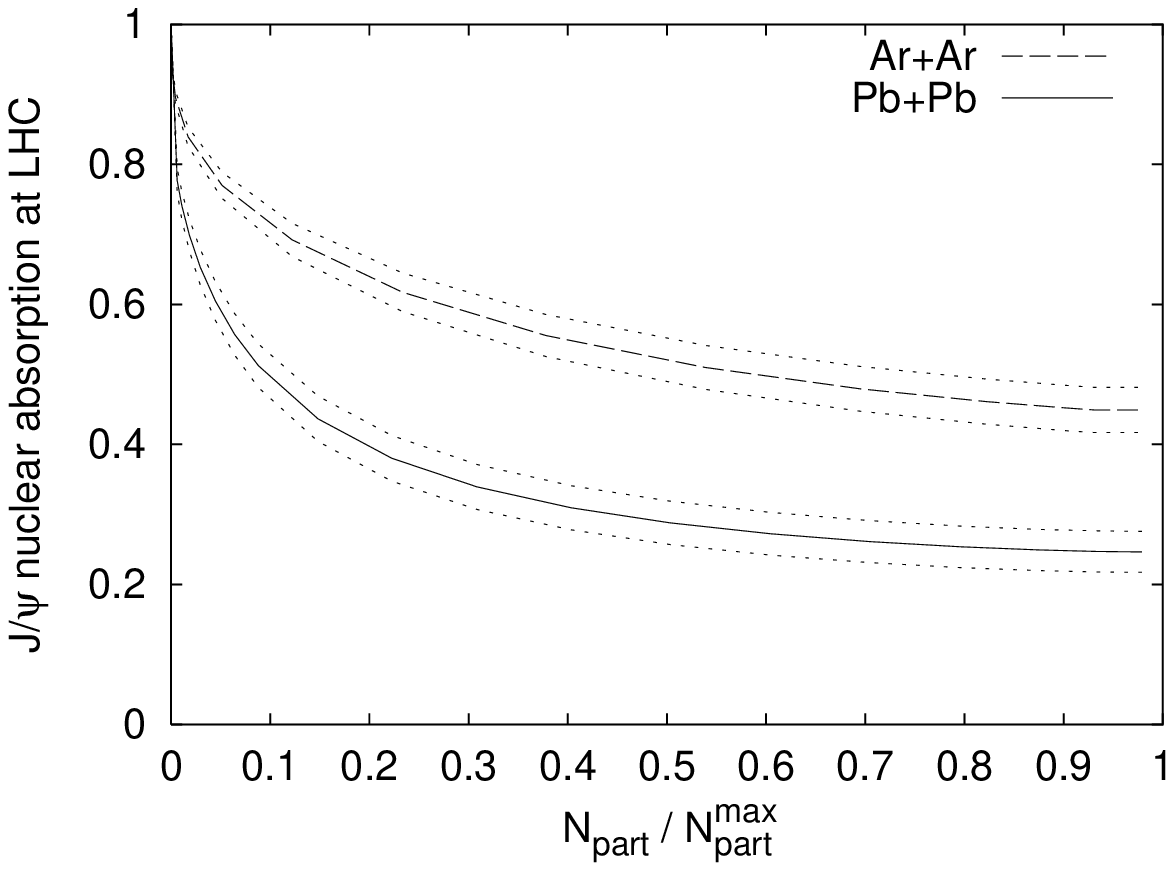}}
\caption{The centrality dependence of $S_{J/\psi}$ at the LHC in 
Pb+Pb and Ar+Ar collisions, both at $\sqrt{s} =5.5$ TeV. 
In both cases, the survival probabilities in the
central curves are calculated with the median $\sigma_{\rm abs}$ extrapolated
to LHC energies while the upper and lower curves give the uncertainty in 
$S_{J/\psi}$ due to the absorption cross section.
$N_{\rm part}^{\rm max} = 2 A$ is the maximum 
number of participants.} 
\label{kharzeev:fig3}
\end{figure}

Since the impact parameter of the collision cannot be measured directly, 
it is convenient to choose the number of participants, those nucleons that have
participated in at least one inelastic collision, as a measure of centrality.
At least in principle the number of participants, $N_{\rm part}$, can be 
inferred from the measurements of the 
number of spectator nucleons, $N_{\rm spect}$, going forward since 
$N_{\rm part} = A - N_{\rm spect}$. The survival probabilities 
in Pb+Pb and Ar+Ar collisions at the LHC are 
shown in Fig.~\ref{kharzeev:fig3} as a function of $N_{\rm part}/N_{\rm 
part}^{\rm max}$ where $N_{\rm part}^{\rm max} = 2A$. 

\begin{figure}[htb]
\begin{center}
\epsfig{file=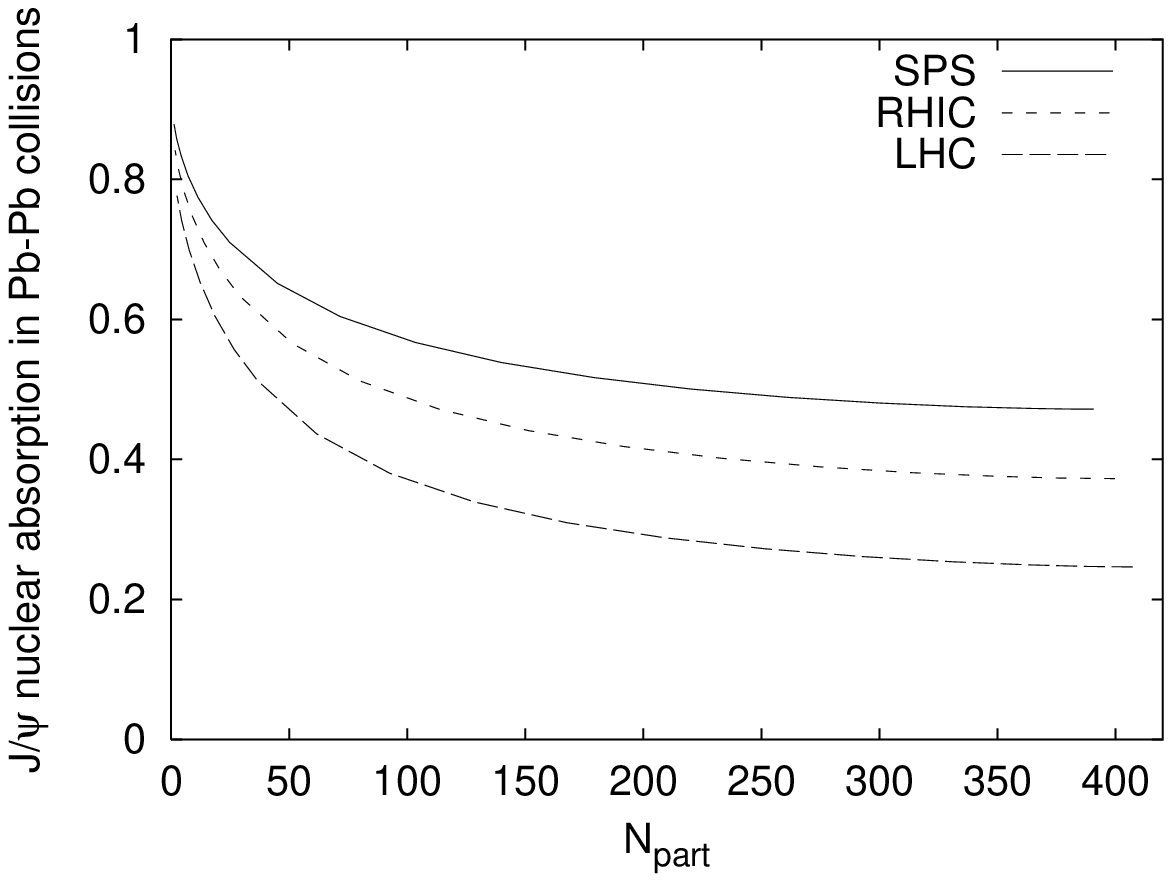, width=10cm}
\caption{The dependence of $S_{J/\psi}$ on $N_{\rm part}$ for Pb+Pb collisions
at SPS, RHIC and LHC energies.}
\label{kharzeev:fig4}
\end{center} 
\end{figure}

Figure~\ref{kharzeev:fig4} the compares charmonium survival 
probabilities in Pb+Pb as a function of $N_{\rm part}$ at SPS, RHIC, and the 
LHC.  The suppression due to nuclear absorption alone at the LHC is 
expected to be much stronger than at the SPS.

\begin{figure}[htb]
\begin{center}
\epsfig{file=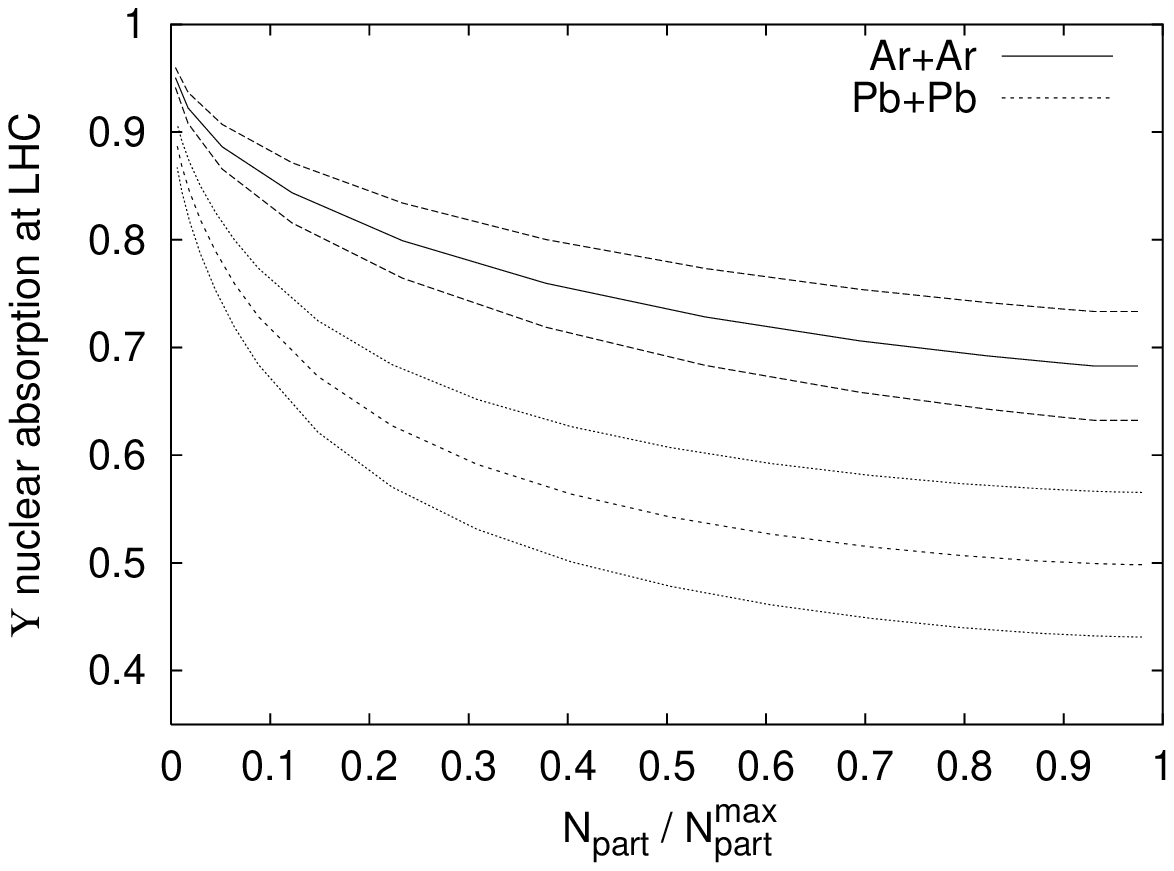, width=10cm}
\caption{The centrality dependence of the $\Upsilon$ survival probability 
at the LHC in Pb+Pb and Ar+Ar collisions. 
In both cases, the survival probabilities in the
central curves are calculated with the median $\sigma_{\rm abs}$ extrapolated
to LHC energies while the upper and lower curves give the uncertainty in 
$S_{J/\psi}$ due to the absorption cross section.
$N_{\rm part}^{\rm max} = 2 A$ is 
the maximum number of participants.} 
\label{kharzeev:fig5}
\end{center}
\end{figure}

Finally we note that at the LHC, bottomonium production will be abundant enough
for study of their suppression in hot matter to be feasible. 
Because of their smaller size and larger binding energy, the $\Upsilon$ states
will provide valuable information complementary to the study of charmonium. 
The available E772 data on $\Upsilon$ production 
at $\sqrt{s} = 39$ GeV \cite{Alde:1991sw,Mike} 
imply an effective $b \overline b$ 
absorption cross section of $\sigma_{\rm abs} = 
2.5 \pm 0.5$ mb. Assuming the energy dependence 
of this effective absorption cross section is the same as that of 
Eq.~(\ref{endep1}), we find $\sigma_{\rm abs} = 4.6 \pm 0.9$ mb at 5.5 TeV. 
The corresponding $\Upsilon$ survival probabilities are presented in
Fig.~\ref{kharzeev:fig5}.    

\subsection[A Note on the $J/\psi$ Strong Couplings]
{A Note on the $J/\psi$ Strong Couplings~\protect
\footnote{Authors: A.~Deandrea, G.~Nardulli, A.~D.~Polosa.}}
\label{section:como}

\begin{figure}[htbp]
\epsfysize=10truecm \centerline{\epsffile{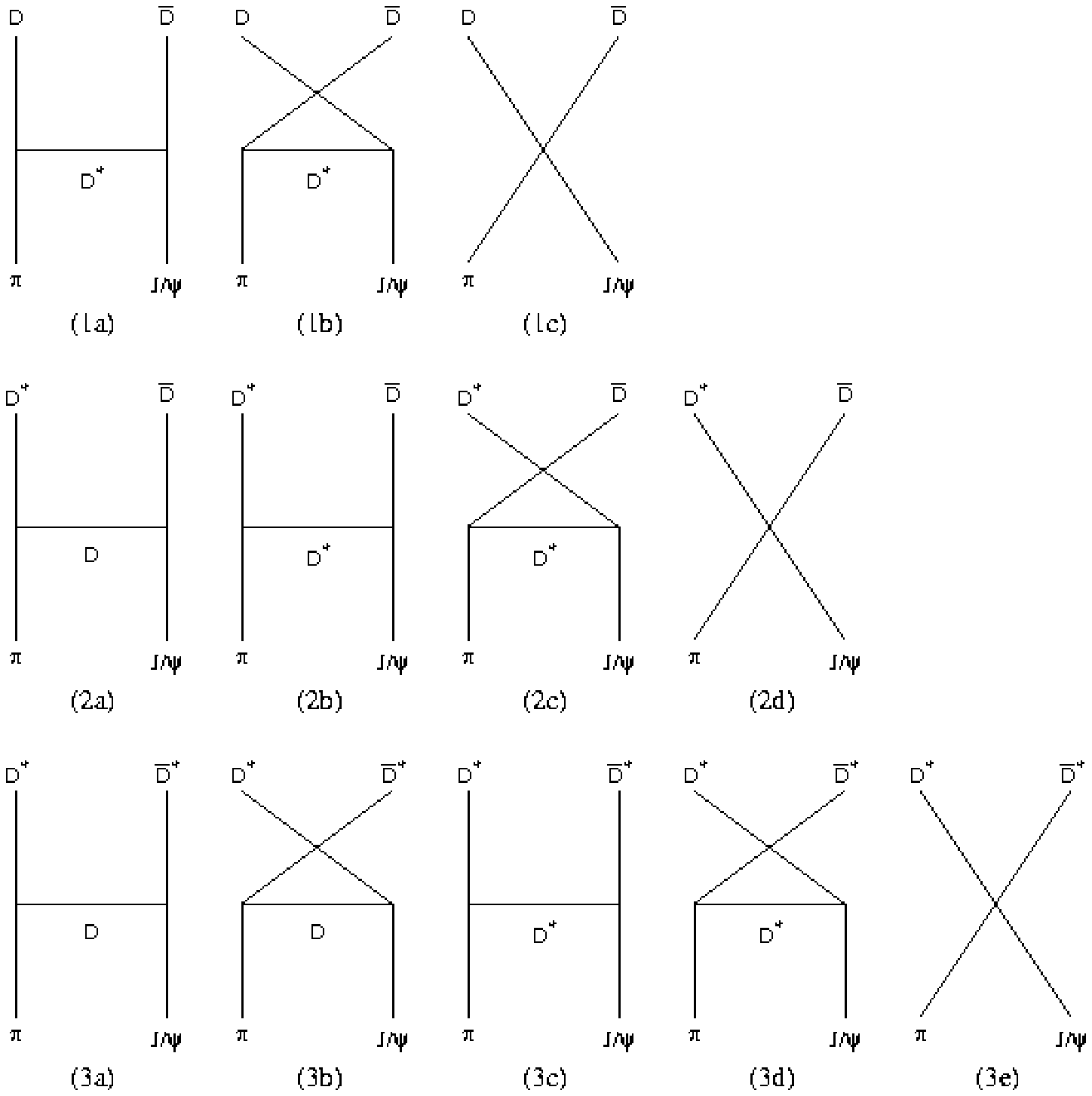}}
\caption{\label{gypsy:fig1} Feynman diagrams for $J/\psi$
absorption by the pion: (1) $J/\psi \pi \to D{\overline D}$, (2)
$J/\psi \pi \to {\overline D}D^*$ and $J/\psi \pi \to {\overline D}^* D^*$.}
\end{figure}

In this section, we present a preliminary report on a study of $J/\psi$ 
absorption due to its interaction with the hot hadronic
medium formed in relativistic heavy-ion collisions. We will give
the full analysis elsewhere~\cite{noi}. Here we concentrate on the strong 
couplings of the $J/\psi$ to low-mass 
charm mesons and pions. In the calculation of the relevant absorption
cross sections, tree-level diagrams such as those
depicted in  Fig.~\ref{gypsy:fig1} are encountered. 
Previous studies of these effects can be
found in \cite{Lin:1999ad,Haglin:2000ar,Oh:2000qr,Qiu:1998rz}. 
Besides the $DD^*\pi$
couplings, for which  both theoretical 
\cite{Colangelo:1994jc,Colangelo:1994es,Casalbuoni:1996pg}
and experimental \cite{Ahmed:2001xc} results are available, in 
Fig.~\ref{gypsy:fig1} the
$JD^{(*)}D^{(*)}$ and $JD^{(*)}D^{(*)}\pi$ couplings also appear
(in this section, we shall often use $J$ to indicate the $J/\psi$). These
couplings have been estimated by different methods, that are, in our
opinion, unsatisfactory. For example, the use of
SU(4) symmetry puts the charm quark and
the light quarks on the same footing, at odds with the results obtained
within the Heavy Quark Effective Theory (HQET), which treats the
charm as infinitely massive, $m_c\gg \Lambda_{\rm QCD}$ (see 
Ref.~\cite{Casalbuoni:1996pg}). 
Similarly, approaches
based on Vector Meson Dominance (VMD) should be considered critically
given the large extrapolation involved, $p^2=0\to m_{J/\psi}^2$.
A different evaluation, based on QCD Sum Rules, 
presents the typical theoretical
uncertainties of this method \cite{Matheus:2002nq,Duraes:2002px}. 
Here, we take
another approach, based on the Constituent Quark Model (CQM),
a quark-meson model which explicitly takes into account the
HQET symmetries.  For more details on the CQM see 
Ref.~\cite{Deandrea:1998uz,Polosa:2000ym}.

\begin{figure}[htbp]
\begin{center}
\epsfig{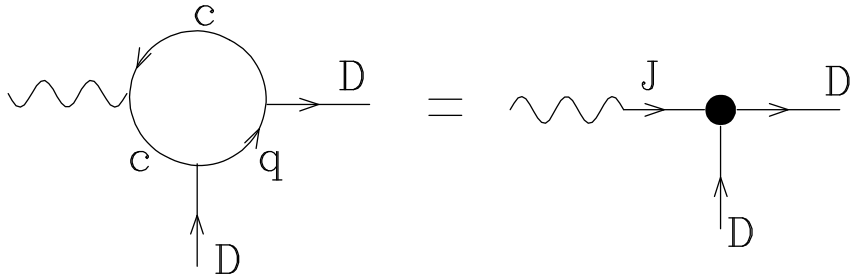}
\caption{\label{gypsy:fig2} The Vector Meson Dominance
equation for the coupling of the $J/\psi$ with $D$ and $D^*$ in terms of
the Isgur-Wise function $\xi$. The function $\xi$ on the left-hand side is
computed by a diagram with a quark loop. The coupling of each
$D^{(*)}$ meson to quarks is given by $\sqrt{Z_H m_D}$.}
\end{center}
\end{figure}

The CQM is particularly suitable for 
studies of exclusive heavy meson decays. Since its Lagrangian
contains the Feynman rules for the heavy-light vertices formed by a heavy
meson, transition amplitudes are
computable via simple constituent quark loop diagrams where
mesons enter as external legs. The model is relativistic and
incorporates the chiral SU(2) symmetry of the light quark sector
as well as the heavy quark symmetries. The calculation of the
$DD^*\pi$ coupling in the CQM can be found in
Ref.~\cite{Deandrea:1998uz,Polosa:2000ym}. Here we shall consider the 
calculation of the $JD^{(*)}D^{(*)}$ and $JD^{(*)}D^{(*)}\pi $ vertices 
within the CQM. This will be done by treating the $J/\psi$ with VMD, as
depicted in Fig.~\ref{gypsy:fig2}, and computing the Feynman diagrams 
prescribed by the CQM.

In the CQM, the evaluation of the loop diagram depicted on the
left-hand side of Fig.~\ref{gypsy:fig2} amounts to the calculation
of the Isgur-Wise function \cite{Deandrea:1998uz,Polosa:2000ym}. The result is 
\begin{equation} 
\xi(\omega)=Z_H\left[ \frac{2}{1+\omega}
I_3(\Delta_H)+
\left(m+\frac{2\Delta_H}{1+\omega}I_5(\Delta_H,\Delta_H,\omega)\right)\right]\
, \end{equation} 
where $m$ is the light constituent mass, $Z_H$ is a constant arising from the
$D^{(*)}$ coupling to their constituent quarks so that the coupling constant
is $\sqrt{Z_Hm_D}$ \cite{Deandrea:1998uz,Polosa:2000ym}, and the $I_i$ 
integrals are listed in sect.~\ref{DNPappendix}. 
Here $\omega = v \cdot v'$, the product of the two heavy quark
velocities, $v$ and $v'$, assumed to be equal to the $D^{(*)}$ velocities
in the infinite heavy quark mass limit.  The hadron-quark mass difference is 
$\Delta_H = m_D - m_c \approx
0.3-0.5$ GeV \cite{Deandrea:1998uz,Polosa:2000ym}.  This form of 
$\xi(\omega)$ for the $JDD$ coupling arises from the loop integral 
\begin{equation} \label{integral}
m_D Z_H\frac{iN_c}{16\pi^4}\int
d^4\ell \frac{{\rm Tr}\left[(\gamma\cdot\ell+m) \gamma_5
(1+\gamma\cdot v^\prime)\gamma_\mu (1+\gamma\cdot
v)\gamma_5\right]}{4 (\ell^2-m^2)(v\cdot
\ell+\Delta_H)(v^\prime\cdot\ell-\Delta_H)}\ , \end{equation} 
where $N_c = 3$ and $(1+\gamma\cdot v)/2 v\cdot k$ 
is the heavy quark propagator in HQET.  To calculate couplings for $D^*$ rather
than $D$, the factor $(-\gamma_5)$ in Eq.~(\ref{integral}) must be replaced by
$\gamma \cdot \epsilon$ where $\epsilon$ is the $D^*$ polarization.

The Isgur-Wise function obeys Luke's theorem, $\xi(1)=1$, 
arising from  the flavor symmetry of HQET, corresponding to the 
Ademollo-Gatto theorem~\cite{Ademolo:sr,Luke:1990eg} for light flavours.
The definition of the Isgur-Wise form factor is
\begin{equation} 
\langle H(v^\prime)|\overline c \gamma_\mu c
|H(v)\rangle=-\xi(\omega){\rm Tr}\left(\overline H\gamma_\mu H\right)\
\end{equation} 
where $H$ is the multiplet containing both the $D$ and the
$D^*$ mesons \cite{Casalbuoni:1996pg}, 
\begin{equation} 
H=\frac{1+\gamma\cdot v}{2}(-P_5\gamma_5+ \gamma\cdot P)\ \,\, ,
\end{equation} 
where $P_5$ and $P^\mu$ are charm meson
annihilation operators.  For example,
the transition between two pseudoscalar $D$ mesons is
\begin{equation} 
\langle D(v^\prime)|\overline c \gamma_\mu c
|D(v)\rangle=m_D\xi(\omega)(v+v^\prime)_\mu. 
\end{equation}
The Isgur-Wise function can be calculated in the CQM for any value of 
$\omega$, not only in
the region $\omega>1$, which is experimentally accessible via
semi-leptonic $B\to D^{(*)}$ decays.  We note that $\omega$ is related to
the meson momenta by
\begin{equation}
\omega=\frac{p_1^2+p_2^2-p^2}{2 \sqrt{p_1^2p_2^2}} \, \, ,
\end{equation}
where $p_1$ and $p_2$ are the momenta of the two $D$'s.

We now consider the right-hand side of Fig.~\ref{gypsy:fig2}.
We use the matrix element
\begin{equation} 
\langle 0|\overline c \gamma^\mu c
|J(q,\eta)\rangle=f_Jm_{J/\psi}\epsilon^\mu 
\end{equation} 
for the coupling of the $J/\psi$ to the current with $f_J=0.405\pm
0.014$ GeV.  The strong couplings $JD^{(*)}D^{(*)}$ are given in 
the following effective lagrangians, 
\begin{eqnarray} 
{\cal L}_{JDD}&=&ig_{JDD}\left(\overline{D}
{\stackrel{\leftrightarrow}{\partial}}_{\nu}D\right)J^\nu\ , \cr
{\cal L}_{JDD^*}&=&ig_{JDD^*}\epsilon^{\mu\nu\alpha\beta}J_{\mu}
\partial_{\nu}\overline D\partial_{\beta}D^*_\alpha\ ,\cr&&\cr
{\cal L}_{JD^*D^*}&=& ig_{JD^*D^*}\Big[ \overline{D}^{*\mu}\left(
{\partial}_{\mu}D^*_\nu\right)J^\nu - {D}^{*\mu} \left(
{\partial}_{\mu}\overline D^*_\nu\right)J^\nu 
- \left(
 \overline{D}^{*\mu}{\stackrel{\leftrightarrow}{\partial}}_{\nu}
D^*_\mu\right)J^\nu \Big]\ .
\end{eqnarray}

\begin{figure}[htbp]
\begin{center}
\epsfig{height=4truecm,figure=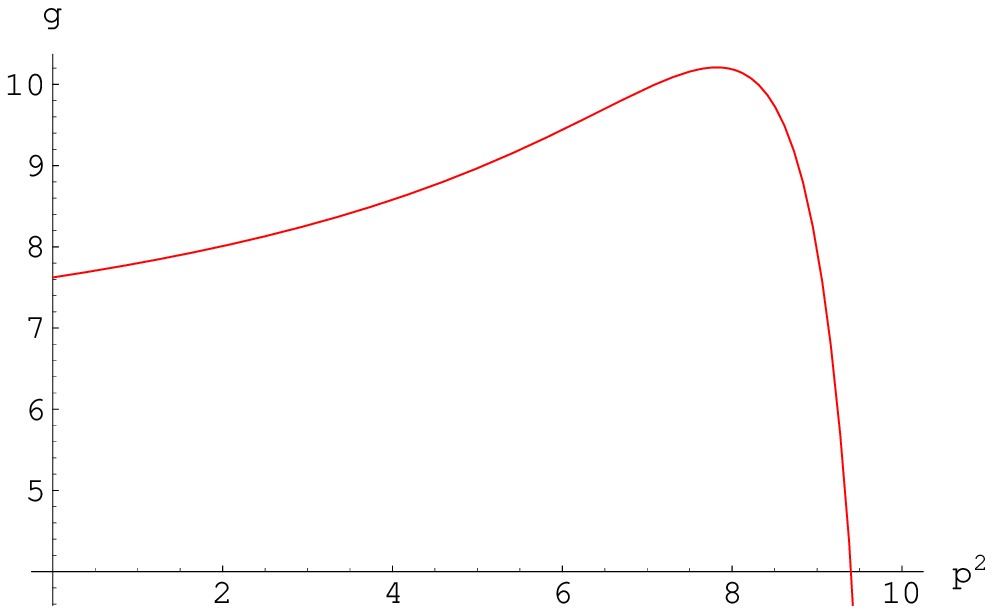}
\caption{\label{gypsy:fig3} The $p^2$ dependence of
$g=g_{JDD}(m_D^2,\,m^2_D,\,p^2)$, showing the almost complete
cancellation between the pole of $\xi(\omega)$ and the
kinematic zero.  The units of $p^2$ are GeV$^2$.}
\end{center}
\end{figure}

As a consequence of the HQET spin symmetry, we have \begin{eqnarray}
g_{JD^*D^*}&=&g_{JDD}\ ,\cr &&\cr
g_{JDD^*}&=&\frac{g_{JDD}}{m_D}\,\, .
\end{eqnarray} 
On the other hand, the VMD ansatz gives
\begin{equation} \label{eq:coupling1}
g_{JDD}(p_1^2,\,p_2^2,\,p^2)=\frac{m^2_{J/\psi}-p^2}{f_J
m_{J/\psi}} \, \xi(\omega) \ .
\end{equation} 
Since $g_{JDD}$ has no zeros, Eq.~(\ref{eq:coupling1}) shows that $\xi(\omega)$
must have a pole at $p^2=m^2_{J/\psi}$, as expected from
dispersion relation arguments. The CQM evaluation of $\xi$ does
show a strong peak at $p^2\approx(2m_c)^2$ even though, due to
${\cal O} (1/m_c)$ effects, the
location of the singularity is not exactly at $p^2=m^2_{J/\psi}$.
This is shown in Fig.~\ref{gypsy:fig3} where we plot
$g_{JDD}(p_1^2,\,p_2^2,\,p^2)$ for on-shell $D$ mesons as a
function of $p^2$ using $\Delta_H=0.4$~GeV and
$Z_H=2.36/$GeV. 
For $0<p^2<4$ GeV$^2$,
$g_{JDD}$ is almost constant, 
\begin{equation} 
g_{JDD}=8.0\pm 0.5\
.\label{gjdd}
\end{equation} 
For larger values of $p^2$ the method is
unreliable due to the incomplete cancellation
between the kinematical zero and the pole.  The distorted shape
around the $J/\psi$ pole suggests that the contribution of the
nearby $\psi(2S)$ pole could also be relevant. Therefore, we
extrapolate the smooth behavior of $g_{JDD}$ in the small $p^2$
region up to $p^2=m^2_{J/\psi}$ and assume the value of $g_{JDD}$ in 
Eq.~(\ref{gjdd}) for on-shell $J/\psi$.  On the
other hand, the behavior with $p_1^2$ and $p^2_2$ is smooth, 
compatible with that of a smooth form factor.
We finally note that Eq.~(\ref{gjdd}) agrees with
the result of the QCD sum rule analysis \cite{Matheus:2002nq,Duraes:2002px}. 
This is not surprising 
since QCD sum rules involve a perturbative part and a suppressed 
nonperturbative contribution.  The perturbative term has its
counterpart in the CQM loop calculation of Fig.~\ref{gypsy:fig2} with an
overall normalization that should agree with the CQM as a consequence of Luke's
theorem.

We now turn to the $JD^{(*)}D^{(*)}\pi$ couplings. As
discussed in Refs.~\cite{Casalbuoni:1996pg,Falk:1990yz}, the leading
contributions to the current matrix element, $\langle
H(v^\prime)\pi |\overline c \gamma^\mu c|H(v)\rangle$, in the soft pion
limit (SPL) are the pole diagrams. Technically,
in the SPL, the reducing action of a pion derivative in the matrix
element is compensated in the polar diagrams by the effect of the
vanishing denominator in the combined $q_\pi\to
0,\,m_c\to\infty$ limit. Since the effect of the pole diagrams is
explicitly taken account in Fig.~\ref{gypsy:fig1}, we do not include any
further contributions.  To be definite, we consider the coupling $g_{JDD\pi}$
which can be obtained  by a VMD ansatz similar to Fig.~\ref{gypsy:fig2}. 
Now the left-hand side is modified by the insertion of a soft pion 
on the light quark line with the 
coupling $q_\pi^\mu/f_\pi \gamma_\mu\gamma_5$. 
We call $\xi^\pi(\omega)$ the analogous form factor in the SPL,
\begin{equation} 
\xi^\pi(\omega)=Z_H\left[
\frac{4m+2\Delta_H}{1+\omega}I_4(\Delta_H)-
\left(m^2+\frac{2\Delta_H^2+4m\Delta_H}{1+\omega}\right)
\frac{\partial I_5(\Delta_H,\Delta_H,\omega)}{\partial m^2}
\right] \, \, .
\end{equation} 
The VMD ansatz of Fig.~\ref{gypsy:fig2} gives 
\begin{equation} 
{\cal L}_{JDD\pi}=ig_{JDD\pi}\epsilon^{\mu\nu\alpha\beta}J_{\mu}
\partial_{\nu}D\partial_{\alpha}\overline{D}\partial_{\beta}\pi
\end{equation}
with 
\begin{equation}
\label{eq:coupling2}
g_{JDD\pi}(p_1^2,\,p_2^2,\,p^2)=\frac{(m_J^2-p^2)\xi^{\pi}
(\omega)}{f_\pi f_J m_D m_J} \, \, . \end{equation}

\begin{figure}[htbp]
\begin{center}
\epsfig{height=4truecm,figure=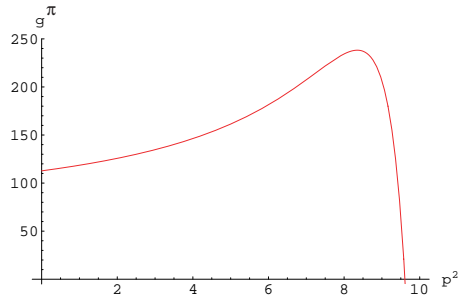}
\caption{\label{gypsy:fig4} The $p^2$ dependence of
$g^\pi=g_{JDD\pi}(m_D^2,\,m^2_D,\,p^2)$. As in Fig.~\protect\ref{gypsy:fig3} 
there is an almost
complete cancellation between the pole of the form factor and the
kinematic zero. The units on $p^2$ are GeV$^2$ and GeV$^{-3}$ for
$g_{JDD\pi}$.}
\end{center}
\end{figure}

In Fig.~\ref{gypsy:fig4} we plot our result for $g_{JDD\pi}$ with 
on-shell $D$ mesons.
By the same arguments used to determine $g_{JDD}$, we find,
with all mesons on mass-shell, 
\begin{equation}
g_{JDD\pi}=125\pm 15~{\rm GeV}^{-3} \, \, . \label{gddjp} 
\end{equation} 
We now compare this result with the effective $JDD\pi$ coupling obtained 
by a polar diagram with an intermediate $D^*$ state.  In this case,
\begin{equation}
g_{JDD\pi}^{\rm polar}\approx\frac{(g_{JDD^*})(g_{D^*D\pi})}{2q_\pi\cdot p_D}
\ \, .
\end{equation} 
Since all the calculations are valid in
the SPL, only pion momenta of up to a few hundred MeV should be considered. 
Using the result
$g_{D^*D\pi}=2m_D/f_\pi g $, with $g=0.59\pm 0.01\pm 0.07$ 
\cite{Ahmed:2001xc}, we find
$g_{JDD\pi}^{\rm polar}\approx 393$, 196, and 98 GeV$^{-3}$ for $|q_\pi| = 
50$, 100, and 200 MeV respectively. This analysis shows
that, within the region of validity of the model,  in spite of the
rather large value of the coupling in Eq.~(\ref{gddjp}), the diagrams
containing this coupling are in general suppressed. 
Similar conclusions are
reached for $D^*$'s.

Finally, we discuss the kinematical limits of our approach. To
allow the production of a $D^{(*)}\overline D^{(*)}$ pair, as shown in 
Fig.~\ref{gypsy:fig1}, we must extend the region of validity of the model 
beyond the SPL since the $D^{(*)}\overline D^{(*)}$ threshold is $|\vec
q_\pi|=700-1000~$MeV. The CQM, along with other models, is a chiral model 
which puts limits on the pion momenta. Therefore it is necessary to 
include a form factor enhancing the
small pion momenta region, for example 
\begin{equation} 
f(|\vec q_\pi|)=\frac{1}{1+ (|\vec q_\pi|/m_\chi)} \, \, .
\label{f}
\end{equation} 
A similar form factor is considered in Ref.~\cite{Lin:1999ad} but
with a different motivation. Here we introduce it to ensure the
validity of our approach.  In this sense, the cross sections we can
compute may be considered as lower bounds.
Since the main effect of Eq.~(\ref{f}) should be to reduce
contributions from pion momenta larger than a few hundred MeV, we
expect $400< m_\chi < 600$ MeV. This
choice implies that the direct couplings in Fig.~\ref{gypsy:fig1}, diagrams 1c,
2d and 3e, should not dominate the final result since their
contribution is larger where the form factor is more effective.

\subsubsection{Definitions of the integrals}
\label{DNPappendix}

We list the expressions used to compute the integrals
$I_i$ used in the text. The ultraviolet cutoff $\Lambda$, the
infrared cutoff, $\mu$, and the light constituent mass, $m$, are 
$\Lambda=1.25$~GeV, $\mu=0.3$~GeV and
$m=0.3$~GeV \cite{Deandrea:1998uz,Polosa:2000ym}.

\begin{eqnarray} I_3(\Delta) &=& - \frac{iN_c}{16\pi^4} \int^{\mathrm
{reg}} \frac{d^4k}{(k^2-m^2)(v\cdot k + \Delta +
i\epsilon)}\nonumber \\ &=&{N_c \over {16\,{{\pi }^{{3/2}}}}}
\int_{1/{{\Lambda}^2}}^{1/{{\mu }^2}} {ds \over {s^{3/2}}} \; e^{-
s( {m^2} - {{\Delta }^2} ) }\; [1 + {\mathrm {erf}}
(\Delta\sqrt{s})] \\
I_4(\Delta)&=&\frac{iN_c}{16\pi^4}\int^{\mathrm {reg}}
\frac{d^4k}{(k^2-m^2)^2 (v\cdot k + \Delta + i\epsilon)}
\nonumber\\ &=&\frac{N_c}{16\pi^{3/2}}
\int_{1/\Lambda^2}^{1/\mu^2} \frac{ds}{s^{1/2}} \;
e^{-s(m^2-\Delta^2)} \; [1+{\mathrm {erf}}(\Delta\sqrt{s})] \\
I_5(\Delta_1,\Delta_2,\omega) &= & \frac{iN_c}{16\pi^4}
\int^{\mathrm {reg}} \frac{d^4k}{(k^2-m^2)(v\cdot k + \Delta_1 +
i\epsilon ) (v'\cdot k + \Delta_2 + i\epsilon )} \nonumber \\
 & = & \int_{0}^{1} dx \frac{1}{1+2x^2 (1-\omega)+2x
(\omega-1)} \nonumber\\ 
&& \times \Big[
\frac{6}{16\pi^{3/2}}\int_{1/\Lambda^2}^{1/\mu^2} ds~\sigma_\Delta \;
e^{-s(m^2-\sigma_\Delta^2)} \; s^{-1/2}\; [1+ {\mathrm {erf}}
(\sigma_\Delta \sqrt{s})] \nonumber\\
&&+ \frac{6}{16\pi^2}\int_{1/\Lambda^2}^{1/\mu^2} ds \;
e^{-s\sigma_\Delta^2}\; s^{-1}\Big] \, \, , \end{eqnarray} 
where
\begin{equation}
\sigma_\Delta \equiv 
\sigma(x,\Delta_1,\Delta_2,\omega)={{{\Delta_1}\,\left( 1 - x
\right)  + {\Delta_2}\,x}\over {{\sqrt{1 + 2\,\left(\omega -1
\right) \,x + 2\,\left(1-\omega\right) \,{x^2}}}}} \, \, .
\end{equation}

\subsection[Heavy Quark Energy Loss in QCD Matter]
{Heavy Quark Energy Loss in QCD Matter~\protect
\footnote{Author: D. Kharzeev.}}
\label{section:eloss}

The study of heavy quark production in nuclear collisions 
will allow extraction of valuable information on the gluon densities of 
the colliding nuclei. 
In addition, since the heavy quarks propagate through the 
quark--gluon matter created in such collisions, they can be used to 
probe the properties of the dense matter. The 
practical importance of heavy quark energy loss to the
charmed hadron and lepton spectra has been clearly established
\cite{Shuryak:1996gc,Lin:1997cn,Lin:1998bd,Mustafa:1997pm,Srivastava:2002kg}.
   
The idea of heavy quark energy loss has attracted considerable
attention. In particular, collisional energy loss in quark--gluon
plasma has been evaluated \cite{Braaten:1991we,Thoma:1990fm,Baier:1999dz} 
using finite--temperature QCD, see Ref.~\cite{Blaizot:2001nr} for a 
comprehensive review. However, the energy loss of fast partons in medium is
dominated by gluon radiation
\cite{Gyulassy:1993hr,Baier:1994bd,Zakharov:1996fv}.  
Radiative energy loss by heavy quarks was
evaluated in Ref. \cite{Dokshitzer:2001zm}.  
It was shown \cite{Dokshitzer:2001zm} that the ``dead cone effect'' in 
the radiation of a heavy quark significantly 
suppresses its energy loss while making the calculation more reliable by 
reducing the sensitivity to the infrared region. 

In fact, the first measurements of charm production in Au+Au 
collisions at RHIC \cite{Adcox:2002cg,Averbeck:2002nz} indicated very small, 
if any, nuclear effect on the shape of charm transverse momentum 
distributions.  The smaller loss predicted 
in Ref.~\cite{Dokshitzer:2001zm} provides 
a plausible and economical explanation of the observed phenomenon 
\cite{Gallmeister:2002tv,Batsouli:2002qf}.  
The ``dead cone'' effect has been taken into account 
in the studies of charm and bottom production at the LHC \cite{Lokhtin:2002wu}.

In the following we give a brief and qualitative explanation of the results of 
Ref.~\cite{Dokshitzer:2001zm}. 
We first recall the basic features of gluon radiation caused by
propagation of a fast parton (quark) through a QCD medium.

As pointed out in Ref.~\cite{Baier:1994bd}, the accompanying radiation is
determined by multiple rescattering of the radiated gluon in the
medium.  The gluon, during its formation time,
\begin{equation} 
   t_{\rm form} \simeq \frac{\omega}{k_T^2}\,,
\label{form}
\end{equation}
accumulates a typical transverse momentum
\begin{equation} 
   k_T^2 \simeq \mu^2 \ {t_{\rm form} \over \lambda}, \label{walk}
\end{equation}
where $\lambda$ is the 
mean free path and $\mu^2$ the characteristic momentum transfer
squared in a single scattering.  This is the random walk pattern with
an average number of scatterings given by
$t_{\rm form}/\lambda$.

Combining Eqs.~(\ref{form}) and (\ref{walk}) we obtain
\begin{equation}
\label{Ncoh}
  N_{\rm coh}=\frac{t_{\rm form}}{\lambda} = 
\sqrt{\frac{\omega}{\mu^2\,\lambda}}\,,
\end{equation}
the number of scattering centres which participate, {\em
coherently}, in the emission of the gluon with a given energy
$\omega$.  For sufficiently large gluon energies,
$\omega>\mu^2\lambda$, when the coherent length exceeds the mean free
path, $N_{\rm coh}>1$.  In this situation, the standard Bethe-Heitler
energy spectrum per unit length describing {\em independent}\/
emission of gluons at each scattering centre is suppressed:
\begin{equation}\label{spec}
 \frac{dW}{d\omega dz} = \frac{1}{N_{\rm coh}} \left(
\frac{dW}{d\omega dz}\right)^{\rm BH} 
= \frac{\alpha_s C_R}{\pi\omega\,\lambda} \sqrt{\frac{\mu^2\,\lambda}{\omega}} 
=  \frac{\alpha_s C_R}{\pi\omega} \sqrt{\frac{\hat{q}}{\omega}} \, \, .
\end{equation}
Here $C_R$ is the ``colour charge'' of the parton projectile
($C_R=C_F=(N_c^2 - 1)/2 N_c = 4/3$ for quarks).

In Eq.~(\ref{spec}) we have substituted the characteristic ratio 
$\mu^2/\lambda$ by the gluon 
{\em transport coefficient}\/
\cite{Baier:1996sk},
\begin{equation}
 \hat{q} \equiv 
 \rho \ \int {d \sigma \over dq^2}\ q^2\ dq^2 \, \, , \label{qhat}
\end{equation}
proportional to the density $\rho$ of the scattering centres
in the medium.  The transport coefficient characterizes 
the typical momentum transfer in gluon
scatterings off these centres.

An important feature of medium-induced radiation is the relation
between the transverse momentum and the energy of the emitted gluon.
Indeed, from Eqs.~(\ref{form}) and (\ref{walk}), see also Eq.~(\ref{qhat}),
we derive
\begin{equation}
  k_T^2 \simeq \sqrt{\hat{q}\ \omega} \label{star} \, \, .
\end{equation} 
Thus the angular distribution of gluons with a given energy
$\omega$ is concentrated at a characteristic energy- and medium-
dependent emission angle,
\begin{equation}\label{angle}
\theta \simeq \frac{k_T}{\omega} 
\sim \left(\frac{\hat{q}}{\omega^3}\right)^{1/4}. 
\end{equation}

Gluon bremsstrahlung off a heavy quark differs from the case of a
massless parton produced at the same hardness scale
in one respect: gluon radiation is suppressed at angles smaller than
the ratio of the quark mass $m_Q$ to its energy $E$.  
Indeed, the distribution of soft gluons radiated by a heavy quark is
given by
\begin{equation}
dP_{\rm HQ}  = {\alpha_s\ C_F \over \pi}\ 
{d\omega \over \omega}\ {k_T^2 \,dk_T^2\over 
(k_T^2 + \omega^2 \theta_0^2)^2} \, \, , \qquad
\theta_0\equiv\frac{m_Q}{E} \, \, ,
\label{dist}
\end{equation}
where 
the strong coupling constant $\alpha_s$ should be evaluated at the scale
determined by the denominator of Eq.~(\ref{dist}).  Equating 
$k_T$ with $\omega\theta$ in the
small-angle approximation, we
conclude that Eq.~(\ref{dist}) differs from the standard
bremsstrahlung spectrum,
\begin{equation}
 dP_0\> \simeq \> \frac{\alpha_s\,C_F}{\pi}
 \frac{d\omega}{\omega}\,\frac{dk_T^2}{k_T^2}
\> =\> \frac{\alpha_s\,C_F}{\pi} \frac{d\omega}{\omega}\,
\frac{d\theta^2}{\theta^2} \, \, ,
\end{equation}
by the factor 
\begin{equation}\label{factor}
 dP_{\rm HQ} = dP_0 \left( 
1+\frac{\theta_0^2}{\theta^2}\right)^{-2} \, \, .
\end{equation}
This effect is known as the ``dead cone'' phenomenon.  Suppression of
small-angle radiation has a number of interesting implications, such
as the perturbative calculability of, and nonperturbative $\Lambda/m_Q$
corrections to, heavy quark fragmentation
functions~\cite{Dokshitzer:1995ev,Nason:1996pk}, 
multiplicity and energy spectra of light particles accompanying hard
production of a heavy quark~\cite{Schumm:1992xt,Dokshitzer:fc}.

In the present context we should compare the angular distribution
of gluons induced by the quark propagation in the medium with the
size of the dead cone.
To this end, for the sake of a semi-quantitative estimate, we 
substitute the characteristic angle, Eq.~(\ref{angle}), into 
the dead cone suppression factor, Eq.~(\ref{factor}), and combine it with
the radiation spectrum, Eq.~(\ref{spec}), to arrive at
\begin{equation}
I(\omega) = \omega {d W \over d \omega} = {\alpha_s\ C_F \over \pi}
\sqrt{{\omega_1 \over 
\omega}} \ {1 \over (1 + (\ell\, \omega)^{3/2})^2} \, \, , 
\label{eq:spechq}
\end{equation} 
where
\begin{equation}
  \ell \equiv  \hat{q}^{-1/3}\ \left({m_Q \over E}\right)^{4/3}. \label{apar}
\end{equation}

To determine whether the finite quark mass essentially affects the medium-%
induced gluon yield, we need to estimate the product $\ell\omega$ for
the maximum gluon energy, $\omega\simeq\omega_1$, achievable in
Eq.~(\ref{spec}),
\begin{equation}
 \ell\omega_1= \hat{q}^{-1/3}\ \left({m_Q \over E}\right)^{4/3}
\hat{q}L^2
= \left(\frac{E_{\rm HQ}}{E}\right)^{4/3} \, , \qquad 
E_{\rm HQ} \equiv m_Q\sqrt{\hat{q}L^3} \, \, .
\end{equation}
where $L$ is the path length through the medium.
Thus the quark mass becomes irrelevant when its
energy exceeds the characteristic value, $E_{\rm HQ}$, dependent on the
size of the medium and on its ``scattering power'' embodied in the
transport coefficient $\hat{q}$.

Which regime is realized in the experiments on heavy quark 
production in nuclear collisions? 
Taking $m_c = 1.5$ GeV, we estimate \cite{Dokshitzer:2001zm}
\begin{eqnarray}
\label{EHQc}
E_{\rm HQ}^{\rm cold} &=& 
\sqrt{\hat q_{\rm cold}}\ L^{3/2}\ m_c \simeq \ 20\, {\rm GeV}
\left({L \over 5\, {\rm fm}}\right)^{3/2} \, , \\
\label{EHQh}
E_{\rm HQ}^{\rm hot} &=& \sqrt{\hat q_{\rm hot}}\ L^{3/2}\ m_c \simeq \ 
92\,  {\rm GeV} \left({L \over 5\, {\rm fm}}\right)^{3/2},
\end{eqnarray}
for cold and hot matter, respectively.  We observe that, in practice,  
$E \ll E_{\rm HQ}$ for the
transverse momentum (energy) distributions of heavy mesons, especially in the
hot medium.  We thus conclude that the pattern of medium-induced gluon
radiation appears to be {\em qualitatively different for heavy and
light quarks}\/ in the kinematic region of interest.

The issue of in-medium quenching of inclusive particle spectra was
addressed in Ref.~\cite{Baier:2001yt}.  The $p_T$ spectrum is
given by the convolution of the transverse momentum distribution in an
elementary hadron--hadron collision, 
evaluated at a shifted value $p_T+\epsilon$, with the 
distribution $D(\epsilon)$ in the energy $\epsilon$ lost 
by the quark to the medium-induced gluon radiation:
\begin{equation}
{d \sigma^{\rm med} \over d p_T^2} = \int d \epsilon \ D(\epsilon)\ 
{d \sigma^{\rm vac}(p_T + \epsilon ) \over d p_T^2}
\equiv {d \sigma^{\rm vac}(p_T) \over d p_T^2} \, Q(p_T),
  \label{defpt}
\end{equation}
where $Q(p_T)$ is a medium-dependent {\em quenching factor}.  Because in the
region of interest, $\epsilon \ll p_T$, when the vacuum cross section is a 
steeply-falling function, the calculation of the quenching factor $Q$ can be
simplified by approximating the $\epsilon$-integral in
Eq.~(\ref{defpt}) as an exponential,
\begin{equation}\label{Qfint}
 Q(p_T) \simeq \int d \epsilon \ D(\epsilon)\ \exp\left\{ 
\frac{\epsilon}{p_T}  {\cal{L}} \right\} \, \, , \qquad 
{\cal{L}} \equiv \frac{d}{d\ln p_T} 
\ln \left[ {d \sigma^{\rm vac}(p_T) \over d p_T^2} \right] \, \, .
\end{equation} 

\begin{figure}[h]
\begin{center}
\epsfig{file=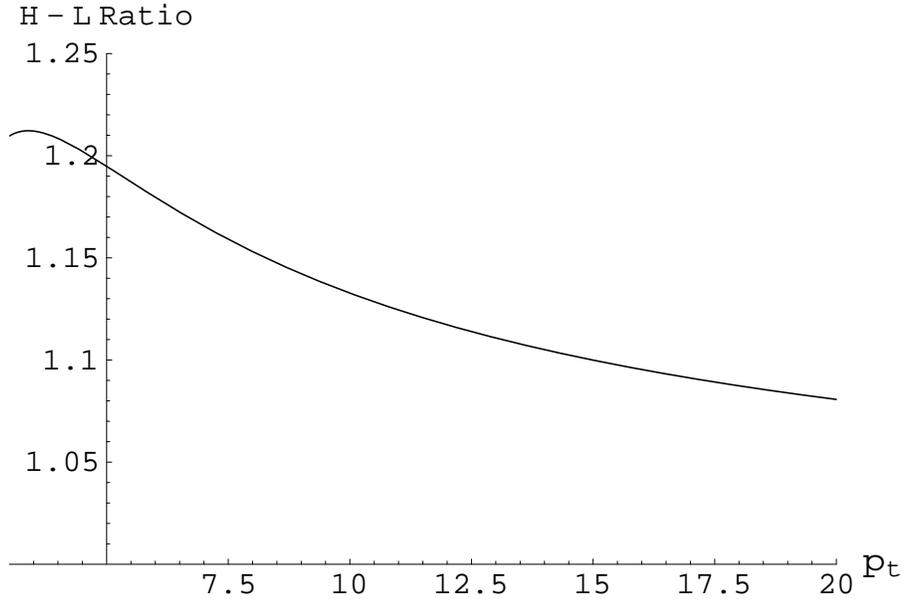, width=12cm}
\end{center}
\caption{The ratio of quenching factors $Q_H(p_T)/Q_L(p_T)$ 
for charm and light quarks in cold nuclear matter
with $\hat{q}=0.01$ GeV$^3$ and $L=5$ fm.  From Ref.~\cite{Dokshitzer:2001zm}.
}
\label{Fcold}
\end{figure}

According to Eq.~(\ref{EHQh}), we expect a larger quenching ratio for
a hot medium. Indeed, the $D/\pi$
ratio should become significantly enhanced relative to $pp$
collisions.  Assuming a fixed path length traversed by the quarks in the hot
medium, $L=5$ fm, we find a factor of $\sim 2$ enhancement at
$p_{T} \sim 5- 10$ GeV.

The following result for the heavy quark quenching factor 
was derived in Ref.~\cite{Dokshitzer:2001zm}: 
\begin{equation}
Q_H(p_T) \simeq \exp \left[- {2 \alpha_s C_F \over \sqrt{\pi}}\ 
L\,\sqrt{\hat{q}\frac{{\cal{L}}_H}{p_T}} \,
 + \,
{16 \alpha_s C_F \over 9 \sqrt{3}} L
\left( \frac{ \hat{q}\> \> m_Q^2}{m_Q^2+p_T^2}\right)^{1/3}  \right] \, \, .
\label{finres}
\end{equation}
The first term in the exponent in Eq.~(\ref{finres}) represents the
quenching of the transverse momentum spectrum, universal for
light and heavy quarks,
\[
Q_L(p_T) \simeq \exp \left[- {2 \alpha_s C_F \over \sqrt{\pi}}\ 
L\,\sqrt{\hat{q}\frac{{\cal{L}}_L}{p_T}}\, \right] 
\]
modulo the difference in ${\cal{L}}$ 
determined by the vacuum $p_T$ distributions.  The second
term in Eq.~(\ref{finres}) is specific to heavy quarks. 
It has a positive sign so
that the suppression of the heavy hadron $p_T$ distributions is
always smaller than that of the light hadrons. This is a
straightforward consequence of the fact that the heavy quark mass
suppresses gluon radiation. At very high transverse momenta both
terms vanish in accord with the QCD factorization theorem which states
that medium effects should disappear as $p_T
\to \infty$. How fast this regime is approached depends, however, on
the properties of the medium encoded in the transport
coefficient $\hat{q}$ and the path length $L$.

\begin{figure}[h]
\begin{center}
\begin{minipage}{12cm}
\begin{center}
\epsfig{file=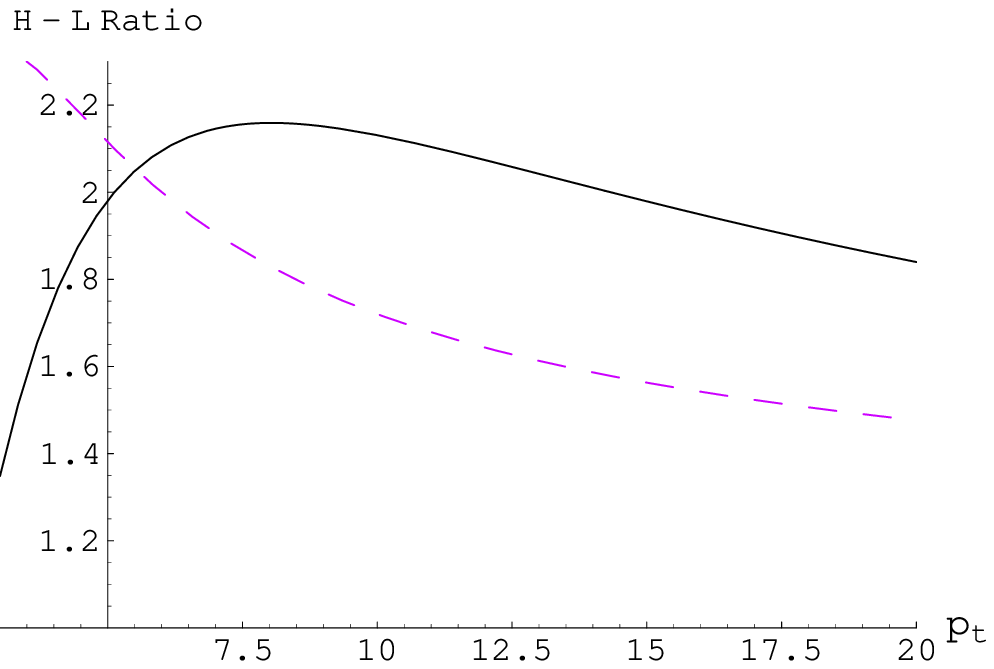, width=11cm}
\end{center}
\end{minipage}

\begin{minipage}{12cm}
\begin{center}
\epsfig{file=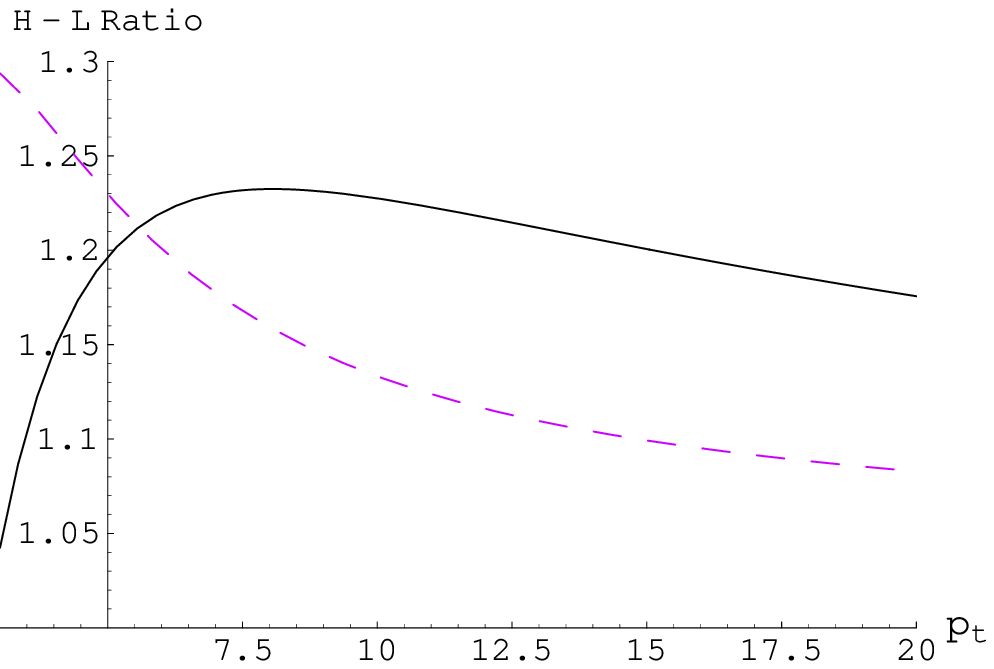, width=11cm}
\end{center}
\end{minipage}
\caption{The ratio of quenching factors $Q_H(p_{T})/Q_L(p_{T})$ 
for charm and light quarks 
in hot matter with $\hat{q}=0.2$ GeV$^3$ ($L=5$ fm upper panel,
$L=2$ fm lower panel).  Solid lines correspond to unrestricted
gluon radiation, while the dashed lines are based on the calculation
with the cut $\omega>0.5$ GeV on gluon energies. From 
Ref.~\cite{Dokshitzer:2001zm}.}
\label{Fhot}
\end{center}
\end{figure}

Constructing the ratio $Q_H$ to $Q_L$, the
heavy-to-light quenching should be enhanced by
\begin{equation}
\frac{Q_H(p_T)}{Q_L(p_T)} \>\simeq\> 
 \exp \left[ {16 \alpha_s C_F \over 9 \sqrt{3}} L
\left( \frac{ \hat{q}\> \> m_Q^2}{m_Q^2+p_T^2}\right)^{1/3}  \right] \, \, .
\label{eq:ratio}
\end{equation}
This simple expression provides a reasonably good approximation to the
more accurate quantitative results presented in Ref.~\cite{Dokshitzer:2001zm}.

As discussed above, the quenching of heavy hadron $p_T$ distributions
caused by QCD matter is much weaker than for
pions because the gluon cloud around the heavy quark is
``truncated'' by the large quark mass in a medium-dependent way. 
This interesting effect can be illustrated
by the transverse momentum dependence of the ratio of hadrons
originating from heavy and light quark fragmentation such as the $D/\pi$ 
ratio in heavy ion collisions.

Fig.~\ref{Fcold} shows the ratio of quenching factors, Eq.~(\ref{eq:ratio}),
for heavy and light quarks in cold nuclear matter, $L=5$ fm, 
relevant to high-$p_T$ particle production in $pA$
collisions.  A small value of the transport coefficient, $\hat q_{\rm cold}
\simeq 0.01$ GeV$^3$, translates into a $\sim$ 15\% enhancement, reduced to
$\approx 1$\% for $L=2$ fm.

We only present the {\em ratio}\/ of quenching factors because, as pointed
out in Ref.~\cite{Baier:2001yt}, the {\em absolute}\/ magnitude of the 
quenching turns out
to be extremely sensitive to gluon radiation in the few-hundred MeV
energy range. Thus the magnitude cannot be quantitatively predicted without
a detailed understanding of the spectral properties of the medium.

The heavy-to-light ratio, however, proves to be much less sensitive to
the infrared region since gluon radiation off heavy and light quarks
is universal in the $x\to 0$ limit, see Eq.~(\ref{eq:spechq}). To
illustrate this point, in Fig.~\ref{Fhot} we show
the ratio of quenching factors calculated with the gluon energies restricted
to $\omega>500$ MeV.  We see that, in the
$5-10$ GeV range of $p_T$, the ratio is modified by
$20-30$~\%.\footnote{The quenching factors themselves change
(increase) by an order of magnitude when the radiation of gluons with
energies smaller than 500 MeV is vetoed.}

Clearly, detailed calculations have to be performed before a
reliable estimate of the magnitude of the $D/\pi$ enhancement can be
presented.  Nevertheless, the $D/\pi$ and $B/\pi$ ratios appear to be extremely
sensitive to the density of colour charges in QCD matter. 
Of special interest is the $B/D$ ratio, for which these calculations
become even less sensitive to the infrared region and thus more stable.

%% file: kinetic.tex
\section[QUARKONIUM FORMATION FROM UNCORRELATED QUARK-ANTIQUARK PAIRS]
{QUARKONIUM FORMATION FROM UNCORRELATED QUARK-ANTIQUARK 
PAIRS~\protect\footnote{Author: R.~L.~Thews.}}
\label{sec:thews}
\subsection{Introduction}
\label{thews.intro}

The goal of this section is to assess the possibility that
quarkonium  production rates
may be enhanced in nucleus-nucleus interactions at the LHC
relative to that predicted by extrapolation of processes thought
to be dominant at lower energy.  This enhancement could follow from the
effects of incoherent
recombination mechanisms involving uncorrelated pairs of heavy
quarks and antiquarks which result from multiple pair production.
Two different approaches have been considered:
statistical hadronization and kinetic formation. Updated predictions
relevant to Pb+Pb collisions at the LHC are given.

The utility of heavy quarkonium
production rates in nuclear collisions as
a signature of color deconfinement
was proposed more than 15 years ago \cite{Matsui:1986dk}.
Since one expects that the long-range color confining potential will be
screened in a deconfined medium, the quark and antiquark constituents of
bound states will be liberated.  As the system expands and cools, these
constituents will, in general, diffuse away from each other
to separations  larger than typical hadronic dimensions.
When the confining potential reappears, a given heavy
quark will not be able to ``find'' its heavy antiquark partner and form heavy
quarkonium.  It must then bind with one of the antiquarks within range
at hadronization.  Since these antiquarks are predominantly the lighter
$u$, $d$, and $s$ flavors, the final hadronic states will preferentially be
those with ``open'' heavy flavor.  The result will be a
decreased population of heavy quarkonium relative to that which would
have formed if a region of deconfinement had not been present.  This
scenario as applied to the charm sector is known as $\J$ suppression.

At LHC energy, perturbative QCD estimates predict that
hundreds of pairs of charm-anticharm quarks will be produced
in a central lead-lead collision.
This situation provides a ``loophole'' in the Matsui-Satz
argument~\cite{Matsui:1986dk} 
since there will be  copious numbers of heavy antiquarks in the
interaction region with which any given heavy quark may combine.  
In order for this to
happen, however, one must invoke a physical situation in which
quarkonium states can be formed from {\em all combinations}
of heavy quarks and antiquarks.
This of course
would be expected to be valid in the case that a space-time region
of color deconfinement is present, but it is not necessarily limited to
this possibility.  

One can make a model-independent estimate of how
such a ``recombination" mechanism would depend on nuclear collision
observables. For a given charm quark,
the probability $\cal{P}$ to form a $\J$ is
proportional
to the number of available anticharm quarks relative to the number of
light antiquarks,
\begin{equation}
{\cal{P}} \propto \frac{\Ncbar}{N_{\overline u, \overline d, \overline s}}
\propto \frac{\Nccbar}{N_{\rm ch}} \, \, .
\end{equation}
In the second step, we have replaced the number of available anticharm
quarks by the total number of pairs initially produced, assuming that
the total number of bound states formed remains a small fraction of the total
$c \overline c$ production.
We normalize the number of light antiquarks by the number of
produced charged hadrons.
Since this probability  is generally very small, one can simply multiply by the
total number of charm quarks, $\Nc$, to obtain the number of
$\J$ expected in a given event,
\begin{equation}
\NJ \propto \frac{{\Nccbar}^2}{N_{\rm ch}} \, \, ,
\label{eqquadratic}
\end{equation}
where the use of the initial values $\Nccbar = \Nc = \Ncbar$ is again justified
by the relatively small number of bound states formed.

The essential property of this result is that the growth of $\NJ$, quadratic
in the total number of charm quarks, with energy \cite{Gavai:1994gb}
is expected to be much faster than
the growth of total particle production in heavy ion collisions
\cite{Bazilevsky:2002fz}.   Without this quadratic mechanism, $\J$ production 
is typically some small energy-independent fraction of total initial charm
production \cite{Gavai:1994in}.  We thus anticipate that the quadratic
formation will become dominant at sufficiently high energy.
Generic estimates of the significance of this type of formation
process can be made \cite{Thews:2001hy}.  Here we
look at specific predictions of two models -- statistical hadronization
and kinetic formation, considered in sects.~\ref{thews.stat} 
and~\ref{thews.kinetic} respectively -- which share the above
properties, and we update the expectations to LHC energies.

\subsection{Statistical Hadronization}
\label{thews.stat}

The statistical hadronization model is 
motivated by the successful fits of relative abundances
of light hadrons produced in high energy heavy ion interactions
according to a hadron gas in chemical and thermal equilibrium
\cite{Braun-Munzinger:2001ip}.  Extension of the model 
to hadrons containing heavy quarks
underpredicts the observed abundances.  This effect may be
attributed to the long time scales associated with
thermal production and annihilation of heavy quarks.
The statistical hadronization model as first formulated 
for charm quarks \cite{Braun-Munzinger:2000px} 
assumes that the $c \overline c$ pairs produced in the initial hadronic
interactions survive until their subsequent hadronization, at which time they
are distributed into hadrons according
to the same thermal equilibrium parameters that fit the light
hadron abundances.  Chemical equilibrium abundances are 
adjusted by a factor $\gamma_c$ which accounts for 
the non-thermal 
heavy quark density. One power of this factor multiplies a given 
thermal hadron population for each heavy quark or antiquark
contained in the hadron.  Thus the relative abundance of the
$\J$  to that of $D$ mesons, for example, may be enhanced in this model.

The value of $\gamma_c$ is determined by conservation of the heavy quark
flavor.
For the charm sector, the conservation constraint relates
the number of initially-produced $c\bar{c}$ pairs $\Nccbar$ to their
distribution into open and hidden charm hadrons,
\begin{equation}
\Nccbar = {1\over 2}\gamma_c N_{\rm open} + {\gamma_c}^2 N_{\rm hidden},
\label{eqgcstat}
\end{equation}
where $N_{\rm open}$ is the number of hadrons containing one $c$ or
$\overline c$ quark and $N_{\rm hidden}$ is the number of hadrons containing
a $c \overline c$ pair. 
For most applications, $N_{\rm hidden}$
(and also multi-charm hadrons) can be neglected compared with
$N_{\rm open}$ due to the mass differences.  Thus the charm
enhancement factor is simply
\begin{equation}
\gamma_c = {2\Nccbar\over N_{\rm open}},
\end{equation}
leading directly to the quadratic dependence of the hidden charm hadron
population on $\Nccbar$.
One can then express the total number of $\J$ in terms of
the various thermal densities, $n_i$, and the total number of $c \overline
c$ pairs, $\Nccbar$.
One factor of system volume $V$ remains implicit here.  It is generally
replaced by the ratio of number to density for total charged
hadrons, $n_{\rm ch}/N_{\rm ch}$. Then the number of $\J$ produced obeys the
generic form anticipated in Eq.~(\ref{eqquadratic}).
\begin{equation}
\NJ = 4 {n_{\rm ch} n_{\J}\over {n_{\rm open}}^2} {\Nccbar^2\over N_{\rm ch}}
\label{eqstathad}
\end{equation}                                                                 

For collider experiments such as those at the LHC and RHIC, 
relating the corresponding central rapidity densities will be
more relevant.  Since Eq.~(\ref{eqstathad}) is homogeneous in
the total particle and
quark pair numbers, it will also be valid if these 
are replaced by their rapidity densities.  To get an order of
magnitude estimate,    
we choose a ``standard'' set of thermal
parameters, $T = 170$ MeV and $\mu_B \approx 0$, for which the thermal
density ratio is approximately $0.5$.  For a specific normalization,
we assume $dN_{\rm ch}/dy$ = 2000 for a central
collision at the LHC and take the initial charm rapidity density to be 
$dN_{c \overline{c}}/dy = 25$, roughly corresponding to $\Nccbar = 200$ 
for central collisions $(b=0)$.  
Using these inputs, one predicts 
$d\NJ/dy = 0.625$, indicating that several $\J$ will form through 
statistical hadronization in a  central collision.  
To put this number in perspective,
it is revealing to form the $\J$ to $\Nccbar$ rapidity
density ratio, 0.025 with the same assumptions.
For comparison, one expects the corresponding hadronic production ratio 
to be of $0.01$.  This number would then be significantly reduced
if placed in a region of color deconfinement.  Thus the efficiency of
$\J$ formation via statistical hadronization at the LHC is expected
to be substantial. 

These numbers can be easily adjusted to other charm and
charged particle densities using Eq.~(\ref{eqstathad}).  
Variations of the thermal parameters can also be investigated.
For example, if the hadronization
temperature is decreased to 150 MeV, the prefactor
combination of thermal densities increases by approximately
a factor of two.  

The centrality dependence is controlled by the behavior
of $\Nccbar$ and $N_{\rm ch}$.  The former should be 
proportional to the nuclear overlap function $T_{AA}(b)$ but is 
generally recast in terms of the dependence on the
number of nucleon participants, $N_{\rm part}$.  The calculation of 
$N_{\rm part}$ requires a model
calculation dependent on the total inelastic cross section, $\sigma_{\rm in}$, 
as well as the nuclear geometry.  We parameterize the 
expected behavior as a power-law $\Nccbar \propto 
{N_{\rm part}}^{4/3}$.  However, there will be deviations from this
behavior for the larger values of $\sigma_{\rm in}$
expected at the LHC \cite{Andronic:2002pj}.  The centrality dependence
of $N_{\rm ch}$ at RHIC is also consistent with a power-law with
exponent $\approx$ 1.2
\cite{Bazilevsky:2002fz,Kharzeev:2001yq}.  We will use the same dependence for
our estimates at the LHC.  
It is clear that for sufficiently peripheral collisions one will encounter
situations in which the average number of initially produced $c \overline c$
pairs is of order unity or less.  At this point, one must revisit 
the assumptions of the original statistical
hadronization model which assumed a grand canonical ensemble.  The grand
canonical approach is valid only when $\Nccbar$ is large enough for the
fluctuations about the average value to be negligible.   Thus for
peripheral collisions, one must recalculate the statistical results
in the canonical approach where the charm number is exactly 
conserved, as noted in \cite{Gorenstein:2000ck}.  Charm conservation can be
implemented via a correction factor \cite{Cleymans:1990mn},
\begin{equation}
\Nccbar = {1\over 2} \gamma_c N_{\rm open} {{I_1(\gamma_c N_{\rm open})}\over 
{I_0(\gamma_c N_{\rm open})}} + {\gamma_c}^2 N_{\rm hidden}.
\end{equation} 
In the limit of large $\gamma_c N_{\rm open}$, 
the ratio of Bessel functions $I_i$ 
approaches unity and the grand canonical result
is recovered.  In the opposite limit when
$\gamma_c N_{\rm open} \rightarrow 0$, the ratio of Bessel functions approaches
${1\over 2} \gamma_c N_{\rm open}$.  In this limit, the dependence on
$\Nccbar$ in Eq.~(\ref{eqstathad}) changes from quadratic to linear.
At the LHC this effect will not be relevant until one reaches
very peripheral events, but at lower energies it can be 
significant over a much larger range of centralities \cite{Gorenstein:2001xp}.
\begin{figure}[htb]
\begin{center}
\includegraphics[width=12.5cm]{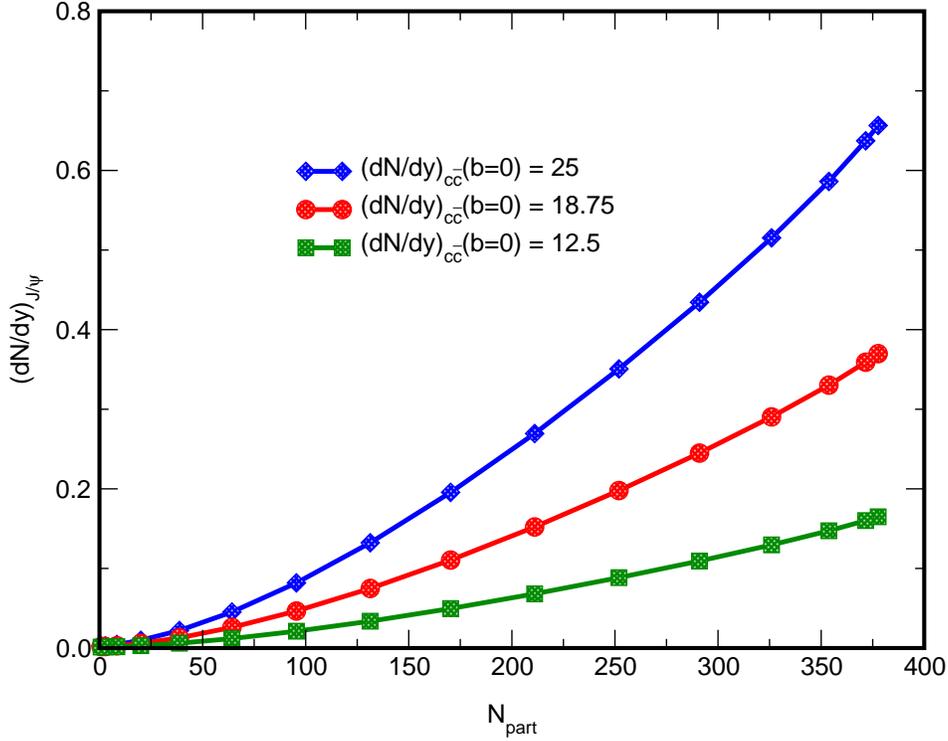}
\caption{Statistical hadronization results for $\J$ production as a function 
of $N_{\rm part}$ at the LHC.}
\label{lhcstatcharm}
\end{center}
\end{figure}

The results for $d\NJ/dy$ as a function of $N_{\rm part}$ at the LHC are 
shown in Fig.~\ref{lhcstatcharm}.
The results are shown for three different values of $dN_{\ccbar}/dy (b=0)$, 
corresponding to $\Nccbar(0) \approx 200$, 150, and 100.  
There is a rapid increase with
centrality due to the quadratic dependence of $\NJ$ on
$\Nccbar$. 

\begin{figure}[hbt]
\begin{center}
\includegraphics[width=12.5cm]{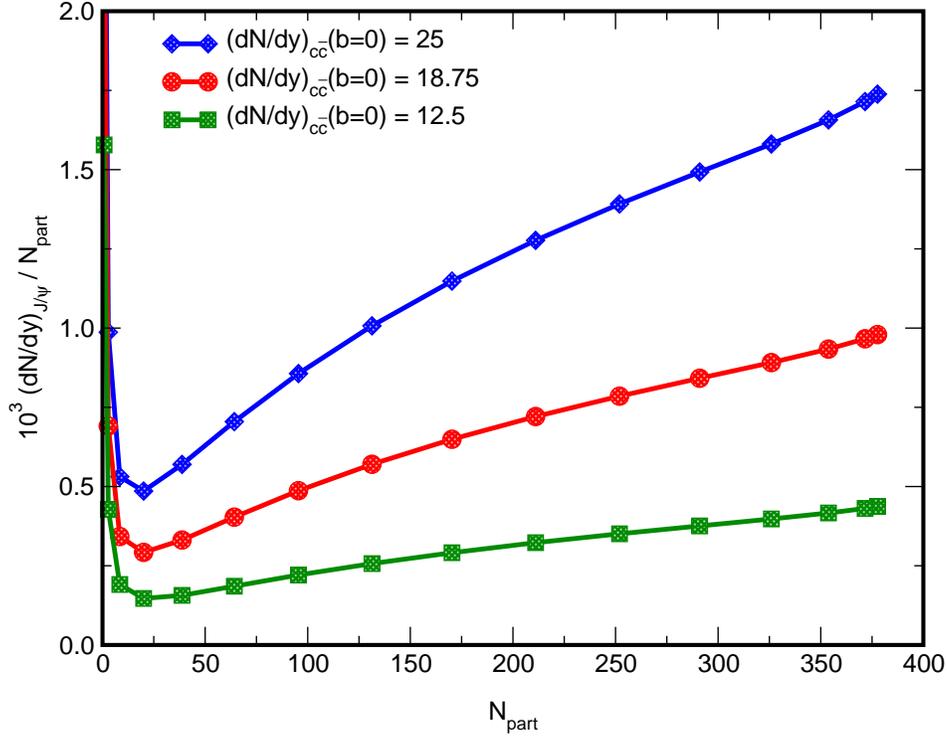}
\caption{The statistical hadronization results of Fig.~\ref{lhcstatcharm}
divided by $N_{\rm part}$ as a function of $N_{\rm part}$.}
\label{lhcstatcharmovernp}
\end{center}
\end{figure}

\begin{figure}[hbt]
\begin{center}
\includegraphics[width=12.5cm]{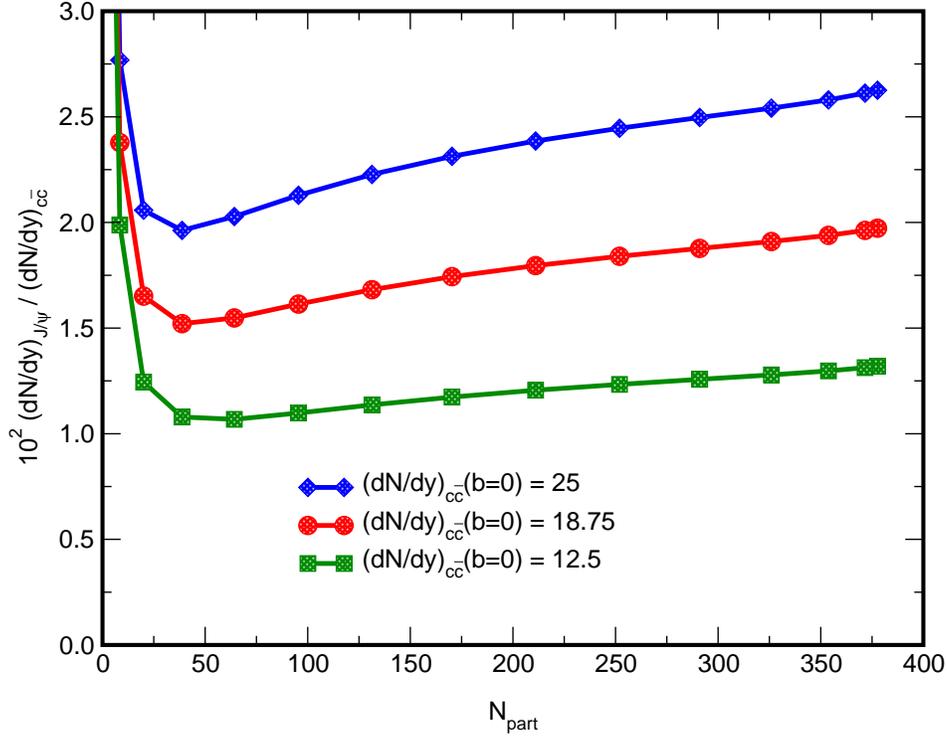}
\caption{The statistical hadronization results for $J/\psi$ production
at the LHC, divided by the open charm multiplicity, $d\Nccbar/dy$, as a
function of $N_{\rm part}$.}
\label{lhcstatcharmovercharm}
\end{center}
\end{figure}

It is also interesting to look at these results normalized by $N_{\rm part}$, 
shown in Fig.~\ref{lhcstatcharmovernp}.
This ratio also increases with centrality, providing a signature
for the statistical hadronization process that is less dependent on $\Nccbar$
for the overall normalization.   The corresponding
results when normalized by $d\Nccbar/dy$ are shown in 
Fig.~\ref{lhcstatcharmovercharm}.  The same general behavior is seen
but the increase with centrality is less pronounced since the
$d\Nccbar/dy$ is assumed to vary with a larger power, ${N_{\rm part}}^{4/3}$.
All of these ratios are at the percent level for central collisions and
hence are larger than expected if the total
$\J$ population were due to initial production followed by 
suppression in a deconfined medium.

The region of very peripheral collisions deserves some separate
comments.  First, there is a rise at low $N_{\rm part}$ in both 
Figs.~\ref{lhcstatcharmovernp} and \ref{lhcstatcharmovercharm} due 
to the onset of corrections from the canonical ensemble treatment.  However, 
the extremely large values of the ratios as $N_{\rm part} \rightarrow 0$ 
are an artifact of the decreasing
interaction volume, $V \rightarrow 0$.  
This calculation must be cut off before $N_{\rm part} = 2$, i.e. only one 
interacting pair. It is also in this region where one must 
take into account the remaining $\J$ from initial production.  Since
the survival probability is maximum for very peripheral collisions
and the statistical hadronization process is least effective in this
same region, there will be a crossover in the relative importance of
the two mechanisms.  Some studies have already been performed for this
situation at SPS and RHIC energies 
\cite{Grandchamp:2002iy,Grandchamp:2002wp,Grandchamp:2001pf}.

Finally, there is another lower cutoff in centrality for 
the statistical hadronization results, 
needed to avoid a contradiction with the $\psi\prime/(J/\psi)$ ratio at the
SPS.  Since both of these states
receive identical factors of $\gamma_c$, their ratio must be that
predicted for chemical equilibrium in the absence of any charm
enhancement or suppression.  Although the measured ratio appears
to be consistent for more central collisions
\cite{Sorge:1997bg}, there is an indication that it rises sharply for
more peripheral collisions.  Most treatments have
thus inserted a cutoff of $N_{\rm part}$ = 100, below which model predictions
become inconsistent \cite{Braun-Munzinger:2000px}.

The numerical values for $d\NJ/dy$ are tabulated as a function of
impact parameter in Table~\ref{thews.tab1} for the
three choices of initial charm multiplicity density and the default values
of all other quantities.


\subsection{Kinetic Formation in a Deconfined Region}
\label{thews.kinetic}

The kinetic model has been developed \cite{Thews:2000rj,Thews:2001em}
to investigate the possibility that $\J$ may form directly in
a deconfined medium.  This formation takes advantage of the mobility
of the initially-produced charm quarks in a deconfined region.
In order to motivate this view, consider the ``standard'' physical picture
of deconfinement in which quarkonium is suppressed by collisions with free
gluons in the medium \cite{Kharzeev:1994pz}.  Then the formation process,
in which a $c$ and $\overline c$ in
a relative color octet state are captured into a color-singlet
quarkonium bound state and emit a color octet gluon,
is simply the
inverse of the the breakup reaction responsible for
the suppression.  This is an inevitable consequence of the suppression picture.

The proper time evolution of the $\J$ population
is given by the rate equation
\begin{equation}\label{eqkin}
\frac{d\NJ}{d\tau}=
  \lambda_{\mathrm{F}} \frac{N_c\, N_{\overline c}}{V(\tau)} -
    \lambda_{\mathrm{D}} \NJ\, \rho_g\, \, ,
\end{equation}
where $\rho_g$ is the gluon number density
and $V(\tau)$ is the time-dependent volume of the deconfined spatial region.
The reactivities $\lambda_{\rm F,D}$ are
the reaction rates, $\langle \sigma v_{\mathrm{rel}} \rangle$,
averaged over the momentum distributions of the initial
participants, i.e. $c$ and $\overline c$ for $\lambda_{\rm F}$ and
$\J$ and $g$ for $\lambda_{\rm D}$.                             

The solution of Eq.~(\ref{eqkin}) grows quadratically
with $\Nccbar$, as long as $\NJ \ll \Nccbar$.  In
this case, we have
\begin{equation}
\NJ(\tau_f) = \epsilon(\tau_f) \left[\NJ(\tau_0) +
\Nccbar^2 \int_{\tau_0}^{\tau_f} \, d\tau
\lambda_{\mathrm{F}}\, [V(\tau)\, \epsilon(\tau)]^{-1} \right] \, \, .
\label{eqbeta}
\end{equation}
The function $\epsilon(\tau_f) =
\exp(-\int_{\tau_0}^{\tau_f} d\tau \lambda_{\mathrm{D}}\, \rho_g)$
would be the suppression factor if formation were neglected.  
                                                                        
The quadratic factor $\Nccbar^2$ is present, as expected, for
the additional formation process.  The
normalization factor of $N_{\rm ch}$ is not immediately evident, but
is implicit in the system volume factor.
This volume is now time-dependent, accounting for the decreasing charm quark
density during expansion.  
Here the transverse area of the deconfined region is determined not just by the
nuclear geometry but by the dynamics which determine the extent of the 
deconfined region.  This area is modeled by the energy density in terms of 
the local participant density in the transverse plane, $n_{\rm part}(b, s=0)$. 
The transverse area is defined by the ratio of the participant number 
to the local participant density.  Note that the maximum local density is
at $s=0$.  Thus,
\begin{equation}
A_T(b) = A_T(0) [N_{\rm part}(b) 
n_{\rm part}(0, s=0)/N_{\rm part}(0) n_{\rm part}(b, s=0)]
\end{equation}                                      
These area effects will be more explicit when the centrality
dependence is considered.    

The numerical results depend on a number of parameters, including the
initial volume and temperature, the time expansion profile, the
reaction cross sections, the behavior of the quarkonium 
masses and binding energies in the deconfined region, and the charm quark
momentum distributions.  For specifics, see Ref. \cite{Thews:2002jg}.
Our previous results have used initial values $\Nccbar = 200$, 150, and 
100, spanning a reasonable range of expectations \cite{Vogt:2001nh}.
The results are very sensitive to the initial charm quark momentum 
distributions, as may be expected. We assume the charm $p_T$ distributions
are Gaussian and the charm rapidity distributions are flat 
over a plateau of variable 
width, $\Delta y$.  The range $1 < \Delta y < 7$ spans the range between an 
approximate thermal momentum distribution, $\Delta y \approx 1$, to a 
distribution similar to that of the initial pQCD production, $\Delta
y \approx 7$.

\begin{figure}[htb]
\begin{center}
\includegraphics[width=12.5cm,clip]{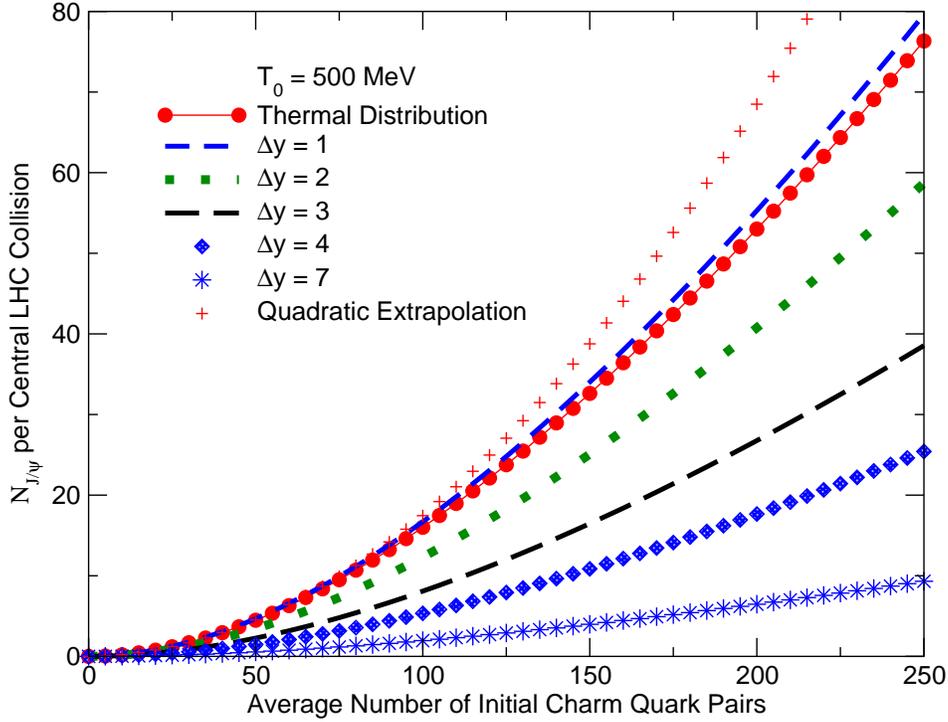}
\caption{The $J/\psi$ production per central LHC collision in the kinetic
model as a function of the initial number of $c \overline c$ pairs.}
\label{lhcbzero}
\end{center}
\end{figure}

The results as a function of the initial number of $c \overline c$ pairs 
produced in central collisions are shown in Fig.~\ref{lhcbzero}.
There is a rapid decrease in formation with increasing $\Delta y$.
The quadratic dependence on $\Nccbar$ is evident, but there is also 
a substantial linear component in some of the curves.  This linear contribution
arises because the final $\J$ formation by this mechanism
is large enough for exact charm conservation to reduce the number of $c$
and $\overline c$ quarks available to participate
in the formation process.  The curve labeled
``Quadratic Extrapolation'' uses a quadratic
dependence derived from a fit 
valid only for low $\Nccbar$.  Note that the result for a
thermal distribution is very similar to the assumption $\Delta y =1$.

\begin{figure}[htb]
\begin{center}
\includegraphics[width=12.5cm,clip]{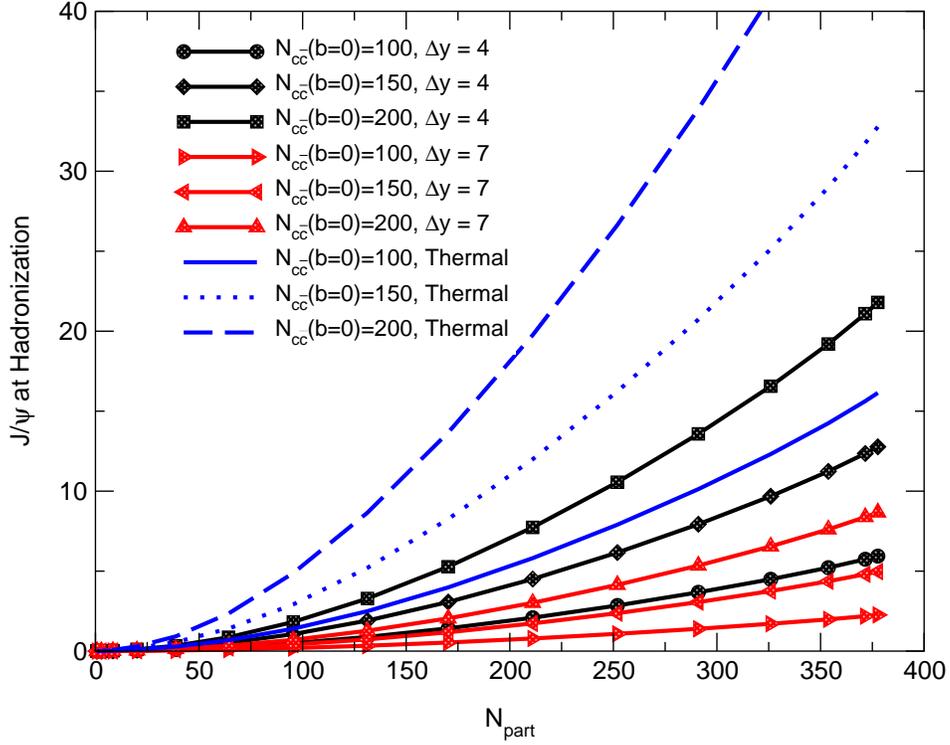}
\caption{The centrality dependence of $J/\psi$ production in the kinetic
model.}
\label{lhcb}
\end{center}
\end{figure}

The corresponding centrality dependence is presented in 
Fig.~\ref{lhcb}, where we give $\NJ$ at hadronization
for three different initial charm quark momentum distributions, thermal,
$\Delta y = 4$ and $\Delta y = 7$, as well as for our three choices of 
$\Nccbar(b=0)$.  

\begin{figure}[h]
\begin{center}
\includegraphics[width=12.5cm,clip]{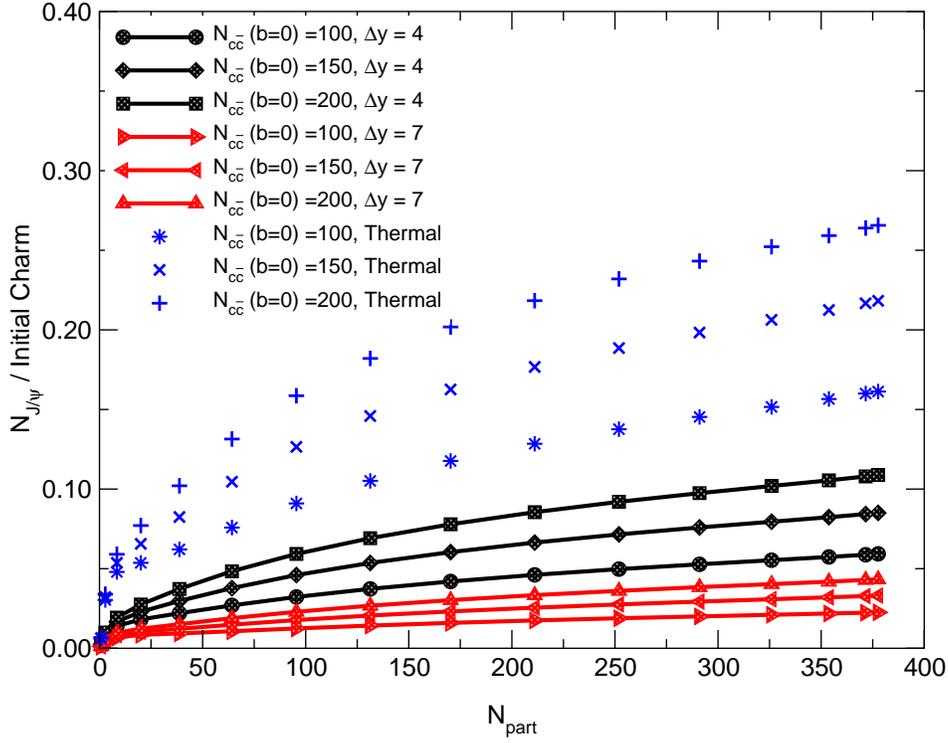}
\caption{The ratio of the number of produced $J/\psi$'s in the kinetic model
to the initial number of $c \overline c$ pairs as a function of $N_{\rm
part}$.} 
\label{lhcjpsiovercharm}
\end{center}
\end{figure}

Finally, the ratio of final $\J$ to initial charm production is shown
in Fig. \ref{lhcjpsiovercharm} using 
the same parameters as in Fig.~\ref{lhcbzero}.  These ratios are most
easily compared to either initial production or
suppression.  There is a substantial variation
in the predictions and it is evident that a simultaneous
measurement of open charm will be required for an
interpretation.  However, the centrality dependence is
opposite to that expected in any pure suppression scenario.

\begin{table}[htb]
\begin{center}
\caption{Comparison of $J/\psi$ production at the LHC by the statistical
hadronization (left-hand side) and kinetic formation (right-hand side) models.}
\label{thews.tab1}
\begin{tabular}{|c|c|c|c||c|c|c|} \hline \hline
 & \multicolumn{3}{|c||}{$dN_{\J}/dy$ (Statistical)} &
\multicolumn{3}{|c|}{$N_{\J}$ (Kinetic, LO Charm)}\\ 
 & \multicolumn{3}{|c||}{$dN_{c\overline c}(0)/dy$} &
\multicolumn{3}{|c|}{$N_{c \overline c}(0)$} \\ \hline \hline
$b$ (fm) & 25 & 18.75 & 12.5 & 200 & 150 & 100 \\ \hline \hline
0&0.656&0.370&0.165&4.0&2.26&1.03 \\ \hline
1&0.637&0.359&0.160&3.85&2.19&1.00\\ \hline
2&0.586&0.330&0.147&3.51&2.00&0.91\\ \hline
3&0.515&0.290&0.130&3.04&1.73&0.79\\ \hline
4&0.434&0.245&0.109&2.50&1.43&0.65\\ \hline
5&0.351&0.198&0.088&1.97&1.12&0.51\\ \hline
6&0.270&0.152&0.068&1.46&0.84&0.38\\ \hline
7&0.196&0.110&0.050&1.01&0.58&0.27\\ \hline
8&0.132&0.075&0.034&0.65&0.38&0.18\\ \hline
9&0.082&0.046&0.021&0.38&0.22&0.10\\ \hline
10&0.045&0.026&0.012&0.20&0.12&0.057 \\ \hline
11&0.022&0.013&0.0061&0.087&0.054&0.028 \\ \hline
12&0.0097&0.0058&0.0029&0.034&0.022&0.012 \\ \hline
13&0.0045&0.0029&0.0016&0.011&0.0075&0.0041 \\ \hline
14&0.0028&0.0019&0.0012&0.0021&0.0013&$6.8 \times 10^{-4}$ \\ \hline
15&0.0025&0.0018&0.0012&$1.8 \times 10^{-4}$&$1.0 \times 10^{-4}$ & $5.1 \times
10^{-5}$\\ \hline \hline
\end{tabular}
\end{center}

\end{table}

We have updated the calculations to include the
charm quark momentum distribution from a leading order pQCD calculation
\cite{thewsinprep}.  The rapidity distribution has a somewhat larger 
effective $\Delta y$ and the $p_T$ distribution does not fall as fast 
as a simple Gaussian.  As a result, the formation efficiency is further 
reduced.  Such distributions may be most relevant, given
preliminary results from RHIC \cite{Adcox:2002uc,Frawley:2002vz}.

\begin{table}[htb]
\begin{center}
\caption{Kinetic $J/\psi$ formation at the LHC assuming both
thermal charm momentum (left-hand side) and
$\Delta y = 4$ (right-hand side).}
\label{thews.tab2}
\begin{tabular}{|c|c|c|c||c|c|c|} \hline \hline
 &\multicolumn{3}{|c||}{$N_{\J}$ (Thermal)}&
\multicolumn{3}{|c|}{$N_{\J}$ ($\Delta y = 4$)} \\ 
 & \multicolumn{3}{|c||}{$N_{c \overline c}(0)$} &
\multicolumn{3}{|c|}{$N_{c \overline c}(0)$} \\ \hline \hline
$b$ (fm) & 200 & 150 & 100 & 200 & 150 & 100 \\ \hline \hline
0&52.7&32.5&16.4&17.5&10.8&5.48\\ \hline
1&50.5&31.2&15.8&16.8&10.4&5.25\\ \hline
2&44.8&27.7&14.0&14.9&9.21&4.65\\ \hline
3&37.0&22.9&11.5&12.3&7.62&3.82\\ \hline
4&28.6&17.7&8.73&9.54&5.89&2.90\\ \hline
5&20.7&12.7&6.05&6.90&4.23&2.01\\ \hline
6&13.8&8.32&3.72&4.61&2.77&1.24\\ \hline
7&8.28&4.71&2.14&2.76&1.57&0.71\\ \hline
8&4.10&2.36&1.10&1.36&0.79&0.37\\ \hline
9&1.78&1.04&0.50&0.59&0.35&0.16\\ \hline
10&0.65&0.39&0.19&0.22&0.13&0.064\\ \hline
11&0.19&0.12&0.063&0.065&0.040&0.021\\ \hline
12&0.048&0.032&0.018&0.016&0.010&0.006\\ \hline
13&0.011&0.0078&0.0049&0.0037&0.0026&0.0016\\ \hline
14&0.0026&0.0019&0.0012&$8.6 \times 10^{-4}$&$6.3 \times 10^{-4}$&
$4.1 \times 10^{-4}$ \\ \hline
15&$5.9 \times 10^{-4}$ &$4.4 \times 10^{-4}$ & $2.9 \times 10^{-4}$&
$2.0 \times 10^{-4}$& $1.5 \times 10^{-4}$&$9.7 \times 10^{-5}$\\ \hline \hline
\end{tabular}
\end{center}
\end{table}

The numerical values for $\NJ$ are compared with the statistical hadronization 
model results for $dN_{\J}/dy$ in Table~\ref{thews.tab1}.
The overall magnitudes are comparable, although
the centrality dependences differ somewhat.  Thus details such as the 
resulting $\J$ momentum distributions will be required to differentiate 
between these two models \cite{thewsinprep}.
For completeness, $\NJ$ for the thermal distributions and the assumption 
$\Delta y = 4$ are presented in Table~\ref{thews.tab2}.

\subsection{Conclusions}
\label{thews.conclus}

The "smoking gun" signature of the quarkonium formation mechanism
is the quadratic dependence on the total number of charm quarks.  For central
collisions at the LHC one
expects that this feature will lead to a total $\J$ rate greater than
that produced by an incoherent superposition
of the initial nucleon-nucleon collisions, even including
suppression due to deconfinement effects.  
In addition, the centrality dependence can be used to identify the 
quadratic dependence on charm assuming that the initial charm
production scales with the number of binary collisions.  Binary scaling 
leads to an 
increase of the ratio of $\J$ to initial charm as the collision
centrality increases, independent of specific parameters which control
the overall magnitudes.  A simultaneous measurement
of total charm will be essential for such conclusions to be drawn.

Uncertainties in the absolute magnitude of the formation process are
inherent in the model parameters.  For statistical hadronization,
one can constrain the thermal parameters to within a factor of
two using the observed hadron populations.  There is some additional
uncertainty related to the lower cutoff on centrality needed to ensure 
the quarkonium ratios are consistent with an overall thermal picture.  There 
is also the possibility that the correction for canonical ensemble effects 
will involve a thermal volume parameter not necessarily equal to the
total system volume \cite{Redlich:2001kb}.  In addition, the formation 
mechanism could be limited to those charm quarks whose phase space separation 
is within some maximum value, introducing another as yet
unconstrained parameter \cite{Grandchamp:2002wp}.  With kinetic 
formation, a similar set of uncertainties exist. 
There are uncertainties in the space-time properties of the deconfinement
region. In addition, possible variations of charmonium binding energies 
and reaction cross sections in a deconfined region are at present not well
understood.  There are indications that the efficiency of the 
formation mechanism is considerably reduced when included in a partonic
transport calculation \cite{Zhang:2002ug}.  

The primary uncertainty in both models is still the initial number of charm
quarks and their momentum distributions.  The tabulated
$\J$ results should be regarded in this
light.  Thus numbers may be only an order of magnitude estimate.  
However, the variation with centrality 
and total initial charm 
should provide experimental signatures which are largely independent
of the overall magnitudes.

\noindent
{\em Acknowledgments} 
My thanks to Anton Andronic for discussions on the Statistical Model and
Martin Schroedter for updates on the Kinetic Model calculations.

%% file: hotqcd.tex
\section[QUARKONIA IN HOT QCD MATTER: DISSOCIATION RATES]
{QUARKONIA IN HOT QCD MATTER: DISSOCIATION RATES~\protect
\footnote{Authors: D.~Blaschke, D.~Kharzeev, P.~Petreczky, H.~Satz.}}
\label{sec:lattice}

\subsection{Introduction}
\label{lattice.intro}

Quarkonium suppression was long ago suggested as a signal of
deconfinement \cite{Matsui:1986dk}. Due to
their small size, quarkonia can, in principle, survive the deconfinement phase 
transition. However, because of color screening, no bound state can exist at 
temperatures $T>T_D$ when the screening radius, $1/\mu_D(T)$, becomes 
smaller than the typical bound-state size \cite{Matsui:1986dk}.
Later it was realized that dramatic changes in the gluon momentum 
distributions at the deconfinement phase transition result in a sharp 
increase in the quarkonium dissociation 
rates \cite{Kharzeev:1994pz,Kharzeev:1995kz,Xu:1995eb}. Both the 
magnitude \cite{Shuryak:1978ij} and the energy dependence 
\cite{Bhanot:1979vb} of charmonium 
dissociation by gluons result in a significant suppression   
of the $c \overline c$ states even for $T<T_D$ but higher than 
deconfinement transition temperature, $T_c$. Moreover, close to 
$T_D$ the thermal activation mechanism is expected to dominate
\cite{Kharzeev:1995ju,Wong:2001uu}. The relative importance of 
gluon dissociation and thermal activation is governed by the ratio of the 
quarkonium binding energy $\varepsilon(T)$ and the temperature $T$, 
$X(T) \equiv \varepsilon(T)/T$ 
\cite{Kharzeev:1996se}.  At $X(T) \ll 1$ thermal activation 
dominates while for $X(T) \gg 1$ the dominant mechanism is ``ionization'' 
by gluons. 

The study of quarkonium dissociation via color screening 
provides an upper bound on the temperature at which 
quarkonium bound states can exist.  This temperature is also
referred to as the Mott temperature \cite{Ropke:zz}.  We note, however,
that resonances may still exist above the Mott temperature \cite{fehrenbach}.

The dissociation temperature due to color screening was studied using potential
models with different parameterizations of the heavy quark potential
\cite{Digal:2001ue,Ropke:zz,Karsch:1987pv,Hashimoto:1987hf}. 
All these studies predicted 
that excited charmonium states ($\chi_c$, $\psi'$) will essentially dissolve
at $T_c$ while the ground state $J/\psi$ will dissociate at 
$1.1~T_c$--$1.3~T_c$.
Some potential models also predicted strong change in the binding energy,
see e.g. Ref.~\cite{Karsch:1987pv}. Recently, charmonium properties
were investigated using lattice calculations 
\cite{Umeda:2002vr,Datta:2002ck} which 
indicate that the ground states exist with essentially unchanged properties at 
temperatures around $1.5 T_c$. This issue will be 
discussed in section~\ref{lattice.in-med}. 
Lattice investigations may indicate that at low temperatures, $T<1.5 T_c$, 
screening is not efficient and therefore gluon dissociation may be the 
appropriate source of quarkonium suppression, discussed in section 
\ref{lattice.sigma}. Finally the equilibrium
survival probabilities  are estimated in section \ref{lattice.abund}.

One should keep in mind that nonequilibrium effects in the very
early stages of a heavy-ion collision, when the energy density is highest,
should be also considered for quarkonium suppression. 
Not much is known about these effects.  However,
they may be an even more important source of quarkonium suppression than
those of the thermalized system, see e.g. Ref.~\cite{Digal:2002bm}.

\subsection{Investigation of In-medium Quarkonium Properties on the Lattice}
\label{lattice.in-med}

Temporal meson correlators $\Delta(\tau,\vec{p})$
may allow direct study of in-medium quarkonium 
properties since they are related to the $T \ne 0$ spectral functions 
$\sigma(\omega,\vec{p},T)$ by
\begin{equation}
\Delta(\tau,\vec{p})=\int_0^{\infty} d \omega \, \sigma(\omega,\vec{p},T)
\frac{\cosh(\omega (\tau-1/(2T)))}{\sinh(\omega/(2 T))}.
\end{equation}
Because at finite temperature, $T \ne 0$, the extension of the time
direction is limited by the inverse temperature, the only way to
study meson properties is to reconstruct the spectral function\footnote{At 
$T=0$ meson properties such as the mass, decay constants etc.
can be extracted from the large distance behavior of meson correlators.}.
The reconstruction of the spectral function with any reasonable 
discretization of the $\omega$ -interval  involves reconstruction
of several hundred degrees of freedom from $N_{\tau} \sim 10$ data 
points on $\Delta(\tau,\vec{p})$  where $N_{\tau}$ is the number
of time-slices. In general, this can be
done only using the so-called {\em Maximum Entropy Method}
(MEM) \cite{Nakahara:1999vy,Asakawa:2000tr}. 
For a review see Ref.~\cite{Asakawa:2000tr} and references therein.
The method has also been quite successful at finite
temperature \cite{Karsch:2002wv,Asakawa:2002xj}.

If we consider point sources, the correlators are defined as 
\begin{equation}
\Delta(\tau,\vec{p})=\langle O(\tau,\vec{p}) O(0,\vec{p}) \rangle,
\end{equation}
where 
\begin{equation}
O(\tau,\vec{p})=\sum_{\vec{x}} e^{i \vec{p} \vec{x}} 
\overline q(\tau,\vec{x}) \Gamma q(\tau,\vec{x}),
\end{equation}
and $\Gamma=1$, $\gamma_5$, $\gamma_{\mu}$ and $\gamma_{\mu} \gamma_5$ for the
scalar, pseudo-scalar, vector and axial-vector channels. By checking
the quantum numbers one realizes that the $^1S_0$, $^3S_1$, $^3P_0$ and
$^3P_1$ states actually correspond to the pseudo-scalar, vector, scalar and
axial-vector channels. 

One can also consider correlators of extended 
operators (sources)
\begin{equation}
O^{\rm ext}(\tau,\vec{p})=\sum_{\vec{x}} e^{i \vec{p} \vec{x}}
\sum_{\vec{y}} \phi(\vec{y}) \overline q(\tau,\vec{x}+\vec{y})
\Gamma q(\tau,\vec{x}),
\end{equation}
where $\phi(\vec{y})$ is some trial quarkonium
wave function which typically has the form $\exp(-a y^p)$ 
\cite{Umeda:2002vr,Umeda:2000ym}. Extended operators have a better 
overlap with the ground state than a point source. Though the spectral
functions of extended operators are easier to reconstruct due to their
simplified structure, typically a single pole, they have the disadvantage of
introducing a bias by the choice of some specific trial wavefunction.
They also do not correspond to any physically observable 
quantity such as the dilepton rate. The only information the study of such
correlators can provide is the position and
width of the resonance, provided that the corresponding structure of
the spectral function can be interpreted as a narrow resonance.

The following discussion is primarily 
based on the finite temperature charmonium
analysis of Ref.~\cite{Datta:2002ck}.  In this study, the nonperturbative 
clover action for quarks \cite{Luscher:1996jn}
was used with isotropic lattices, those with the same spacing in the temporal 
and spatial directions, with inverse lattice spacing, $a^{-1}$, between $4.04$
GeV and $9.72$ GeV.  The quark mass was varied in a range corresponding to the
pseudo-scalar $\eta_c(^1S_0)$ mass between $2.4$ GeV and $4.1$ GeV.
\begin{table}[htb]
\begin{center}
\caption{The lattice parameters used in the studies where $g^2$ is the lattice
coupling, $T_c$ is the critical temperature for deconfinement.  The 
lattice spacing $a$ is obtained from the string tension. 
The hopping parameter $\kappa$ is related 
to the quark mass by $m_q a=(1/2\kappa-1/2\kappa_c)/u_0$
where $u_0$ is the average value of the link field and 
$\kappa_c(g^2)$ was calculated in Ref.~\cite{Luscher:1996jn}.}
\label{tbl.lattices}
\begin{tabular}{|ccccc|} 
\hline
$6/g^2$ & $a^{-1}$ (GeV) & Size & $T / T_c$ & $\kappa$ \\
\hline
6.499 & 4.042 & $48^3 \times 16$ & 0.93 & 0.1300,~0.1234 \\
6.499 & 4.042 & $48^3 \times 12$ & 1.25 & 0.1300,~0.1234 \\
6.499 & 4.042 & $48^3 \times 10$ & 1.5 & 0.1300,~0.1234 \\
6.640 & 4.860 & $48^3 \times 16$ & 1.1 & 0.1290 \\
6.640 & 4.860 & $48^3 \times 12$ & 1.5 & 0.1290 \\
7.192 & 9.720 & $64^3 \times 24$ & 1.5 & 0.13114 \\
7.192 & 9.720 & $48^3 \times 12$ & 3.0 & 0.13114 \\ \hline
\end{tabular} 
\end{center}
\end{table}
The parameters of the simulations are summarized in Table \ref{tbl.lattices}.
The results of this study are compared to another
study \cite{Umeda:2002vr} using a standard Wilson action for quarks with
anisotropic lattices and extended operators. The temporal lattice spacing
in this study was four times smaller than the spatial one,
$1/(2.03 \, {\rm GeV})$, allowing more points in the time direction while 
keeping the spatial volume reasonably large. On the other hand, the large
spacing in the spatial direction may result in many unwanted lattice artifacts.
Both studies were done in the quenched approximation, neglecting the effects 
of quark loops. This approximation was used because 
such studies need quite small lattice spacings both because of the
heavy quark mass and the need to have many points in
the time direction, $N_\tau$, while keeping the temperature $1/(a  N_{\tau})=T$
fixed. Going beyond this approximation would require more than 10 Teraflops of 
computational power.

\begin{figure}
\centerline{\includegraphics[width=3in]{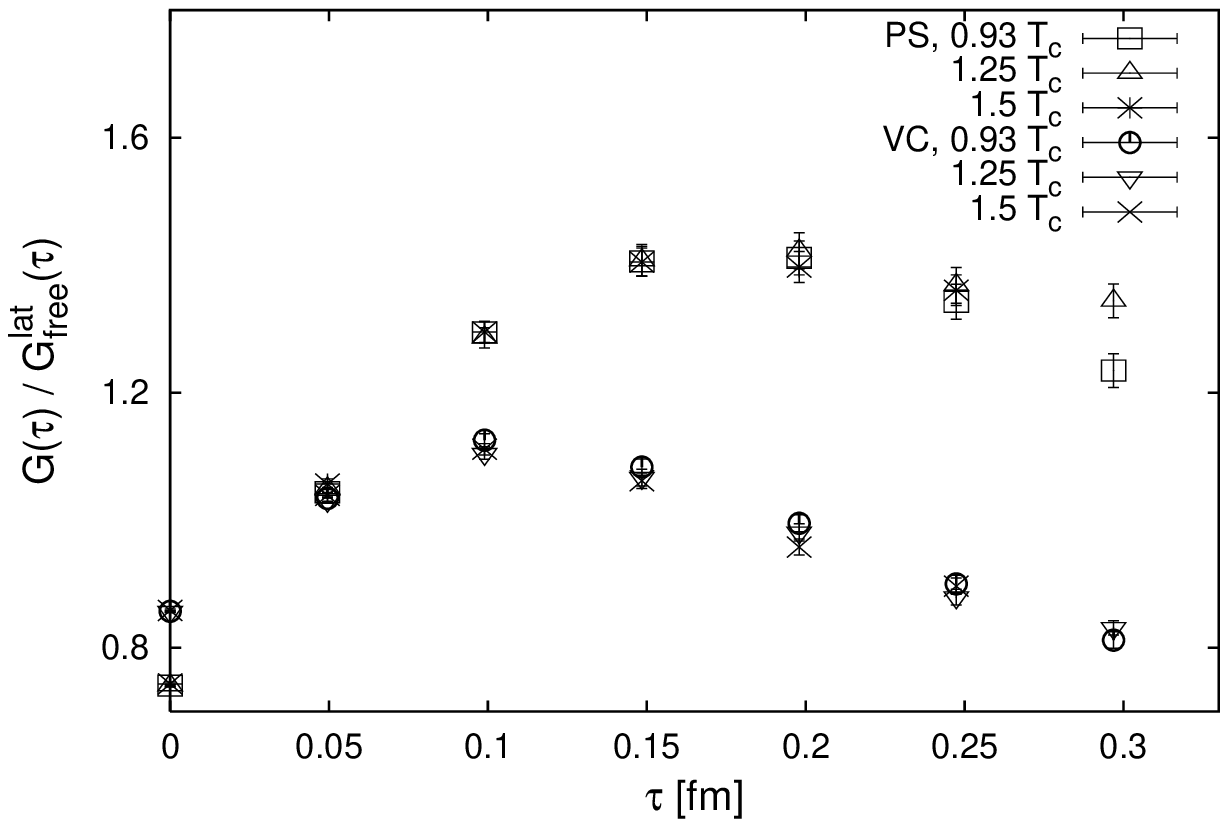}
\hspace*{0.2cm}
\includegraphics[width=3in]{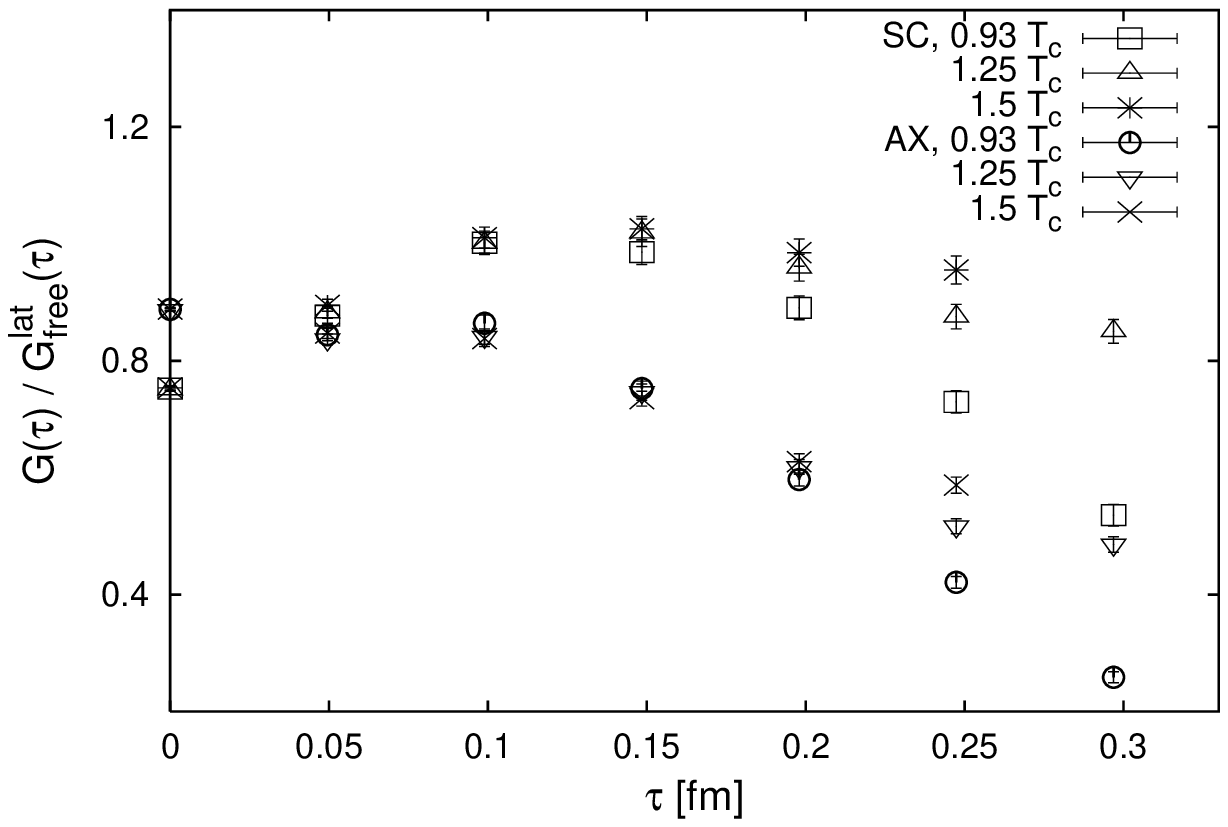}}
\caption{The temporal correlators for different channels at $6/g^2 = 6.499$ 
and $\kappa = 0.13$.  The left-hand side
shows the pseudo-scalar and vector channels, the right-hand side, the scalar 
and axial-vector channels.}
\label{tcorr}
\end{figure}

\begin{figure}
\centerline{\includegraphics[width=3in]{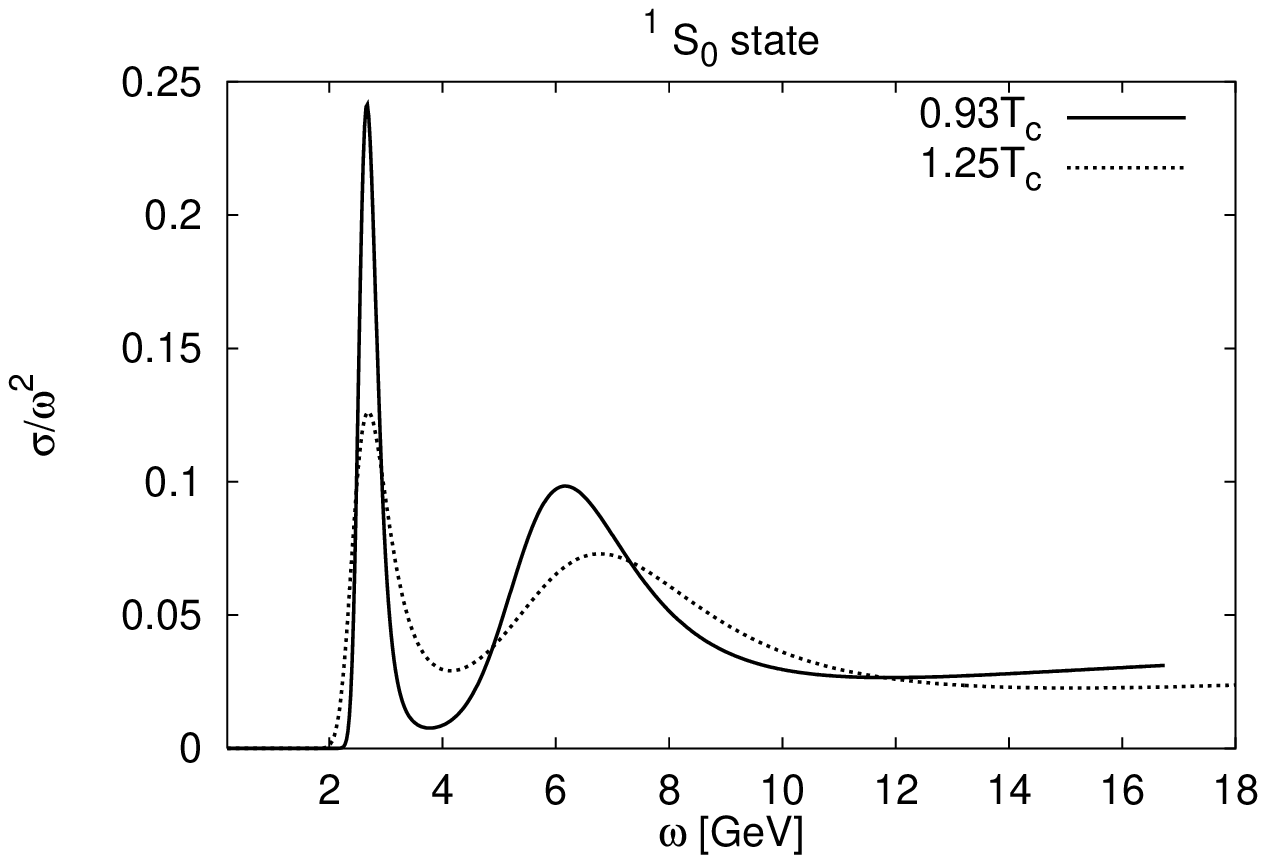}
\hspace*{0.2cm}
\includegraphics[width=3in]{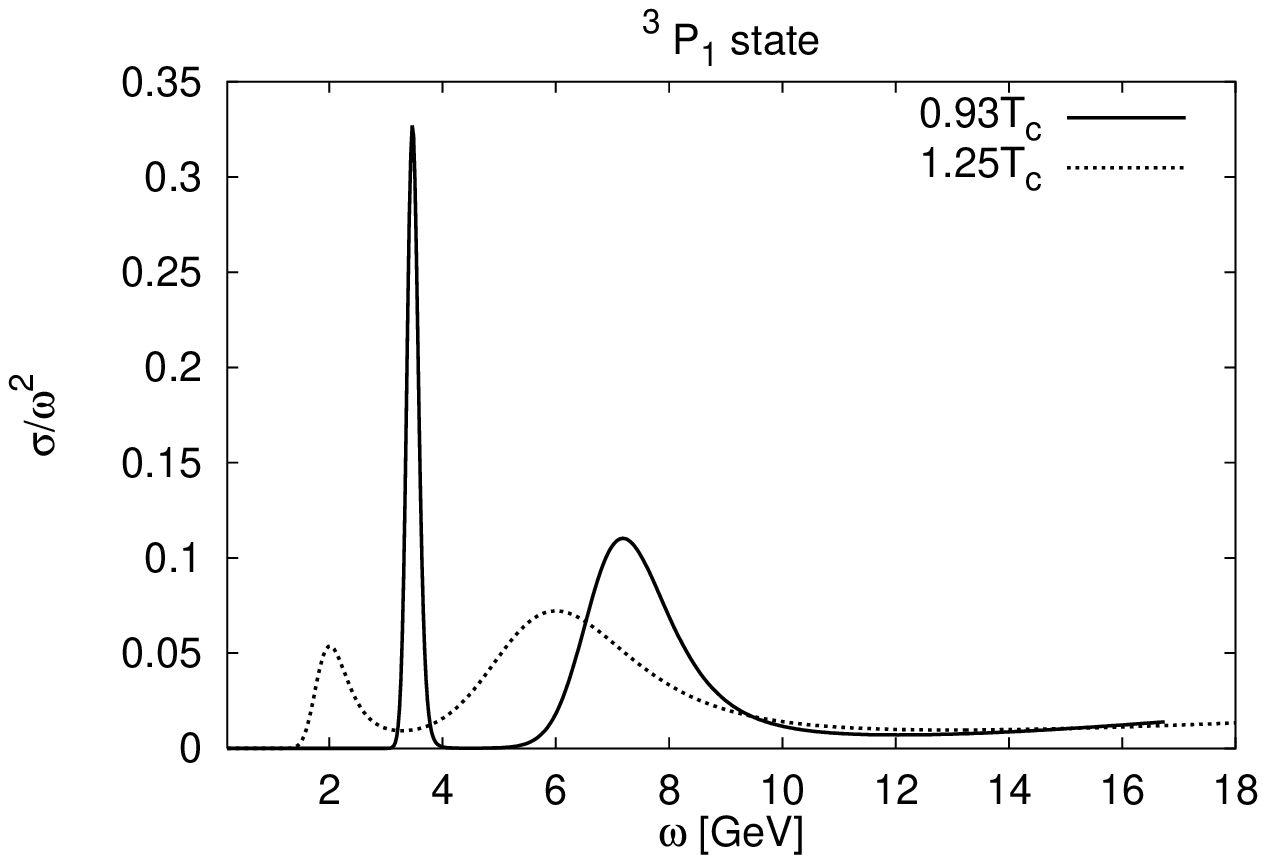}}
\caption{The spectral function for the pseudo-scalar
(right) and axial-vector (left) channels at $0.93T_c$ and $1.25T_c$ and 
for $6/g^2=6.499$ and $\kappa=0.1300$.}
\label{spfqq}
\end{figure}

The temperature dependence of the temporal correlators alone 
can provide some information on the change of the meson spectral
properties. In Fig.~\ref{tcorr} we show the ratio of the correlators
at different temperatures to the free correlators calculated on the lattice
for quark masses corresponding to a $^1S_0$ mass of $2.4$ GeV.
As seen in the figure, the correlators in the pseudo-scalar
and vector channels show only a very small temperature dependence
across $T_c$ while  in the scalar and axial-vector channels there
are quite noticeable changes. This difference may imply that the ground
state charmonia are not affected by deconfinement while the excited
$1P$ states are. 
This conclusion seems to be supported by the analysis of the spectral 
functions shown in Fig.~\ref{spfqq}.   While the spectral function in 
the pseudo-scalar channel peaks at almost the same
position at $0.93T_c$ and $1.25T_c$, in the axial-vector channel the peak 
present below $T_c$ is completely absent above $T_c$.
The same results apply to the vector and scalar channels.
Thus it seems that $1S$ ground state quarkonia are present
above $T_c$ with essentially unchanged masses but the excited
$1P$ states disappear already at $T \sim 1.25T_c$. The behavior of the
correlators in Fig.~\ref{tcorr} suggests that situation is similar at
$1.5T_c$. To see what actually happens at temperatures of $1.5T_c$ or
higher one has to consider simulations with lattice spacings
$a^{-1}=4.86$ GeV and $a^{-1}=9.72$ GeV with $N_{\tau}\ge 12$.
The analysis of the spectral functions in the pseudo-scalar and
vector channels calculated with
$a^{-1}=4.86$ GeV shows that the ground state peak survives at least until
$1.5T_c$ and the spectral functions are essentially unchanged
for $1.1T_c\leq T \leq 1.5T_c$ while the scalar and
axial-vector spectral functions are similar to those at $1.5T_c$.
This conclusion is also supported by the analysis of the spectral
functions at $1.5T_c$ with $a^{-1}=9.72$ GeV.
However, at $3T_c$ the spectral functions do not show any peak-like structures.
On the other hand, the correlators
show quite sizable changes between $1.5T_c$ and $3T_c$.

Finally we discuss the results obtained using extended
operators \cite{Umeda:2002vr}. Here only the pseudo-scalar and
vector channels were considered.  The simulations were done
at two temperatures, $T=0.88T_c$ and $1.08T_c$. The spectral
functions were first extracted using the MEM. Due to use of extended
sources, the spectral functions essentially consist of a 
single peak with a temperature-independent position. 
Thus the authors of Ref. \cite{Umeda:2002vr}
used a $\chi^2$ fit to a Breit-Wigner function to determine
the thermal width $\Gamma(T)$, obtaining
\begin{eqnarray}
^1S_0: \Gamma(T=1.08 \, T_c) & = & 0.12 \pm 0.03 \, {\rm GeV}, \nonumber \\
^3S_1: \Gamma(T=1.08 \, T_c) & = & 0.21 \pm 0.03 \, {\rm GeV}. \nonumber
\end{eqnarray}

The picture which emerges from the results of the lattice
studies described above can be summarized as follows. The 
ground state quarkonia survive up to $\sim 1.5T_c$.
Moreover, its mass does not change significantly relative
to its $T=0$ mass, contrary to potential model predictions which show 
significant mass shifts as well as dissociation of ground state charmonia well
below $1.5T_c$ \cite{Digal:2001ue,Karsch:1987pv}. Thus 
screening is not an efficient mechanism for charmonium
dissociation at temperatures close to $T_c$. 
However, there are indications that the 
ground state charmonia acquire a thermal width, implying that collisions 
with gluons may be the dominant mechanism for quarkonium dissociation at 
low temperatures. As the temperature increases,
screening becomes more and more important and will lead
to the disappearance of any bound state. The present study 
only indicates that in the absence of dynamical fermions 
this dissociation will happen somewhere between $1.5T_c$ and $3T_c$.
The excited $1P$ states seems to dissolve already at
$\sim 1.25T_c$, probably because the $1P$ state is larger than the ground 
state and therefore screening may be sufficient to dissolve it.

We now discuss whether these results can be interpreted within the present
understanding of finite temperature QCD.  First we note that the detailed 
picture of screening depends on the quantity one considers.  Universal
screening can be understood in terms of a simple
(perturbative) picture only at astronomically high temperatures
due to additional infrared problems arising at finite
temperature.  For a review, see e.g. Ref.~\cite{Philipsen:2000qv}.  
The most relevant quantity as far as the finite temperature properties of
quarkonium are concerned is probably the color singlet free 
energy\footnote{The gauge invariant definition of the color singlet free
energy was suggested only very recently \cite{Philipsen:2002az}. So far,
most lattice studies only considered the color-averaged free
energy which also includes the octet contribution.}, $F_1(r,T)$, studied in 
Ref.~\cite{Kaczmarek:2002mc}. This quantity shows quite 
complex short distance behavior and cannot be parameterized
by any simple form. Nevertheless, for $T_c < T < 1.5 T_c$, a rough estimate of
the screening length is $0.4 -0.6$ fm. From the small value of the 
screening length, it is possible to conclude that the charmonium ground state 
should also dissolve below $1.5T_c$.
However, the free energy also contains an entropy contribution,
$S(r,T)=-\partial F_1(r,T)/\partial T$ and therefore is not directly 
related to the potential between static
charges.  In fact, at large distances the entropy
is positive.  Therefore the free energy, $F_1(r,T)=V_1(r,T)-TS(r,T)$, is 
always smaller than the potential energy $V_1(r,T)$ \cite{Kaczmarek:2002mc}. 
If this situation persists over all distances
it would imply that screening in $F_1$ is stronger 
than in $V_1$, and could possibly explain the discrepancy between
the lattice studies of the charmonium spectral functions and potential
models.

\subsection{Quarkonium Dissociation Cross Sections and Rate Coefficients}
\label{lattice.sigma}

In this section, we evaluate the dissociation rates of 
Coulombic bound states by collisions in a quark-gluon plasma. 
For the first step, the plasma is represented by an ideal gluon gas 
with thermal distribution functions
$n_g(\omega)=g_g [\exp(\omega_g(p)/T) - 1]^{-1}$,
where $T$ is the temperature, $\omega_g(p)$ is the dispersion 
relation and $g_g=2(N_c^2-1)$ is the gluon degeneracy factor with $N_c = 3$.
We focus here on the $^3S_1$ quarkonium ground states, $J/\psi$ and
$\Upsilon$, which we consider to be Coulombic bound states.
Their masses are expected to be rather independent of the plasma temperature,
as discussed in section~\ref{lattice.in-med}.  These results
are consistent with experimental results for excitonic states in an 
electron-hole plasma \cite{fehrenbach} and with dynamical 
screening \cite{Ropke:zz} 
as opposed to static Debye screening \cite{Hashimoto:1987hf}.   
The effects of lowering the continuum threshold as well as temperature 
dependent modifications of the heavy quark potential will be neglected in
this first step.

The quarkonium breakup cross section by gluon impact
can be estimated using the Bhanot--Peskin formula \cite{Bhanot:1979vb},
\begin{equation}
\sigma_{(Q\overline Q) g} (\omega)= 
\frac{2^{11}}{3^4}\alpha_s \pi a_0^2
\frac{(\omega/\varepsilon(0) - 1)^{3/2}}{(\omega/\varepsilon(0))^5} 
\Theta(\omega - \varepsilon(0)) \, \, ,
\label{cross}
\end{equation}
where $\varepsilon(0)$ is the $T=0$ binding energy of the $1S$ 
quarkonium state with a Coulombic rms radius 
$\sqrt{\langle r^2\rangle_{1S}}=\sqrt{3} a_0=
2\sqrt{3}/(\alpha_s m_Q)$, $m_Q$ is the heavy-quark mass and $\omega$ is
the energy of the incoming gluon. 
The values of $m_Q$ and $\varepsilon(0)$ used in 
our further calculations are given in Table~\ref{tbl.hqp}.
\begin{table}[htb]
\begin{center}
\caption{Values of binding energy, $\varepsilon(0)$, and heavy quark mass, 
$m_Q$, \cite{Arleo:2001mp} used in our further calculations.  
}
\label{tbl.hqp}
\begin{tabular}{|ccccc|} 
\hline
 & \multicolumn{2}{c}{Set (i)} & \multicolumn{2}{c|}{Set (ii)} \\ \hline
System & $\varepsilon(0)$ (GeV) & $m_Q$ (GeV) & $\varepsilon(0)$ 
(GeV) & $m_Q$ (GeV) \\
\hline
bottomonium & 0.75 & 5.10 & 1.10 & 5.28 \\
charmonium & 0.78 & 1.94 & 0.62 & 1.86 \\
\hline
\end{tabular} 
\end{center}
\end{table}

The Bhanot--Peskin formula is derived using the 
operator product expansion, keeping the leading color--electric dipole 
operators to leading order in $1/N_c$, so that the final-state colour octet
interactions of heavy-quark pair are negligible. The 
corresponding expression for the Wilson coefficients (chromo--electric 
polarizabilities) of the excited $1P$ states is given in 
Ref.~\cite{Kharzeev:1995ij}.  Since
quarks have been found to be less efficient in dissociating quarkonium states
\cite{Kharzeev:1995kz}, we only include the effects of gluons here.

\begin{figure}
\centerline{\includegraphics[width=2.5in,angle=-90]{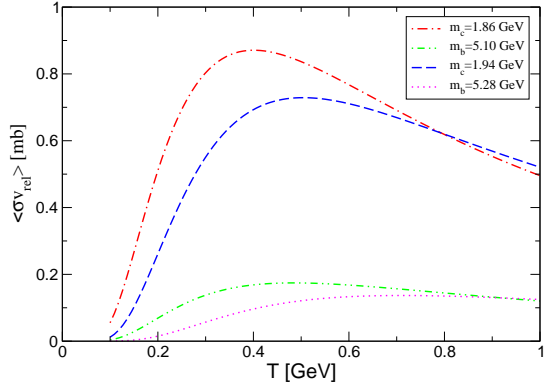}}
\caption{Thermal averaged cross sections for heavy quarkonium dissociation
by gluon impact as a function of temperature.}
\label{sigma_qgp}
\end{figure}

The cross section in Eq.~(\ref{cross}) 
has a steep rise at threshold with a drop 
after the maximum.  We give the thermal--averaged cross sections in 
Fig. \ref{sigma_qgp}.
A straightforward extension to the $p_T$ dependence of the quarkonium 
state in the plasma has been studied in Ref.~\cite{Xu:1995eb}.

\begin{figure}
\centerline{\includegraphics[width=2.5in,angle=-90]{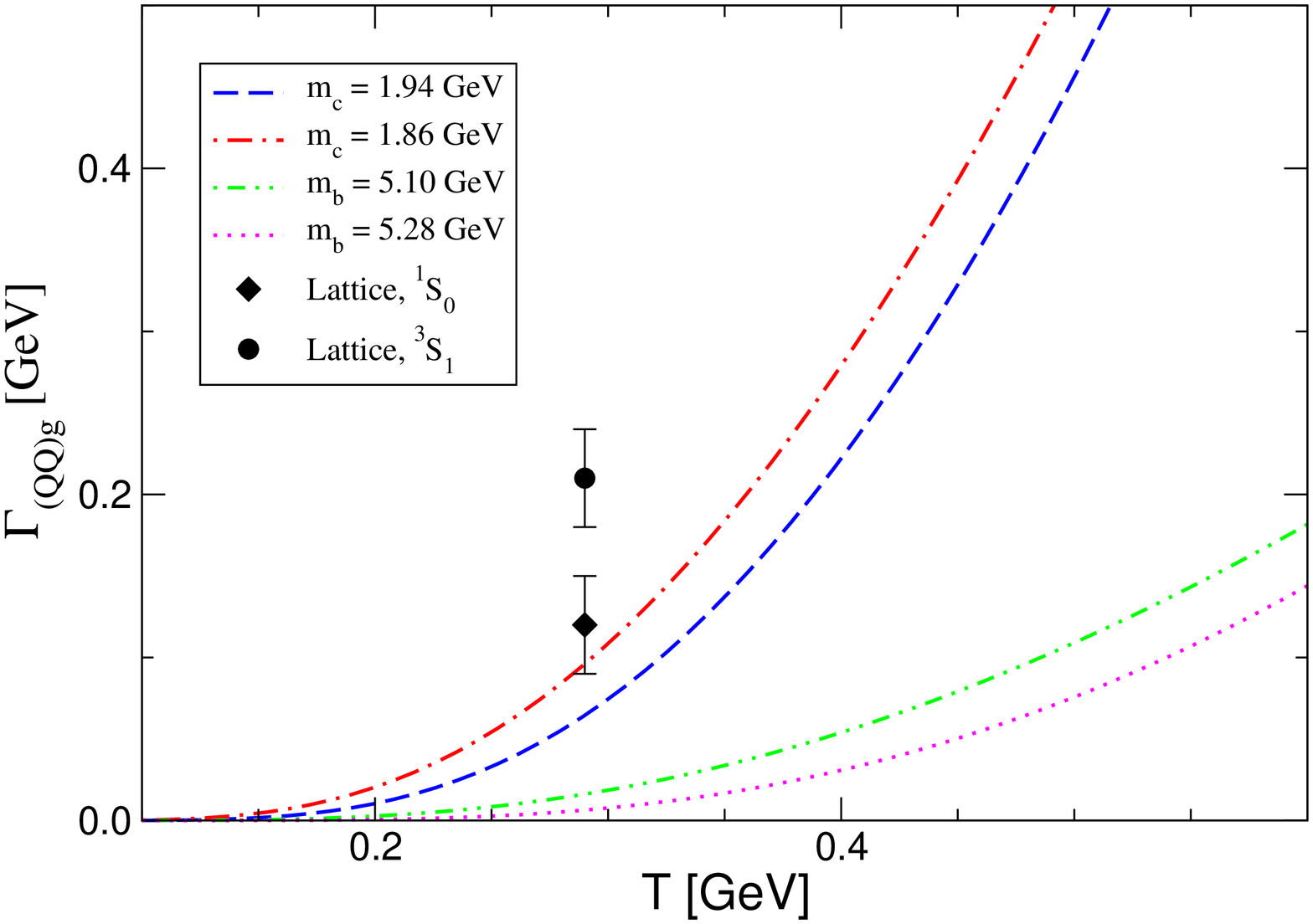}
\includegraphics[width=2.5in,angle=-90]{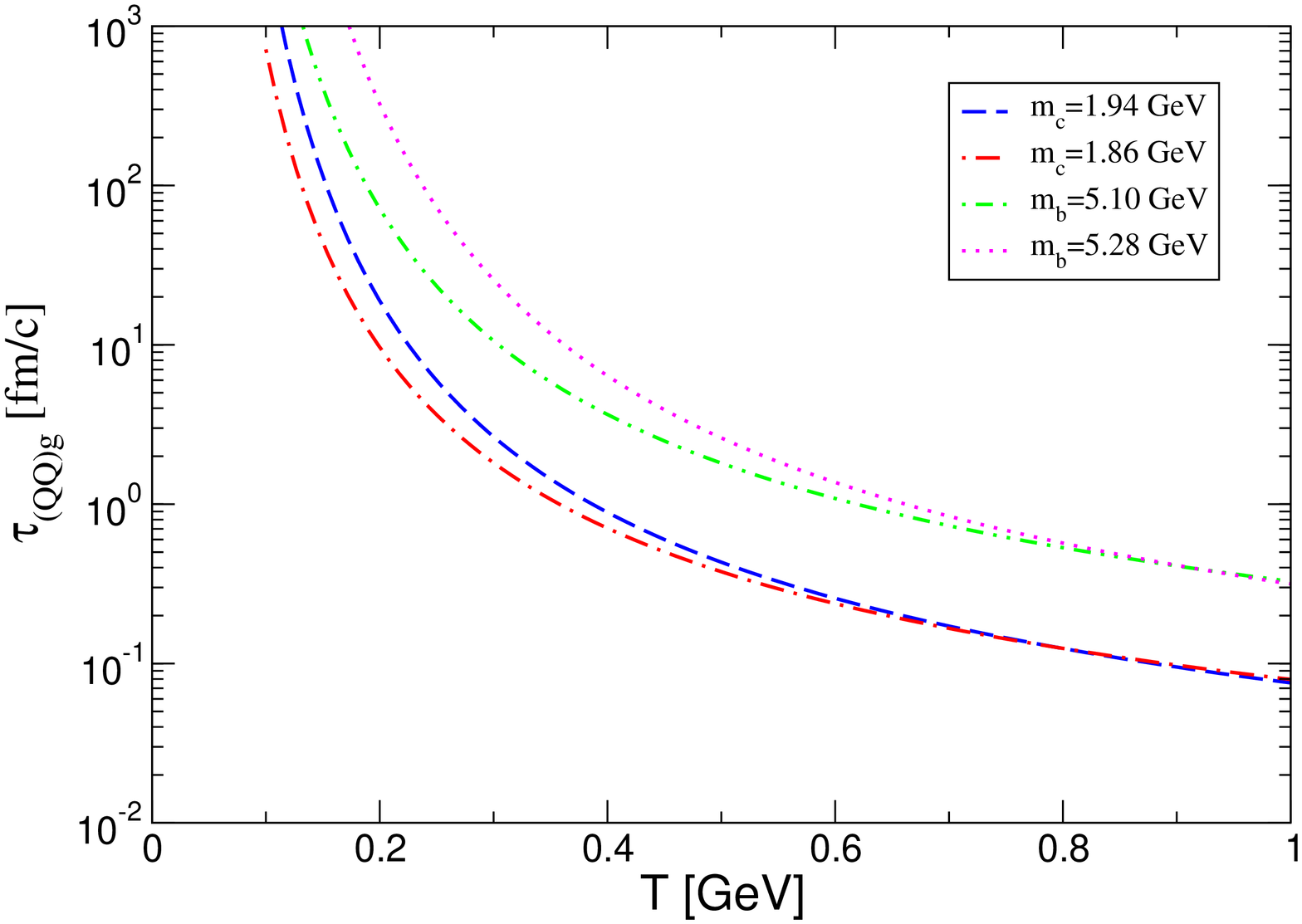}}
\caption{The rate coefficients 
$\Gamma_{(Q\overline Q)g}=\tau_{(Q\overline Q) g}^{-1}$
for heavy quarkonium dissociation by gluon impact as a function of the
gluon plasma temperature (left panel) and the corresponding lifetime 
$\tau_{(Q\overline Q) g}$ (right panel).
For comparison, the Breit-Wigner fits to the $^1S_0$ and
$^3S_1$ charmonium spectral widths from lattice simulations are shown.}
\label{gamma-tau-g}
\end{figure}

The dissociation rate for a quarkonium $1S$
state at rest in a heat bath of massless gluons 
at temperature $T$ distributed according to $n_{g}(\omega)$ is
\begin{equation}
\tau_{(Q\overline Q) g}^{-1}(T) \equiv \Gamma_{(Q\overline Q) g}(T)=
\langle \sigma_{(Q\overline Q) g}(\omega)n_g(\omega)\rangle_T~.
\end{equation}
Results for the partial widths $\Gamma_{(Q\overline Q) g}$
due to collisions, together with their 
inverses corresponding to partial 
lifetimes, are shown in Fig.~\ref{gamma-tau-g} for gluon breakup.
It is interesting to note that there is close correspondence between the 
charmonium decay widths due to gluon impact, the gluonic E1 transition, and 
the widths extracted from recent lattice simulations in the previous section. 

\subsection{Quarkonia Abundancies and Observable Signatures}
\label{lattice.abund}

\begin{figure}
\centerline{\includegraphics[width=2.5in,angle=-90]{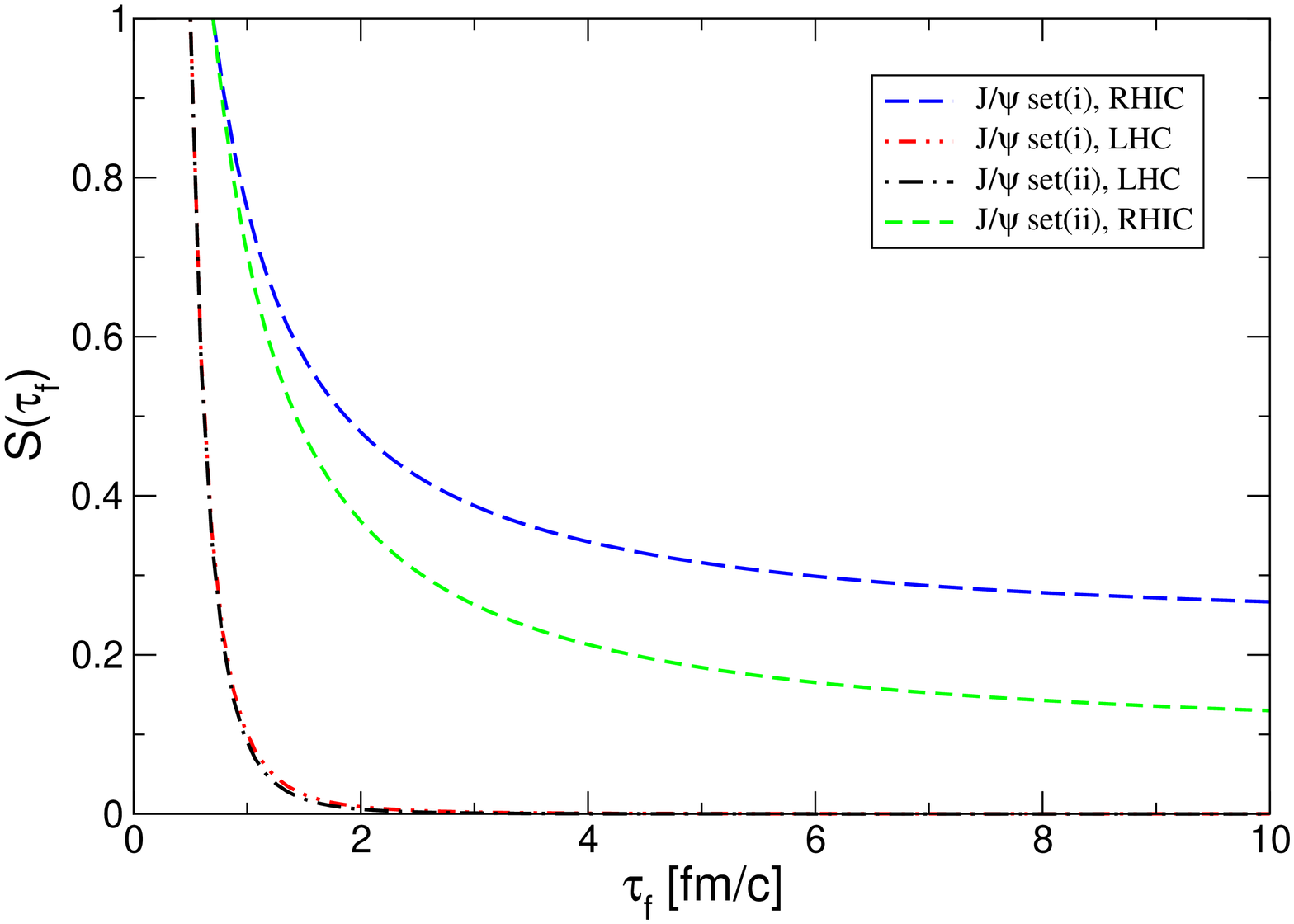}
\includegraphics[width=2.5in,angle=-90]{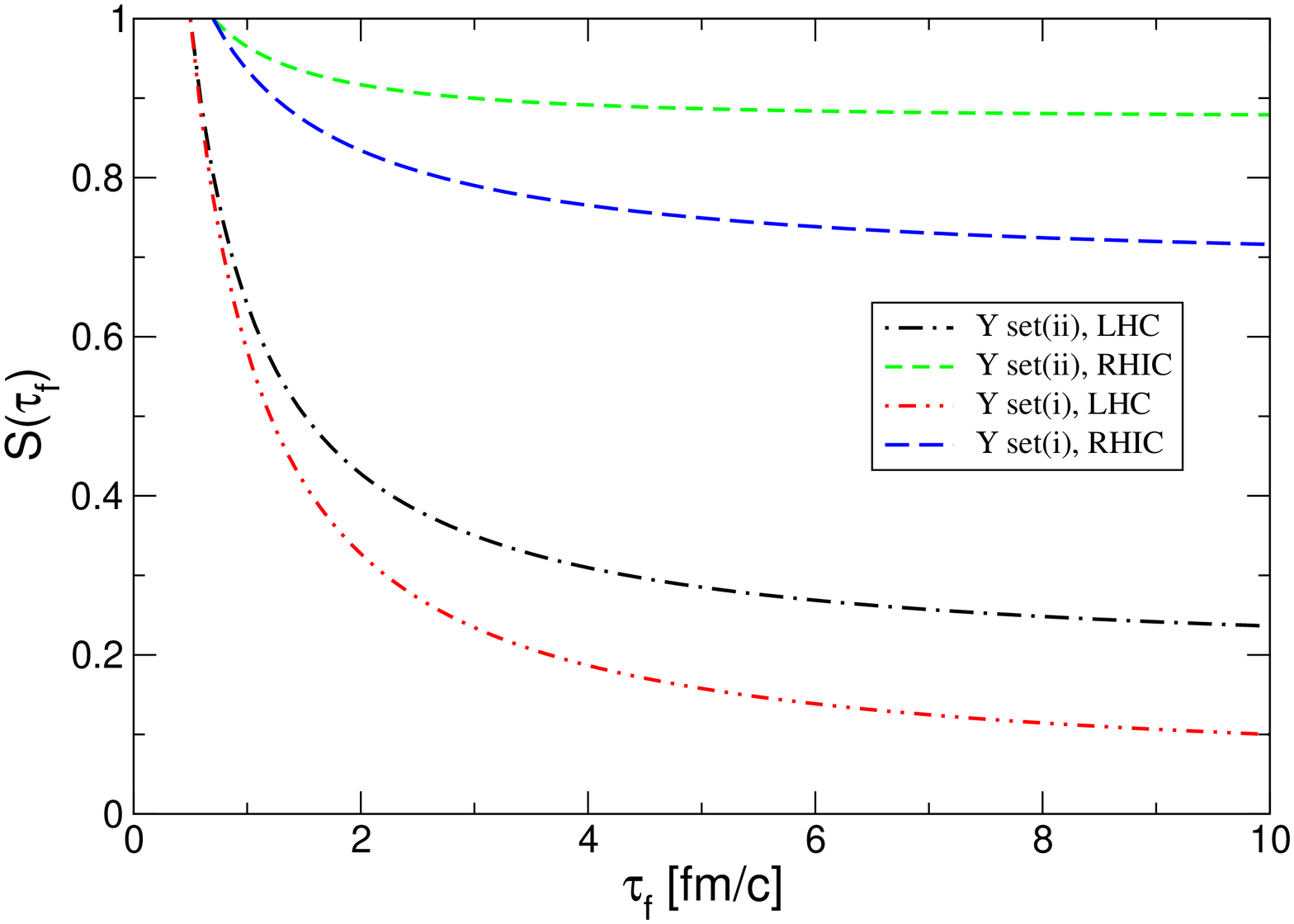}}
\caption{The survival probabilities of heavy quarkonia in a 
longitudinally expanding gluon plasma as a function of the plasma lifetime
using the initial conditions of Table~\ref{tbl.HIC} and the heavy quark masses
and binding energies of Table~\ref{tbl.hqp}.  On the left-hand side are the
$J/\psi$ results and on the right-hand side are the $\Upsilon$ results.}
\label{suppr}
\end{figure}

In order to study observable signatures, we adopt the Bjorken 
scenario \cite{Bjorken:1982qr} of plasma evolution, i.e. a longitudinal 
expansion with conserved entropy: $T^3 \tau = T_0^3 \tau_0={\rm const}$.
The initial temperatures and proper times are give in Table~\ref{tbl.HIC}. 
The RHIC and LHC values are set 3 from Ref.~\cite{Xu:1995eb}, assuming the 
highest initial fugacities.  The SPS values are typical of those used in
Ref.~\cite{Bjorken:1982qr}.
\begin{table}[htb]
\begin{center}
\caption{Initial conditions for plasma evolution.  The results for LHC and RHIC
are from Ref.~\cite{Xu:1995eb}.}
\label{tbl.HIC}
\begin{tabular}{|cccc|} 
\hline
 & LHC & RHIC & SPS  \\
\hline
$T_0$ [GeV]   & 0.72 & 0.4 & 0.25 \\
$\tau_0$ [fm/$c$] & 0.5  & 0.7 & 1.0 \\ \hline
\end{tabular} 
\end{center}
\end{table}
We calculate the quarkonium survival probability neglecting nucleon
absorption, hadronic comovers and hadronization effects,
\begin{equation}
S(\tau_f)=\exp\left(-\int_{\tau_0}^{\tau_f} ~d\tau~
\tau_{(Q\overline Q) g}^{-1}(T) \right)~.
\end{equation}
At the freeze-out time, $\tau_f$, the collisions stop changing the number of 
$J/\psi$ and $\Upsilon$. Using Bjorken scaling, $\tau_f$ 
can be translated into a 
freeze-out temperature. In Fig. \ref{suppr} we show the survival probability
of $J/\psi$ (left panel) and $\Upsilon$ (right panel) due to gluon impact
using the initial conditions from Table \ref{tbl.HIC}.  Note that set (i) gives
the lowest $\Upsilon$ survival probability due to its lower values of the
binding energy and bottom quark mass while set (ii), with its lower charmonium
binding energy and charm mass, has the lower survival probability for the
$J/\psi$. 
The simple estimates presented here for the LHC suggest that none of the
initially produced $J/\psi$'s will survive their passage through the plasma.
An analysis of the $p_T$ dependence of the suppression pattern, as in
Ref.~\cite{Xu:1995eb}, should provide a more complete picture.
The $\Upsilon$, however, will be a good probe of
the plasma lifetime as well as its temperature.  

\subsection{Conclusions}
\label{lattice.end}

We have used recent lattice data 
on the mesonic spectral functions to constrain predictions 
of the temperature dependence of the two-particle quarkonium spectra.
We have evaluated the collisional broadening of quarkonia states in a 
quark-gluon plasma using the Bhanot-Peskin cross section.
The $J/\psi$ partial widths in a gluon gas are in good 
agreement with lattice results for the spectral widths without dynamical 
fermions. 

Neglecting secondary quarkonium production we have evaluated the time 
dependence of the survival probability of $J/\psi$ and $\Upsilon$ for 
different initial conditions.  The results suggest that measurements of 
$\Upsilon$ suppression at the LHC can provide rather robust information about 
the initial temperature or
lifetime of the QGP \cite{Ropke:bx}. 

\noindent
{\em Acknowledgments} 
D.B. would like to thank Yuri Kalinovsky and Gerhard Burau for useful
discussions. 

%% file: montecarlo.tex
\section[HEAVY QUARK PRODUCTION IN MONTE CARLO GENERATORS]
{HEAVY QUARK PRODUCTION IN MONTE CARLO GENERATORS~\protect
\footnote{Authors: N.~Carrer, A.~Dainese, H.~Niemi.}}\label{sec:monte}

\subsection{Introduction}

This section presents a study of the predictions of Monte Carlo 
parton shower programs for the generation of charm and bottom
quarks in heavy-ion collisions at LHC energies. 
In section~\ref{sec:NLOxsec} the production cross sections in nucleon-nucleon
collisions, calculated at the next-to-leading order (NLO), and their 
extrapolations to Pb+Pb collisions are reported.
The NLO kinematical distributions are then compared with
the results obtained with \textsc{pythia}~\cite{Sjostrand:2000wi,
Sjostrand:1993yb} and \textsc{herwig}~\cite{Corcella:2000bw,Corcella:2001wc,
Marchesini:1991ch} in sect.~\ref{sec:pythia} and sect.~\ref{sec:herwig}
respectively.
Section~\ref{sec:hijing} presents a short description of 
\textsc{hijing}~\cite{Wang:1991ht,Gyulassy:ew},
a generator for nucleus-nucleus collisions, used to
simulate the charm- and bottom-hadron production background in $AA$ collisions.

\subsection{NLO heavy flavour production cross sections at the LHC}
\label{sec:NLOxsec}

The cross sections for charm and bottom production in $pp$
collisions at LHC energies, $\sqrt{s}=5.5\ \TeV$ and 14 TeV, 
have been calculated at NLO using a program by Mangano, Nason and 
Ridolfi~\cite{Mangano:jk} (referred to as MNR hereafter).
Two sets of parton distribution functions, MRST \cite{Martin:1998sq} and 
CTEQ5M1 \cite{Lai:1999wy}, have been used. 
All the following MNR calculations have been performed using the same 
values for the heavy quark masses ($m_c$ and $m_b$) and the 
factorization and renormalization scales ($\mu_F$ and $\mu_R$) as in
Ref.~\cite{Vogt:2001nh}:
\begin{eqnarray}
\label{eq:paramCharm}
m_c & = & 1.2\,\, \mathrm{GeV}\, ,
\,\,\,\,\,\,\,\,\,\,\,\,\,\,\,\,\,\,\,\,\,\,\,\,\,\,\,\,\,\,\,\,\,\,\,\,\,\,\,
\mu_F = \mu_R = 2\, \mu_0 \, , \\
\label{eq:paramBottom}
m_b & = & 4.75\, \, \mathrm{GeV}\, ,
\,\,\,\,\,\,\,\,\,\,\,\,\,\,\,\,\,\,\,\,\,\,\,\,\,\,\,\,\,\,\,\,\,\,\,\,\,\,\,
\mu_F = \mu_R = \mu_0 \, ,
\end{eqnarray}
where $\mu_0 = m_Q$ in calculations of the total cross section and
$\mu_0 = \sqrt{(p_{T,Q}^2+p_{T,\overline{Q}}^2)/2+m_Q^2}$ for differential
cross sections.
The $pp$ results are reported in Table~\ref{tab:lhcXsec}. The difference 
due to the choice of the parton distribution functions is relatively small,
$\sim 20-25$\% at 5.5~TeV and slightly less at 14~TeV. 
In $AA$ collisions, the cross section per nucleon is modified by shadowing
which we implement using the EKS98 
parameterization~\cite{Eskola:1998iy,Eskola:1998df}.

A discussion of the inclusive heavy quark $p_T$ and $y$ spectra and 
the $Q\bQ$ correlations is presented in the following section. The NLO
predictions are also compared to results obtained using {\sc pythia}.

\begin{table}[hb]
\caption{NLO calculation~\cite{Mangano:jk} of the total $c\overline{c}$
  and $b\overline{b}$
  cross sections in $pp$ collisions at 5.5 and 14~TeV, using the MRST
  and CTEQ5M1 parton distribution functions.}
\label{tab:lhcXsec}
\begin{center}
\begin{tabular}{|c|c|c|c|}
\hline
 &  & \multicolumn{2}{c|}{$\sigma_{pp}$ (mb)}\\ \hline
$\sqrt{s}$ (TeV) & $Q\bQ$ & MRST & CTEQ5M1 \\
\hline
& $c\overline{c}$ & 5.86 & 7.42 \\
\cline{2-4}
\raisebox{1.5ex}[0cm][0cm]{5.5} & $b\overline{b}$ & 0.19 & 0.22 \\
\hline\hline
& $c\overline{c}$ & 10.28 & 12.07 \\
\cline{2-4}
\raisebox{1.5ex}[0cm][0cm]{14} & $b\overline{b}$ & 0.46 & 0.55 \\
\hline
\end{tabular}
\end{center}
\end{table}

\subsection{Heavy flavour production in {\sc pythia}}
\label{sec:pythia}

The MNR code is not well suited for simulation purposes since it is not 
an event generator. Also, at LHC
energies the NLO corrections are large.  Therefore still higher order
corrections may be important because of, e.g., multiple gluon emission. 
On the other hand, Monte Carlo event
generators like {\sc pythia} and {\sc herwig} offer a different set
of approximations. They are based on LO pQCD matrix elements and  
parton showers which simulate real corrections to LO
QCD processes. For example, a final state heavy quark can further
radiate a gluon or an initial-state gluon can split into a
quark-antiquark pair. The parton shower approach to 
contributions beyond LO is not exact even at NLO but
it catches the leading-log aspects of multiple-parton 
emission.

In the simulations described here, {\sc pythia} 6.150 was used. We
will indicate light quarks ($u$, $d$ or $s$) by $q$ and
charm or bottom quarks by $Q$.

\subsubsection{Comparison with LO MNR results}

\begin{figure}[!bh]
  \begin{center}
    \includegraphics[width=\textwidth]{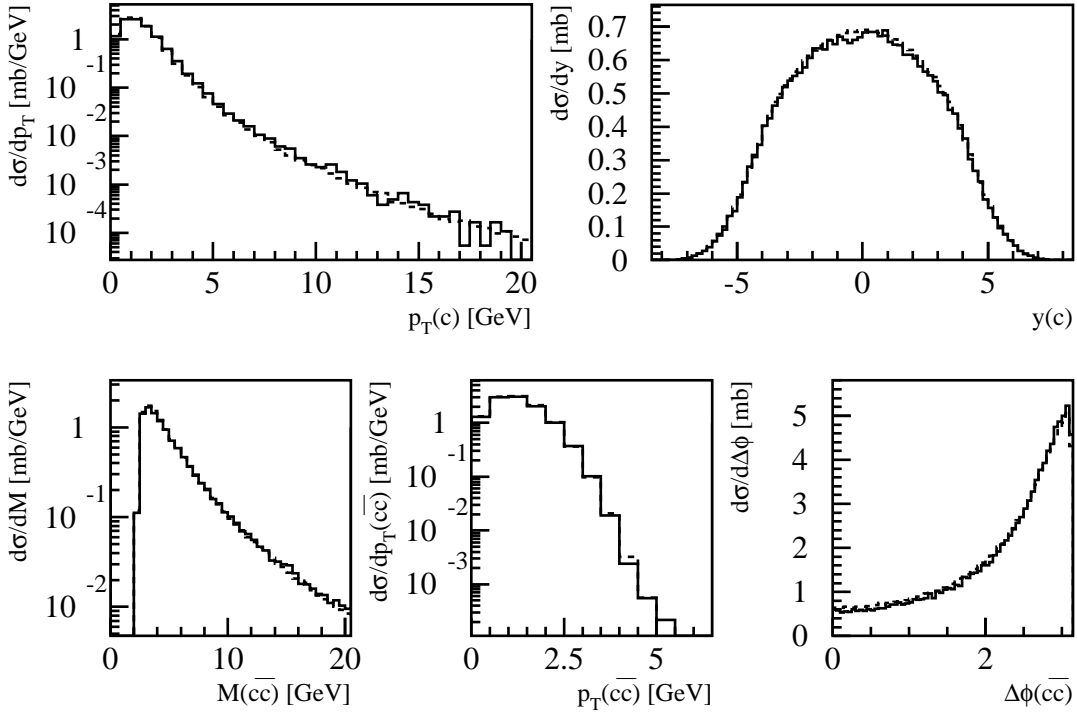}
  \end{center}
  \caption{Comparison between {\sc pythia} results (solid histogram) for
    $gg \rightarrow c \overline{c}$, without parton showers, and the
    MNR calculation of the same process at LO (dashed histogram). 
    The centre of mass energy is $\sqrt{s}=5.5\ \TeV$.}
  \label{fig:cmpCharmLO}
\end{figure}

On general grounds, one expects that the LO description of $Q\bQ$ production 
should be the same in {\sc pythia} and MNR when parton showers are
turned off in {\sc pythia}. 
Thus before comparing the results of the full {\sc pythia}
simulation with the ones obtained with the NLO MNR calculation, we first
check if the basic LO subprocesses are handled correctly in {\sc pythia}.
We compare the {\sc pythia} calculations without initial and final-state
showers to the MNR LO results in $\sqrt{s} = 5.5$ TeV $pp$ collisions for the 
following observables:
\begin{itemize}
\item inclusive $\pt$ and rapidity distribution of the $Q$ ($\bQ$);
\item pair $\pt$, defined as the projection on the plane normal
  to the beam axis of the $Q\bQ$ total momentum
  $\mathbf{p}_Q+\mathbf{p}_{\bQ}$;
\item pair mass, $M(Q\bQ) =
  \sqrt{(E_Q+E_{\bQ})^2-(\mathbf{p}_Q+\mathbf{p}_{\bQ})^2}$,
  where $E_Q = \sqrt{m_Q^2 + (\mathbf{p}_Q)^2}$ is the quark energy;
\item angle $\Delta\phi$ between the $Q$ and the $\bQ$ in the
  plane normal to the beam axis.
\end{itemize}
The results for charm production via gluon fusion\footnote{In {\sc pythia}, 
$gg \rightarrow c \overline c$ is called by ISUB=82.}, $gg \rightarrow c
\overline c$, are shown in
Fig.~\ref{fig:cmpCharmLO}. For a meaningful comparison the same input
parameters were used in {\sc pythia} and in MNR.  We use the CTEQ4L 
parton densities with a two-loop evaluation of $\alpha_s$, $m_c = 1.2$ GeV,
$\mu_F = \mu_R = \mu = 2\sqrt{(p_T^{\rm hard})^2 + m_c^2}$ where 
$p_T^{\rm hard}$ is the quark transverse momentum in the lab frame.
Parton intrinsic momentum smearing is modeled by a Gaussian with
$\langle k_T^2 \rangle = 1$ GeV$^2$.
Multiple interactions and initial and final-state parton showers
were switched off in {\sc pythia} for this comparison. 

The level of agreement between the two calculations is excellent. Thus the
LO calculations in both methods gives the same results when the same 
parameters are used.  Since at LO the pair is produced
back-to-back, the $p_T$ of the pair should be zero and
$\Delta\phi = \pi$. The deviation from the strict back-to-back
configuration is due to the intrinsic $k_T$ of the incoming partons.
The $k_T$-kick is implemented differently in the two calculations. Thus
the fact that MNR and {\sc pythia}
give the same results for this value of $\langle k_T^2 \rangle$ is 
quite remarkable. However, this fact should be considered as accidental,
and one should not expect it to hold in general, for example for larger
values of $\langle k_T^2 \rangle$.

\subsubsection{Parton shower and process classification in 
\textsc{pythia}}

When the parton shower is included, the partonic interaction can be
subdivided in different stages.  The event generation starts with the
LO process and then the parton shower is used to generate different
histories for the incoming and outgoing particles that take part in
the LO process. For example, outgoing particles from the LO process
can emit gluons or an incoming charm quark could come from the
splitting of a gluon into a $Q\bQ$ pair. Therefore the
parton shower will shift the kinematics of the produced particles
and at the same time can increase the multiplicity of final state. The
parton showers do not affect the probabilities of the LO
processes but increase the total rate of heavy quark production
through gluon splitting into $Q\bQ$ pairs.

\begin{figure}[!b]
  \begin{center}
    \includegraphics[width=\textwidth]{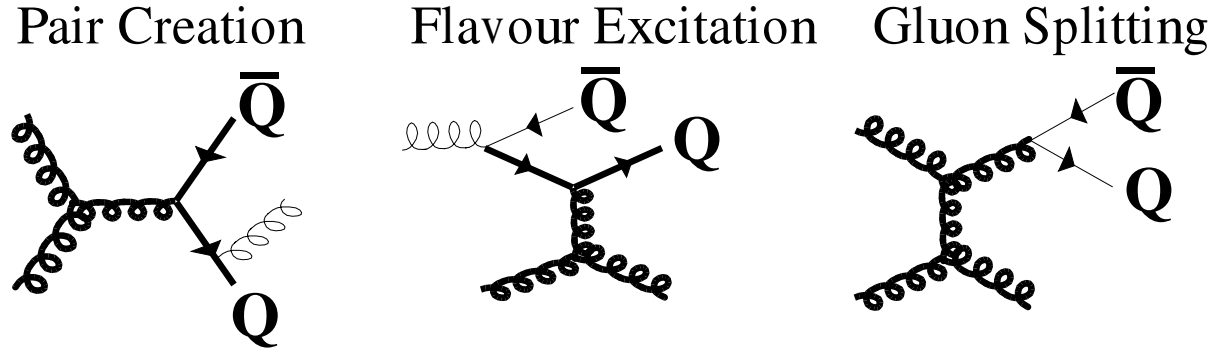}
  \end{center}
  \caption{Examples of pair creation, flavour
    excitation and gluon splitting. The thick lines correspond to the
    hard process, the thin ones to the parton shower.}
  \label{fig:processes}
\end{figure}

In {\sc pythia} the processes are divided in the following three classes,
according to the number of heavy quarks in the final state of the hard
process, where the hard process is defined as the one with the highest
virtuality:
\begin{description}
\item[pair creation:] The hard process produces two heavy quarks in the final
  state through the reactions $gg \rightarrow Q\bQ$ and
  $q \overline{q}\rightarrow Q\bQ$; these are the reactions which constitute
  the LO contribution in MNR;
\item[flavour excitation:] A heavy quark in one incoming hadron
  is put on mass shell by scattering off a parton from the opposite hadron,
  $qQ\rightarrow qQ$ or $gQ\rightarrow gQ$. The incoming heavy quark is
  assumed to come
  from a $g \rightarrow Q\bQ$ splitting in the initial-state 
  shower.  This hard process is characterized by one heavy quark in 
  the final state;
\item[gluon splitting:] No heavy quark is involved in the hard
  scattering.  Instead a $Q\bQ$ pair is produced in the initial or
  final-state showers from a $g \rightarrow Q\bQ$
  branching.
\end{description}
This classification might be misleading since all three classes
produce pairs at $g \rightarrow Q\bQ$ vertices when showers are
turned on. However, the
requirement that the hard scattering should be more virtual than the
shower avoids double-counting~\cite{Norrbin:2000zc}.
Fig.~\ref{fig:processes} shows some topologies belonging to the
processes specified above.

Flavour excitation and gluon splitting contributions are calculated
in the approximation of massless quarks. As a consequence, these cross
sections diverge as $\pt^{\rm hard}$ vanishes. These
divergences are usually regularized by putting a lower bound on
the allowed values of $\pt^{\rm hard}$. However, the choice of 
the minimum $\pt^{\rm hard}$ is arbitrary and is usually considered to be
a parameter. The lower bound on $\pt^{\rm hard}$ has a
strong influence on the heavy flavour cross section at low $\pt$.
Experiments with acceptances starting at several GeV, such as CDF and D\O~
at the Tevatron and ATLAS and CMS at the LHC, are not very 
sensitive to $p_T^{\rm hard}$. On the other hand, ALICE 
has been designed for good acceptance down to very low $\pt$ 
since the measurement of the total heavy flavour cross section in nuclear 
reactions is of great interest. Therefore, the choice of the minimum 
$\pt^{\rm hard}$ in the simulations of heavy quark production 
has to be carefully evaluated. The criteria used in this choice are 
explained in the next section.

\subsubsection{Comparison with MNR at NLO and tuning of \textsc{pythia} 
  parameters}

Because of the divergence at low $p_T$, {\sc pythia} does not offer a
valid baseline for the calculation of the total cross section and of
the kinematical distributions in the low-$\pt$ region. For this reason
our approach has been to adopt the NLO calculation as the baseline
for our simulations and to tune the {\sc pythia} parameters to
reproduce the NLO predictions as closely as possible. The main parameter
tuned is the lower bound of $\pt^{\rm hard}$. The simulations of Pb+Pb
collisions at $\sqrt{s}=5.5$ TeV/nucleon includes shadowing using the
EKS parameterization \cite{Eskola:1998iy,Eskola:1998df}, 
as noted in section~\ref{sec:NLOxsec}.
The complete list of parameters used is shown in
Table~\ref{tab:PythiaParams}. 
See Ref.~\cite{Sjostrand:2000wi,Sjostrand:1993yb} 
for a description
of the {\sc pythia} parameters. The larger value of the
intrinsic $k_T$-kick in Pb+Pb relative to $pp$
collisions ($k_T$ broadening) has been taken from Ref.~\cite{Vogt:2001nh},
based on a $Q^2$-dependent nuclear broadening.

\begin{table}[!b]
\caption{{\sc pythia} parameters used for the generation of charm and
  bottom quarks for Pb+Pb collisions at $\sqrt{s}=5.5$ TeV/nucleon and 
  $pp$ collisions at $\sqrt{s}=14$ TeV. All unspecified parameters
  are {\sc pythia} 6.150 defaults.}
\label{tab:PythiaParams}
\begin{center}
\small
\begin{tabular}{|c|c|c|c|}
\hline
description & parameter & Charm & Bottom \\
\hline
\hline
Process types & MSEL & 1 & 1 \\
\hline
Quark mass & PMAS(4/5,1) & 1.2 & 4.75 \\
\hline
Minimum $p_T^{\rm hard}$ & CKIN(3) & 2.1 & 2.75 \\
\hline
CTEQ4L     & MSTP(51) & 4032 & 4032 \\
Proton PDF & MSTP(52) & 2 & 2 \\
\hline
Switch off & MSTP(81) & 0 & 0 \\
multiple & PARP(81) & 0 & 0 \\
interactions & PARP(82) & 0 & 0 \\
\hline
Initial/Final parton & MSTP(61) & 1 & 1 \\
shower on & MSTP(71) & 1 & 1 \\
\hline
2$^{\mathrm{nd}}$ order $\alpha_s$ & MSTP(2) & 2 & 2 \\
\hline
QCD scales & MSTP(32) & 2 & 2 \\
for hard scattering & PARP(34) & 1 & 1 \\
and parton shower & PARP(67) & 1 & 1 \\
& PARP(71) & 4 & 1 \\
\hline
& MSTP(91) & 1 & 1 \\
                & PARP(91) & 1.304 (Pb+Pb) & 2.035 (Pb+Pb) \\
Intrinsic $k_T$ &          & 1     ($pp$)   & 1     ($pp$)   \\
                & PARP(93) & 6.52 (Pb+Pb)  & 10.17 (Pb+Pb) \\
                &          & 5    ($pp$)    & 5     ($pp$)   \\
\hline
\end{tabular}
\end{center}
\end{table}

\begin{figure}
  \begin{center}
    \includegraphics[width=.93\textwidth]{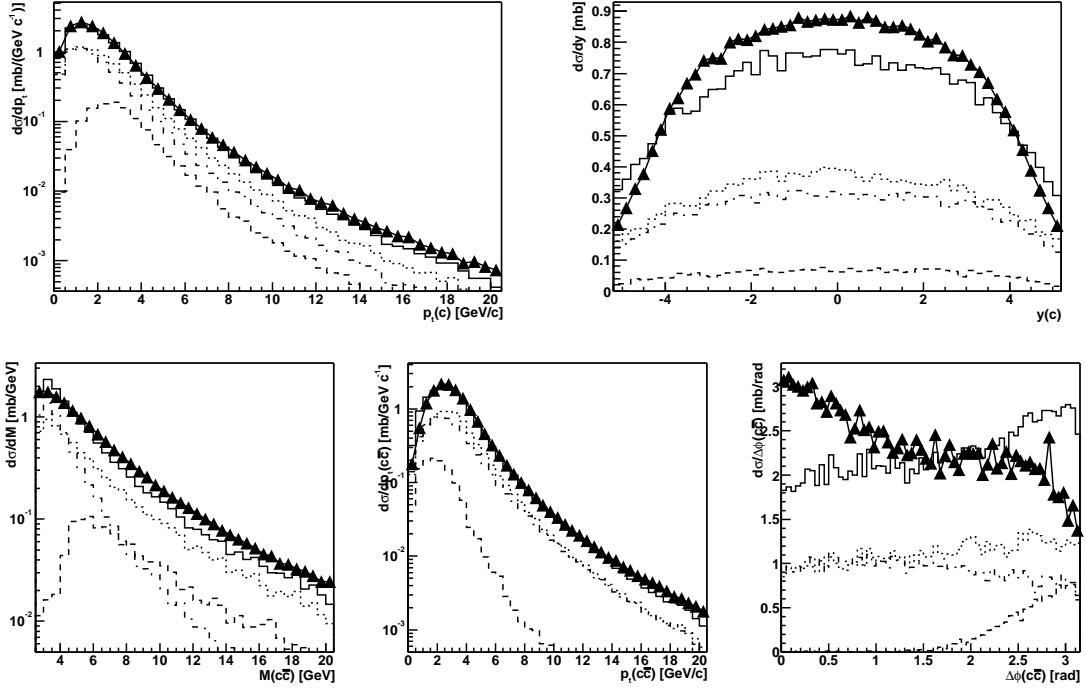}
  \end{center}
  \caption{Comparison between charm production in the NLO calculation
    by Mangano, Nason and Ridolfi and in {\sc pythia} with parameters tuned as
    described in the text for Pb+Pb collisions at 
    $\sqrt{s}=5.5$ TeV/nucleon. 
    The triangles show the NLO calculation, the
    solid histogram corresponds to the {\sc pythia} total production. The
    individual {\sc pythia} contributions are pair production (dashed),
    flavour excitation (dotted) and gluon splitting (dot-dashed).}
  \label{fig:charmPyMNR}
\end{figure}
\begin{figure}
  \begin{center}
    \includegraphics[width=.93\textwidth]{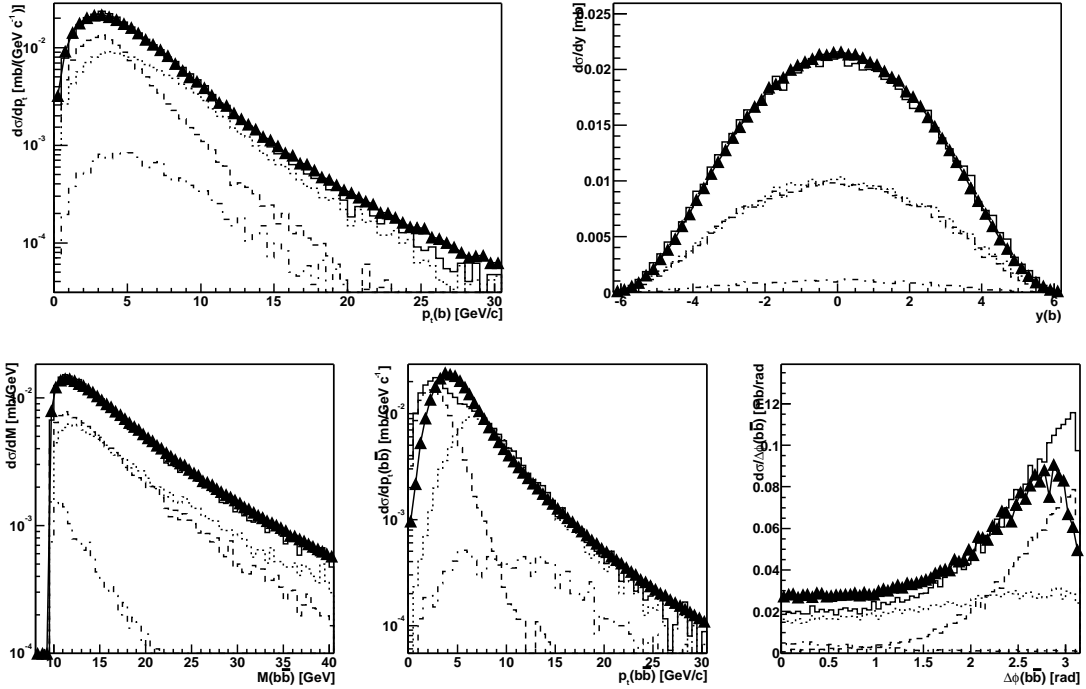}
  \end{center}
  \caption{Same as Fig.~\ref{fig:charmPyMNR}, but for bottom production.}
  \label{fig:bottomPyMNR}
\end{figure}
The results of this tuning for Pb+Pb collisions at 5.5 TeV/nucleon
are shown in Fig.~\ref{fig:charmPyMNR}
for charm and in Fig.~\ref{fig:bottomPyMNR} for bottom. The
{\sc pythia} and NLO distributions are compared. The
{\sc pythia} results are scaled by factors of 0.62 and 0.57 respectively to
give the same total cross section as the NLO calculation. Despite the
fundamental differences between the two models, the agreement is
relatively good. However, significant discrepancies are present,
especially in the $\Delta\phi$ distribution of the charm quarks.

\begin{figure}[!t]
  \begin{center}
    \includegraphics[width=0.9\textwidth]{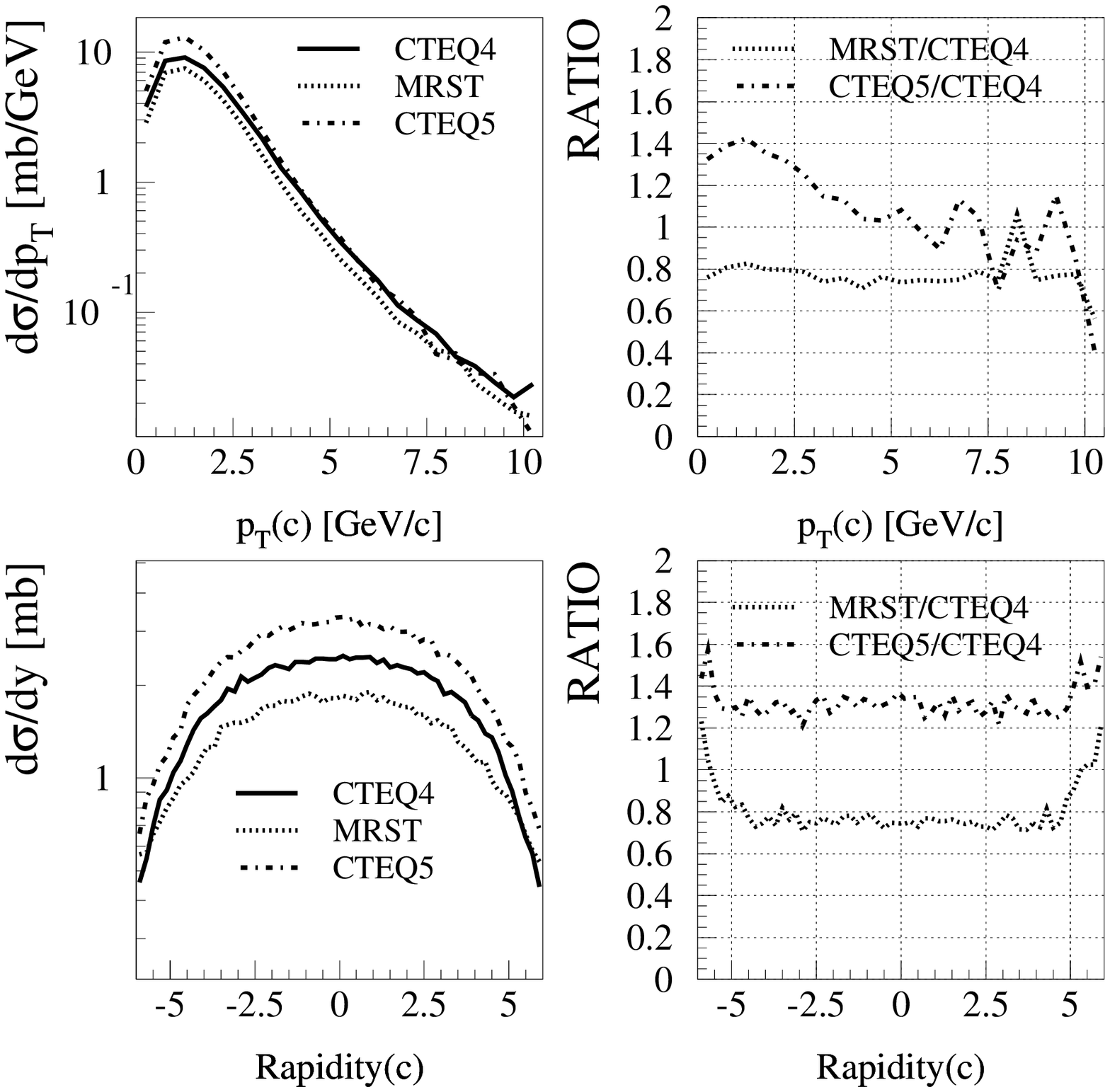}
  \end{center}
  \caption{Inclusive $p_T$ and
    rapidity distributions for charm with the CTEQ4L, CTEQ5L and MRST
    structure functions. The distributions are obtained using the MNR code
    for Pb+Pb collisions at 5.5 TeV/nucleon.}
  \label{fig:charmPyPDFsets}
\end{figure}

We note that the CTEQ4L parameterization used here is rather outdated.
However, we verified with the MNR code, that these results lie 
in between those
obtained with CTEQ5 and MRST for all the relevant kinematical
quantities.  The sensitivity to the parton
distribution functions does not exceed 30\% for charm and 20\% for
bottom. Figure~\ref{fig:charmPyPDFsets} shows the inclusive $p_T$ and
rapidity distributions for charm with the CTEQ4L, CTEQ5L and MRST
distributions\footnote{A direct comparison with the results in
Fig.~\ref{fig:charmPyMNR} is not possible since a slightly
different set of parameters was used.}.

\begin{figure}[!th]
  \begin{center}
    \includegraphics[width=.82\textwidth]{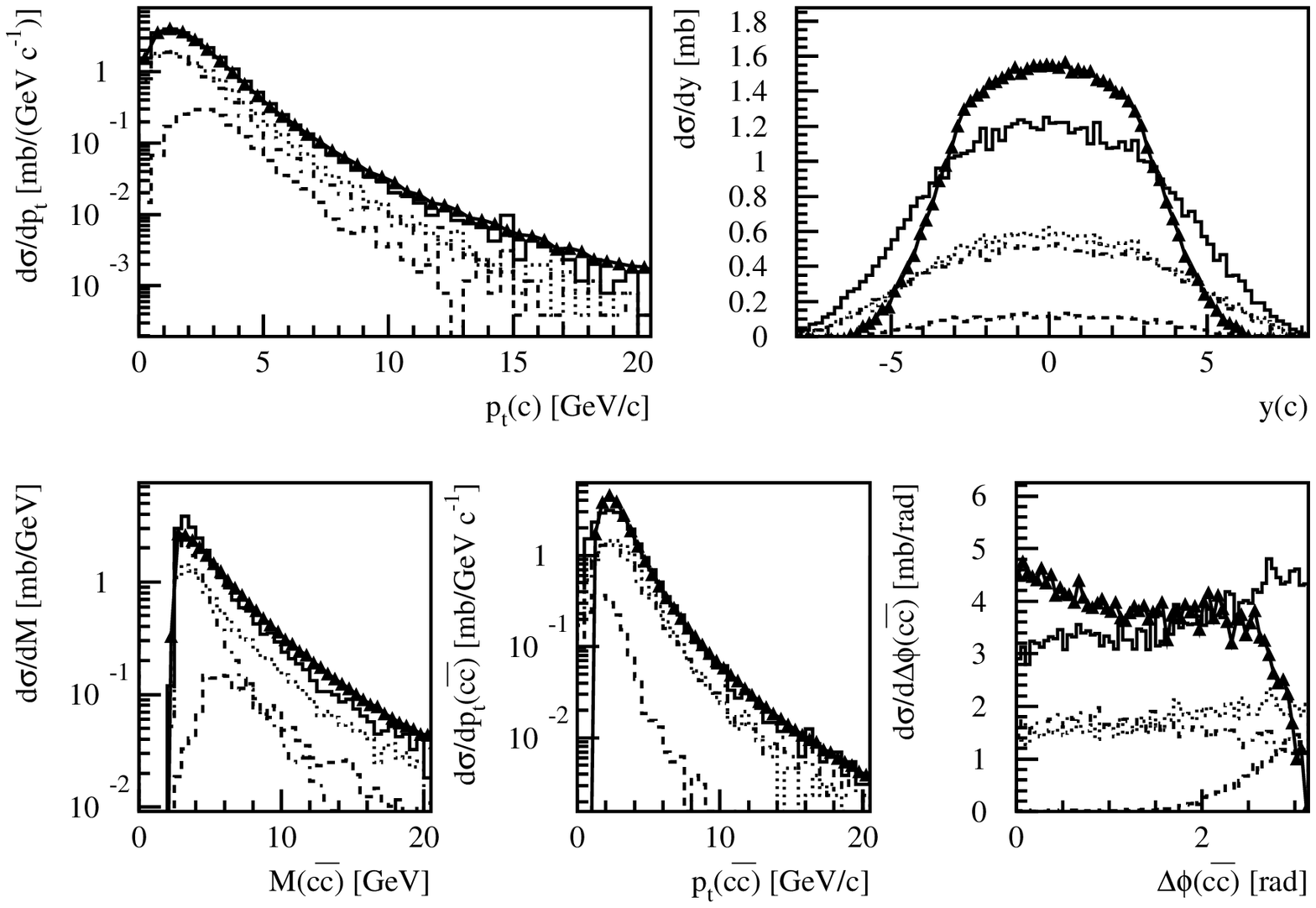}
  \end{center}
  \caption{Comparison between charm production in 14 TeV $pp$ interactions
    with the MNR code
    and {\sc pythia} with parameters of Table~\protect\ref{tab:PythiaParams}.
    The curves are the same as those in Fig.~\ref{fig:cmpCharmLO} but now the 
    MNR results are shown by the solid curves.}
  \label{fig:charmPyPpMNR}
  \begin{center}
    \includegraphics[width=.82\textwidth]{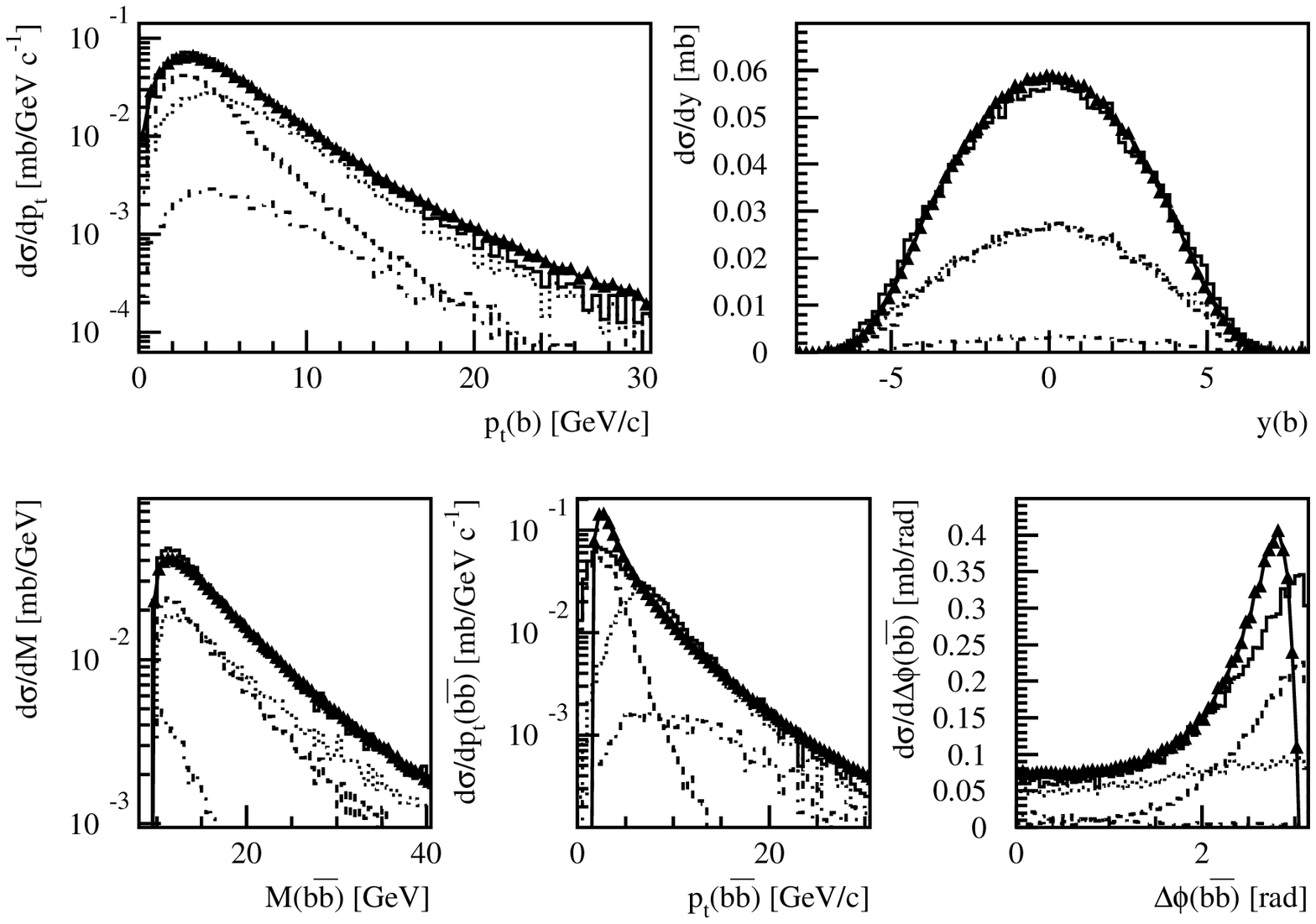}
  \end{center}
  \caption{Same as Fig.~\ref{fig:charmPyPpMNR}, but for bottom production.}
  \label{fig:bottomPyPpMNR}
\end{figure}

To simulate heavy quark production for the 14 TeV $pp$ run, the same
parton distribution functions and values of $m_Q$ and $\mu$ are used,
see Table~\ref{tab:PythiaParams}. The comparison of the {\sc pythia}
spectra with the MNR calculation is shown in Fig.~\ref{fig:charmPyPpMNR} 
for charm and in Fig.~\ref{fig:bottomPyPpMNR} for bottom. The largest
difference is a rather poor description of the charm quark 
rapidity distribution due to the range of validity of the 
parton distribution functions.
Most widely used parton distributions, including CTEQ4, are valid only down 
to $x \sim 10^{-5}$.  Below this $x$, the behaviour depends on the
implementation of the parton density but has no physical meaning. For example,
the CTEQ4 gluon density is fixed to its value at 
$x=10^{-5}$.
The rapidity range over which the evolution of the set of parton densities
is reliable depends on the energy.  For charm production at $\sqrt{s}=5.5$ and
14 TeV this range corresponds to $|y|<4.3$ and $|y|<3.4$ respectively.  

\subsubsection{Fragmentation and decay}

We use the default {\sc pythia} quark fragmentation and hadron decay 
parameters.  Table~\ref{tab:partcomp} summarizes the total yield and the 
rapidity density, $dN/dy$, averaged over $|y|<1$ for heavy flavour hadrons.

\begin{table}[!b]
  \caption{Total yield and average rapidity density for $|y|<1$ for
    hadrons with charm and bottom produced in Pb+Pb collisions at
    $\sqrt{s} = 5.5$~TeV/nucleon.}
  \label{tab:partcomp}
  \begin{center}
  \begin{tabular}{|c|c|c||c|c|c|}
\hline
  Particle & Yield & $\langle dN/dy \rangle_{|y|<1}$ & Particle & Yield &
    $\langle dN/dy \rangle_{|y|<1}$ \\\hline
$D^0$&   68.9&   6.87&$B^0$&   1.86&  0.273\\
$\overline{D}^0$&   71.9&   6.83&$\overline{B}^0$&   1.79&  0.262\\
$D^+$&   22.4&   2.12&$B^+$&   1.82&  0.251\\
$D^-$&   22.2&   2.00&$B^-$&   1.83&  0.270\\
$D_s^+$&   14.1&   1.30&$B_s^0$&   0.53&  0.077\\
$D_s^-$&   12.7&   1.22&$\overline{B}_s^0$&   0.53&  0.082\\
$\Lambda_c^+$&    9.7&   1.18&$\Lambda_b^0$&   0.36&  0.050\\
$\overline{\Lambda_c}^-$&    8.2&   0.85&$\overline{\Lambda_b}^0$&
    0.31&  0.047\\
\hline
  \end{tabular}
  \end{center}
\end{table}

\begin{figure}[!b]
  \begin{center}
    \includegraphics[width=\textwidth]{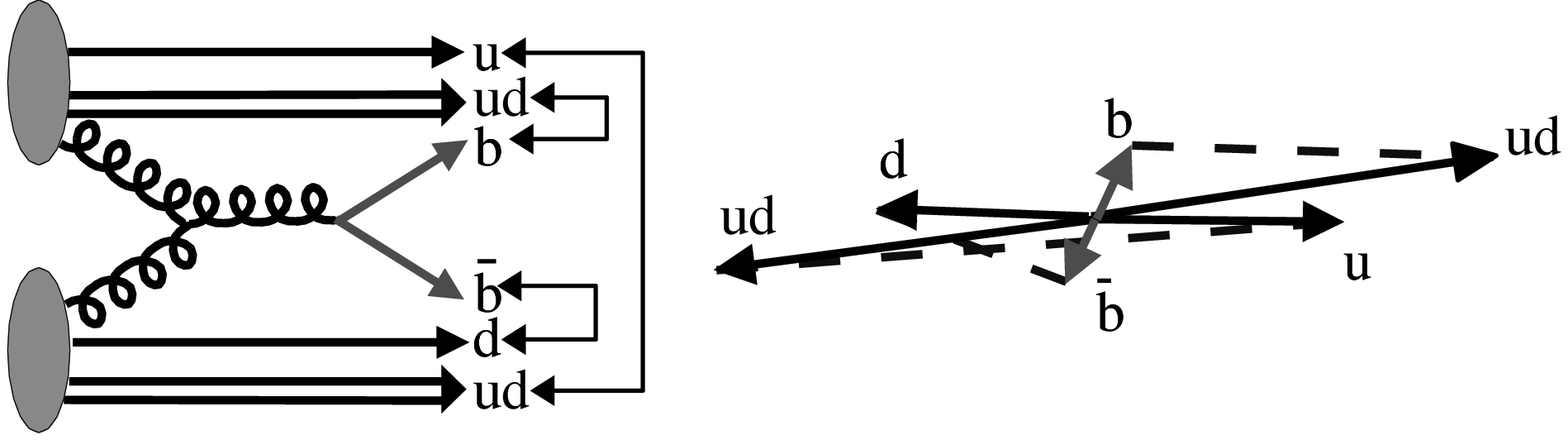}
  \end{center}
  \caption{Sketch of the string structure for the LO graph $gg
    \rightarrow b \overline{b}$. Left: the heavy quark is attached
    to a beam remnant (quark or diquark). Right: the same view in
    momentum space (not to scale).  The dashed lines
    represent strings.}
  \label{fig:fragmSketch}
\end{figure}

\begin{figure}
  \begin{center}
    \includegraphics[width=.6\textwidth]{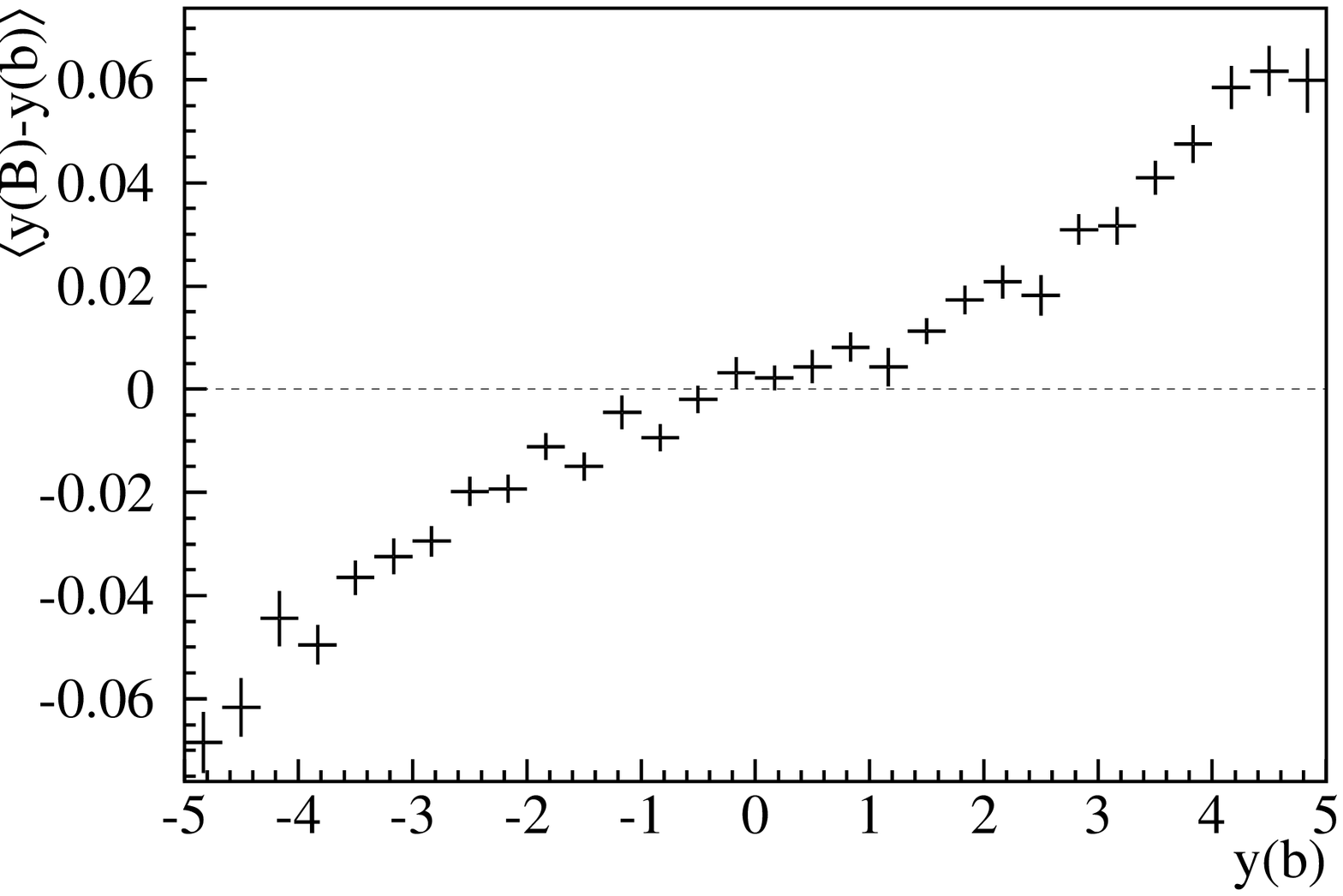}
  \end{center}
  \caption{Average difference between the rapidity of the B meson and
    the one of the corresponding b quark, as a function of the quark
    rapidity, for interactions with $p_T^{\rm hard} > m_b$.}
  \label{fig:bottomDeltaY}
\end{figure}

The default fragmentation scheme in {\sc pythia} is Lund string
fragmentation. The string topology is derived from the colour flow in
the hard process, determining the partonic partners in the colour
singlet state forming the string endpoints. Let us consider, for
example, one of the LO graphs contributing to
$g g \rightarrow b \overline{b}$, see
Fig.~\ref{fig:fragmSketch}.  The colour representation of the beam
remnants is always described by a quark in a colour triplet
state and a diquark in a colour anti-triplet state which
thus behaves like an antiquark. Therefore, in
this simple situation, there are only three strings, as shown in
Fig.~\ref{fig:fragmSketch}.  The $b$ quark is colour connected to a
diquark, the $\overline{b}$ is connected to a quark and the last
string connects the remaining quark-diquark system.  When we include
the parton shower, the number of quarks in the final state increases
and therefore more colour singlet strings can be produced. The strings
then fragment according to the longitudinal Lund symmetric
fragmentation function. For strings with heavy quark endpoints, the
Lund fragmentation function is too soft to reproduce the charm and bottom
meson fragmentation data. The Lund
fragmentation function is therefore modified through the {\sc pythia} default
settings to take into account the harder spectrum of heavy flavour
mesons using the Bowler parameterization. The fact that the heavy quark is
usually connected with one of the beam remnants shifts the rapidity of
the heavy quark towards the rapidity of the projectile or target 
\cite{Norrbin:2000zc}, see Fig.~\ref{fig:bottomDeltaY}. 

\begin{figure}
  \begin{center}
    \includegraphics[width=.8\textwidth]{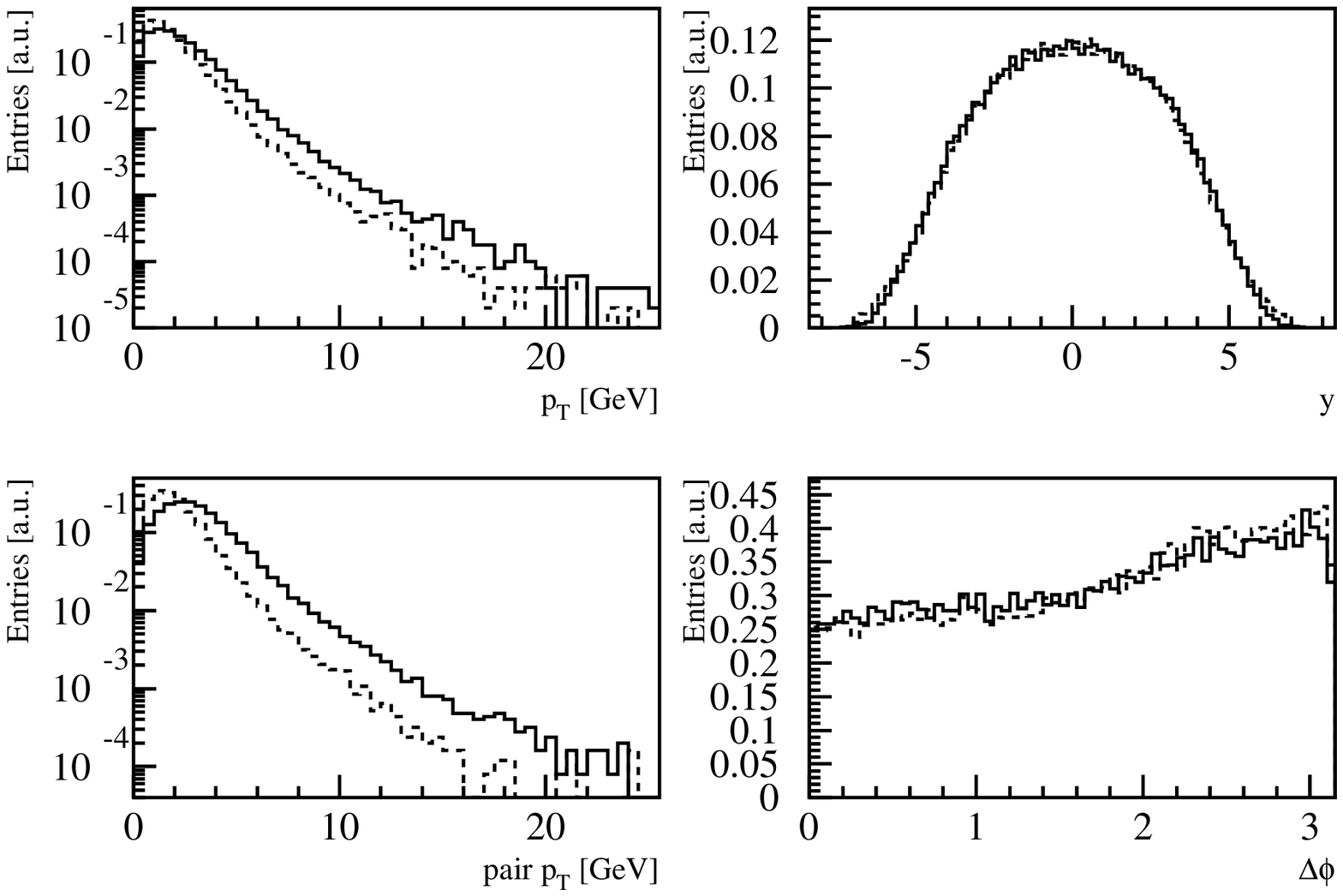}
  \end{center}
  \caption{Distributions for charm quarks (solid histogram) and 
    $D$ mesons (dashed histogram). The normalization is arbitrary.}
  \label{fig:FragmCharm}
\end{figure}

\begin{figure}
  \begin{center}
    \includegraphics[width=.8\textwidth]{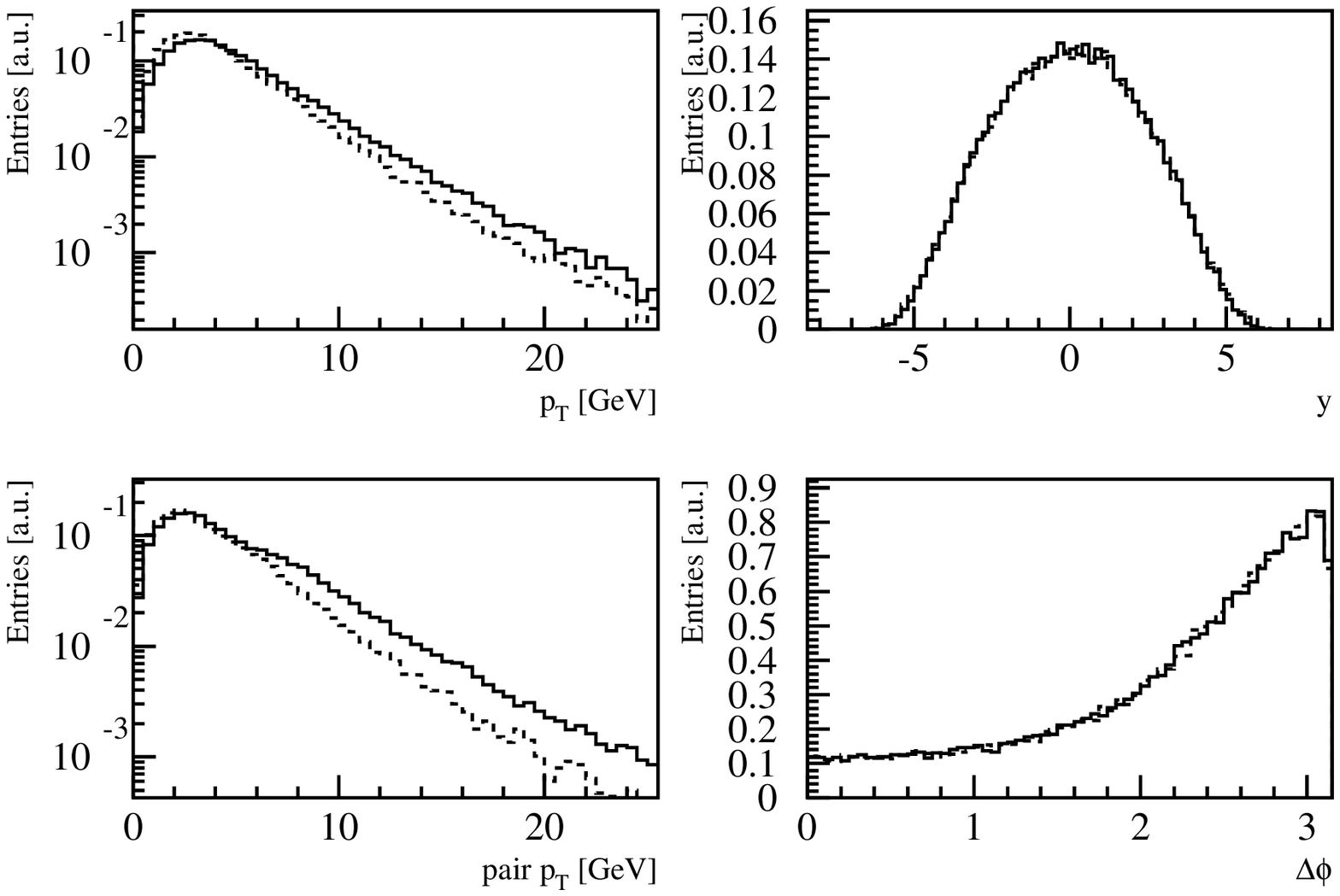}
  \end{center}
  \caption{Distributions for bottom quarks (solid
    histogram) and $B$ mesons (dashed histogram). The normalization is
    arbitrary.}
  \label{fig:FragmBottom}
\end{figure}

A detailed description of the heavy quark string fragmentation
can be found in Ref.~\cite{Norrbin:2000zc}. Here we report only how
it affects the kinematics. Figure~\ref{fig:FragmCharm}
shows the charm $p_T$ and rapidity as well as the $p_T$ and $\Delta\phi$
distributions of the $c\overline c$ pairs.  The solid histograms represent
quarks and the dashed histograms, the charm mesons.  In this figure we 
consider only the mesons into which the quarks hadronize.  The decay of heavy
flavor resonances is not included.  Figure~\ref{fig:FragmBottom} shows the
equivalent plots for bottom.

The most important effect is a reduction of the transverse momenta,
directly reflecting the magnitude of the longitudinal
fragmentation function. The fragmentation reduces the
momentum of charm mesons by 25\% and of bottom mesons by 14\% on average. The
rapidity shift shown in Fig.~\ref{fig:bottomDeltaY} is much smaller
than the rapidity range of the interactions.  Therefore it does not strongly
affect the inclusive rapidity distribution.  The azimuthal
correlation is also not significantly affected by fragmentation.

\subsection{A study of heavy flavour production in \textsc{herwig}}
\label{sec:herwig}

Like {\sc pythia}, 
{\sc herwig}~\cite{Corcella:2000bw,Corcella:2001wc,Marchesini:1991ch} 
is a general-purpose
event generator which uses parton showers to simulate higher-order
QCD effects. The main differences with respect to {\sc pythia} are 
the treatment of the parton showers, which accounts more correctly
for coherence effects, and the hadronization model, based on cluster
rather than string fragmentation. The version used for the analysis
presented here is {\sc herwig}~6.4.
Here we investigate the main features of heavy
flavour production in {\sc herwig} rather than trying to produce agreement
between {\sc herwig} and the NLO calculation.
The following simulations have been done for $pp$ collisions at $\sqrt{s} = 
5.5$ TeV.

\subsubsection{Comparison with LO MNR results}

As with {\sc pythia}, we first checked the description of the LO processes 
in {\sc herwig} by comparing to the LO results of the MNR code. For the LO 
calculations, we used the same parameters in both calculations.  We used the 
CTEQ4L parton distributions with $\Lambda_{\rm QCD}^{(4)} = 236$ MeV, $m_c 
= 1.2$ GeV, $m_b = 4.75$ GeV and default {\sc herwig} factorization and 
renormalization scales.  Good agreement was found when the {\sc herwig}
parton showers were switched off.

\subsubsection{Parton showers in \textsc{herwig}: gluon splitting and flavour 
excitation}

Since gluon splitting and flavour excitation cannot be considered
separately in the NLO code, as in event generators, we cannot compare the
two results at this level.  Instead we  can compare {\sc herwig} with
{\sc pythia} to see if there are
fundamental differences in the implementation of these processes.

\begin{figure}
  \begin{center}
    \includegraphics[width=.9\textwidth]{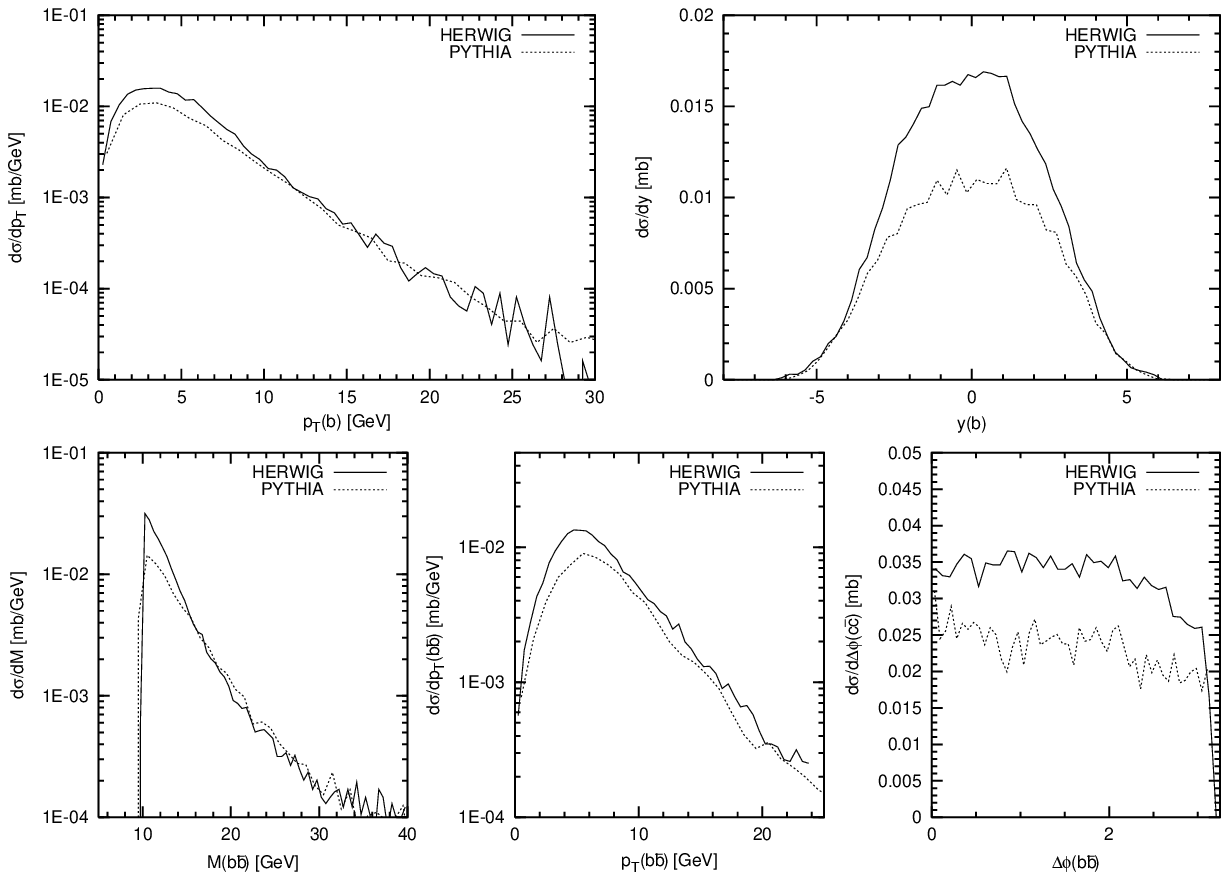}
    \caption{Gluon splitting results in $b\overline{b}$ production,
 obtained with {\sc herwig} (solid) and {\sc pythia} (dotted).}
    \label{gsbottom}
  \end{center}
\end{figure}

The gluon splitting results for bottom production are compared 
in Fig.~\ref{gsbottom}.
The shape of the distributions from the two generators
agree at high $p_T$. Since different scales are used in {\sc
  herwig} and {\sc pythia} we do not expect the result to be exactly
the same, especially at low $p_T$.

\begin{figure}
  \begin{center}
    \resizebox{160mm}{!}{\includegraphics{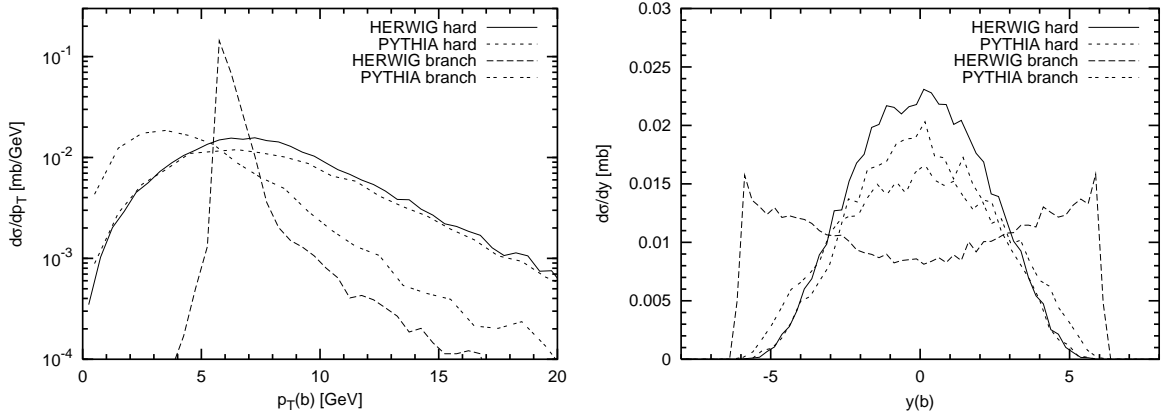}}
    \caption{Comparison of flavour excitation in {\sc herwig} and
      {\sc pythia}. The distributions are divided into two parts, a
      heavy quark coming from hard scattering (hard) and from gluon
      splitting (branch), see Fig.~\ref{fig:processes}.}
    \label{febranch}
  \end{center}
\end{figure}

On the other hand, flavour excitation is clearly different in the two
generators.  The problem is apparent in the single quark $p_T$
spectrum where a large peak is present at $p_T \sim m_b$ when {\sc herwig}
is used. We studied the origin of this peak by comparing the
single quark spectrum for the quark coming from the hard scattering
and for the quark coming from the initial-state gluon splitting, the middle
diagram in Fig.~\ref{fig:processes}.  We have performed this separation
and we present the comparison in Fig.~\ref{febranch}.  We can see that, 
while the hard scattering part is compatible for the two models, {\sc herwig} 
fails to describe the initial-state evolution of the gluon splitting into a
$Q\bQ$ pair. This may be due to the fact that
initial-state shower variables in {\sc herwig} are chosen for light
parton showers and mass effects are not considered~\cite{WebberComm}.
We found the same kind of behaviour for charm production although the peak in
$p_T$ spectrum is shifted to a value closer to $m_c$.
Given that \textsc{herwig} does not correctly describe flavour
excitation in the low-$\pt$ region, it should not be used for 
simulations in such a region. 

\subsection{\textsc{hijing}: event generator for nucleus-nucleus collisions}
\label{sec:hijing}

\textsc{hijing} (Heavy Ion Jet INteraction Generator) 
is widely used to simulate the background for  
feasibility studies of heavy flavour detection.
\textsc{hijing} combines a QCD-inspired model of 
jet production~\cite{Wang:1991ht,Gyulassy:ew} with the Lund model 
for jet fragmentation. Hard or semi-hard parton scatterings with transverse 
momentum of a few GeV are expected to dominate high energy heavy ion 
collisions. The \textsc{hijing} model has been developed with special 
emphasis on the role of minijets in $pp$, $pA$ and $AA$ interactions at
collider energies.

The Lund \textsc{fritiof}~\cite{Andersson:1986gw} model and the Dual Parton 
Model~\cite{Capella:yb} have guided the formulation of \textsc{hijing} 
for soft 
nucleus-nucleus interactions at low energies,
$\sqrt{s}\simeq 20$ GeV/nucleon. The hadronic collision model has been 
inspired by the successful implementation of perturbative QCD processes in 
\textsc{pythia}.

Two important features of nuclear collisions in \textsc{hijing} are the 
simulation of jet quenching and nuclear shadowing.
The charged-particle rapidity-density given by \textsc{hijing} with jet 
quenching for central Pb+Pb collisions at $\sqrt{s}=5.5$ TeV/nucleon
is $dN_{\rm ch}/dy\simeq 6200$, about a factor of 2 larger than that
predicted using recent analyses of RHIC results~\cite{Eskola:2001bf}. 
However, such a large multiplicity can be considered as a ``safety factor'' 
for the feasibility studies.  \textsc{hijing} 
also provides the relevant 
background sources such as direct leptons and strange particles.  

\subsection{Conclusions}
\label{sec:concl}

We have studied heavy quark production in nucleon-nucleon collisions
at LHC energies with the most widely used Monte Carlo
generators---\textsc{pythia} and \textsc{herwig}.  The
resulting distributions were compared 
to those obtained in perturbative QCD to NLO using
the MNR code \cite{Mangano:jk} to establish a baseline simulation of heavy
flavour production.
We found that by tuning the main \textsc{pythia} parameters, satisfactory
agreement with the NLO bare quark transverse momentum distributions was
obtained for both charm and bottom production in $pp$ collisions at
$\sqrt{s} = 5.5$ and 14 TeV.  The most important parameters tuned in the 
comparison were the heavy quark masses, the minimum $p_T^{\rm hard}$ of the
heavy quarks in the lab frame, and the width of the transverse momentum
smearing of the incoming partons, $\langle k_T^2 \rangle$.  However, we found
that \textsc{herwig} does not treat flavour excitation correctly at small
transverse momenta, making it unsuitable for simulations of heavy quark 
production. 

This problem is solved by merging the NLO computations with parton showers, 
as advocated in MC@NLO~\cite{Frixione:2003ei,Frixione:2002ik}. The 
implementation of heavy flavour hadroproduction, Ref.~\cite{Frixione:2003ei},
became available after the completion of the work reported in this
section.  We point out that, in the context of MC@NLO, no additional
tuning of the Monte Carlo parameters is needed to reproduce the NLO results
in the relevant regions since this is guaranteed by the
formalism.

%% file: alice_s.tex
\section[QUARKONIA AND HEAVY FLAVOUR DETECTION WITH THE ALICE 
DETECTOR]{QUARKONIA AND HEAVY FLAVOUR DETECTION WITH THE ALICE 
DETECTOR~\protect\footnote{Authors: P.~Crochet, A.~Dainese, E.~Vercellin.}}
\label{sec:alice}

\subsection{Introduction}

ALICE is the dedicated heavy-ion experiment at the LHC.
The apparatus will investigate strongly interacting
matter at extreme energy densities by comprehensive measurements of a 
large variety of observables~\cite{:1995pv,ALICEPPR}.
Among the most promising observables, heavy flavours are especially relevant 
since they provide an essential probe of the earliest stage of heavy ion 
collisions.
They further give precious information on the properties of the medium 
produced at longer time scales.

Quarkonium states can be pertinent signatures of the Quark 
Gluon Plasma (QGP).
From the early predictions of charmonium suppression by Debye screening in a 
deconfined medium~\cite{Matsui:1986dk}, to the recent results from the NA50 
collaboration at SPS~\cite{Abreu:2000ni}, much effort has been devoted to the 
subject (for reviews see Refs.~\cite{Satz:2000bn,Vogt:cu}).
While at SPS energies only charmonium states are experimentally accessible,
the much higher LHC (and RHIC) energies make bottomonium measurements 
feasible, thus providing an additional probe for QGP studies.
In fact, since the $\Upsilon$(1S) state only dissolves significantly above the 
critical temperature~\cite{Digal:2001ue,Wong:2001uu}, 
at a value which might only be 
reachable at energies above that of RHIC, the 
spectroscopy of the $\Upsilon$ family at the LHC should reveal unique 
characteristics of the QGP~\cite{Gunion:1996qc}.
We note also that quarkonium measurements at RHIC and LHC are complementary 
in terms of the accessible $x$ range.

The study of quarkonium states at the LHC is significantly different from 
those at the SPS and RHIC.
First of all, in addition to prompt charmonia produced by hard 
scattering, significant indirect charmonia can be produced 
by $B$ meson decays~\cite{Groom:in}, $D\overline{D}$ 
annihilation~\cite{Braun-Munzinger:2000dv}, 
and by coalescence mechanisms which 
could result in enhancement rather than suppression, see 
section~\ref{sec:thews}.
Then, in the environment of a heavy-ion reaction, in-medium effects 
such as shadowing and heavy quark energy loss may
substantially modify the final yields and spectra, see 
section~\ref{sec:dima}.
It is therefore obvious that an understanding of the QGP requires
systematic investigations.
More precisely, quarkonium must be studied:
\begin{itemize}
\item{as a function of centrality to identify suppression/enhancement 
patterns;}
\item{as a function of system size to vary the energy densities and thus 
disentangle normal and anomalous suppression;}
\item{for all species because the survival probability
reflects the temperature;}
\item{as a function of $p_T$ to disentangle models;}
\item{with good vertex resolution to distinguish between prompt and 
secondary production;}
\item{with respect to the orientation of the reaction plane to unravel 
Glauber and comover absorption;}
\item{together with other QGP signals to explore correlations;}
\item{together with open charm and bottom, the most natural normalization of
the quarkonium signals.}
\end{itemize}
This physics program should be achievable with the ALICE detector.
Indeed, quarkonium states will be identified both in the dielectron and in 
the dimuon channels. 

The study of open charm and bottom in
heavy ion collisions at the LHC is another important issue
addressed by ALICE since it probes the mechanisms 
of heavy quark production in the hot and dense medium formed in the 
early stage of the collision. In this sense, comparison with 
$pp$ and $pA$ interactions will be essential for establishing the 
baseline production cross sections and nuclear shadowing effects.

If a QGP is formed, secondary parton scattering may provide an 
additional source of charm quarks~\cite{Muller:xn,Geiger:1993py}.
In such a scenario, the number of heavy quarks produced would strongly 
depend on the initial temperature and lifetime of the plasma, providing
information on the early stage of the collision.
 
Plasma formation would not only modify the total production
cross section but would also affect the kinematic distributions of
the produced heavy quarks.  Elastic collisions of the heavy quarks with 
partons in the plasma and radiative loss in the medium may reduce the heavy 
quark momenta, see e.g. Ref.~\cite{Lin:1998bd}.
Energy loss by $D$ and $B$ mesons, originating from $c$ and $b$
quarks, is particularly relevant because heavy flavor energy loss is 
expected to be significantly less than for light hadron production,
dominated by light parton 
fragmentation at LHC energies, see Section~\ref{section:eloss}
and the jet chapter of this report for more discussion.

The present experimental picture of heavy flavor production 
in heavy ion collisions is quite unclear. The dimuon spectrum measured by
NA50 \cite{Abreu:2000oc} at the SPS may indicate enhanced charm production in 
central Pb+Pb collisions.  On the other hand, PHENIX measurements of $D$ 
production through the electron yield at RHIC suggest no charm enhancement 
in Au+Au collisions \cite{Adcox:2002cg}.

ALICE is equipped with dedicated subdetectors for the identification of 
secondary vertices with a displacement of $\sim 100~\mu{\rm m}$ from the 
primary interaction vertex.  These displaced vertices are the primary 
signature of heavy flavour decays.  Inclusive $D$ and $B$ meson production
will be measured through their semi-leptonic decays. 
Moreover, the exclusive reconstruction of hadronic $D$ meson decays 
will provide a direct measurement of the $D$ $p_T$ 
distribution.

The expected detector performance in the detection of hidden and 
open heavy flavours is presented here.

\subsection{The ALICE Detector}

\begin{figure}[hbt]
  \centering{\epsfig{file=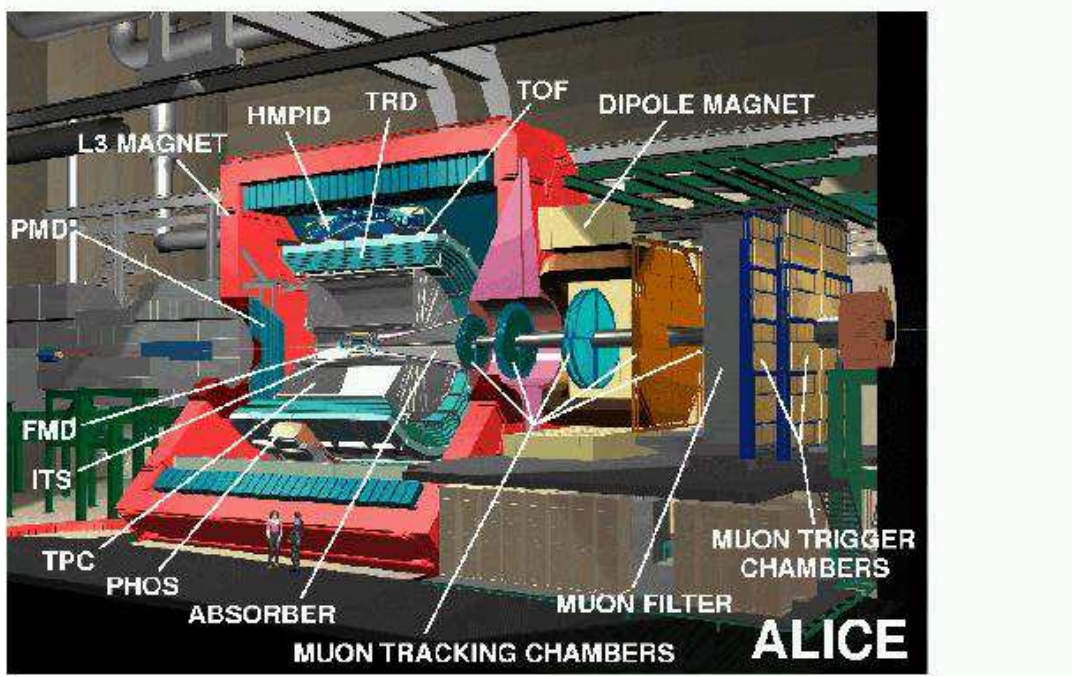,width=0.9\linewidth}}
  \caption{Schematic view of the ALICE detector.}
  \label{alice}
\end{figure}

The ALICE detector, shown in Fig.~\ref{alice}, is designed to cope with large 
particle multiplicities which, in central Pb+Pb collisions, are expected
to be between 2000 and 8000 per unit rapidity at midrapidity.
It consists of a central part, a forward muon
spectrometer and forward/backward small acceptance detectors.
The central part of ALICE consists of four layers of detectors 
placed in the small, $< 0.5$ T, solenoidal field provided by the LEP L3 magnet.
From the inner side to the outer side, these detectors are the Inner Tracker 
System (ITS), the large Time Projection Chamber (TPC), the Transition 
Radiation Detector (TRD) and the Time of Flight system (TOF).
They provide charged particle identification in the pseudorapidity range 
$|\eta|<0.9$, with full azimuthal coverage and a broad $p_T$ acceptance.
These large area devices are complemented by smaller acceptance detectors: 
the High Momentum Particle Identification (HMPID), the PHOton Spectrometer
(PHOS) and the Photon Multiplicity Detector (PMD).
In the forward/backward region, the charged multiplicity and the zero
degree energy will be measured by additional detectors (T0, V0, FMD, ZDC)
which will allow fast characterisation and selection of the events.
Finally a forward muon spectrometer covering the pseudorapidity range 
$2.5 < \eta <4$ is placed on the right side of the central part. 
It makes use of the usual techniques of muon identification at small angle
and consists of a front absorber, a dipole magnet, ten high-granularity
tracking chambers, a muon filter and four large area trigger chambers.

\subsection{Accessible $x$ Range}
\label{Ali_x1x2}

The LHC can probe both the nucleon parton distribution functions, in $pp$
collisions, and their modifications in the nucleus, using heavy ion collisions,
down to unprecedented low $x$ values
where $x$ is the fraction of the nucleon momentum carried by the 
interacting parton. In this section, the $x$ range of heavy flavour production
accessible to ALICE is quantitatively assessed.

The $x$ range probed depends on the center of mass energy
per nucleon pair $\sqrt{s}$, the $Q \overline Q$ pair invariant mass, 
$M_{Q\overline{Q}}$, and on the pair rapidity, $y_{Q\overline{Q}}$. 
We consider the leading order process, $gg\rightarrow Q\overline{Q}$ 
in the collision of two ions with mass and charge $(A_1,Z_1)$ and $(A_2,Z_2)$.
The square of the pair invariant mass is equal to the 
square of the center of mass energy of the initial gluons, $\hat{s}$,
\begin{equation}
M^2_{Q\overline{Q}}=\hat{s}=x_1\,x_2\,s =x_1\,\frac{Z_1}{A_1}\,x_2\,\frac{
Z_2}{A_2}\,s_{\rm pp},
\end{equation}
where $x_1$ and $x_2$ are the gluon momentum fractions 
and $s_{\rm pp} = (14~{\rm TeV})^2$ is the square of the $pp$ center of mass
energy at the LHC.
The longitudinal rapidity of the pair in the laboratory is given by:
\begin{equation}
\label{ypair}
y_{Q\overline{Q}}=\frac{1}{2}\ln \left[\frac{E+p_z}{E-p_z}\right]=\frac{1}{2}\
\ln \left[ \frac{x_1}{x_2} \frac{Z_1 A_2}{Z_2 A_1} \right].
\end{equation}
From the pair mass and rapidity, the dependence of $x_1$ and $x_2$ on 
$A$,$Z$, $M_{Q\overline{Q}}$ and $y_{Q\overline{Q}}$ can be derived, obtaining
\begin{equation}
x_1 = \frac{A_1}{Z_1}\frac{M_{Q\overline{Q}}}{\sqrt{s_{\rm pp}}}\exp\left
({y_{Q\overline{Q}}}\right)~~~~~~~~~~~~~~~~ 
x_2 = \frac{A_2}{Z_2}\frac{M_{Q\overline{Q}}}{\sqrt{s_{\rm pp}}}\exp\left
({-y_{Q\overline{Q}}}\right). 
\end{equation}

We first focus on Pb+Pb and $pp$ reactions.
At central rapidities, $x_1\simeq x_2$ with the magnitude determined 
by the ratio $M_{Q \overline Q}/\sqrt{s}$.  The $x$ values for $Q \overline Q$
production at threshold, 
$M_{c\overline{c}} =2m_c=2.4$~GeV and $M_{b\overline{b}} =2m_b=9$~GeV, are
given in Table~\ref{tab:x}.
\begin{table}[!h]
  \begin{center}
  \caption{The $x$ values for $Q \overline Q$ pair production at threshold
for central rapidities in Pb+Pb and $pp$ collisions.}
  \label{tab:x}
  \begin{tabular}{|c|c|c|}
  \hline
system ($\sqrt{s}$) & $x$ $(c \overline c)$ & $x$ $(b \overline b)$ \\
  \hline
  \hline
Pb+Pb (5.5 TeV) & $4.3\times 10^{-4}$ & $1.6\times 10^{-3}$ \\
$pp$ (14 TeV) & $1.7\times 10^{-4}$ & $6.4\times 10^{-4}$ \\ \hline
  \end{tabular}
  \end{center}
\end{table}

Because of the lower mass, charm probes smaller $x$ values than bottom. 
The capability for charm and bottom measurements in the forward 
rapidity region, $y\simeq 4$, gives access to $x$ regimes about 
2 orders of magnitude lower, down to $x\sim 10^{-6}$. 

\begin{figure}[!ht]
  \begin{center}
    \includegraphics[width=.75\textwidth]{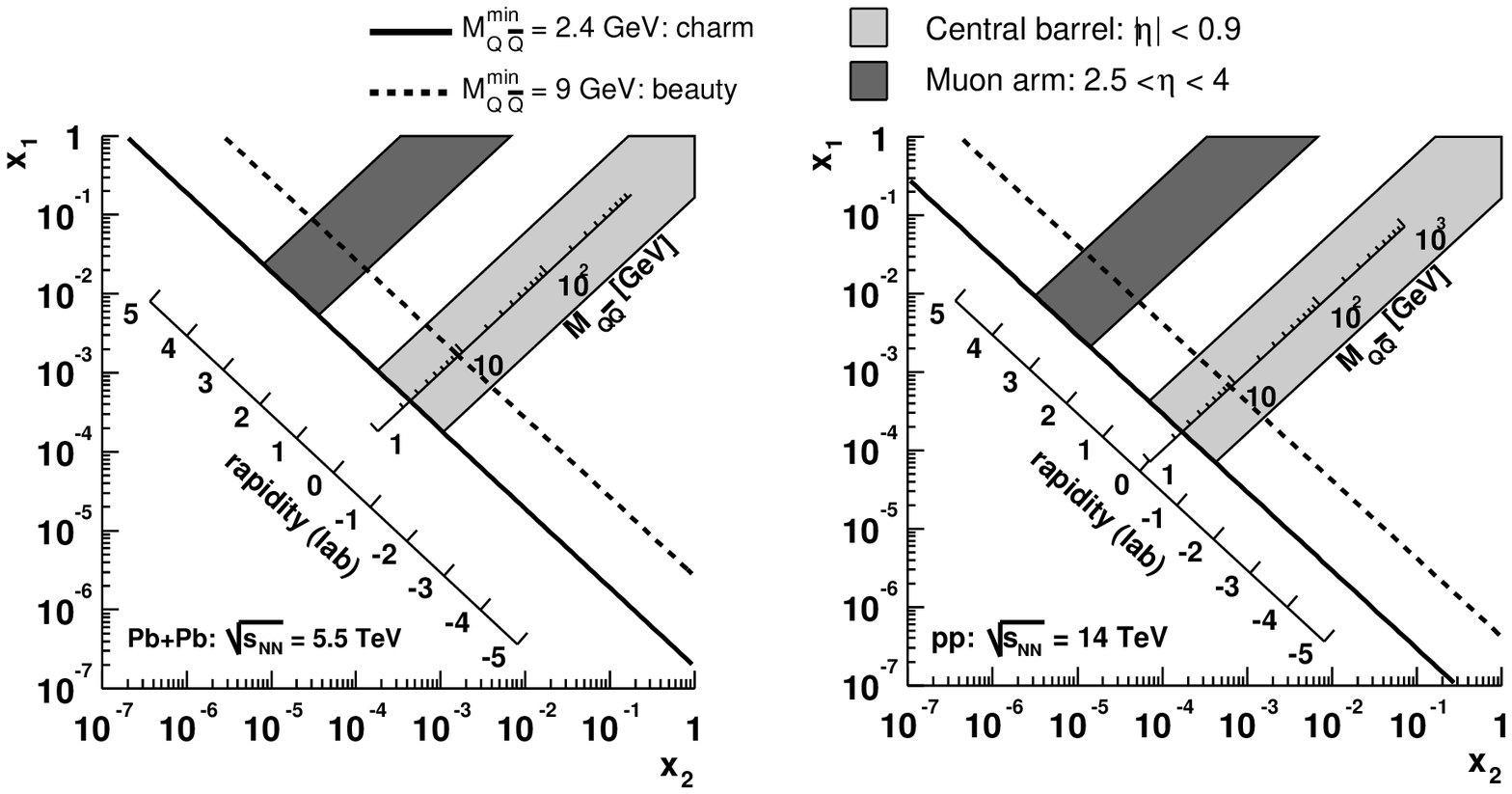}
  \end{center}
  \caption{ALICE heavy flavour acceptance in the ($x_1$, $x_2$) plane for  
Pb+Pb (left) and $pp$ (right) collisions.}
  \label{fig:x1x2_AApp}
  \begin{center}
    \includegraphics[width=.75\textwidth]{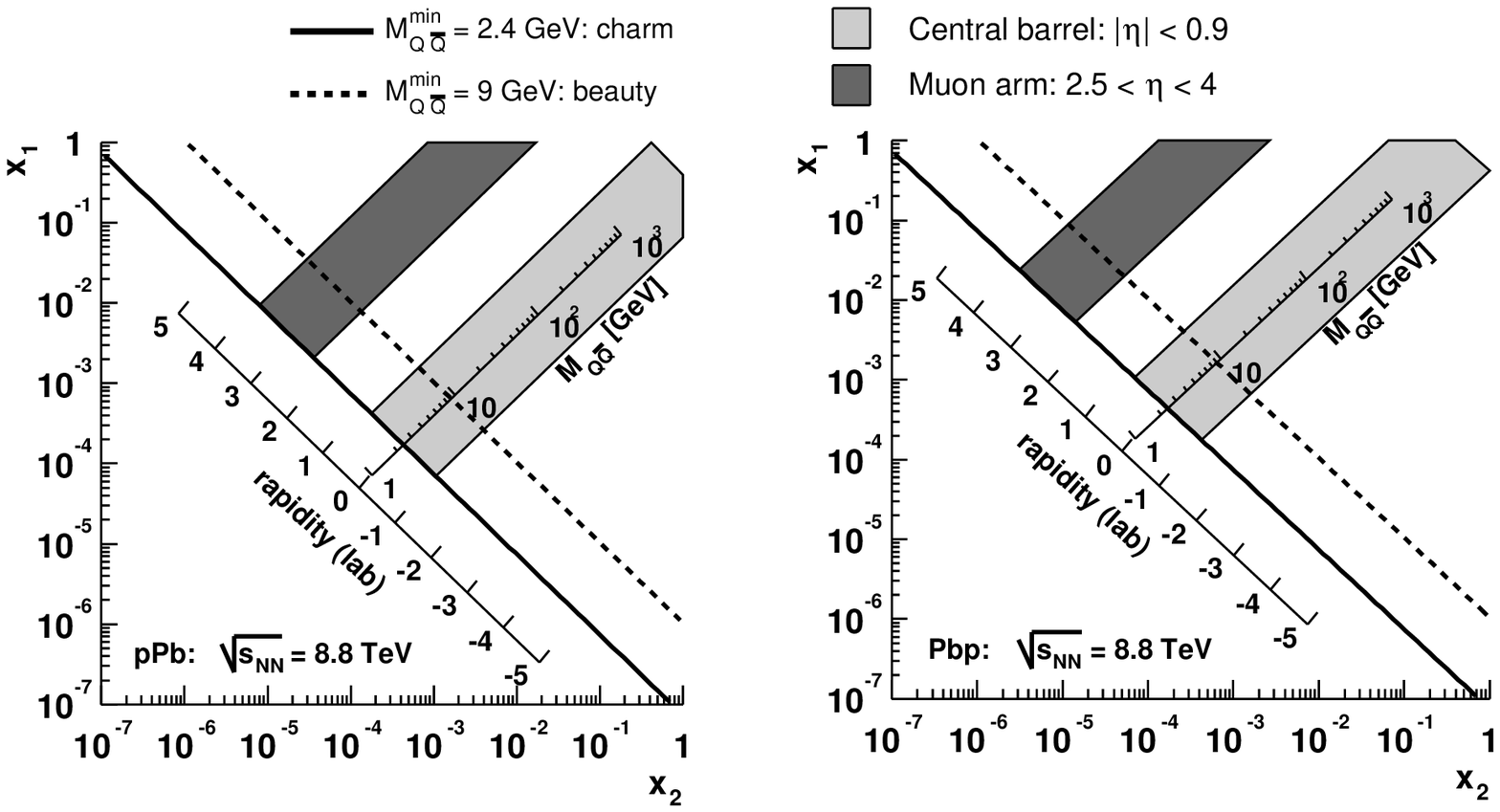}
  \end{center}
  \caption{ALICE heavy flavour acceptance in the ($x_1$, $x_2$) plane for 
$p$Pb (left) and Pb$p$ (right) collisions.}
  \label{fig:x1x2_pAAp}
  \begin{center}
    \includegraphics[width=.75\textwidth]{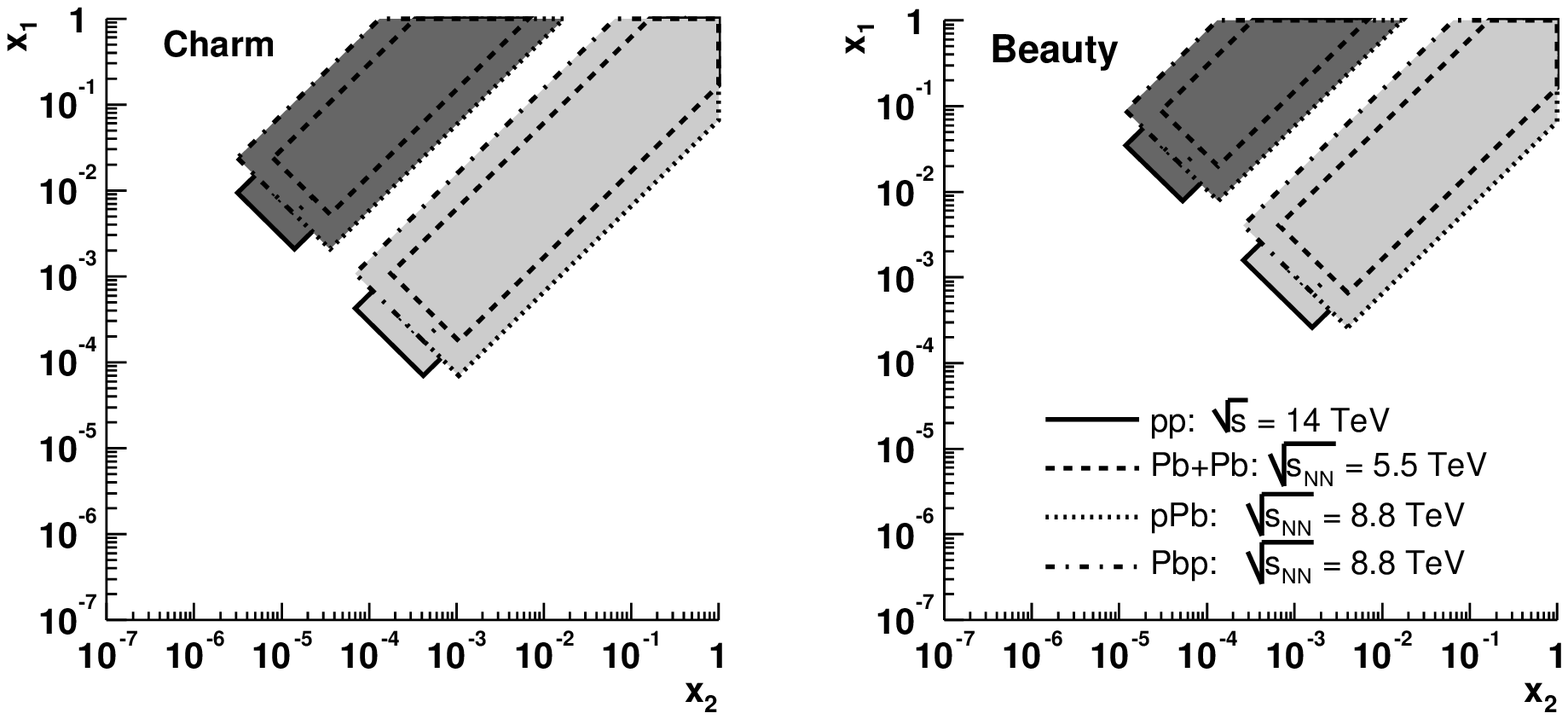}
  \end{center}
  \caption{Compiled ALICE acceptance in the ($x_1$, $x_2$) plane for 
charm (left) and bottom (right) in $pp$, Pb+Pb, $p$Pb and Pb$p$ collisions.}
  \label{fig:x1x2_global}
\end{figure}

In Fig.~\ref{fig:x1x2_AApp}, the regions of the ($x_1$, $x_2$) plane 
for charm and bottom measurements in the ALICE acceptance are shown for Pb+Pb 
collisions at 5.5~TeV 
and $pp$ collisions at 14~TeV. 
In this plane, points of constant invariant 
mass lie on hyperbolae, $x_1=M^2_{Q\overline{Q}}/(x_2\,s)$, 
which are straight lines on the log-log scale.  The solid and dashed
lines correspond to $c\overline{c}$ and $b\overline{b}$ pair production 
at threshold.  Points with constant rapidity lie
on straight lines, $x_1=x_2\exp(-2y_{Q\overline{Q}})$. The shaded regions 
show the acceptance of the ALICE central barrel, $|\eta|<0.9$, and 
of the muon arm, $2.5<\eta<4$.

In the case of asymmetric collisions, $Z_1 \neq Z_2$ and $A_1\neq A_2$, 
the $Q \overline Q$ rapidity is shifted by:
\begin{equation}
\Delta y_{Q \overline Q} = \frac{1}{2}\,\ln
\left(\frac{Z_1 A_2}{Z_2 A_1}\right) \, \, ,
\end{equation}
as seen in Eq.~(\ref{ypair}).
For $p$Pb (Pb$p$) collisions, $\Delta y=+0.47~(-0.47)$.
Therefore, combining $p$Pb and Pb$p$ runs will allow the 
largest $x_1$ and $x_2$ coverage with the central barrel and the muon arm. 
Figure~\ref{fig:x1x2_pAAp} shows the $p$Pb and Pb$p$ acceptances
while Fig.~\ref{fig:x1x2_global} compares the $pp$, Pb+Pb, $p$Pb and Pb$p$
coverages for charm (left) and beauty (right). 
  
Note that these figures only give a qualitative idea of the $x$ regions 
accessible to ALICE since the leading order definitions of $x$ were used
and the rapidity cuts were applied to the $Q \overline Q$ pair and not
the detected particles. In addition, no minimum $p_T$ cuts 
were applied.  Such cuts increase the minimum accessible value of 
$M_{Q\overline{Q}}$, correspondingly increasing the minimum accessible $x$. 
These approximations, however, are not too drastic since 
there is a very strong rapidity correlation between the 
initial $Q\overline{Q}$ pair and the produced heavy flavour hadrons
and, as shown in the following sections, the minimum $p_T$ 
cut will be quite low, lower than the hadron mass, for most of the 
channels studied at ALICE.

\subsection{Muons}

The ALICE forward muon spectrometer is designed to detect 
heavy quarkonia in the $\mu^+\mu^-$ decay channel. 
The $\mu^+\mu^-$ continuum will be measured together with the resonances
allowing the study of open charm and open bottom.

\subsubsection{Quarkonia}

\begin{figure}[hbt]
  \centering\mbox{\epsfig{file=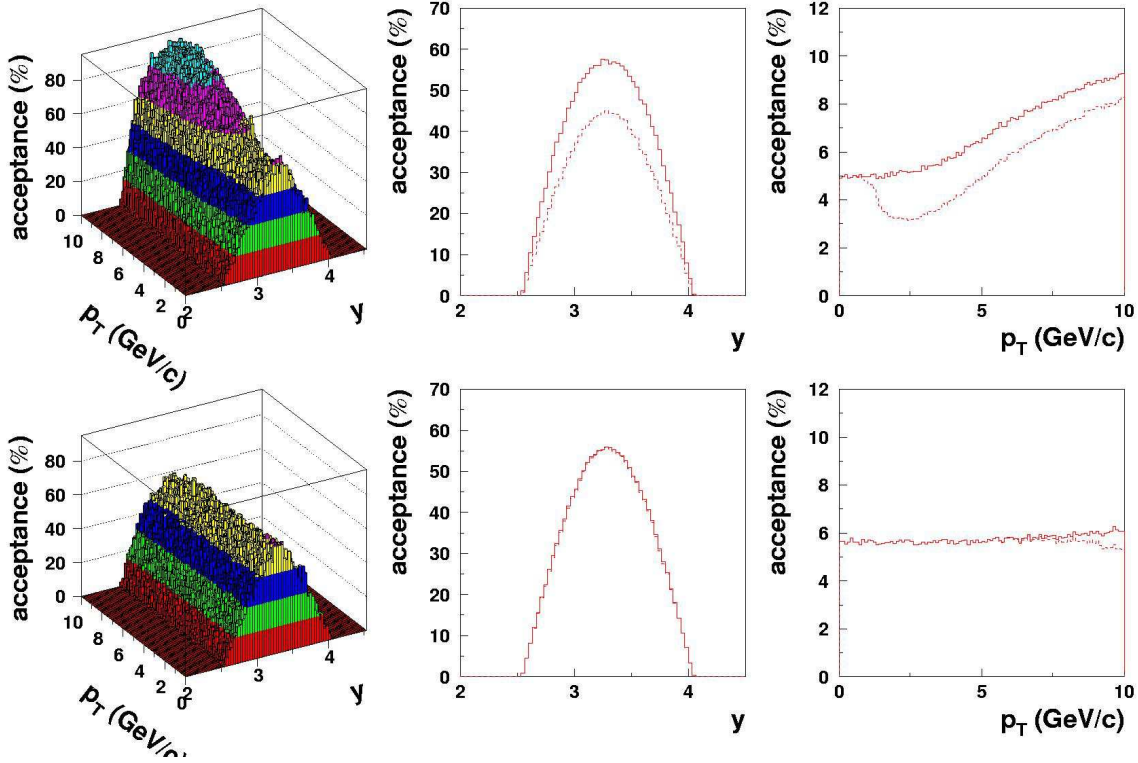, width=1.0\textwidth}}
  \caption{Geometrical acceptance for $J/\psi$ (top) and $\Upsilon$ (bottom).
    The left panels show the geometrical acceptance versus rapidity and 
    transverse momentum. The middle and right panels show the acceptance
    as a function of rapidity and transverse momentum, respectively.
    In the middle and right panels, the acceptance is shown without (solid
    histograms) and with (dashed histograms) the trigger cut on single muon 
    $p_T$. From Ref.~\cite{ALICEPPR}.}
  \label{dimu_accep}
\end{figure}

Muons are detected in the pseudo rapidity interval 
$2.5< \eta <4$.
To eliminate the huge low-$p_T$ background from $\pi$ and $K$ decays, 
a muon $p_T$ threshold is applied to single muons at the trigger level.
The resulting geometrical acceptance is shown in Fig.~\ref{dimu_accep}
for $J/\psi$ and $\Upsilon$.

The projection of the 3-D plots on the $y$ and $p_T$ axes are 
shown in the middle and right panels of the same figure.
Together with these projections, the ones obtained after applying a sharp
$p_T$ cut of 1 (2)~GeV/$c$ for $J/\psi$ ($\Upsilon$) are also shown to 
give an idea of the trigger effects.
As can be seen, only the $J/\psi$ acceptance is slightly 
affected by the trigger $p_T$ cut,
implying that $J/\psi$ and $\Upsilon$ can be detected down to $p_T = 0$.

\begin{figure}[hbt]
  \centering\mbox{\epsfig{file=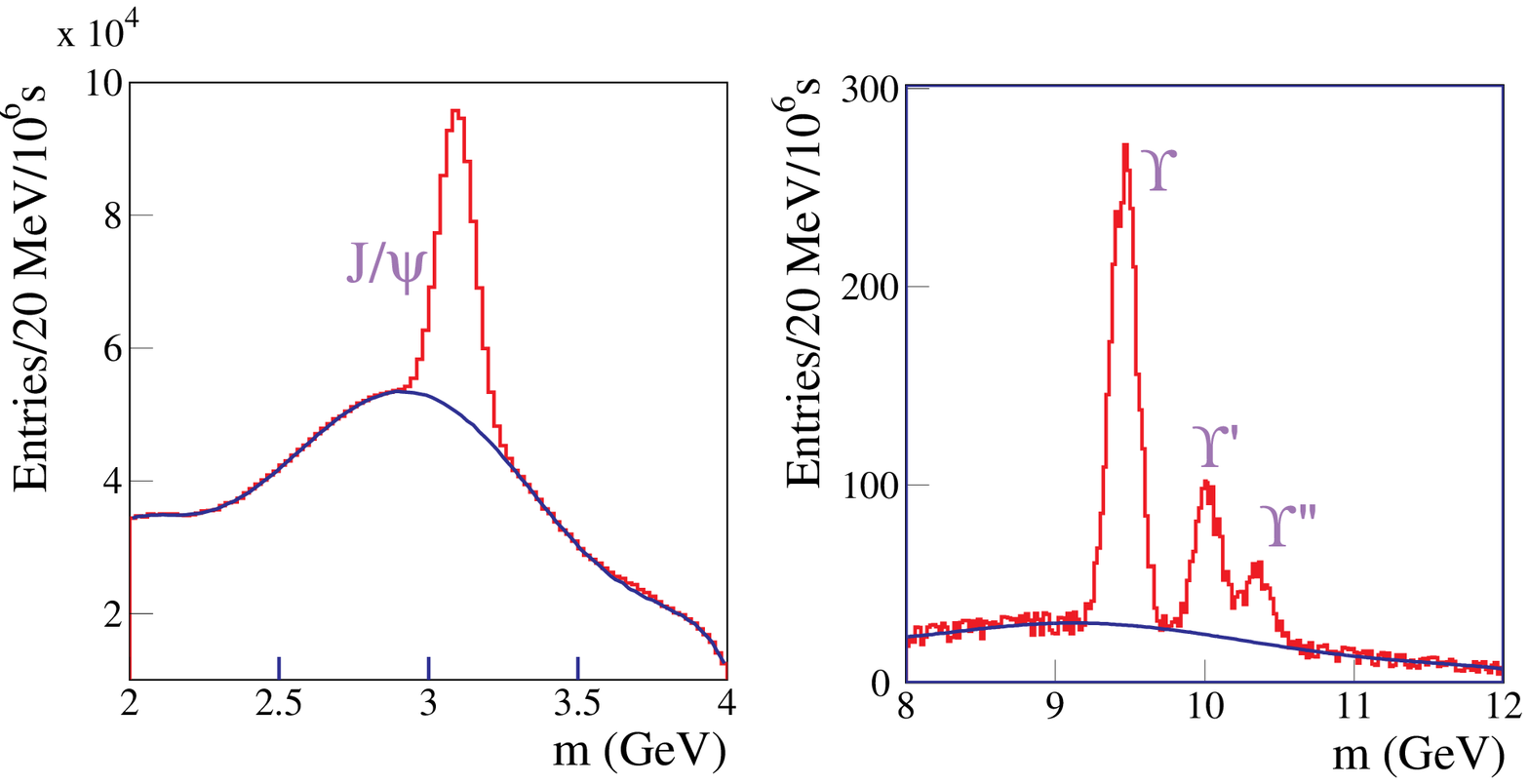, width=0.85\textwidth}}
  \caption{Opposite sign dimuon mass spectra in the region 
    $2 < m < 4~{\rm GeV}$ 
    (left) and $8 < m < 12~{\rm GeV}$ (right) for the $10\%$ most central
    Pb+Pb reactions. 
    The spectra were obtained by means of fast simulations including 
    acceptance cuts and detector efficiencies and resolutions.
    They correspond to a luminosity of 
    $5\times 10^{26}~{\rm cm}^{-2}{\rm s}^{-1}$ and a running time of 
    $10^6~{\rm s}$. From Refs.~\cite{Beole:1996yp,Morsch:xf}.}
  \label{dimu_resol}
\end{figure}

An important characteristic of the spectrometer is that its mass resolution
is expected to be of about $70$ ($100~{\rm MeV}/c^2$) for 
$J/\psi$ ($\Upsilon$).
As shown in Fig.~\ref{dimu_resol}, such resolution allows separation of
the $\Upsilon$ states.
It should then be possible to disentangle scenarios of quarkonium suppression 
by studying the $p_T$ dependence of quarkonium ratios~\cite{Gunion:1996qc}.

The statistics expected in a $10^6$ s run, roughly corresponding to one month 
of data taking, are summarized in Table~\ref{dimu_stat} 
for different colliding systems.  (Note that since Ca has been replaced by Ar
in the list of accepted nuclei, the Ca+Ca results are indicative of the 
expected rates for an intermediate mass system.)
The signal-to-background ratios and the significances
for detecting the various resonances are also given.
We note that these numbers were obtained by applying 
analysis cuts which are not fully optimized. 
Therefore they should be considered as indicative numbers only.
All numbers refer to primary production of the resonances and do not include 
any specific in-medium effects such as enhancement or suppression mechanisms.
The total quarkonium production cross sections 
in proton-proton collisions were taken from Ref.~\cite{Gavai:1994in}.  
These cross sections have been scaled to central ($10\%$) nucleus-nucleus 
reactions according to $A^2/(2\cdot 2)$ 
where the factors of $2$ represent the 
reduction for central relative to minimum bias collisions 
and nuclear shadowing~\cite{Beole:1996yp}.
The $p_T$ distributions have been parameterized according to 
CDF results (see Ref.~\cite{Beole:1996yp} for more details).

\begin{table}[htb]
\begin{center}
\caption{Expected background and signal rates, signal-to-background ratios 
and significance in pp and in central Pb+Pb and Ca+Ca collisions for 
charmonium and bottomonium. 
The numbers correspond to an interval of $\pm 1\sigma$ 
around the resonance mass. All rates and ratios are for a $10^6$s run.
The luminosities for the Pb+Pb, Ca-Ca and pp system
are $5\times 10^{26}$, $1\times 10^{29}$ and 
$1\times 10^{31}~{\rm cm}^{-2}{\rm s}^{-1}$, 
respectively. From Ref.~\cite{Beole:1996yp}.}
\label{dimu_stat}
\begin{tabular}{|c|c|c|c|c|c|} \hline
 system  & state & B ($\times 10^3$) & S ($\times 10^3$) & S/B & 
 S/$\sqrt{\rm S+B}$ \\
\hline
\hline
              & $J/\psi$                   & 320 & 230 & 0.72 & 310  \\
              & $\psi^\prime$              & 150 & 4.6 & 0.03 & 12   \\
Pb+Pb         & $\Upsilon$                 & 0.25 & 1.8 & 7.1 & 39   \\
              & $\Upsilon^\prime$          & 0.22 & 0.54 & 2.5 & 19  \\
              & $\Upsilon^{\prime\prime}$  & 0.18 & 0.26 & 1.5 & 12  \\
\hline
              & $J/\psi$                   & 760 & 2000 & 2.7 & 1200 \\
              & $\psi^\prime$              & 360 & 41 & 0.11 & 64    \\
Ca+Ca         & $\Upsilon$                 & 1.5 & 16  & 10.7 & 120  \\
              & $\Upsilon^\prime$          & 1.3 & 4.8 & 3.6 & 61    \\
              & $\Upsilon^{\prime\prime}$  & 1.1 & 2.3 & 2.0 & 39    \\
\hline
              & $J/\psi$                   & 64 & 850 & 13 & 890     \\
              & $\psi^\prime$              & 32 & 17 & 0.53 & 76     \\
$pp$          & $\Upsilon$                 & 0.49 & 6.5 & 13 & 78    \\ 
              & $\Upsilon^\prime$          & 0.45 & 2.0 & 4.4 & 40   \\
              & $\Upsilon^{\prime\prime}$  & 0.38 & 0.94 & 2.5 & 26  \\
\hline
\end{tabular}
\end{center}
\end{table}

As shown in the table, the high statistics expected for $J/\psi$ will
permit detailed studies as a function of both $p_T$ and centrality.
It is then also possible to investigate the azimuthal distribution of $J/\psi$
with respect to the reaction plane~\cite{DIMURANI}.
For the $\Upsilon$, the statistics are obviously reduced but are 
nevertheless sufficient to investigate the $p_T$ and centrality 
dependence.
Note particularly the large expected signal-to-background
ratio (about 10) and good significance.

\subsubsection{Open Charm and Open Bottom}
\label{dimu_open}

\begin{figure}[hbt]
  \centering\mbox{\epsfig{file=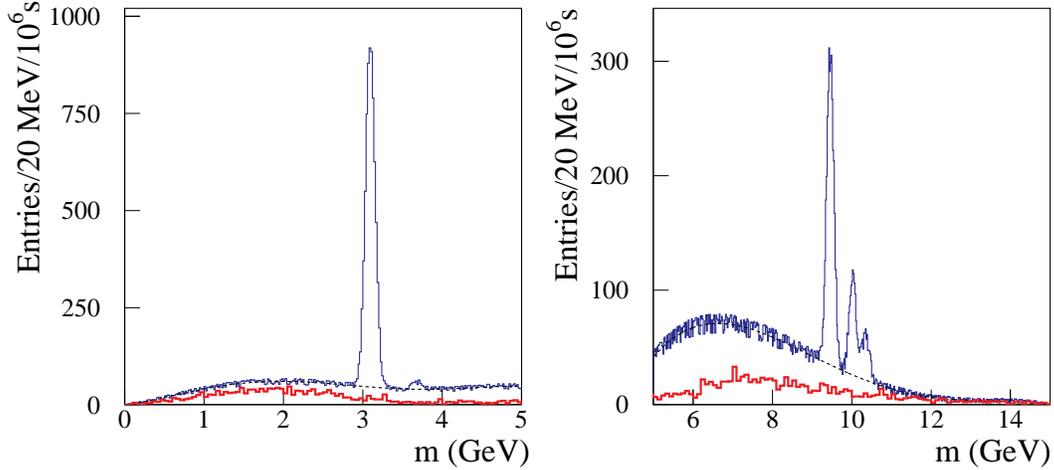, width=0.87\textwidth}}
  \caption{Opposite sign dimuon mass spectra in the region 
    $0 < m < 5~{\rm GeV}$ 
    (left) and $5 < m < 15~{\rm GeV}$ (right) for central Pb+Pb 
    reactions. 
    A $p_T$ threshold of $3~{\rm GeV}/c$ is applied on each single muon.
    The upper and lower histogram correspond to the total combinatorial 
    background and to the correlated signal from bottom decay, respectively.
    The spectra were obtained by means of fast simulations.
    From Ref.~\cite{andreaspriv}.}
  \label{dimu_corel}
\end{figure}

The muon spectrometer should allow measurements of the open bottom 
cross section.
This is possible because the background muons from $\pi$ and $K$ decay 
surviving
the 1~GeV/$c$ trigger $p_T$ cut, as well as those coming from charm decay,
can be efficiently removed with a higher software $p_T$ cut.
In fact, as illustrated in Fig.~\ref{dimu_corel}, when applying a $p_T$ 
cut of $3~{\rm GeV}/c$ on each muon, the correlated signal from bottom decays 
represent a significant fraction of the total dimuon yield both at 
low ($1 < m < 3~{\rm GeV}$) and high ($6 < m < 9~{\rm GeV}$) 
invariant mass.
Therefore, after subtraction of the uncorrelated combinatorial background
(evaluated, for instance, by event mixing) from the total 
dimuon yield, one should
be able to extract the unlike-sign correlated bottom decay signal 
in these two invariant mass regions.

It is important to note that the unlike-sign correlated signal from bottom 
decay has two different origins.
In the high invariant mass region (Fig.~\ref{dimu_corel}, right) each muon 
comes from the direct decay of a $B$ meson.
In the low invariant mass region (Fig.~\ref{dimu_corel}, left), the correlated 
signal from bottom is dominated by the so-called $B$-chain channel 
where both muons come from the decay of a single $B$ via a $D$.
This implies that the bottom cross section can be estimated 
independently (at least to some extent) by a study of the low and high
invariant mass regions.
Note that similar measurements of the charm cross section should be achievable
with appropriate analysis strategies.

In addition, it is important to remark that the bottom cross section can be 
also determined from the like-sign dimuon yield. 
In fact, $B$ meson oscillation as well as specific $B$ meson decay chains
represent a source of correlated like-sign muon pairs.
This component can be extracted by subtracting the uncorrelated signal 
from the like-sign spectrum~\cite{Crochet:2001qd}.

Note finally that the open heavy flavor cross sections can also
be extracted from the single muon spectra, as recently shown by the PHENIX
collaboration with single electron spectra~\cite{Adcox:2002cg}.

\subsection{Electrons}

The electrons, measured in the central region, will also give access to 
heavy quarkonia together with open charm and open 
bottom from the dielectron continuum. 
The key detector is the TRD which provides 
electron identification for $p_T>1$~GeV/$c$ and an electron trigger
for $p_T>3$~GeV/$c$.
It is operated in conjunction with the other ALICE central detectors
for precise momentum measurements.
In addition, the vertex capabilities of the ITS can distinguish
between primary and secondary $J/\psi$.
While the identification of the primary $J/\psi$ is of crucial importance
for QGP studies, the identification of secondary $J/\psi$ can lead to a direct 
measurement of the $B$ meson production cross section.
Furthermore, single high-$p_T$ electrons with displaced vertices give
access to the inclusive heavy flavor cross sections.

\subsubsection{Quarkonia}

\begin{figure}[htb]
  \centering\mbox{\epsfig{file=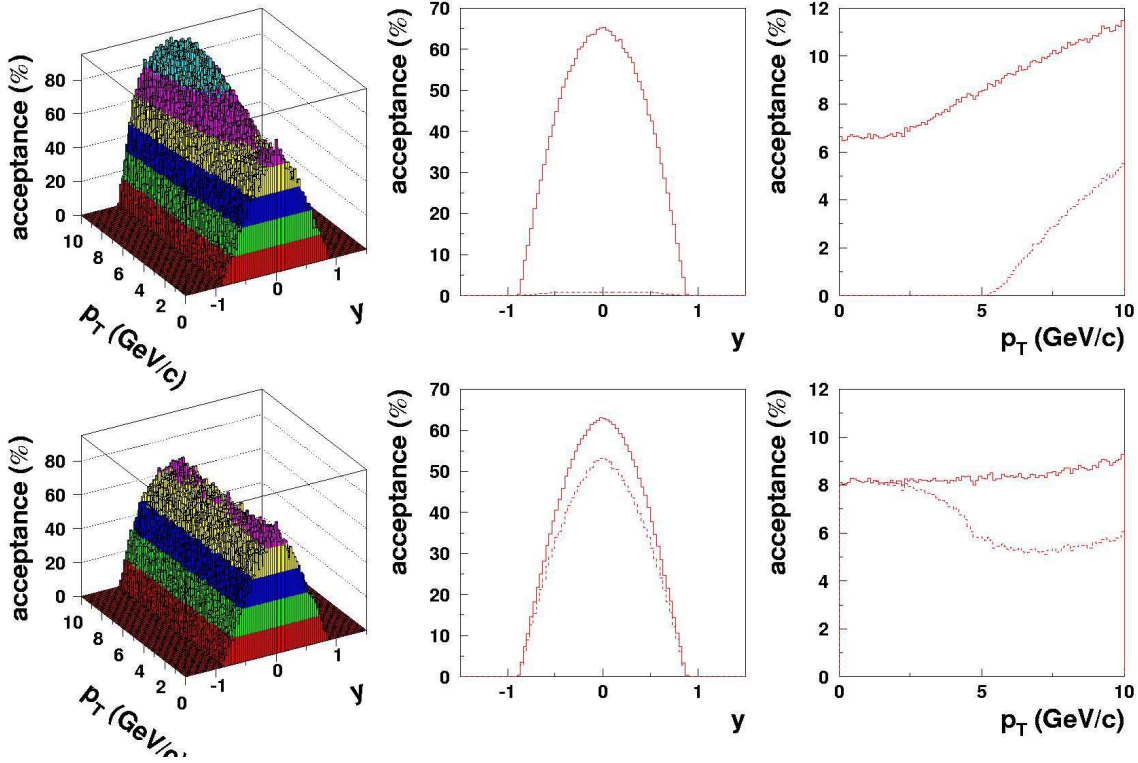,width=1.0\textwidth}}
  \caption{Same as Fig.~\ref{dimu_accep} for the $J/\psi$ (top) and $\Upsilon$ 
    (bottom) in the dielectron channel.}
  \label{diel_accep}
\end{figure}

The $J/\psi$ and $\Upsilon$ acceptances in the dielectron channel are 
displayed in Fig.~\ref{diel_accep}.
The effects of the TRD-L1 trigger, simulated by requiring a sharp 
$p_T$ threshold of 3~GeV/$c$ on both decay electrons, are shown 
in the middle and right panels of the figure.

The effects of the $p_T$ cut on the $\Upsilon$ are similar
to those observed for the $J/\psi$ in the dimuon channel, see 
Fig.~\ref{dimu_accep}. 
The cuts have a much stronger effect on the $J/\psi$ distribution.
It is only at rather large $p_T$ that the decay kinematics allow
both electrons to have $p_T >3~{\rm GeV}/c$. 
Therefore, while the $\Upsilon$ can be measured down to $p_T=0$, 
there is no $J/\psi$ acceptance below $p_T\sim 5.2$~GeV/$c$ because of 
the trigger condition.
As in the case of the dimuon channel, the expected mass resolution
is of the order of $100~{\rm MeV}/c^2$ for the $\Upsilon$ when the L3 magnet 
is operated at ${\rm B}=0.4~$T, allowing the $\Upsilon$ states to be separated.

The expected statistics are strongly dependent on the trigger efficiency
which decreases with particle multiplicity.
In the worst case anticipated for Pb+Pb, $dN_{\rm ch}/d\eta = 8000$, 
the expected number of $\Upsilon$ 
per month is $\sim 2600$ in minimum-bias collisions.
This number is expected to be significantly larger for lower charged particle 
multiplicities~\cite{TRDTDR}.

The current $J/\psi$ simulations indicate that the statistics will be 
significantly lower than that achieved in the dimuon channel, 
mainly due to the high $p_T$ cut of the electron trigger.
In a scenario with a smaller charged 
particle multiplicity, the electron trigger $p_T$ threshold could 
be lowered, leading to better $J/\psi$ detection efficiency.

\subsubsection{Open Charm and Open Bottom}

\begin{figure}[hbt]
  \centering\mbox{\epsfig{file=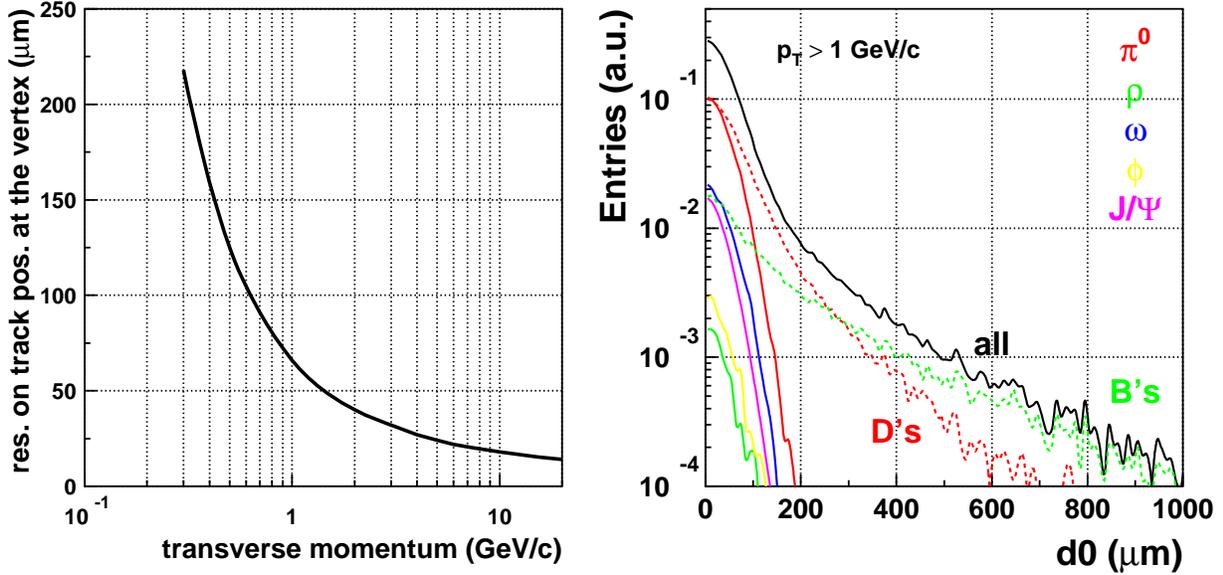, width=1.0\textwidth}}
  \caption{Left: the transverse momentum dependence of the bending 
    projection (d0) of the impact parameter 
    resolution.
    Right: the d0 distribution of electrons originating from different 
    parent particles. From Ref.~\cite{TRDTP}.}
  \label{diel1}
\end{figure}

The same studies described in section~\ref{dimu_open} for the muon channel
will be also carried out in the electron channel using similar methods.
In addition, the electron identification in the TRD can be used with 
the very good vertexing capabilities provided by the ITS.
Due to the finite $c\tau$ of $D$ and $B$ mesons, their decay
electrons have impact parameters\footnote{The impact parameter d0 is
defined as the distance of closest approach of the particle trajectory,
projected in the plane orthogonal to the beam axis, 
to the interaction vertex.} 
d0~$\simeq 100-300~\mu{\rm m}$ and d0~$\simeq 500~\mu{\rm m}$, respectively.
Figure~\ref{diel1} shows the resolution of the track position at the vertex
achieved in ALICE\footnote{To obtain the impact parameter resolution, the
track position resolution must be quadratically combined with the error on the
primary vertex position, negligible in Pb+Pb collisions, $\simeq 15$ $\mu$m.} 
(left) and the d0 distributions for electrons 
coming from $D$ and $B$ mesons and from other parent particles (right).
The expected rates (per unit rapidity at midrapidity) for semileptonic decays
of $D$ and $B$ 
mesons in central Pb+Pb collisions (5\% $\sigma_{\rm tot}$)
at $\sqrt{s}=5.5~{\rm TeV}$ are 2 and 0.09 per event, respectively.

\begin{figure}[!ht]
  \begin{center}
  \includegraphics[width=0.49\textwidth]{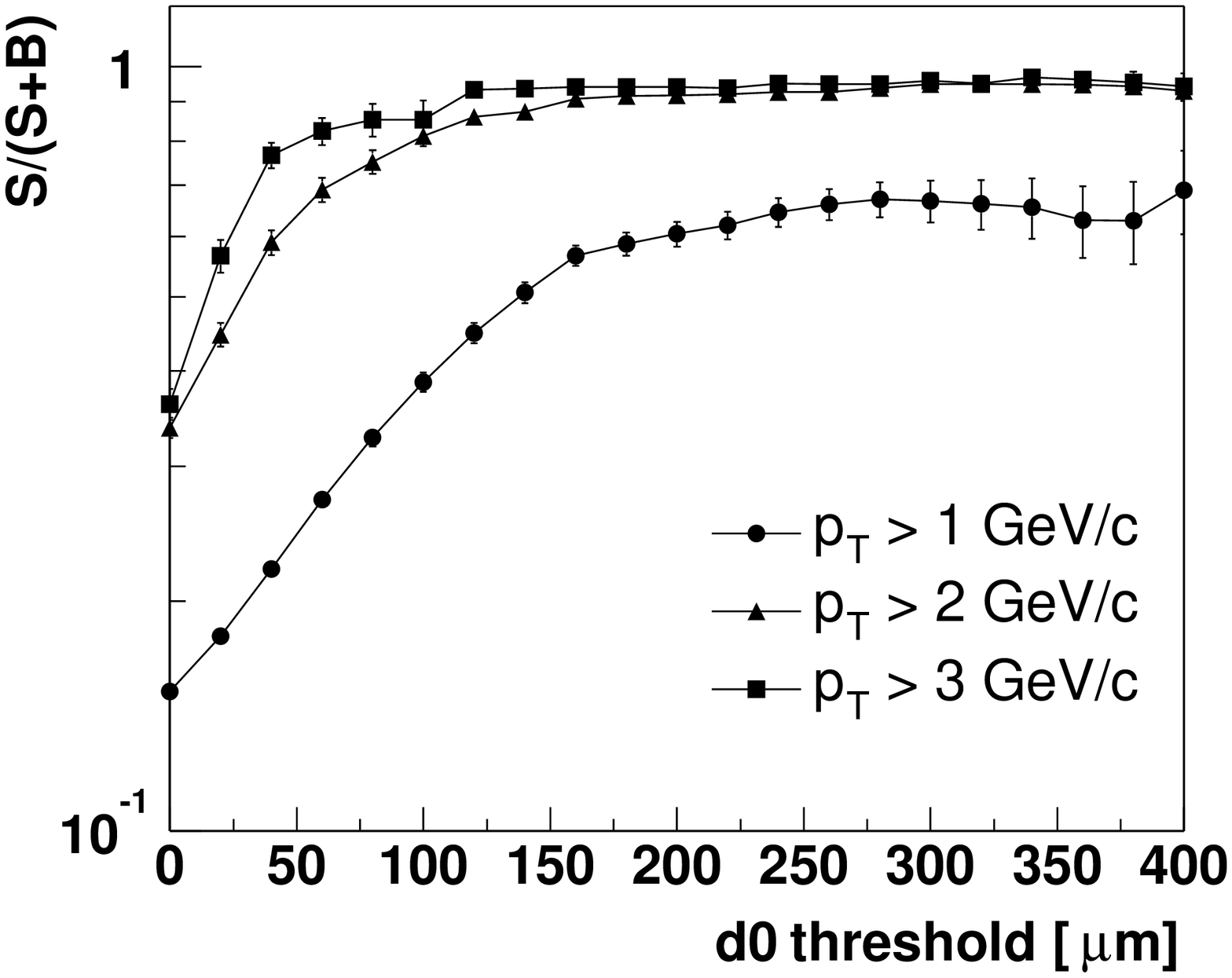}
  \includegraphics[width=0.49\textwidth]{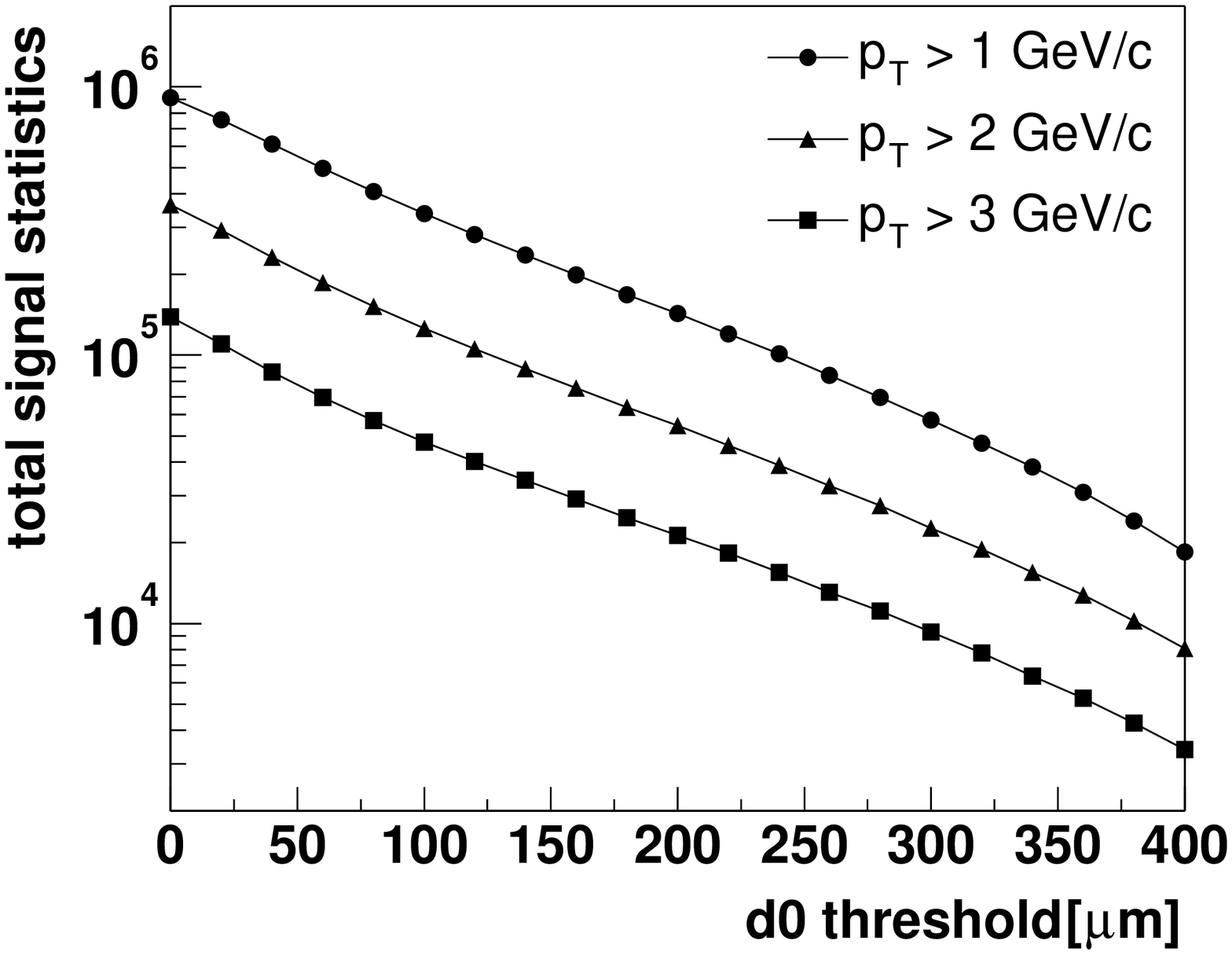}
  \caption{Detection of $B\to e+X$: ${\rm S/(S+B)}$ ratio (left) and 
           signal statistics
           for $10^7$ central Pb+Pb events (right) as a function of the 
           d0 threshold and for three values of the $p_T$ threshold.
           From Ref.~\cite{noteBeauty}.}
  \label{resultsB}
  \end{center}
\end{figure}

The performance of ALICE for open bottom detection in the 
$B\to e+X$ channel has been studied using a detailed simulation of the 
apparatus and including all relevant background sources (direct-charm 
decays, pair production due to photon conversion in the detector materials,
Dalitz decays of light and strange mesons, pions misidentified as 
electrons)~\cite{noteBeauty}. A very pure electron sample is obtained 
using the transition radiation technique (TRD) in conjunction with the 
$dE/dx$ measurement  in the TPC.  The fraction of pions misidentified as 
electrons is $\sim 10^{-4}$. The electrons from $B$ decays have 
harder $p_T$ spectra and, as already mentioned, 
broader d0 distributions.  Therefore, a sample of 
electrons from open bottom decays can be selected using suitable $p_T$ 
and d0 thresholds.  The left-hand side of Fig.~\ref{resultsB} presents the 
signal/(signal+background) ratio for three values of the electron $p_T$ cut 
as a function of the d0 threshold. 
As an example, a condition of $p_T >2$ GeV/$c$ and 
d0~$>180~\mu {\rm m}$ gives a sample of $90~\%$ purity. The expected 
number of $B$ mesons detected in $10^7$ central Pb+Pb events 
(1 month of ALICE data taking) is $\simeq 80000$ without 
any specific electron trigger. A more complete view of the 
attainable statistics is shown on the right-hand side of Fig.~\ref{resultsB} 
as a function of the d0 threshold for three different $p_T$ 
thresholds.

If the $p_T$ and d0 thresholds are lowered to $\sim 1$ GeV/$c$
and $\sim 100~\mu {\rm m}$, the contribution of electrons from 
open charm decays becomes comparable with that of open beauty. 
A method for disentangling the two contributions is currently under study. 

\begin{figure}[htb]
  \centering\mbox{\epsfig{file=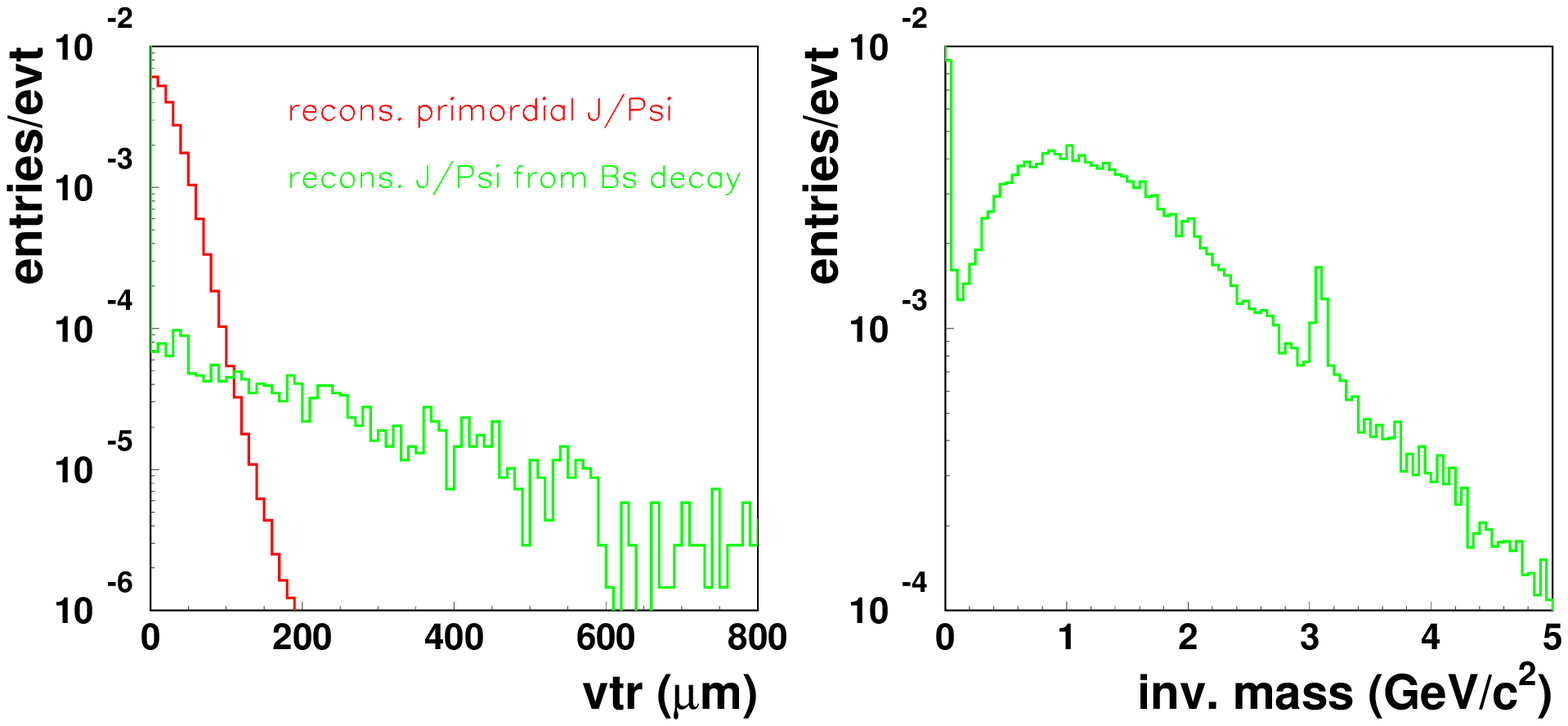,width=1.0\textwidth}}
  \caption{Left: dependence of the $J/\psi$ signal on the distance to the
    primary vertex.  Right: invariant mass distribution of pairs with 
    displaced vertices. From Ref.~\cite{TRDTP}.} 
  \label{second_jpsi}
\end{figure}

The second way to measure the open bottom cross section is based on the
measurement of secondary $J/\psi$'s from $B$ decays. 
These secondary $J/\psi$'s are produced at large distances from the primary
vertex and can therefore be selected by identifying $e^+e^-$ pairs with 
displaced vertices, as shown on the left-hand side of 
Fig.~\ref{second_jpsi}.  The resulting invariant mass spectrum is presented on
the right-hand side of Fig.~\ref{second_jpsi}. 

\subsection{Electron-Muon Coincidences}

The correlated $c\overline{c}$ and ${b}\overline{b}$ 
cross sections can be measured in 
ALICE using opposite sign electron-muon pairs.
The electron is identified in the central part and the muon is detected
in the forward muon spectrometer.
This channel is the only leptonic channel that gives direct access to 
correlated ${c}\overline{c}$ and ${b}\overline{b}$ pairs.
Indeed, in contrast to the $e^+e^-$ and $\mu^+\mu^-$ channels,
resonance decays, direct dilepton production and thermal production 
cannot produce correlated $e\mu$ pairs.
Within ALICE, the $e\mu$ channel has the additional advantage that the 
rapidity distribution of the corresponding signal extends in  $1 < y < 3$, 
bridging the central and the forward acceptances~\cite{Lin:1998bd}.
Measurements of $e\mu$ coincidences were successful
in $pp$ interactions at $\sqrt{s}=60~{\rm GeV}$~\cite{Chilingarov:1979ur} 
and in $pN$ interactions at $\sqrt{s}=29~{\rm GeV}$~\cite{Akesson:1996wf}. 
They are planned for heavy ion collisions with the PHENIX detector 
at RHIC~\cite{phe93}.
Preliminary simulations have shown that such 
measurements might be possible with ALICE~\cite{TRDTP,:1999kg,MUONTDR}.

\subsection{Hadrons}

In the central part of ALICE, heavy mesons and baryons can be fully 
reconstructed from their charged particle decay products in the ITS, 
TPC and TOF.
Thus, not only their integrated yields, but also their
$p_T$ distributions can be measured.
The most promising open charm decay channel is 
$D^0 \rightarrow K^-\pi^+$ (and its charge conjugate) with
a branching ratio of $\sim 3.9$\% and $c\tau=124~\mu{\rm m}$.
The expected rates (per unit rapidity at midrapidity) 
for $D^0$ (and $\overline{D}^0$) mesons, decaying into $K^\mp\pi^\pm$ pairs, 
in the 5\% most central 
Pb+Pb collisions at $\sqrt{s}=5.5~{\rm TeV}$ and in $pp$ 
collisions at $\sqrt{s}=14~{\rm TeV}$ are $5.3\times 10^{-1}$ and 
$7.5\times 10^{-4}$ per event, respectively.

\begin{figure}[hbt]
  \centering\mbox{\epsfig{file=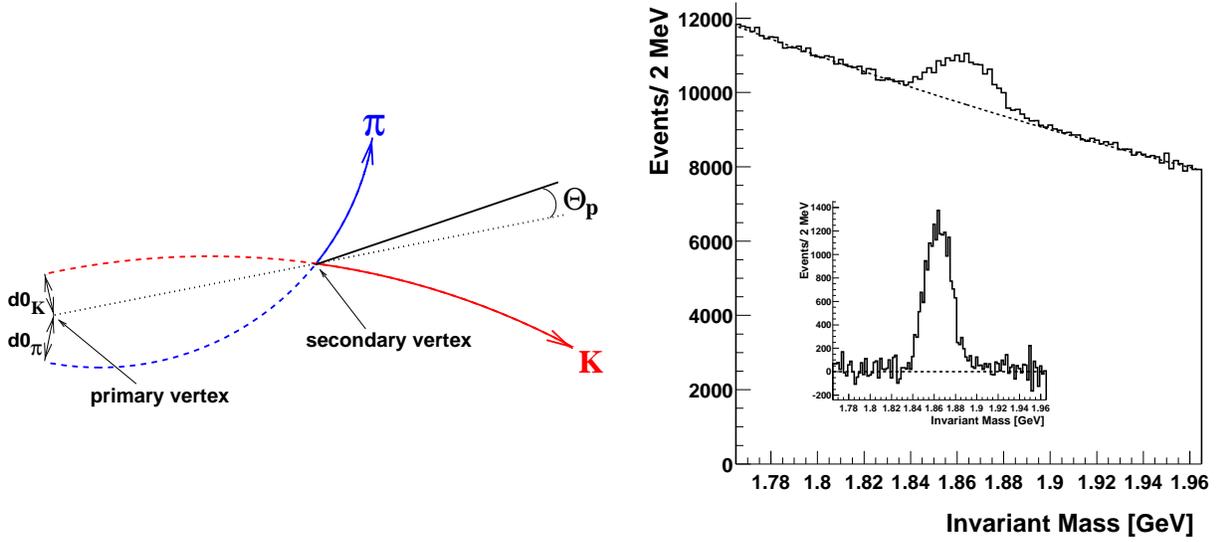,angle=0,width=1.0\linewidth}}
  \caption{Left: schematic representation of the $D^0 \rightarrow K\pi$
    decay showing the impact parameters of the kaon (d0$_K$)
    and of the pion (d0$_\pi$) and the pointing angle 
    ($\Theta_{\rm p}$).
    Right: $K\pi$ invariant mass distribution for $10^7$ central 
    Pb+Pb events.
    The background-subtracted distribution is shown in the inset.
    From Ref.~\cite{Carrer:aj}.}
  \label{D0combined}
\end{figure}

This decay channel allows direct identification of 
the $D^0$ particles by computing the invariant mass of fully-reconstructed 
topologies originating from displaced secondary vertices.  The left-hand side
of Fig.~\ref{D0combined} sketches the decay.  The main feature 
of this topology is the presence of two tracks with impact parameters 
of order $100~\mu{\rm m}$. 

The capabilities of the ALICE central barrel for $D^0$-meson searches have
been investigated both for Pb+Pb collisions, see 
Refs.~\cite{ALICEPPR,andreanote,Carrer:aj},
and for $pp$ collisions.  All the relevant background 
sources have been included in these studies, along with a detailed simulation 
of the detector response. In particular, for $pp$, the 
position of the interaction vertex will have to be reconstructed
event-by-event using the tracks\footnote{The transverse size of the LHC 
Pb beams will
be only $15~{\rm \mu m}$. In the $pp$ runs, the beams will be defocused at 
the ALICE interaction point
in order to bring the luminosity down to $\sim 10^{30}~{\rm 
cm^{-2}s^{-1}}$, see Section~\ref{running}.  Thus, the transverse 
size of the beams may be increased up to $\simeq 100-200~{\rm \mu m}$.
The resolution of the interaction vertex position in the transverse 
plane achieved using the tracks is
$\simeq 50~{\rm \mu m}$.}. 
 
In Table~\ref{tab:D0initial} the statistics for signal, S, and 
background, B, per event and the signal-to-background ratio 
after reconstruction are presented.  The only selection is on pair invariant 
mass, $|M_{K\pi} - M_{D^0}|<3~\sigma = 36~{\rm MeV}$.
\begin{table}[!h]
  \begin{center}
  \caption{Statistics for signal and background and signal-to-background ratio 
including only selection on pair invariant mass, 
$|M_{K\pi}-M_{D^0}|<3~\sigma = 36~{\rm MeV}$.}
  \label{tab:D0initial}
  \begin{tabular}{|c|c|c|c|}
  \hline
system ($\sqrt{s}$) & S/event & B/event & S/B \\
  \hline
  \hline
Pb+Pb (5.5 TeV) & $1.3\times 10^{-1}$ & $2.8\times 10^4$ & $4.5\times 10^{-6}$
\\
$pp$ (14 TeV) & $2.4\times 10^{-4}$ & $1.1\times 10^{-1}$ & 
$2.3\times 10^{-3}$ \\
  \hline
  \end{tabular}
  \end{center}
\end{table}

The most effective selection in order to extract the charm signal out of 
the large combinatorial background of opposite-charge track pairs 
is based on the requirement to have two tracks with large impact parameters
and a good pointing of the reconstructed $D^0$ momentum to the collision 
point ({\it i.e.} the pointing angle $\Theta_p$ between the $D^0$ momentum 
and its flight-line should be close to 0, as shown on the left-hand side of
Fig.~\ref{D0combined}). The selection strategy is described in detail in
Refs.~\cite{ALICEPPR,andreanote,Carrer:aj}. In Table~\ref{tab:D0final} 
the statistics after selection, together with the signal-to-background ratio
and the significance, are presented. 
The expected statistics are $\simeq 13000$ reconstructed $D^0$ in $10^7$ 
central Pb+Pb events and $\simeq 19000$ in $10^9$ $pp$ events.  The
(relatively) poorer quality of the $pp$ results is due to the larger error on
the primary vertex position with respect to Pb+Pb.

\begin{table}[!h]
  \begin{center}
  \caption{Statistics for signal and background, signal-to-background ratio 
           and significance after selections. The significance is computed 
           for the number of events collected in a one month run,
           $10^7$ for Pb+Pb and $10^9$ for $pp$.}
  \label{tab:D0final}
  \begin{tabular}{|c|c|c|c|c|}
  \hline
  system ($\sqrt{s}$) & S/event & B/event & S/B (\%) & 
  ${\rm S}/\sqrt{{\rm S+B}}$\\
  \hline
  \hline
  Pb+Pb (5.5 TeV) & $1.3\times 10^{-3}$ & $1.2\times 10^{-2}$ & 11 & 
  37~~~($10^7$ events)\\
  $pp$ (14 TeV) & $1.9\times 10^{-5}$ & $1.7\times 10^{-4}$ & 11 & 
  44~~~($10^9$ events) \\
  \hline
  \end{tabular}
  \end{center}
\end{table}

\begin{figure}[!t]
  \begin{center}
  \includegraphics[width=\textwidth]{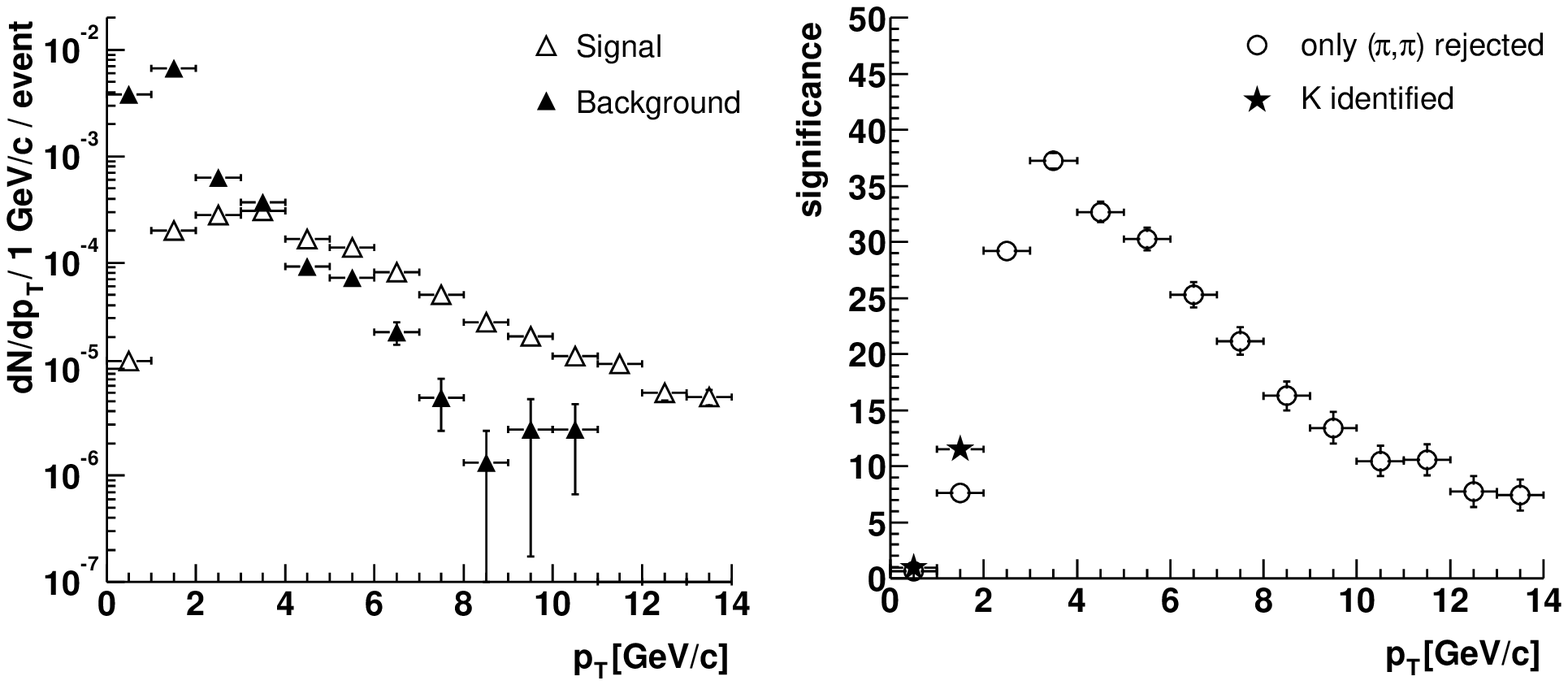}  
  \caption{Left: the transverse momentum distribution of selected 
    $D^0$ candidates in Pb+Pb collisions at $\sqrt{s}=5.5$ TeV.
    The signal and background are 
    normalized to one central Pb+Pb event.
    Right: the corresponding significance for $10^7$ events. 
    At low $p_T$ the values obtained requiring kaon identification 
    are shown by the stars.}
  \label{D0ptPbPb}
  \end{center}
\end{figure}
\begin{figure}[!h]
  \begin{center}
  \includegraphics[width=\textwidth]{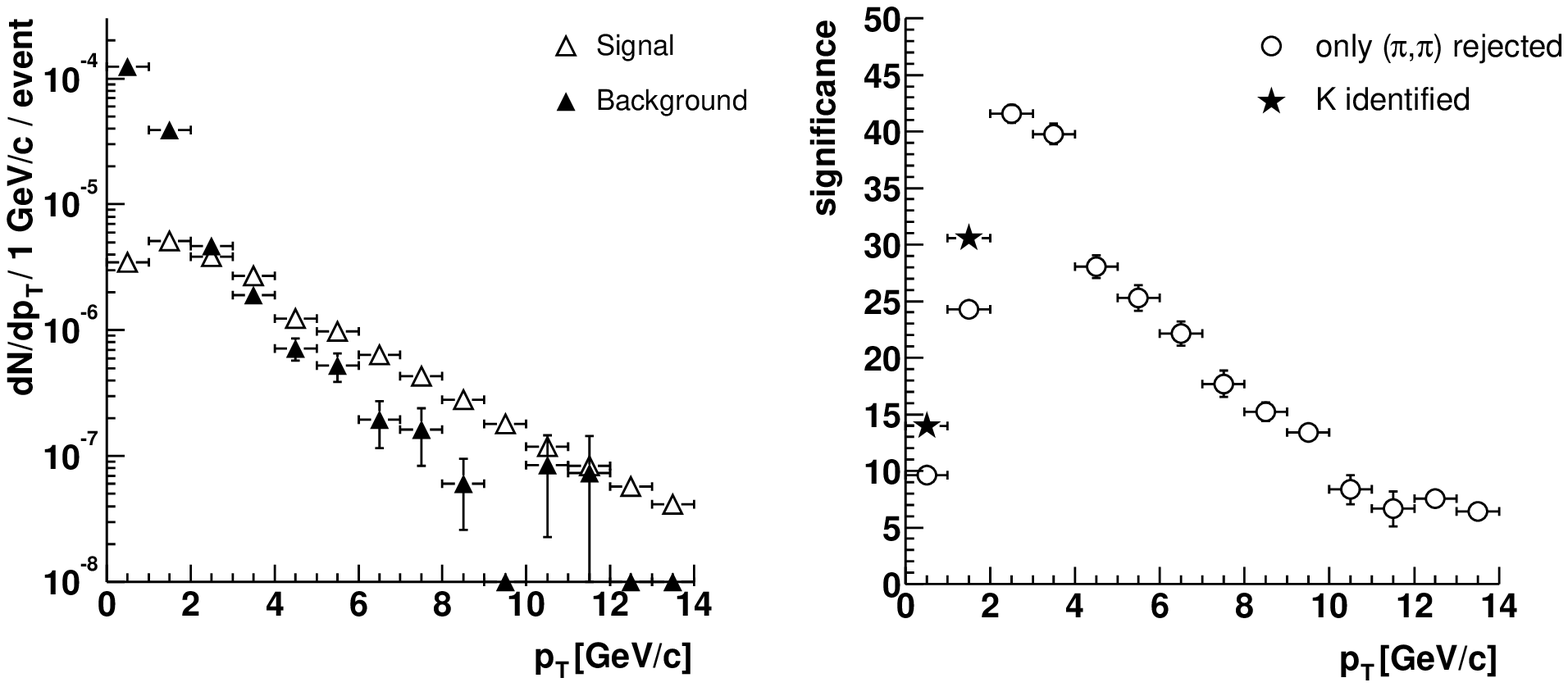}  
  \caption{Left: the transverse momentum distribution of selected
    $D^0$ candidates in $pp$ collisions at $\sqrt{s}=14$ TeV.
    The signal and background are 
    normalized to one minimum bias $pp$ event.
    Right: the corresponding significance for $10^9$ events.
    At low $p_T$ the values obtained requiring kaon identification 
    are shown by the stars.}
  \label{D0ptpp}
  \end{center}
\end{figure}

A typical example of the $D^0$ candidates invariant mass distribution 
in central Pb+Pb reactions, after selections, is shown on the right-hand side 
of Fig.~\ref{D0combined} before and after background subtraction.

Figures~\ref{D0ptPbPb} and~\ref{D0ptpp} show the $p_T$ 
distribution of the signal and 
the background (normalized to one event) and the corresponding significance
in Pb+Pb and $pp$ collisions respectively.
Note that for $p_T \geq 7-8$ GeV/$c$, even if the simulated background
statistics are insufficient, the significance is dominated by the
signal statistics.
The significance is larger than 10 for up to $p_T \approx 10$ GeV/$c$, 
both in Pb+Pb and in $pp$ collisions, if a 1 GeV/$c$ $p_T$ bin size is used.
As shown in Ref.~\cite{Carrer:aj}, the significance at low $p_T$ can be 
further improved by selecting sub-samples of $D^0$ candidates for which the 
kaon is identified. The values obtained with this method are shown with 
different symbols in Figs.~\ref{D0ptPbPb}-\ref{D0ptpp}. In this case
the $D^0$ production cross 
section can be measured down to $p_T \simeq 1$ GeV/$c$ in Pb+Pb 
collisions and down to almost $p_T=0$ in $pp$ collisions.

\subsection{Proton-nucleus Interactions}
\label{pa}

Proton-nucleus collisions are an important part of the ALICE physics 
program~\cite{ALICEPPR}.
The main motivation is to bridge the results obtained in $pp$ and $AA$ 
interactions to unravel initial and final state
medium effects.

More precisely, $pA$ studies of heavy flavour production
will provide essential measurements of cold nuclear matter effects such
as nuclear shadowing effects on heavy quarks \cite{Vogt:2001nh} and
nuclear absorption of quarkonium.  Nuclear absorption effects have
been demonstrated to be of crucial importance for
disentangling ``normal nuclear absorption'' from
``anomalous suppression'' of quarkonium at the 
SPS~\cite{Abreu:2000ni,Satz:2000bn,Vogt:cu}.
Similarly, the $pA$ open charm and bottom yields will 
provide valuable references for the onset of
thermal charm production in nuclear collisions and for the modification of
the heavy flavour spectra due to energy loss in the QGP.

The ALICE $pA$ program, foreseen as dedicated $p$Pb, $d$Pb or
$\alpha$Pb runs, is expected during the first few years of LHC
operation~\cite{alice_ppr_ch2}.
As in $pp$ and $AA$ collisions, quarkonia and heavy flavor
mesons will be measured 
through their (semi-)electronic decay channels in the ALICE central barrel.
Also, $D$ mesons will be fully reconstructed using exclusive hadronic 
decay channels.  The detector performance for
$D^0\to K^-\pi^+$ (and charge conjugate) 
decays in $p$Pb collisions is expected to be similar 
to that of $pp$ since the multiplicity of background tracks
will not be much larger than in $pp$ interactions. 
The asymmetric muon spectrometer allows all these studies in both direct
and inverse kinematics since both
$pA$ and ``$Ap$" operation is, in principle, possible.
Detailed simulations investigating the expected detector performance
in $pA$ collisions are underway.

\subsection{Run Plan}
\label{running}

The present ALICE plans for data taking during the LHC heavy ion runs, in
addition to participation in the 14 TeV $pp$ runs, are now
briefly described, see Ref.~\cite{alice_ppr_ch2} for more details.
In the first 5-6 years of ion runs, we propose:
1-2 years Pb+Pb; 1-2 years Ar+Ar; and 1 year $p$Pb-like collisions ($p$Pb,
$d$Pb or $\alpha$Pb).
Later options depend on the outcome of the early runs:
dedicated $pp$ or $pp$-like ($dd$ or $\alpha \alpha$) collisions;
other $AA$ systems such as O+O, Kr+Kr or Sn+Sn;
other $d (\alpha)A$ systems; or a low energy Pb+Pb run.

The maximum Pb+Pb luminosity for the ALICE central detectors
is $L=10^{27}~{\rm cm}^{-2}{\rm s}^{-1}$, imposed by the pile-up event rate
in the TPC and the DAQ bandwidth, 1~kHz, of the muon spectrometer.
Nevertheless, to achieve better muon statistics, 
dedicated high luminosity runs 
of up to $L=2\times 10^{28}~{\rm cm}^{-2}{\rm s}^{-1}$
involving the muon spectrometer without the
TPC could be foreseen with a high $p_T$ muon trigger, primarily for 
$\Upsilon$ decays.

To study the results as a function of energy density, at least one intermediate
mass system, most likely Ar+Ar, will be studied in the first years.
If more energy density points are needed, other systems such as Sn+Sn, Kr+Kr
and O+O may be studied at a later stage.  Luminosities similar to those used
for the Pb+Pb runs will be used for the intermediate mass ion runs.

Although ALICE will take $pp$ data at 14 TeV in parallel with the other 
experiments, the luminosity must be reduced to
$\approx 3\times 10^{30}~{\rm cm}^{-2}{\rm s}^{-1}$
to keep the pile up rate in the TPC and the ITS at an acceptable level.
Higher luminosity runs with the muon spectrometer alone are possible with
luminosities up to $5\times10^{31}~{\rm cm}^{-2}{\rm s}^{-1}$.

In addition, lower energy $pp$ collisions, closer to the Pb+Pb center
of mass energy, $5.5-7$~TeV, might be needed for reference data.

\subsection{Summary}

The capabilities of ALICE for heavy flavour physics at the LHC will provide a
comprehensive understanding of open and hidden heavy flavour production 
at low $x$ where strong nuclear gluon shadowing is expected.
The large number of accessible channels will permit detailed investigations
of the properties of the nuclear medium.
Open heavy flavour measurements will provide information on the primary 
production mechanisms, parton energy loss and possible
secondary production of heavy quarks.
The complete spectroscopy of the quarkonium states will be used to probe the
deconfined medium and to pin down its temperature. 
The simultaneous measurement of open and hidden heavy flavours will provide a 
powerful tool to study quarkonium suppression/enhancement in the QGP.
Thanks to the large acceptance of the detector, the signals will be 
reconstructed over a broad transverse momentum range.
This is particularly important for establishing the 
reference production rates with high accuracy and to investigate 
the characteristics of the QGP.
The measurements will be performed for a variety of 
heavy and light nucleus-nucleus systems as well as for $pp$ and $pA$ reactions.
The latter are mandatory since, besides probing shadowing effects,
they will provide the normalization for understanding the properties of 
the nuclear medium in $AA$ collisions.

In addition to the measurements discussed here, further exciting possibilities 
should be opened with, for example, the reconstruction of hadronic decays of
charged $D$ and $B$ mesons, reconstruction of double charm baryons 
by combining hadronic and leptonic channels, and dilepton measurements at very
high invariant mass.

%% file: cms.tex
\section[QUARKONIA AND HEAVY FLAVOUR DETECTION WITH THE CMS 
DETECTOR]{QUARKONIA AND HEAVY FLAVOUR DETECTION WITH THE CMS 
DETECTOR~\protect\footnote{Authors: M.~Bedjidian, O.~L.~Kodolova,
R.~Kvatadze, I.~P.~Lokhtin.}}
\label{sec:cms}


\subsection{Introduction}
\label{CMS.intro}

The interest in quarkonium production at the LHC emerged from the CERN SPS 
results~\cite{Abreu:1999qw,Abreu:2000ni} which showed a strong 
anomalous suppression in 
\jpsi\ production in \PbPb\ collisions at $\sqrt s = 17.3$ GeV. 
RHIC will study \jpsi\ production and 
suppression in detail at $\sqrt{s} = 200$ GeV since the temperature of the
system produced at RHIC should be high enough for direct \jpsi\ suppression. 
Although the \ups\  
production cross section is large enough to be observed at RHIC, albeit 
with limited statistics, its suppression is not 
expected until the higher initial temperatures at the LHC are reached. 
CMS is particularly well suited to study the \ups\ family, the
continuum up to the $Z^0$ mass and, to a lesser extent, the \jpsi\ and $\psi'$.

Open heavy flavour production is important for studying 
the behaviour of massive colour charges in a dense medium at the LHC. The 
charm and bottom production cross sections are much larger than at RHIC. 
Systematics studies of heavy flavours, especially $B$ mesons, can be 
performed with CMS. In particular, in-medium gluon radiation and collisional 
energy loss of heavy quarks can result in the suppression and modification 
of the high-mass dilepton~\cite{Lin:1998bd,Lokhtin:2001nh,Lokhtin:ay} 
and secondary $B \rightarrow J/\psi$ decay 
spectra~\cite{Lokhtin:2001nh,Lokhtin:ay}.
Finite quark mass effects can suppress
medium-induced radiation of heavy quarks, enhancing e.g. the $B/\pi$
ratio~\cite{Dokshitzer:2001zm}. 

We briefly describe the CMS detector in section \ref{CMS.cmsdet}. 
Then we provide the resonance cross sections and 
the background description used in our simulations in section \ref{CMS.qdet}. 
Section \ref{CMS.detres} describes the detector response, 
the acceptance and the dimuon reconstruction algorithm.  The 
results of the simulations are presented in section \ref{CMS.masdis}, 
including the signal/background 
ratios and the dimuon invariant mass distributions. In the last section, 
\ref{CMS.flav}, the CMS capabilities for open heavy flavour 
and $Z^0$ measurements in the mass range above the \ups\ are discussed.

\subsection{The CMS Detector}
\label{CMS.cmsdet}

The CMS detector is designed to identify and measure muons, 
electrons, photons and jets over a large energy and rapidity range. It 
offers the widest 
muon acceptance centered at midrapidity. 
CMS is particularly well suited to study the \ups\ family, the
continuum up to the $Z^0$ mass and, to a lesser extent, the \jpsi\ and $\psi'$.
The CMS dilepton capability allows systematic studies of heavy flavour physics.
The impact parameter (centrality) of the collision can be determined from 
measurements of transverse energy 
production over the range \abseta~$<$~5. 

CMS has a high-field solenoidal 
magnet with a uniform 4 T field, leading to a compact detector. The 
first absorber, the electromagnetic calorimeter, is 1.3~m 
from the interaction point, eliminating a large fraction of the hadronic 
background. The powerful tracking system provides good track 
reconstruction efficiency for dimuons even for very large charged particle 
densities, $dN_{\rm ch}/dy\le$~8000, providing excellent dimuon mass 
resolution. 

\begin{figure}
\begin{center}
\includegraphics[width=12.5cm]{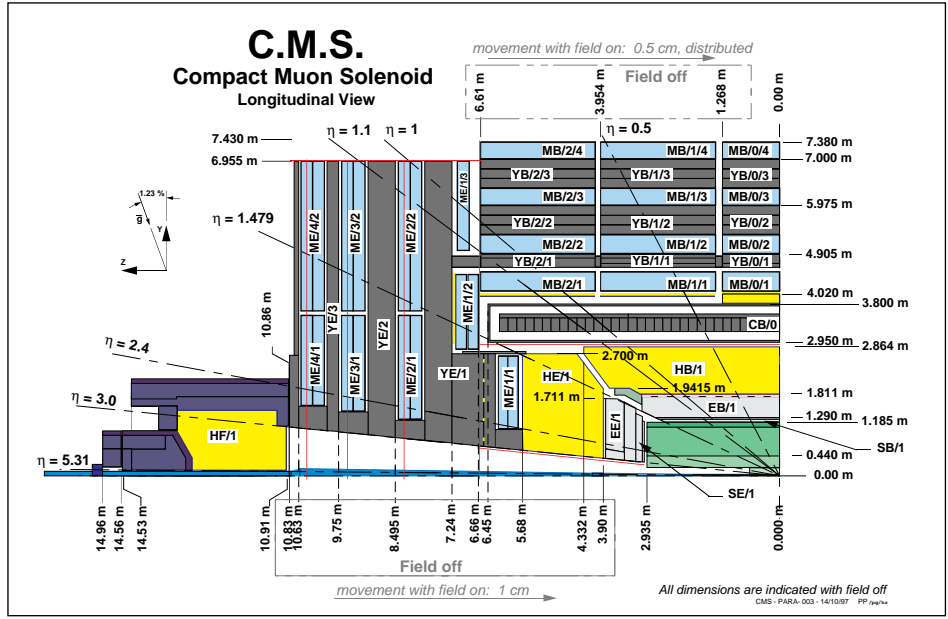} 
\caption{\small A longitudinal view of the CMS detector.}
\label{geom}
\end{center}
\end{figure}

A detailed description of the detector elements can be found in the 
corresponding Technical Design Reports~\cite{Htdr,Mtdr,Etdr,Ttdr}. 
A longitudinal view 
of the detector is shown in Fig.~\ref{geom}. The central element of CMS is 
the magnet, a 13~m long, 6~m diameter, high-field solenoid with an internal 
radius of $\approx$3~m.
The tracker and muon chambers cover the pseudorapidity region $|\eta|<2.4$, 
while the ECAL and HCAL calorimeters reach $|\eta|=3$. A pair of quartz-fiber 
very forward (HF) calorimeters, located $\pm 11$ m from the interaction point,
cover the region $3<|\eta|<5$ and complement the energy measurement. 
The tracker is composed of pixel layers and silicon strip counters.
The barrel part of tracker, $|\eta| <0.8$, consists of three pixel layers 
4, 7, and 11 cm radially from the beam line and 9 layers of 
silicon strip counters.
The endcap part of tracker, $\eta >0.8$, is more complex.
There are two pairs of pixel 
layers symmetrically located 34 cm and 45 cm from the geometrical center of the
detector. Outside this, two symmetric sets of 12 silicon layers are situated 
up to 265 cm from the center of the detector.
The electromagnetic calorimeter is made of almost 83000 scintillating 
PbWO$_4$ crystals.  The hadronic calorimeter consists of scintillator 
inserted between copper absorber plates. 

The CMS muon stations consist of drift tube chambers (DT) in the barrel
region (MB), $|\eta|< 1.2$, cathode strip chambers (CSCs) in the endcap 
regions (ME), $0.9 < |\eta| < 2.4$, and resistive plate 
chambers (RPCs) in both barrel and endcaps, for $|\eta|<1.6$. The RPC detector 
is dedicated to triggering, while the DT and CSC detectors, used for precise 
momentum measurements, also have the capability to self-trigger up to 
$|\eta|<2.1$. The muon system can thus reconstruct muons in the range 
$|\eta|<2.4$.

In each of the 12 sectors of the barrel return yoke of the magnet (YB),
covering $30^\circ$ in azimuthal angle $\phi$, there are four muon barrel 
`stations', MB$i$, $i=1,4$ in Fig.~\ref{geom}.  Each station consists of 
one DT and one (or two, in the two innermost stations) RPC chambers. 
In each DT chamber, there are two
"superlayers" of four DT planes measuring the $(r, \phi)$ coordinates 
in the bending plane and another superlayer measuring the $z$ coordinate. 
In the outermost station, MB4, there are only two $(r, \phi)$ superlayers.  
Thus a muon traversing the whole barrel has 44 measurement points (`hits') in
the DT systems and 6 hits in the RPC.  There are a total of 240 stations in 
the barrel system.
Each endcap region has four muon stations of CSCs. Each chamber 
gives six measurements of the $\phi$-coordinate (strips) and six 
measurements of the $r$-coordinate (wires).
RPCs will be added in the barrel and endcaps to provide an 
additional trigger. Six layers of RPCs will be mounted in the barrel 
chambers and four in the endcaps.

\subsection{$AA$ Collisions}
\label{CMS.qdet}

\subsubsection{Resonance cross sections} 

We use the quarkonium cross sections per nucleon calculated in 
section~\ref{quarkon.cem} in our simulations.  The calculations, performed in
the CEM, include nuclear shadowing but do not include absorption by nucleons
and secondaries since the interplay between shadowing and absorption as a
function of $\sqrt{s}$ is unknown.  For convenience, the inclusive $AA$ dimuon
cross sections given in Tables~9 and 10 are reproduced in Table~\ref{resigma}. 

\begin{table}[ht]
\begin{center}
\caption{Total cross sections for quarkonium production in
minimum bias $AA$ collisions.}
\vskip0.2cm
\label{resigma}
\renewcommand{\arraystretch}{1.5}
\begin{tabular}{|c|r|r|r|r|} \hline
System & \PbPb & \SnSn & \KrKr & \ArAr \\ \hline 
$\sqrt{s}$ (TeV/nucleon) & 5.5 & 5.84 & 6.14 & 6.3 \\ \hline
Resonance &\multicolumn{4}{|c|}{$B_{\mu\mu}$ $\sigma_{AA}$ 
($\mu$b)} \\ \hline
 \jpsi                                & 48930 & 17545 & 9327 & 2321 \\ \hline 
 \psip                                &   879 &   315 & 167.6 & 41.7 \\ \hline 
 \ups                                 &   304 &  108.1 & 57.4 & 13.8 \\ \hline 
 \upsp                                &  78.8 &   28 &  14.8 & 3.6  \\ \hline 
 \upspp                               &  44.4 & 15.8 &  8.4 &  2.0 \\ \hline 
\end{tabular}
\renewcommand{\arraystretch}{1}
\end{center}
\end{table}

\subsubsection{Background and kinematic distributions} 

\paragraph{Soft hadronic background.} 

Soft hadron production is the main 
dimuon background. This background is estimated from the charged 
particle multiplicity and the shape of the pion and kaon
transverse momentum distributions. 

The charged particle multiplicity, $dN_{\rm ch}/d\eta$, 
is unknown for the LHC energies. 
Early estimates of this multiplicity in central collisions 
($b = 0$) and midrapidity
($\eta = 0$) were as high as 8000 for \PbPb collisions \cite{cms060}.  
In contrast, the first extrapolations 
from RHIC suggested that the Pb+Pb multiplicity could be as low as 2000
\cite{Kharzeev:2001yq}. We choose two values for our 
simulations: 2500 (low) and 5000 (high). Lower multiplicities
reduce the dimuon background. The track reconstruction algorithm was 
designed to function for a multiplicity of 8000, so
that the current, lower, estimates should result in higher reconstruction 
efficiencies.  The high and low values of the multiplicities used in the 
simulations are shown in Table~\ref{multh}.  The upper numbers for each set 
correspond to the 5\% most central collisions, assuming the multiplicity scales
as $A^{4/3}$ from Pb+Pb.  The lower numbers for each set correspond to
collisions at $b =0$.  In this case, the multiplicity can be higher.

\begin{table}[htb]
\begin{center}
\caption{The high and low multiplicity sets,
$dN_{\rm ch}/d\eta \vert_{\eta=0}$, used in the simulations.  
In both cases, the
results are shows for the 5\% most central collisions and impact parameter
$b=0$.}
\vskip0.2cm
\label{multh}
\renewcommand{\arraystretch}{1.5}
\begin{tabular}{|c|c|r|r|r|r|} \cline{3-6}
 \multicolumn{2}{c}{ } & \multicolumn{4}{|c|}{$dN_{\rm ch}/d\eta 
\vert_{\eta=0}$} \\ \hline
    Set   &   & \PbPb & \SnSn & \KrKr & \ArAr \\ \hline
    High  & 5\% most central  & 5000 & 2400 & 1500 & 550 \\ 
   & $b=0$  & 7500 & 4000 & 2600 & 850 \\ \hline
    Low  & 5\% most central  & 2500 & 1200 & 750 & 280 \\ 
   & $b=0$  & 3800 & 1900 & 1300 & 450 \\ \hline
\end{tabular}
\renewcommand{\arraystretch}{1}
\end{center}
\end{table}

The soft hadronic background is assumed to entirely consist of pions and kaons
produced according to the ratio $K/\pi = 0.12$, independent of $b$.  We use the
SHAKER algorithm \cite{shak} to simulate the pion and kaon momentum
distributions where $\langle p_T \rangle_\pi = 0.47$ GeV/$c$ and
$\langle p_T \rangle_K = 0.67$ GeV/$c$. 
We note that HIJING \cite{Wang:1991ht,Gyulassy:ew} produces pions
and kaons with lower momenta,  $\langle p_T \rangle_\pi = 0.43$ GeV/$c$ and
$\langle p_T \rangle_K = 0.52$ GeV/$c$.  Due to the magnetic field, only
muons with $p_T > 3.5$ GeV/$c$ will reach the barrel 
muon chambers.  This minimum
momentum requirement serves as the primary background rejection criteria.  More
than 99.8\% of the SHAKER hadron background is rejected.  The lower
average momentum of HIJING reduces the probability for acceptance of 
decay muons by a factor of two with respect to SHAKER.  
Track reconstruction will also be an
effective method of background rejection.

\paragraph{Heavy flavour decays.}

The other important source of quarkonium background is $Q \overline Q$ 
pair production and decay. 
The number of pairs produced in \aacol\ collisions as a function of 
impact parameter $b$ is 
$$ N(Q\overline Q)= \sigma(Q\overline Q)  T_{AA}(b) \, \, , $$ 
directly proportional to the nuclear overlap function
$T_{AA}(b)$ where $T_{\rm PbPb}(0)
= 30.4$/mb, $T_{\rm SnSn}(0) = 13.3$/mb, $T_{\rm KrKr}(0) = 8.5$/mb and 
$T_{\rm ArAr}(0)= 2.9$/mb. The NLO $Q \overline Q$ production cross sections 
in 5.5 TeV $pp$ interactions are $\sigma (c\overline c) = 6.3$ mb and 
$\sigma (b \overline b) = 
0.19$ mb (with large uncertainties -- see section~\ref{sec:open}). 
Since the center of mass energies are higher for lighter ions, the
$Q\overline Q$ cross sections per nucleon increase slightly from Pb+Pb to 
Ar+Ar interactions.  The number of $Q \overline Q$ pairs at $b=0$ 
for the collision systems considered is shown in Table~\ref{qqb}.

\begin{table}[htb]
\begin{center}
\caption{Number of $Q \overline Q$ pairs produced at $b$=0.}
\vskip0.2cm
\label{qqb}
\renewcommand{\arraystretch}{1.5}
\begin{tabular}{|c|r|r|r|r|} \cline{2-5}
\multicolumn{1}{c|}{ }   & \PbPb & \SnSn & \KrKr & \ArAr \\ \hline
\ccb\ & 192 & 86 & 57 & 20 \\ \hline
\bbb\ & 6 & 3 & 2 & 1 \\ \hline
\end{tabular}
\renewcommand{\arraystretch}{1}
\end{center}
\end{table}

Although $N(c \overline c) \gg N(b \overline b)$, dimuons
from \cbarc\ decays are not the largest contribution to the background. 
Among all background sources taken into account, \bbarb\ 
decays produce muons with the highest average transverse momentum, making a  
significant contribution to the background. The muon \pt\ and \etta\ 
distributions are extracted from PYTHIA \cite{Sjostrand:1993yb} with 
\lgl$p_{\rm T}^\mu$\rgl$_c = 0.51$~\Gonc\ and \lgl$p_{\rm T}^\mu$\rgl$_b = 
1.2$~\Gonc. 

Only 6\% of single $b$ decays produce two muons and, of these, 2/3 are of 
opposite sign. When the CMS kinematic cuts, $p_{\rm T}^\mu >3.5$ \Gonc\ 
and $| \eta | < 2.4$, are included,
more than 98\% of the single $b$ decay muons do not survive 
the cuts.  The probability of obtaining a dimuon from these decays 
after cuts is $\approx 10^{-4}$,
shown in Table~\ref{qqmu}. About 80\% of the accepted
pairs are of opposite sign 
so that the correlated like-sign muon rate is negligible. 
Thus, to simplify the simulations,
we assume that all muon pairs from \bbarb\ decays are of opposite sign. 

\begin{table}[htb]
\begin{center}
\caption{The probability of $n$-muon signals in $b$ quark decays.}
\vskip0.2cm
\label{qqmu}
\renewcommand{\arraystretch}{1.5}
\begin{tabular}{|c|r|r|r|r|} \cline{2-5}
\multicolumn{1}{c|}{ } & 0 $\mu$ & 1 $\mu$  & 2 $\mu$ & 3 $\mu$  \\ \hline
before CMS cuts & 0.626 & 0.308 & 0.060 & 0.006 \\ \hline
after CMS cuts & 0.987 & 0.0125 & 0.00018 & 0 \\ \hline
\end{tabular}
\renewcommand{\arraystretch}{1}
\end{center}
\end{table}

\subsection{Detector Response}
\label{CMS.detres}

\subsubsection{Acceptances} 

The $p_T$ and $\eta$ distributions of the quarkonium sstates used in the
simulations were obtained from several sources.  The \pt\ distributions of 
the three $\Upsilon(S)$ states 
were extrapolated from CDF data~\cite{Abe:1997ue} while PYTHIA 
was used to generate the $\eta$ distributions.  PYTHIA was also used to obtain
the \jpsi\ and \psip\ distributions.

\begin{figure} [!ht] 
\begin{center}
\includegraphics[width=9cm,height=8cm]{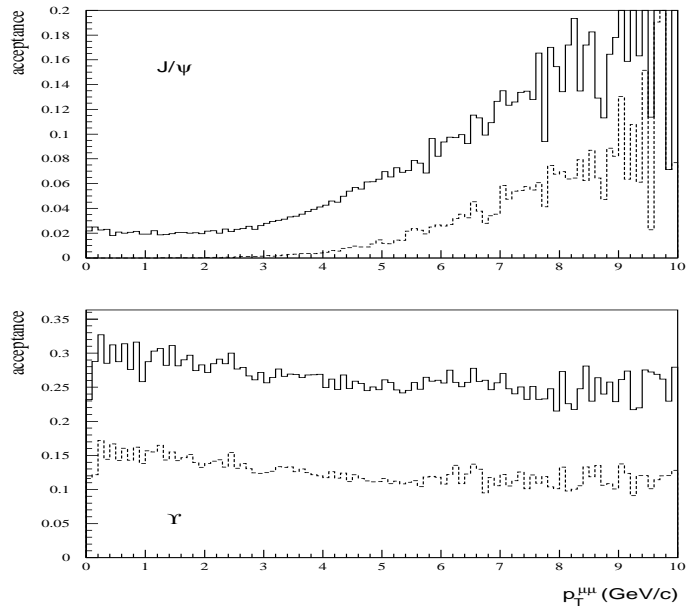}
\caption{\label{ptacc}\small The \pt\ dependence of the dimuon acceptance 
for \jpsi\ (upper) and \ups\ (lower) resonances. In each case both the 
full CMS detector (solid line) and the barrel only (dashed line) have 
been considered.}
\end{center}
\end{figure}

We assume that the primary vertex is at the geometrical centre of the 
detector. Each resonance decay muon is tracked through a 4~T magnetic field by 
GEANT~\cite{geant} using the CMSIM package~\cite{cmsim}. 
A dimuon is accepted if each decay muon of the pair passes through at 
least one muon chamber. We also introduce a $p_T$ cut of 3.5~\Gonc\ 
for each muon, just above the minimum \ptmu\ needed to  reach 
the first muon chambers in the barrel.   This cut on \ptmu\ leads to very
different resonance acceptances.  The \jpsi\ decay muons have
$\langle$\ptmu$\rangle = 1.7$~\Gonc\ so that only 0.8\% of the \jpsi\ 
decays give leptons above the muon threshold.  On the other hand, the \ups\ 
decay muons have $\langle$\ptmu$\rangle = 4.6$~\Gonc\ .  Thus 40\%  
survive the \pt\ cut. The acceptances discussed below and shown in
Fig.~\ref{ptacc} are geometrical only.  The dimuon reconstruction 
efficiencies, the trigger efficiency and the muon $p_{\rm T}$ cut are not
taken into account. 

The \ups\ acceptance is flat, $\approx 27$\% in the full $\eta$ range of the 
detector before the muon $p_T$ cut, dropping to 16\%  
once the muon \pt\ cut is applied.  With the barrel alone, the \ups\ acceptance
is 13\% before the \pt\ cut and 9.5\% after.
The \pt\ dependence
of the accepted \ups\ in Fig.~\ref{ptacc} does not change 
between the full detector and the barrel.  The \pt\ 
distribution starts from zero and is only statistically limited in the high-%
\pt\ region, allowing a comparison between the high- and low-\pt\ events. 

The situation is different for the \jpsi. The integrated 
\jpsi\ acceptance is about ten times lower than the \ups\ acceptance. There is 
an additional factor of 10 reduction when the $J/\psi$ acceptance is limited 
to the barrel due to the natural \ptmu\ cut provided by the material 
located in front of the muon 
chambers. Including a 3.5~\Gonc\ cut diminishes 
the acceptance very strongly. Only muons from high \pt\ \jpsi\ decays 
(\pt\ \jpsi\ > 4 GeV/$c$) can 
reach the barrel muon chambers, as clearly shown in Fig.~\ref{ptacc}. 
However, when the endcap regions are included can the \jpsi\ be detected 
over the whole $p_T$ range.  The acceptance is small but the expected rate is
large.

\subsubsection{Track reconstruction}

Pattern recognition and track reconstruction determine, to a large extent, 
the feasibility of heavy ion measurements with CMS. 
The particles were tracked through
the detector using a full GEANT simulation.  The tracks where followed down
to 100 KeV for electromagnetic processes and 1 MeV for hadronic processes,
an order of magnitude below the default GEANT energy threshold cuts.
The quarkonium studies have
has been carried out for a range of $dN_{\rm ch}/d\eta$, up to an extreme 
value of 8000, far beyond the present upper limit of 5000 for central Pb+Pb 
collisions. The upper limit of $dN_{\rm ch}/d\eta = 8000$
led to occupancies of up to 18\% for the 5 outermost Si-layers of the 
barrel, located more than 70~cm from the beam, and of 50\% or more for the 
innermost Si-strip detectors. Occupancies of  2-4\%~\cite{tn2} are expected 
for the pixel layers due to their very high granularity. Thus, the 
occupancies determine the number of layers used in the
track-reconstruction algorithm.

The goal of the track-finding algorithm is to select pairs of muon tracks 
coming from the interaction point. In the barrel, the algorithm uses 
the two pixel layers at 7 and 11 cm and the four outermost Si-layers.
This is the minimum combination necessary to get good efficiency and purity for
high occupancy rates in the tracker. The usage of the third pixel layer
at 4 cm is under investigation.

\begin{figure}  [!ht]
\begin{center}
\includegraphics[width=12cm]{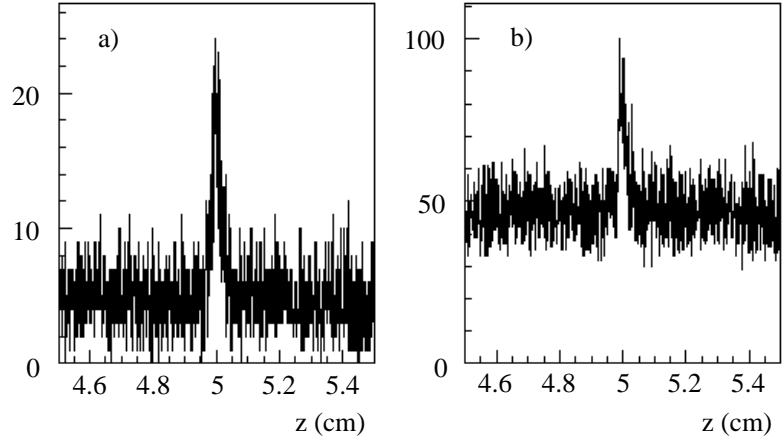}
\caption{\label{vertex}\small Distribution of the $z$ position of the 
primary vertex for $dN_{\rm ch}/d\eta$ of (a) 2500 and (b) 8000. 
The initial collision was generated at $x=y= 5$ $\mu$m and $z= 5$ cm.}
\end{center}
\end{figure}

 
The vertex is determined from a combination of clusters 
in the pixel layers assuming the interaction occurs at the geometrical 
centre of the detector, $(0,0,0)$. Figure~\ref{vertex} shows the results of
the interaction point vertex determination for a simulated
collision at $z = 5$ cm for $dN_{\rm ch}/d\eta=2500$ and 8000. 
The resolution is $\sigma_z = 140$ $\mu$m. 

The track finding algorithm starts from the muon chambers since they are the 
least populated planes. The tracks are propagated within roads in 
transverse ($R/z , \phi$) and longitudinal ($R ,z$) planes with simple 
parameterizations $d\phi = k \, dr / p_{\rm T} $ for the barrel,  
$d\phi = k \, dz / p_{\rm L} $ for the endcaps, and a straight line in 
the ($R , z$) plane. 

The track candidates are then fitted and, after vertex constraints, the 
best quality track is selected. Several dimuon events have been studied:  
homogeneous dimuons from \ups\ decays as well as mixed pairs originating from 
$\pi /K/b$ decays
where one muon comes from a $\pi$ or $K$ decay and the other from a $b$
decay. For the moment, only barrel tracks have been investigated. 
In the region $|\eta| < 0.8$, the efficiency of $\Upsilon$
reconstruction is $\approx$ 90\% for $dN_{\rm ch}/d\eta = 2500$ and 85\% for 
5000 with a purity above 99\%.   
The $\Upsilon$ reconstruction efficiency is 
76\% for $dN_{\rm ch}/d\eta = 8000$.  
The efficiency for $\pi$ and $K$ decays is about six times lower,
leading to good background rejection.  For $0.8 < |\eta| < 1.3$, 
the $\Upsilon$ efficiency is ~50\%.  The high precision 
tracking leads to an excellent dimuon mass resolution of 
46~MeV at the \ups\ peak when both tracks have $|\eta| < 0.8$, increasing to 
60~MeV when one track has $0.8 < |\eta| < 1.3$.

\subsection{Invariant Mass Spectra}
\label{CMS.masdis}

\subsubsection{Signal/background and mass distributions}
\label{CMS.sectsignal}

The simulations are based on the weight method with acceptance tables. 
They consider combinations of muons with $p_T^\mu > 3.5$ GeV/$c$.  Either both
muons originate from $\pi$ and/or $K$ decays, $hh$, or from 
$c \overline c$ and $b \overline b$ decays,
or from a "mixed source", one from $\pi$ or $K$ decays and one from
heavy quark decays, $hb$ and $hc$.

The invariant mass of each pair is calculated. Two mass regions are then
defined, the \ups\ region, 
$8.5<M_{\mu\mu}<11$ \mass\, and the \jpsi\ region, $2<M_{\mu\mu}<4.5$ \mass\ .
The resulting invariant mass distribution is smoothed by a 
Gaussian with a width determined by the expected mass resolution. 

\begin{figure}[!ht]
\begin{center}
\includegraphics[width=7.9cm]{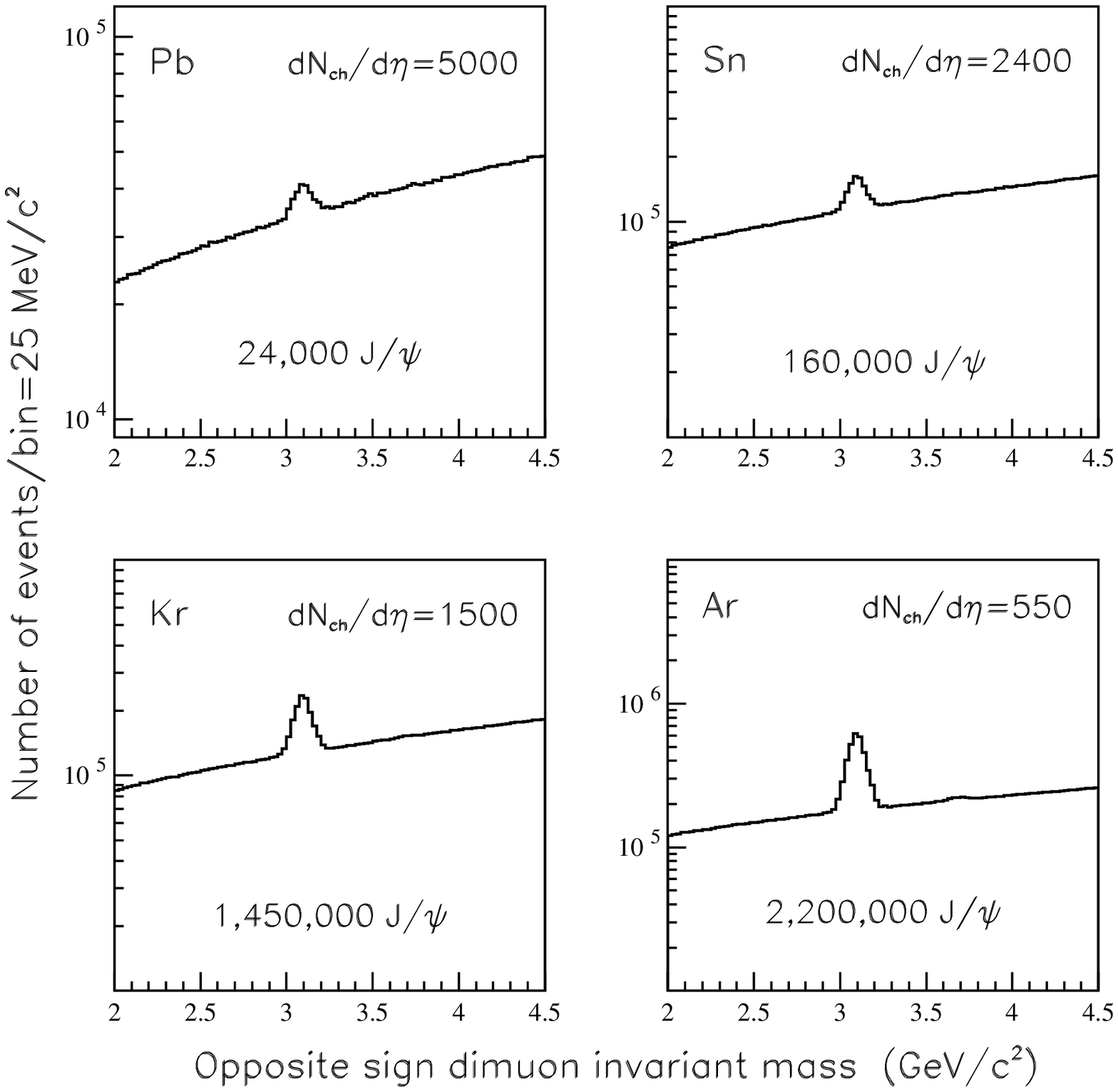}
\includegraphics[width=7.9cm]{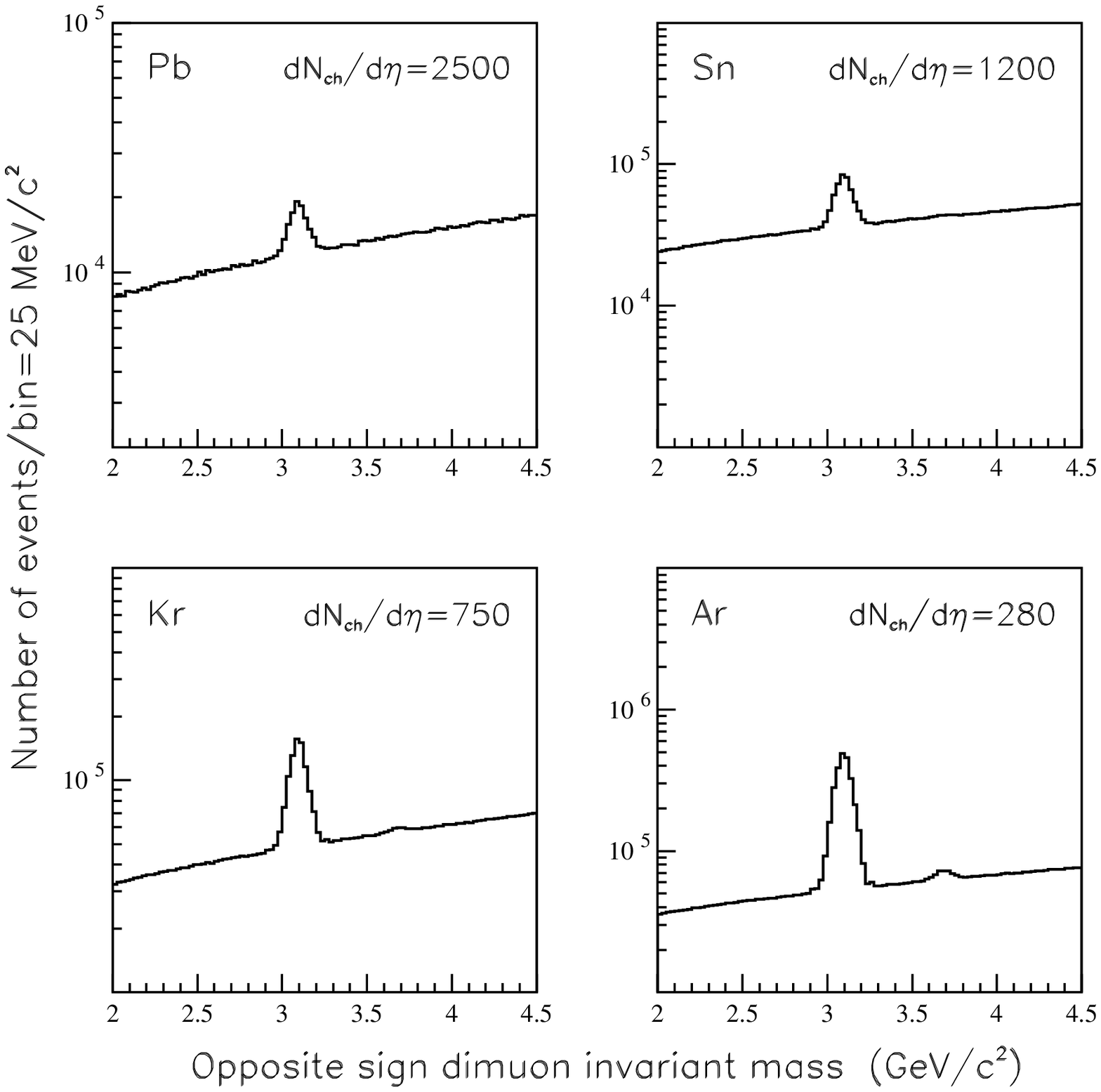}
\caption{\label{jpsi4}\small The opposite sign dimuon invariant mass 
distributions in the \jpsi\ mass region obtained in the 5\% most central
collisions in a one month run with
the high multiplicity set (left) and low multiplicity set (right) as defined
in Table~\protect\ref{multh}.}
\end{center}
\end{figure}

\begin{figure}[!ht]
\begin{center}
\includegraphics[width=7.9cm]{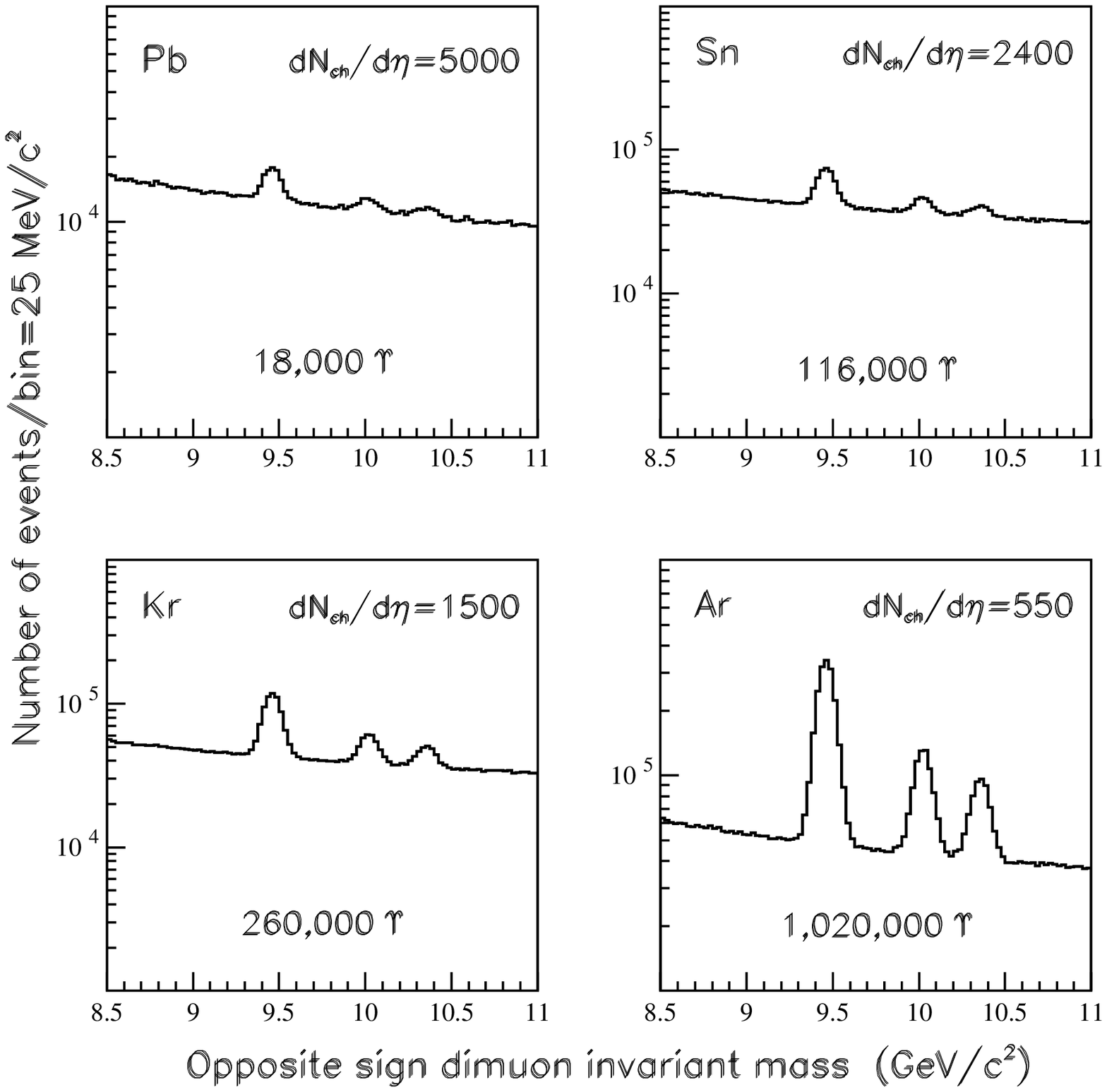}
\includegraphics[width=7.9cm]{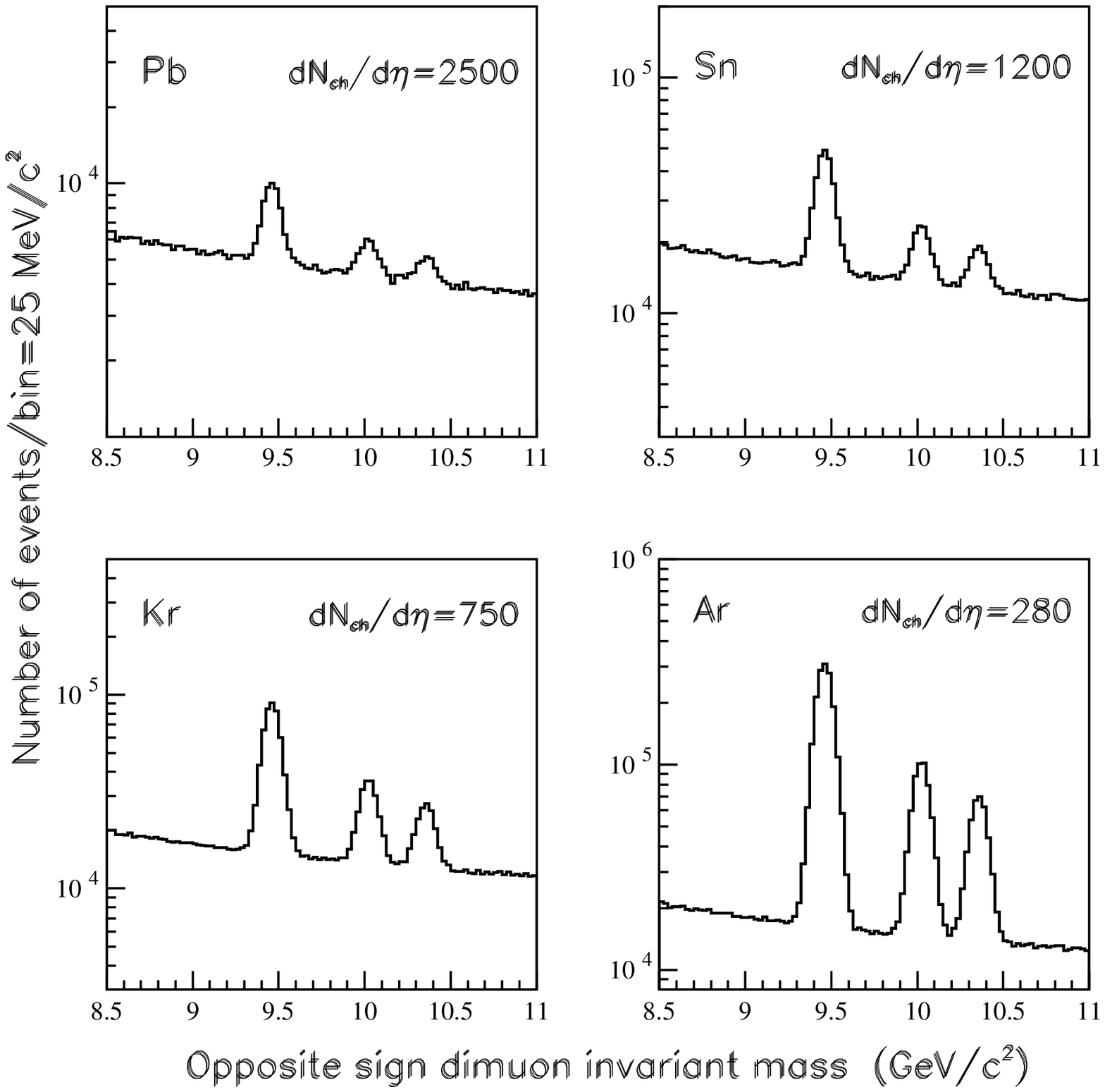}
\caption{\label{upsi4}\small The opposite sign dimuon invariant mass
distributions in the \ups\ mass region obtained in the 5\% most central
collisions in a one month run with
the high multiplicity set (left) and low multiplicity set (right) as defined
in Table~\protect\ref{multh}.}
\end{center}
\end{figure}

The opposite sign dimuon invariant mass distributions obtained in 
Pb+Pb, Sn+Sn, \KrKr\ and \ArAr\ collisions are shown in Fig.~\ref{jpsi4} 
for the \jpsi\ and in Fig.~\ref{upsi4} for the $\Upsilon$. 
Using the high and low multiplicity sets given in Table~\ref{multh}, 
we present results for a one month 
run, $1.3 \times 10^6$ seconds, assuming a 50\% machine efficiency.
The \jpsi\ and \ups\ statistics are given in the 
figures and summarized in Table~\ref{stats}, assuming the average 
luminosities of Ref.~\cite{lumi}.
The 18000 \ups\ measured in a one month \PbPb\ 
run are statistically significant enough to bin with impact parameter. 
Limiting detection 
to the barrel reduces the statistics, but not
unacceptably for intermediate-mass ions. 

The signal-to-background ratio, S/B, between the number of detected 
resonances, \jpsi\ or \ups (1S), 
and the number of opposite sign dimuons under the peak in the mass window 
$M_{\rm res} \pm 50$~MeV, is calculated from Fig.~\ref{jpsi4} 
for the \jpsi\ and from Fig.~\ref{upsi4} for the $\Upsilon$. 
This ratio depends on the 
multiplicity. Table~\ref{soverb} gives the S/B ratio for the $J/\psi$ and 
$\Upsilon$ and the significances, S/$\sqrt{\rm S+B}$, 
for the three $\Upsilon(S)$ states.  The results are shown 
for both the high and low
multiplicity sets of Table~\ref{multh} for each collision system.  The lower
signal-to-background ratios and significances correspond to the larger
background, high
multiplicity set.  The higher
signal-to-background ratios and significances result from the reduced
background of the low multiplicity set.

\begin{table}[htb]
\begin{center}
\caption{\label{stats}The number of resonances detected in a one month, $1.3
\times 10^6$ s, run in the full detector and in the barrel alone 
(in parentheses).
}
\vskip0.2cm
\renewcommand{\arraystretch}{1.5}
\begin{tabular}{|c|r|r|r|r|} \hline
$N_{\rm res}/10^3$ 
& \PbPb & \SnSn & \KrKr & \ArAr \\ \hline
 $J/\psi$    &  24 & 160 & 360 & 1450 \\ 
                       &  (4.7) &(31) & (81) & (275) \\ \hline
 $\Upsilon$  & 18 & 116 & 260 & 1020 \\ 
                       & (7.9) & (51) & (114) & (450) \\ \hline
 $\Upsilon'$ & 5.4 & 35 & 80 & 310 \\ \hline
 $\Upsilon''$ & 3.4 &  22 & 50 & 195 \\ \hline
\end{tabular}
\renewcommand{\arraystretch}{1}
\end{center}
\end{table}

\begin{table}[htb]
\begin{center}
\caption{\label{soverb}The signal-to-background ratios, S/B, and 
the significances, S/$\sqrt{\rm S+B}$ in the 5\% most central collisions for
the two multiplicity sets given in Table~\protect\ref{multh}. 
The smaller numbers correspond to
the high multiplicity set while the larger numbers are obtained with the low 
multiplicity set.}
\vskip0.2cm
\renewcommand{\arraystretch}{1.5}
\begin{tabular}{|c|c|r|r|r|r|} \cline{3-6}
 \multicolumn{2}{c|} { }   & \PbPb & \SnSn & \KrKr & \ArAr \\ \hline
  S/B &\jpsi\ & 0.2-0.5 & 0.4-1.1 & 0.7-1.8 & 2.0-6.8 \\ \cline{2-6}
 & \ups\  & 0.4-0.9 & 0.7-1.9 & 1.5-4.3 & 5.3-15.6 \\ \hline \hline
  &  \ups\   & 69-93 & 220-276 & 396-460 & 925-978 \\ \cline{2-6}
 S/$\sqrt{\rm S+B}$   & \upsp\  &  24-38 & 84-123 & 165-218 &  447-512  
\\ \cline{2-6}  
& \upspp\ &  16-26 & 55-86 & 113-157 &  325-391  \\ \hline
\end{tabular}
\renewcommand{\arraystretch}{1}
\end{center}
\end{table}

\begin{figure}[!ht]
\begin{center}
\includegraphics[width=10cm,height=8cm]{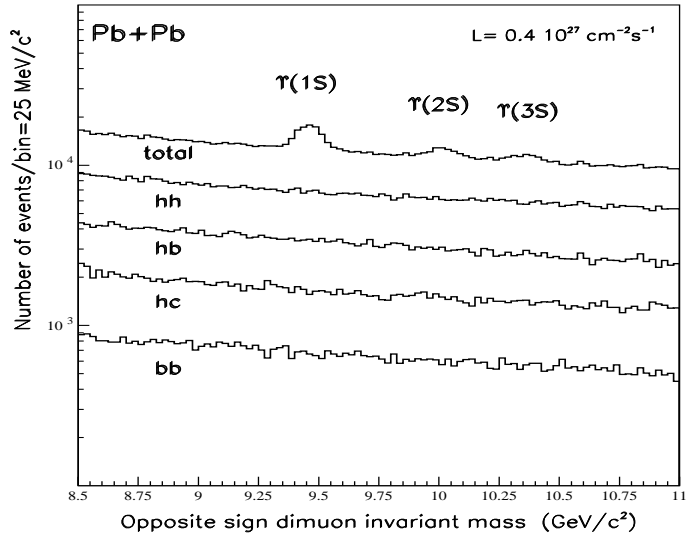}
\caption{\label{upsi1}\small The invariant mass distribution in the 
\ups\ mass region of 
opposite sign dimuons, including the largest background contributions,  
with the high multiplicity set. For $hh$, both muons come from $\pi$ and $K$ 
decays. For $hb$ and $hc$, one muon is from $\pi$ or $K$ decay while the 
second is from a $b$ or $c$ decay.  For 
$bb$, both muons come from \bbb\ decays.}
\end{center}
\end{figure}

Figure~\ref{upsi1}, corresponding to \PbPb\ collisions with the high \
multiplicity set, shows the expected opposite sign dimuon mass distribution 
in the \ups\ mass region in more detail. The most 
important contributions to the background are given. The background is 
typically dominated by $\pi$ and $K$ decays in combination with each other,
$hh$, or with a heavy quark decay, $hb$ and $hc$. 

\begin{figure}[!ht]
\begin{center}
\includegraphics[width=12cm,height=8cm]{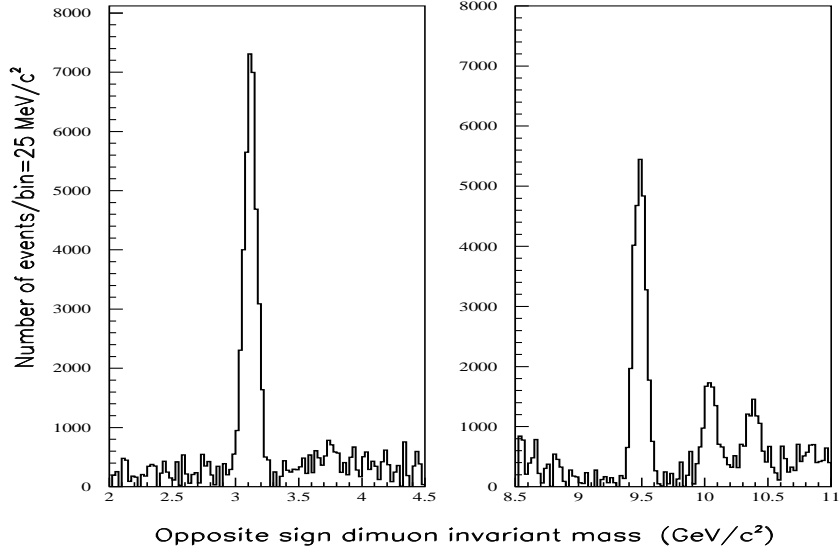}
\caption{\label{signal}\small The signal invariant mass distribution, 
after background subtraction, in 
\jpsi\ (left) and \ups\ (right) mass regions obtained in one month with
the high multiplicity in \PbPb\ collisions.}
\end{center}
\end{figure}

Although $N(b \overline b) \ll N(c \overline c)$, in an $AA$ collision 
the $b \overline b$ contribution is more important.  The background is 
primarily from uncorrelated muon pairs.
This background can be subtracted from the opposite sign 
spectrum using the `like-sign' spectra, as done by the 
SPS NA50 collaboration~\cite{Abreu:1999qw,Abreu:2000ni}. 
In this case, the signal is
$$
N^{\rm Sig} 
= \ N^{+-} - 2 \sqrt {N^{++} \ N^{--} }
$$
where $N^{+-}, N^{++}$ and $N^{--}$ are the
combinations of opposite sign pairs and positive and negative like-sign pairs
respectively in a given mass interval.
Figure~\ref{signal} presents the mass distributions resulting from this 
subtraction. 

\subsection{Open Heavy Flavours} 
\label{CMS.flav}

\begin{figure}[!ht]
\begin{center} 
\includegraphics[width=10cm,height=8cm]{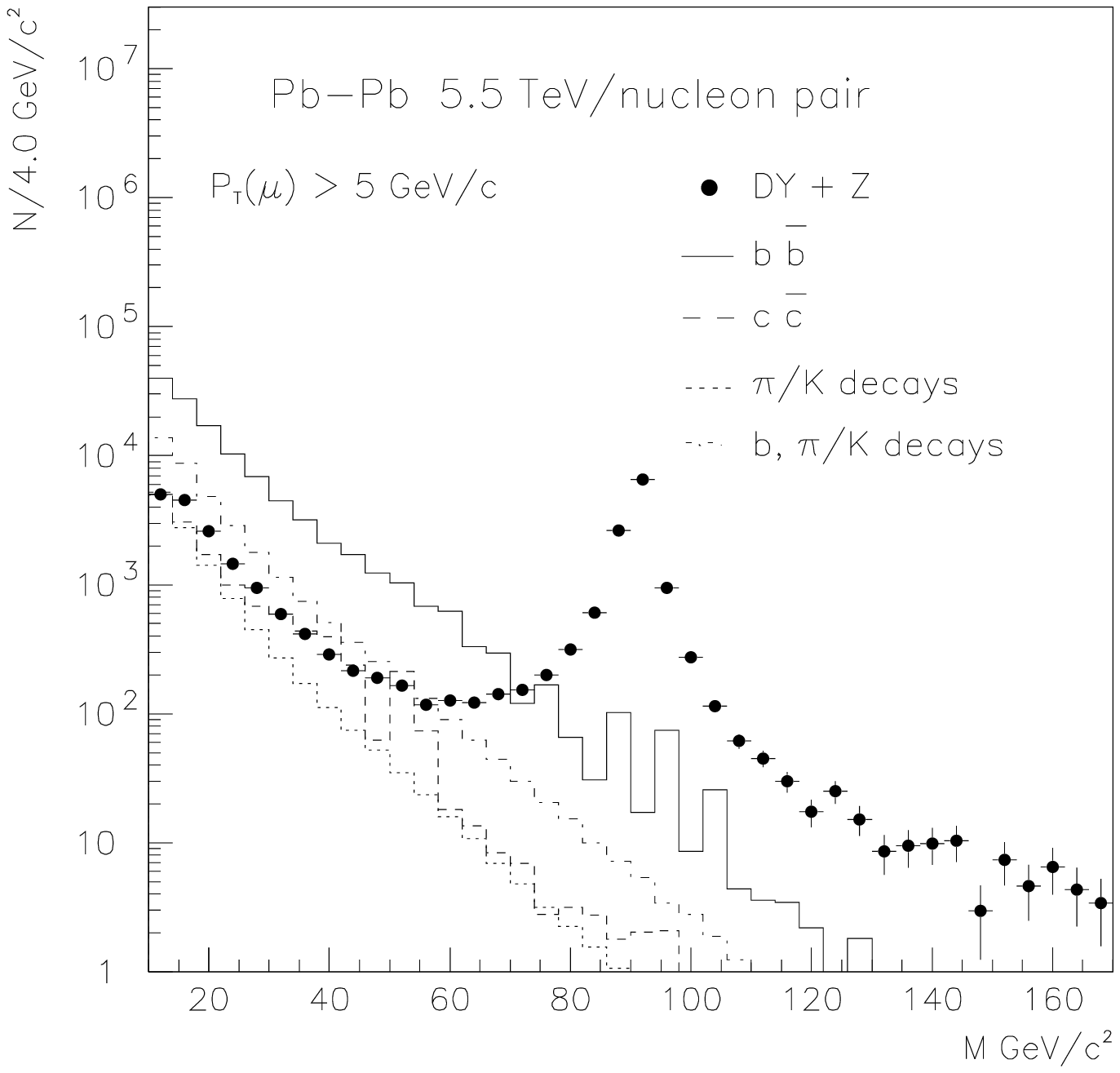} 
\caption{\label{dm:igfig1}\small Invariant mass distribution of 
$\mu^+ \mu^-$ pairs for muons with $p_T > 5$ \Gonc\ in one month Pb+Pb run.} 
\end{center}
\end{figure}

We briefly discuss some preliminary open heavy flavour results in CMS.
The heavy flavour production cross sections in minimum bias  
nucleus-nucleus collisions, as well as those of Drell-Yan and $Z^0$, 
$W$, $WW$, $WZ^0$ and $Z^0Z^0$ decays, were obtained from those in $pp$ 
interactions at $\sqrt{s} = 5.5$ TeV assuming $\sigma_{AA}=A^2 \sigma_{pp}$. 
The $pp$ cross sections were evaluated using PYTHIA with the CTEQ2L parton 
distribution functions.  Since PYTHIA uses leading order matrix elements,
a $K$ factor of two was included
for $c\overline c$ and $b\overline b$ production. Uncorrelated muon pairs from 
hadronic $\pi/K$ decays in minimum bias Pb+Pb collisions were
obtained from HIJING. 
Figure~\ref{dm:igfig1} presents the invariant dimuon mass spectra for 
$p_T^\mu > 5$ \Gonc\ and $|\eta| < 2.4$, corresponding to a one 
month Pb+Pb run, $1.3 \times 10^6$ s, assuming $\epsilon_{\rm machine} = 0.5$
and $\langle {\cal L}_{AA} \rangle = 10^{27}~$cm$^{-2}$s$^{-1}$. 
The dimuon detection efficiencies have been taken into account. In the 
mass range $10 \leq M_{\mu \mu} \leq 70$ \mass\ the dominant contribution 
comes from $b \overline b$ fragmentation \cite{Norrbin:2000zc}. 
Hence, this mass region can be used to 
estimate the in-medium bottom quark energy loss in heavy ion 
collisions~\cite{Lin:1998bd,Lokhtin:2001nh,Lokhtin:ay}. 
The mixed-origin contribution, 
when one muon is from $b \to B$ fragmentation and the other from $\pi/K$ 
decays, is about 16\%. The contribution from 
$c \overline c$ fragmentation and $\pi/K$ decays are 6\% and 5\% respectively. 
Note that the charm energy loss could be significantly larger than the 
bottom loss due to the lower charm quark mass, resulting in an additional 
suppression of the $c \overline c \rightarrow \mu^+\mu^-$ yield. A clear 
signal from $Z^0 \to \mu^+\mu^-$ decays, $\sim 11000$ events within 
$M_Z \pm 10$ GeV/$c^2$, has a background of less than 5\%. Since $Z^0$ 
production is unaffected by final state interactions and is 
assumed to be proportional to the number of
nucleon-nucleon collisions, it can be used as reference 
process to normalize the jet, quarkonium and heavy flavour
rates in $AA$ collisions relative to $pp$ collisions. 
The dimuon rates from other massive sources ($t \overline t$, 
$WW$, $WZ^0$, $Z^0Z^0$) are negligible.

If we consider dimuons from $b\overline{b}$ decays as a signal, two 
kinds of background can be extracted: uncorrelated and correlated. 
The uncorrelated background, random decays of 
pions and kaons and muon pairs of mixed origin, can be estimated from the 
like-sign events and the signal.

\begin{figure}[!ht]
\begin{center} 
\includegraphics[width=10cm]{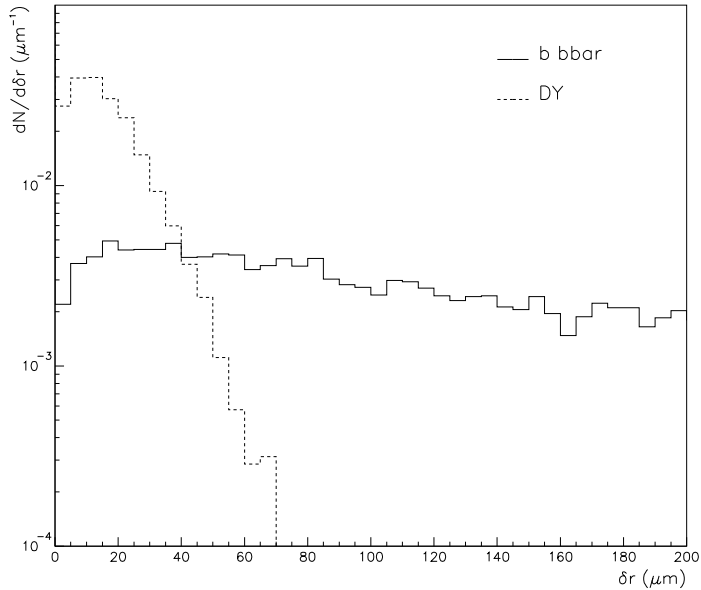} 
\caption{\label{dm:igfig2}\small The distribution 
of $\mu^+\mu^-$ pairs from $b \overline{b}$  decays 
(solid histogram) and from Drell-Yan production (dashed histogram) as a
function of $\delta r$ (see text).} 
\end{center} 
\end{figure}

The main correlated background, Drell-Yan production, 
is unaffected by medium-induced final state interactions. 
These dimuons come directly from the primary nuclear interaction 
vertex while the dimuons from $B$ and $D$ meson decays appear at secondary 
vertices some distance away from the primary vertex. The path length 
between the primary and secondary vertices 
is determined by the lifetime and Lorentz boosts.
A good way to discriminate the dimuons from $B$ mesons from those 
emitted at the primary vertex can 
be the transverse distance, $\delta r$, defined below. 
If $P_{\rm min}$ is defined 
as a track point with minimal distance to the beam axis, $z$, then 
$\delta r$ is the distance in the $x-y$ plane between points $P_{1,~\rm 
min}$ and $P_{2,~\rm min}$ belonging to different muon 
tracks. Muon pairs from $b \overline{b}$ 
decays show a rather flat 
distribution~\cite{Lokhtin:2001nh,Lokhtin:ay} while those
from Drell-Yan production 
sharply decrease, vanishing at $\delta r = 70$ $\mu$m, 
see Fig.~\ref{dm:igfig2}. We estimate the accuracy of 
the track position determination to be 
$\sigma_x =\sigma_y \sim 10$ $\mu$m and 
$\sigma_z\sim 100$ $\mu$m while the accuracy of 
the nuclear interaction point determination is
$\sigma_x =\sigma_y \sim \sigma_z 
\sim 20$ $\mu$m for the CMS tracker~\cite{cms060}. We find 
that, for such a simple simulation of the tracker resolution, a 
$\delta r > 50$ $\mu$m cut suppresses the Drell-Yan rate by two orders 
of magnitude at the price of a $30$\% reduction in signal.  A 
full GEANT-based simulation, 
including the real CMS geometry and dimuon reconstruction algorithm, is 
needed~\cite{Kodolova:2002} for more detailed conclusions. 

Another process which can also carry information about medium-induced 
rescattering and bottom energy loss is secondary $J/\psi$ 
production~\cite{Lokhtin:2001nh,Lokhtin:ay}. The branching ratio 
$B \rightarrow J/\psi X$ 
is $1.15$\%. The $J/\psi$ subsequently 
decays to dimuons with a $5.9$\% branching ratio so that
$gg \rightarrow b\overline{b}\rightarrow B\overline{B}~X\rightarrow 
J/\psi~Y\rightarrow\mu^+\mu^-Y .$ 

From a leading order calculation and assuming
$p^{\mu}_T > 5$ GeV/$c$ and 
$|\eta_{\mu}| < 2.4$, we expect $\approx 13000$ dimuons from secondary 
$J/\psi$ decays in a one month Pb+Pb run. Primary $J/\psi$'s 
produced at the nuclear interaction point can be rejected using tracker 
information on the secondary vertex position. 

Finally, we mention some of the theoretical uncertainties in the $Q \overline
Q$ cross sections and the 
corresponding dimuon spectra.
Nuclear shadowing, implemented with EKS98 \cite{Eskola:1998iy,Eskola:1998df}, 
is not large in this kinematic region, 
an $\approx 15$\% effect on $b \overline b$ and $c \overline c$ 
decays and $\approx 
25$\% for Drell-Yan production. In 
this case, the predicted factor of $\sim 2-4$ $b \overline b$ suppression 
due to final state energy loss can  
clearly be observed over initial-state shadowing. 
However, shadowing has an $\approx 30 $\% effect on secondary $J/\psi$ 
production from $B$ decays. 

\subsection{Conclusions} 

With its $\approx4 \pi$ muon acceptance and calorimetric coverage, CMS can
make very significant and, in some respects, unique contribution to 
heavy ion physics.  Studies of the \ups\ family, from \pp\ 
through Ar+Ar and Pb+Pb, as well as from peripheral to central collisions, is 
likely to be of great interest at the LHC, just as the
\jpsi\ has been for the SPS. 

The key issue for CMS is the muon reconstruction 
efficiency in the tracker with the extreme 
occupancies expected in Pb+Pb collisions. The all-silicon tracker results in 
a 76\% dimuon reconstruction efficiency in $|\eta| <  0.8$ for 
$dN_{\rm ch}/d\eta = 8000$.  

The pixel layers in the CMS tracker play a very important role 
in muon filtering and reconstruction efficiency, especially for the \ups\ 
purity, by suppressing $\pi$/\kaon\ decay muons in 
combination with a beam-line vertex constraint. Furthermore, 
since the Si tracker occupancies could be somewhat lower, 
including deeper Si layers
in the muon reconstruction algorithm than the outermost four could improve 
the muon reconstruction efficiency in the region $0.8 < |\eta| < 1.3$.

With a dimuon mass resolution of $\approx$~50~MeV, CMS is extremely well 
suited for quarkonia detection. 
The large rapidity aperture of the muon detector, as well as the 
precise tracking, result in high statistics and a very good separation 
between the \ups\ states. Significances between 70 for Pb+Pb to near 
1000 for Ar+Ar are found for the $\Upsilon$. Since $p_T^\mu > 3.5$ GeV/$c$ in
the barrel, detection of the higher mass $\Upsilon$ states is favored relative
to the $J/\psi$ where only $J/\psi$'s with $p_T > 5$ GeV/$c$ will pass the muon
$p_T$ cut.  The threshold decreases from 3.5
to ~2.0~\Gonc\ when $1.3< \eta <2.4$ so that muon detection in the 
endcaps would allow CMS to study \jpsi\ 
$\to$ \mup\mum\ over the full \pt\ range.

Above the $\Upsilon$ mass, $Z^0$ detection is a
unique feature of CMS. The independent observation of the \zo\ in the outer 
muon system and in the muon+tracker systems will
calibrate the tracker muon reconstruction efficiency over an extended \pt\  
range. Theoretical uncertainties in the heavy flavour production cross 
sections in nucleon-nucleon collisions at LHC energies are rather large and 
significantly affect the predicted dimuon production rates 
before any energy loss. These uncertainties
arise from the choice of the parton distribution functions, the heavy 
quark mass, next-to-leading order corrections, heavy quark
fragmentation, etc. Thus measurements in $pp$ or $dd$ collisions at 
the same or similar energies per nucleon as those for heavy ions are 
strongly desirable to determine the baseline rate precisely.

%% file: atlas_s.tex
\section[HEAVY FLAVOUR STUDIES IN ATLAS]
{HEAVY FLAVOUR STUDIES IN ATLAS~\protect
\footnote{Authors: B.~Cole, H.~Takai, S.~Tapprogge.}}
\label{sec:atlas}

The ATLAS detector is designed for high-$p_T$ physics in
proton-proton collisions at high luminosity. The detector has a
wide rapidity coverage and enormous potential for
the study of heavy ion collisions. The calorimeter is highly
segmented both longitudinally and transversely and offers a
unique opportunity to detect and measure jet properties. The
standalone muon system behind the calorimeter offers the
possibility of detecting dimuons in a quiet environment.
The ATLAS detector can be used to study the $\Upsilon$
states and $b$-jets. We briefly describe the detector and
discuss possible ways heavy flavour physics
could be studied with ATLAS.

\subsection{The ATLAS Detector}

\begin{figure}[htbp]
\label{fig:atlas}
\begin{center}
\mbox{\epsfig{file=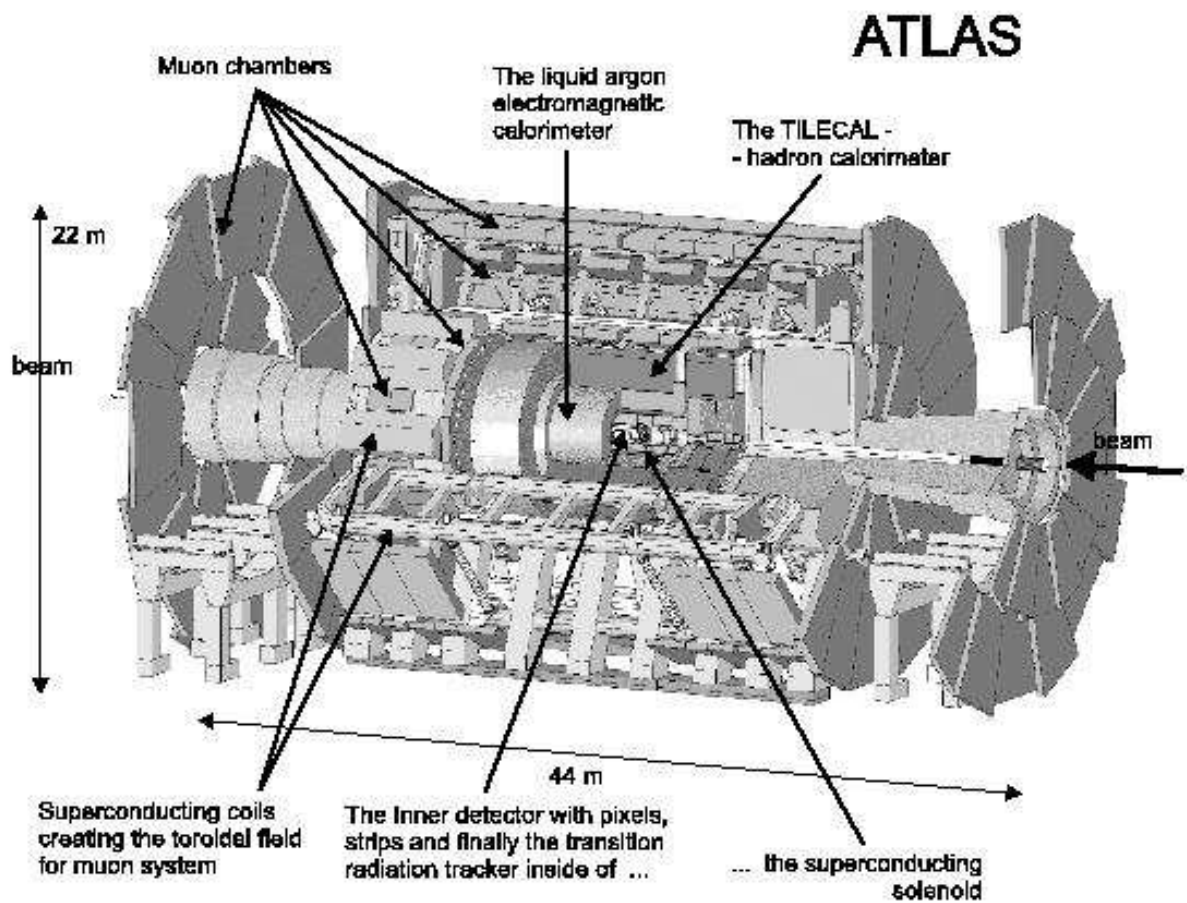,width=16cm}}
\caption{The overall layout of the ATLAS detector.}
\end{center}
\end{figure}

The ATLAS detector is designed to study proton-proton collisions at
the LHC design energy of 14~TeV in the center of mass.  The physics
pursued by the collaboration is vast and includes the Higgs boson search,
searches for SUSY and other scenarios beyond the Standard Model as
well as precision measurements of process within (and possibly beyond)
the Standard Model. To achieve these goals at full machine
luminosity of $10^{34}$ cm$^{-2}$s$^{-1}$, ATLAS will have a precise
tracking system (Inner Detector) for charged particle measurements, an
as hermetic as possible calorimeter system with extremely
fine grained segmentation, and a standalone muon system. An overview of
the detector is shown in Fig.~\ref{fig:atlas}.

The Inner Detector is composed of a finely segmented silicon pixel
detector, silicon strip detectors, Semiconductor Tracker (SCT),
and the Transition Radiation Tracker (TRT).  The segmentation is
optimized for proton-proton collisions at machine design luminosity.
The Inner Detector is designed to cover the pseudorapidity interval
$\mid \eta \mid < 2.5$ and is located inside a 2~T solenoid magnet.

The ATLAS calorimeter system surrounding the solenoid
magnet is divided into electromagnetic and hadronic sections and
covering $\mid \eta \mid < 4.9$. The EM calorimeter is
an accordion liquid argon device and is finely segmented
longitudinally and transversely for $\mid \eta \mid \le 3.1$.  The first
longitudinal segmentation has a granularity of $\Delta \eta \times \Delta \phi
= 0.003 \times 0.1$ in the barrel, becoming slightly coarser in the
endcaps.  The second longitudinal segmentation is composed of $0.025 \times 
0.025$ cells while the last segment has $0.05 \times 0.05$ cells.  In
addition, a finely segmented, $0.025 \times 0.1$, pre-sampler system is
placed in front of the electromagnetic (EM) calorimeter.  The overall
energy resolution of the EM calorimeter determined experimentally is
$10\%/\sqrt{E} \oplus 0.5\%$. The calorimeter also has good pointing
resolution, $60\,  {\rm mrad}/\sqrt{E}$ for photons and a timing resolution
of better than 200 ps for showers with energy greater than 20 GeV.

The hadronic calorimeter is also segmented longitudinally and
transversely.  Except for the endcaps and the forward calorimeters,
the calorimeter is a lead-scintillator tile structure with a granularity of 
$0.1 \times 0.1$.  In the endcaps, the hadronic calorimeter uses
liquid argon technology for radiation hardness and has the
same granularity as the barrel hadronic calorimeter.  The energy
resolution for the hadronic calorimeters is $50\%/\sqrt{E} \oplus 2\%$
for pions.  The very forward region, up to $\eta =4.9$, is covered by
the Forward Calorimeter, an axial drift liquid argon
calorimeter.  The overall performance of the calorimeter system is
described in Ref.~\cite{Armstrong:1994it}.

The ATLAS muon spectrometer is located behind the calorimeters,
thus shielded from hadronic showers. The spectrometer uses several tracking 
technologies along with a toroidal magnet, providing a 4~T field for 
an independent momentum measurement outside the calorimeter volume.  
Most of the spectrometer
volume is covered by Monitored Drift Tubes (MDT).  In the forward region,
where the rate is high, Cathode Strip Chambers are used.
The standalone muon spectrometer momentum resolution is of the order
of $2\%$ for muons with $10 \leq p_T \leq 100$ GeV. The muon
spectrometer covers $\mid \eta \mid < 2.7$.

The ATLAS trigger and data acquisition system is multi-level
and has to reduce the beam crossing rate of 40~MHz to an
output to mass storage rate of $\mathcal{O}(100)$~Hz. The first stage
(LVL1) is a hardware-based trigger, making use of the coarse 
granularity calorimeter data and the dedicated muon trigger chambers only,
reducing the output rate to about 75~kHz, within a maximum latency
of 2.5~$\mu$s. 

The performance results have been obtained using a detailed
full simulation of the ATLAS detector response with GEANT and have 
been validated by an extensive program of test beam measurements of 
all components.

\subsection{Heavy Flavour studies in ATLAS}

Recent theoretical investigations~\cite{Dokshitzer:2001zm}
have indicated that a dense partonic medium
will suppress the gluon radiation from charm and
bottom quarks relative to that from
light quarks.  Thus $b$-jet measurements
in ATLAS would provide an important comparison measurement to light
quark and gluon jets. Such measurements would have important
implications for gluon shadowing and saturation effects. Tagging of
$b$-jets by the associated muon is possible in the $pp$
environment \cite{Armstrong:1994it}. 

It is unlikely that vertexing will be available for these studies due
to the high multiplicity Pb+Pb environment.  Therefore
we are currently studying the possibility of tagging
$b$-jets by matching a measured muon in the muon spectrometer
to the jet measured by the calorimeter.  Vertexing will be available for 
lighter ion and $pA$ collisions.

We will profit from current studies of ATLAS performance for 
soft lepton tagging of the $H \rightarrow b\bar{b}$ decay. Initial studies
of the detector performance for $b$-jet tagging using this
technique can be found in the TDR for high luminosity
$pp$ collisions \cite{:1998fz}.  
We note that the muon momentum should be larger than 
4 GeV to be detected in the spectrometer. The expected muon background
comes from $\pi $, $K$ and $c$  decays. 

Quarkonia suppression is expected in a deconfined medium
due to the screening of the long-range attractive potential. We are
beginning studies of the ATLAS capability for $\Upsilon$ identification. 
The initial evaluation is that the muon system alone
will provide marginal resolution for a clear separation of the three
states. However, use of the SCT and pixel detectors can enhance the
mass resolution. Initial full simulations of Pb+Pb collisions
indicate that the occupancies
in the Inner Detector (with the exception of the Transition Radiation
Tracker) are low enough to allow track reconstruction.
We expect that all of the inner detectors will be available for lighter
ion and $pA$ collisions. The 
expected occupancies obtained from preliminary full
detector simulations using HIJING are listed
in Table~\ref{table:occ} for the barrel detectors in the Inner Detector. 
In the endcaps, the Inner Detector covers $1.4 \leq \eta \leq 2.5$ with average
occupancies of 0.2\% for the silicon pixels and 6\% and 10\% for the SCT, 
depending on the SCT wheel.  There are 9 SCT tracking stations on each endcap
which should allow tracking in the forward direction as well.

\begin{table}[htb]
\caption{Segmentation and occupancies of the Inner Detector (Barrel). Values
were obtained from preliminary full detector simulations using HIJING
with $dN_{\rm ch}/d\eta = 8000$. The $pp$ occupancies are from 
Ref.~\cite{unknown:1997fs}.
The corresponding $pp$ and Pb+Pb luminosities are also quoted.}
\label{table:occ}
\begin{center}
\begin{tabular}{|c|c|c|c|c|c|}
\hline
ID & $\eta$ range & Radius (cm) & Channels & Occupancy & Occupancy\\
&&&&$pp$ ($10^{34})$ & Pb+Pb ($10^{27}$)\\
\hline
Si Pixel 1 & $\pm 2.5$ & 5.0 &  $10^7$         & 0.04  & 0.6     \\
Si Pixel 2 & $\pm 1.7$ & 8.8 & $8 \times 10^7$ & 0.01  & 0.3     \\
Si Pixel 3 & $\pm 1.7$ &12.2 & $8 \times 10^7$ & 0.006 & 0.2    \\
\hline
SCT Layer 1 & $\pm 1.4$ &  30.0 & $3 \times 10^6$ & 0.6 & 15    \\
SCT Layer 1 & $\pm 1.4$ &  37.0 & $3 \times 10^6$ & 0.5 & 10   \\
SCT Layer 1 & $\pm 1.4$ &  44.0 & $3 \times 10^6$ & 0.4 & 7.5   \\
SCT Layer 1 & $\pm 1.4$ &  51.5 & $3 \times 10^6$ & 0.3 & 6.0   \\
\hline
\end{tabular}
\end{center}
\end{table}

The muon spectrometer will benefit from the low nucleus-nucleus luminosity. 
In the high luminosity $pp$ runs a large number of spectrometer hits
come from slow neutrons produced in previous interactions. Initial simulations
using HIJING Pb+Pb events indicate that the muon spectrometer
is sparsely populated. The muon backgrounds are from
$\pi$ and $K$ decays in the calorimeter. However, a substantial number
of these decays are expected to be rejected when matches to Inner Detector
tracks are required. The minimum muon $p_T$ for detection is $\sim 5-6$ GeV. 

\noindent
{\em Acknowledgments} 
The authors would like to thank their collaborators in ATLAS:
S.~Aronson, K.~Assamagan, M.~Dobbs, J.~Dolesji, H.~Gordon,
F.~Gianotti, S.~Kabana, M.~Levine, F.~Marroquim, J.~Nagle, 
P.~Nevski, A.~Olszewski, L.~Rosselet, P.~Sawicki, A.~Trzupek, 
M.~A.~B.~Vale, S.~White, R.~Witt, B.~Wosiek and K.~Wozniak.

%% file: ack.tex
\section{ACKNOWLEDGEMENTS}

\noindent The following sources of funding are acknowledged:

\noindent~~~ Academy of Finland,

Grant 102046: P. Hoyer;

\noindent~~~ 2002 CNRS/SNF Cooperation Program,

Grant 11240: N. Marchal, S. Peign\'e;

\noindent~~~ European Commission IHP program,

Contract HPRN-CT-2000-00130;


\noindent~~~ DFG Grant No. 436 RUS 17/129/02: D. Blaschke;


\noindent~~~ United States Department of Energy, Division of Nuclear Physics,
 
Contract No. DE-AC02-98CH10886: D. Kharzeev, P. Petreczky;

Grant No. DE-FG03-95ER40937: R.L. Thews;

Contract No. DE-AC03-76SF00098: R. Vogt;


\noindent~~~ United States Department of Energy, Division of High Energy 
Physics,
 
Contract No. W-31-109-ENG-38: G.T. Bodwin, Jungil Lee.
